\documentclass[defended,twoside]{new_cit_thesis}
\usepackage{graphicx}
\usepackage{epsfig,pstricks}

\usepackage{latexsym,amsmath,amsfonts,amsthm}
\usepackage[english]{babel}
\usepackage{hyperref}
\input{Qcircuit}

\definecolor{Black}{rgb}{0,0,0}
\definecolor{lred}{rgb}{0.95,0,0}
\definecolor{lblue}{rgb}{0,0,0.9}

\newtheorem{theo}{Theorem}

\newtheorem{defi}{Definition}
\newtheorem{lem}{Lemma}
\newtheorem{props}{Properties}
\newtheorem{propo}{Proposition}

\newcommand{\bs}{$\mathcal{C}_{BS}^{(n)}$ }
\newcommand{\bsc}[1]{$\mathcal{C}_{BS}^{(#1)}$}
\newcommand{\sh}{$\mathcal{C}_{Shor}^{(n)}$ }
\newcommand{\bk}{$\mathcal{C}_{BK}^{(n)}$ }
\begin{document}

%\author{Panos Aliferis}
\author{Panos Aliferis}
%\title{\LARGE{ \sc Rigorous Accuracy Thresholds \\ for Quantum Computation}}
\title{\LARGE{ \sc Level Reduction and \\ the Quantum Threshold Theorem}}
\degreeaward{Doctor of Philosophy}
\university{California Institute of Technology}
\address{Pasadena, California}
\unilogo{cit_logo}
\copyyear{2007}
\date{December 11, 2006}
%\pubnum{}

\dedication{\textit{for Ayis and K., of course}}

\maketitle

%%%%%%%%%%%%%%%%%%%%%%%%%%%%%%%%%%%%%%%%%%%%%%%%%%%%%%%%%%

\begin{frontmatter}
%\makecopyright
\setcounter{page}{2}
{\parindent=0pt 
\null\vspace{2in}
\textbf{Copyright notice:}\par\bigskip
Chapters \ref{ch:syntax} and \ref{ch:threshold} contain some material from \cite{Aliferis05b} and chapter \ref{ch:extensions} contains some material from \cite{Aliferis05c} which are both copyrighted by Rinton Press. Chapter \ref{ch:lower-bounds} contains some material from \cite{Aliferis06c} which is copyrighted by the American Physical Society.  
\par\bigskip
The remaining material is
\vfill
\begin{center}
\copyright 2007\\
Panos Aliferis\\
All Rights Reserved
\end{center}}
\newpage

\makededication

\begin{acknowledgements}

%It took Odysseus ten years to find his way back to Ithaca. It also took Zarathustra ten years of living up on the mountains before he decided to descend down to the people. %Homer's poetry describes Odysseus' long journey as a battle against formidable challenges and symbolic seductions. In a parallel to Homer, Nietzsche describes Zarathustra's battle against vanity and self-contentment and his call for a revaluation of life and an internal metamorphosis.
%
%My Ph.D. journey was similarly long and its conclusion also marks a turning point. % and a call for revaluations. My journey started when I begun studying Electrical Engineering at NTUA in Athens. It continued in the research groups at the Optical Communications Lab at NTUA and later, as a master's student, at the Quantum Nano-optics Lab at the University of Michigan. And it led to the doctorate's program at Caltech and the Institute of Quantum Information (IQI). 
%
%But foremost it was challenging and rewarding. 
%
Living on the open air of the mountains of research and reaching the point of writing this thesis would not have been possible without the guidance, the encouragement, and the support of many people. To them I write the few thanking words that follow.  

First, I would like to thank my advisor John Preskill. He provided the vessel in the form of IQI and the guidance for my education at Caltech. From him I learned much. Most importantly, I learned how the successful scientific spirit combines depth with breadth, seriousness with cheerfulness, abstract proofs with happy dancing. He introduced me to the field of fault-tolerant quantum computation and pointed out interesting problems in this area of research. Many of the results of this thesis came about from my collaboration with him and Daniel Gottesman and would not have been possible without their crucial contributions and insights.

Second, I would like to thank my mentor and friend Debbie Leung. I was fortunate to meet her as a starting graduate student at Caltech and collaborate with her on my first research results. From her I also learned much. I remember her emphasizing in several different occasions the value of developing easily composable frameworks for attacking research problems, and it was inspired by these words that I was guided to the idea of level reduction. Her sincere and active interest about both my personal well-being and education were invaluable in keeping me balanced at many difficult times.

Third, I would like to thank my collaborators David DiVincenzo and Barbara Terhal for sharing many thoughts and  starting new projects with me. My visits to the IBM Quantum Information group in the summers of 2005 and 2006 gave me new ideas and helped me understand in more depth important elements of our common research.

Fourth, I would like to thank my collaborators Andrew Cross and Krysta Svore with whom I exchanged many ideas on various topics of this thesis. Andrew carried out the threshold calculations discussed in chapter \ref{ch:lower-bounds} and welcomed my ideas and suggestions with an enthusiasm that is so characteristic of him. Krysta contributed in the software we used for these calculations and, with her fresh smile and positive attitude, was a source of encouragement and inspiration. % blew forward wind to my sails when new wind was most needed.

Fifth, I would like to thank my fellow graduate students, Kovid Goyal, Hui Khoon Ng, Graeme Smith, and Mike Zwolak, who %, always tolerant of my ignorance, 
spent time discussing their research with me. I also owe thanks to Robin Blume-Kohout, Sergei Bravyi, David Poulin, and Robert Raussendorf whom I met as postdocs at IQI and with whom I have had many insightful discussions.

Finally, I do not forget the influence of my three pre-Caltech mentors: my math mentor in Sparta, Stavros Damopoulos, my advisor at NTUA, Heraklis Avramopoulos, and my advisor at the University of Michigan, Duncan Steel. I also appreciate the help of the members of my defense and candidacy committees, Alexei Kitaev, Hideo Mabuchi, Robert McEliece, and Chris Umans. 

%Odysseus is a heroic character, Zarathustra a tragic one. What this journey will make of me is left to be seen.
\vskip 0.5cm
%\hskip 11cm -- {\it June 2006, Sparta}.
\hskip 11.5cm {\it Pasadena, XII 06}.

\end{acknowledgements}

\begin{abstract}
Computers have led society to the information age revolutionizing central aspects of our lives from production and communication to education and entertainment. There exist, however, important problems which are intractable with the computers available today and, experience teaches us, will remain so even with the more advanced computers we can envision for tomorrow. %And this despite the very rapid pace with which computing power has been advancing, in surprising agreement with Moore's empirical law.

Quantum computers promise speedups to some of these important but classically intractable problems. Simulating physical systems, a problem of interest in a diverse range of areas from testing physical theories to understanding chemical reactions, and solving number factoring, a problem at the basis of cryptographic protocols that are used widely today on the internet, are examples of applications for which quantum computers, when built, will offer a great advantage over what is possible with classical computer technology.

The construction of a quantum computer of sufficient scale to solve interesting problems is, however, especially challenging. The reason for this is that, by its very nature, operating a quantum computer will require the coherent control of the quantum state of a very large number of particles. Fortunately, the theory of quantum error correction and fault-tolerant quantum computation gives us confidence that such quantum states can be created, can be stored in memory and can also be manipulated provided the quantum computer can be isolated to a sufficient degree from sources of noise.

One of the central results in the theory of fault-tolerant quantum computation, the quantum threshold theorem shows that a noisy quantum computer can accurately and efficiently simulate any ideal quantum computation provided that noise is weakly correlated and its strength is below a critical value known as the quantum {\em accuracy threshold}. 
This thesis provides a simpler and more transparent non-inductive proof of this theorem based on the concept of {\em level reduction}. This concept is also used in proving the quantum threshold theorem for coherent and leakage noise and for quantum computation by measurements. In addition, the proof provides a methodology which allows us to establish improved rigorous lower bounds on the value of the quantum accuracy threshold. 
\end{abstract}

\tableofcontents
\cleardoublepage
\end{frontmatter}

%-----------------------------------------%

\chapter{Prologue}
\label{ch:intro}
%--------------------------------------------------------------%
\section{The Quantum Computer}
\label{sec:quantum-computer}

Trying to introduce the concept of quantum computation, it is tempting to begin by saying that quantum computers are computing devices that operate according to the laws of quantum mechanics. But, on second thought, this description is not very informative since quantum mechanics, in so far as it is a complete natural theory, describes every physical computing device and, so, even classical computers. What distinguishes quantum from classical computers is that the operation of the former is based on the two distinctively quantum-mechanical effects of {\em interference} and {\em entanglement} that do not appear in classical physics. So this is how I would describe in one phrase quantum computers to those who have taken a basic course in quantum mechanics: a quantum computer is an interference device of many entangled computation paths. Just as an interference pattern can appear by preparing a particle in a superposition of different geometric paths which are then combined to interfere, the output of a quantum computer is obtained by preparing the quantum bits in a superposition of different classical computation states which are also combined to interfere producing the final computation answer. 

%The problem with noise

This immediately leads to the problem of dealing with noise in quantum computation. If we think of a quantum computer as a big interference device, it is important to ask whether it is practically possible to construct such a device in a way that the arbitrarily complex interference effect will not disappear due to the fact that the components of the computer are not perfectly accurate. After all, we know that the interference pattern in a double-slit experiment disappears if it is {\em in principle} possible to know from which slit the particle has passed. So, if we think of a quantum computer as executing a computation which resembles a very complex interference experiment, it is natural to worry that the complexity of the computation will make it impossible {\em in practice} to maintain the complex superpositions of states and observe an interference effect at the end of the computation.

%Why consider this model of computation? What is the evidence for its power vs the classical computer?   

But before I address the problem of constructing robust quantum computers which is the main topic of this thesis, I  should first mention the indications we have about their greater computational power relative to classical computers. Historically, it was first Feynman \cite{Feynman82,Feynman85} and Deutsch \cite{Deutsch85,Deutsch89} who realized that simulating certain quantum systems on classical computers apparently requires time that scales exponentially with the size of the simulated system (e.g, the number of particles), and suggested that a more natural simulator for quantum systems is a computer which is quantum itself. Mathematical models of quantum computation were also previously discussed by Manin \cite{Manin80}, by Benioff \cite{Benioff82}, and by Peres \cite{Peres85}. The computational power of quantum computers has been investigated extensively since these first proposals. Today, proposed quantum computing applications range from general-purpose quantum simulations and simulators of local quantum field theories (e.g., \cite{Kitaev95,Lloyd96,Zalka98,Freedman02,Byrnes06}), to understanding the physics of condensed-matter systems and computing ground-state energies of complex molecules (e.g., \cite{Abrams97,Abrams99,Aspuru05}). But certainly, the strongest evidence for the power of quantum computers comes from Shor's discovery of a polynomial-time quantum algorithm \cite{Shor94} for finding the prime factors of composite numbers and computing the discrete logarithm. Although there is no proof, it is believed that no polynomial-time classical algorithm exists for these two problems. In fact, the belief in the hardness of these problems has led to their use in the public-key cryptography systems RSA and Diffie-Hellman which are widely used on the internet. More recent results on quantum algorithms include Grover's quantum algorithm \cite{Grover96} for unstructured search problems which offers quadratic speedups over classical search algorithms and Hallgren's quantum algorithm \cite{Hallgren02} for Pell's equation which is exponentially faster than any known classical algorithm.

%what the rest of this section is about

The remaining of this introductory chapter is structured as follows: In \S \ref{sec:QMechanics}, I give a short introduction to quantum computation. I begin by discussing the formalism of quantum mechanics and the theory of physical operations. Then, I introduce the circuit model of quantum computation, including the standard notation for elementary quantum operations and quantum circuit diagrams. Next, I discuss how imperfections in implementing the elementary operations during the execution of a quantum algorithm relate to the accuracy of the final computation result. Finally, I introduce the notion of universal quantum computation and discuss how a quantum computer being able to execute gates from a finite gate set can efficiently simulate any quantum computation to any desired accuracy. In \S \ref{sec:QCoding}, I briefly present the main results in the theory of quantum error correction. I begin by discussing the differences between classical and quantum coding and then I state the formal criteria for quantum error correction. Finally, I introduce the stabilizer formalism that enables a succinct description of a large class of quantum codes and use this formalism to describe CSS quantum codes which are especially advantageous for fault-tolerant quantum computation.

%--------------------------------------------------------------%
\section{Introduction to Quantum Computation}
\label{sec:QMechanics}

Computers are physical devices that run algorithms. Intuitively, an algorithm is a set of instructions that given some input produce as an output the answer to some problem. In fact, the {\em Church-Turing thesis}---really, a hypothesis about the nature of computers---states that for any function that we intuitively consider computable there exists a corresponding algorithm computing the function that runs on a (universal) Turing machine given sufficient time and memory space (see, e.g., Sipser's introductory book on the theory of computation \cite{Sipser97}). The interesting content of this thesis is that it identifies our intuitive notion of what it means to be able to compute a function with the algorithms that Turing machines can run. As a corollary to this thesis, any computer is as computationally powerful as a universal Turing machine if we do not consider any time or memory limitations.

Even though any computer can theoretically run any algorithm, efficiency considerations are important for determining  which algorithms can be executed in practice. To classify algorithms according to their efficiency, one typically considers how the time and memory space for running an algorithm depend on the size of the input to the algorithm, where the size of the input is measured in the {\em natural units} for the problem considered (e.g., number of binary digits when the input is an integer). Another fundamental thesis in the theory of computation, the {\em strong Church-Turing thesis}, states that (probabilistic) Turing machines can efficiently simulate any other realistic model of computation. In other words, any realistic computing device can simulate any other one with a slowdown which scales at most as a polynomial with the size of the input. The word {\em realistic} should be understood as referring to computing devices that can be constructed and they are scalable, i.e., can be extended to receive inputs of any size; e.g., one should probably exclude analog classical computers whose power relies on manipulating real numbers of unlimited precision and which are believed to be extremely vulnerable to inaccuracies and noise.

Quantum computers pose a challenge to the strong Church-Turing thesis. In particular, as I very briefly discussed in the previous section, there are problems for which efficient quantum algorithms exist and no efficient classical algorithms are known. Certainly, at this point one might ask whether quantum computation classifies as a realistic  model of computation. To address this question, I will next first describe the standard quantum circuit model of computation. Then, I will discuss the issue of accuracy and universality in quantum computation. As we shall see, the power of quantum computers stems from the parallelism that characterizes the quantum world, and this power is not lost if we consider the limited precision with which we can execute the elementary operations in a quantum algorithm. 

%--------------------------------------------------------------%
\subsection{Quantum Mechanics and Qubits}
\label{sec:PhysicalOperations}

%But first I will need to review the basic formalism of quantum mechanics and also briefly discuss some results from the theory of physical operations.% which will be useful later in the discussion of quantum error correction. 
%\footnote{Nobody who is not acquainted with the quantum theory should expect to learn it from my thesis---nevertheless, it would be inappropriate to start without introducing the basic formalism and the notation used in quantum mechanics. I advise the uninitiated reader to study the first chapters in Nielsen and Chuang's book \cite{Nielsen00} and, going into more depth, Preskill's lectures notes \cite{Preskill-notes} and Kitaev, Shen and Vyalyi's book \cite{Kitaev-book}. }

Nobody who is not acquainted with the quantum theory and the basic notions of quantum computation should expect to learn them from my thesis. Nevertheless, it would be inappropriate to start without briefly reviewing the basic formalism of quantum mechanics and some basic results from the theory of physical operations---I advise the uninitiated reader to study the first chapters in Nielsen and Chuang's book \cite{Nielsen00} and, going into more depth, Preskill's lectures notes \cite{Preskill-notes} and Kitaev, Shen and Vyalyi's book \cite{Kitaev-book}.

Quantum mechanics is a mathematical theory for the description (and prediction) of outcomes of experiments on physical systems. %\footnote{It should be noted that, although quantum mechanics has been incredibly successful in describing physics at the microscopic level (e.g., elementary particles such as the electron or the photon) or the mesoscopic level (e.g., super-conductivity), it has not been tested at the macroscopic level. It is, in fact, an issue of debate and research why quantum phenomena are not present in the every-day world around us which is apparently classical. Building a full-scale quantum computer can be seen as an interesting endeavour even from the purely theoretical point of view of providing a test of quantum mechanics at the macroscopic, or nearly macroscopic level.} 
Let us first consider {\em closed} physical systems, i.e., systems which can be described in complete isolation from their surroundings. Quantum mechanics associates with such a closed system a {\em quantum state}, $|\psi\rangle$, which is a {\em vector} of unit length in a---finite-dimensional, for our purposes---complex inner-product vector space $\mathcal{H}$ (i.e., a Hilbert space). Overall phases are irrelevant, i.e, $\forall \theta \in \mathbb{R}$, $\forall |\psi\rangle \in \mathcal{H}$, $e^{i\theta}|\psi\rangle$ describes the same state as $|\psi\rangle$. 

Consider the space of bounded operators $\mathcal{B}(\mathcal{H})$ defined in $\mathcal{H}$. To each measured quantity, $x$, there corresponds a self-adjoint (i.e., Hermitian) operator or {\em observable}, $\hat{x}\in \mathcal{B}(\mathcal{H})$, such that the possible outcomes when a measurement of this quantity is performed are given by the eigenvalues, $\{x_i\}$, of $\hat{x}$. The probability of obtaining the $i$th outcome, $x_i$, when a measurement of $x$ is performed on a system prepared in the state $|\psi\rangle$ is given by
\begin{equation}
\label{1.2.1}
{\rm Prob}(x_i) = ||P_{x_i}|\psi\rangle ||^2 = \langle \psi|P_{x_i}|\psi\rangle \; ,
\end{equation}

\noindent where $\langle \psi| \equiv (|\psi\rangle)^\dagger$ (`$^\dagger$' denotes Hermitian conjugation), and
\begin{equation}
\label{1.2.2}
P_{x_i} \equiv \sum\limits_{j:\hat{x}|x_j\rangle = x_i |x_j\rangle} |x_j\rangle \langle x_j| \; 
\end{equation}

\noindent is the projector onto the eigenspace of $\hat{x}$ corresponding to eigenvalue $x_i$. After the measurement, the system is described by the state $P_{x_i} |\psi\rangle$ up to normalization---we say that the measurement  {\em collapses} the initial state $|\psi\rangle$ to the state $P_{x_i} |\psi\rangle$. The requirement that $\hat{x}$ is Hermitian implies that all its eigenvalues are real and its eigenvectors form an orthonormal basis spanning $\mathcal{H}$. 

Given the state $|\psi(t_o)\rangle$ at some time $t_0$, its time evolution is governed by the Schr$\ddot{\rm o}$dinger equation
\begin{equation}
\label{1.2.3}
\partial_t |\psi(t)\rangle = -i H |\psi(t)\rangle  \; , 
\end{equation}

\noindent where $\partial_t$ denotes the partial derivative with respect to time, $H$ is a Hermitian operator (the {\em Hamiltonian} of the system) and $\hbar =1$. Equation (\ref{1.2.3}) implies that the system dynamics is unitary, i.e., the quantum state at time $t$ is $|\psi(t)\rangle = U |\psi(t_0)\rangle$, where $U=\exp(-i H t)$ is a unitary operator.

In analogy with classical computation where the basic carriers of information are bits, in quantum computation the carriers of information are quantum bits or simply {\em qubits}. Qubits are two-dimensional quantum systems such as, e.g., spin $1/2$ particles.\footnote{Depending on the actual experimental situation, a qubit can be formed by any two levels in a multilevel quantum system; e.g., a qubit can be defined by the ground and first excited state of an ion.} A basis for operators in a two-dimensional Hilbert space can be formed by the three {\em Pauli matrices},
\begin{equation}
\label{1.2.4}
X \equiv \sigma_{\rm 1} = \left( \begin{array}{cc} 0 & 1 \\ 1 & 0 \end{array}\right) \;; \;\; Y \equiv \sigma_{\rm 2} = \left( \begin{array}{cc} 0 & -i \\ i & 0 \end{array}\right) \; ; \;\; Z \equiv \sigma_{\rm 3} = \left( \begin{array}{cc} 1 & 0 \\ 0 & -1 \end{array}\right) \; , 
\end{equation} 

\noindent and the identity operator $I \equiv \sigma_{\rm 0} = {\rm diag}(1,1)$. A useful observation is that the Pauli matrices are unitary and Hermitian. The Pauli matrices are, in fact, related to the operators $J_i = {1\over 2}\sigma_{i}$ which generate the rotation group $SU(2)$ in two dimensions: a rotation around the unit-vector  $\vec{n}\in \mathbb{R}^3$ by an angle $\theta\in \mathbb{R}$ is given by $U_{\vec{n}}(\theta) = \exp(-i \; \vec{n} {\cdot} \vec{J}\; \theta) =  \cos({\theta /2}) I - i\; \vec{n} {\cdot} \vec{\sigma}\; \sin({\theta /2})$, where $\vec{\sigma}\equiv (\sigma_1,\sigma_2,\sigma_3)$. 

Unlike the state of a bit which is either 0 or 1, the state of a qubit is a unit vector in a two-dimensional Hilbert space, $\mathcal{H}=\mathbb{C}^2$. According to the standard convention, we can choose a basis for $\mathbb{C}^2$ to be formed by the eigenvectors of the Pauli $Z$ operator: one basis state is the $+1$ eigenvector which we label $|0\rangle$ and the other basis state is the $-1$ eigenvector which we label $|1\rangle$. Then, an arbitrary single-qubit state can be written in the form $|\psi\rangle = \alpha_0 |0\rangle + \alpha_1 |1\rangle$, where $\alpha_0,\alpha_1\in \mathbb{C}$ and $|\alpha_0|^2 + |\alpha_1|^2=1$ in order for $|\psi\rangle$ to be normalized to have unit length. The basis $\{|0\rangle,|1\rangle \}$ (and its obvious generalization to more than one qubits) we will often refer to as the {\em computation basis} since it corresponds to the basis along which classical computation is phrased.

Qubits generalize classical bits since not only can they be in the two orthogonal states $|0\rangle$ and $|1\rangle$, but they can also be in {\em superpositions} of these two states. Similarly, if we consider some number $n>1$ of qubits, the quantum state, $|\psi^{(n)}\rangle$, that describes them is a unit vector in a $2^n$-dimensional Hilbert space; hence, it can in general be written as a superposition of all $n$-bit strings,
\begin{equation}
\label{1.2.5}
|\psi^{(n)}\rangle = \sum\limits_{i_0=0}^{1} \sum\limits_{i_1=0}^{1} \cdots \sum\limits_{i_{n-1}=0}^{1} \alpha_ {i_0, \dots, i_{n-1}} |i_0 i_1 \cdots i_{n-1}\rangle \; ,
\end{equation} 

\noindent where $\alpha_ {i_0, \dots, i_{n-1}}\in \mathbb{C}$ are arbitrary {\em amplitude} coefficients subject to the constraint that $|\psi^{(n)}\rangle$ is normalized. Quantum states such as $|\psi^{(n)}\rangle$ allow us to talk about {\em quantum parallelism} since the $n$ qubits appear to be in all $2^n$ possible classical bit configurations {\em in parallel}. And since both the norms and phases of the amplitude coefficients are important to determine a quantum state, we often say that $|\psi^{(n)}\rangle$ is a {\em coherent} superposition of the different computation-basis states (as opposed to a probabilistic superposition that could result by picking each computation-basis state at random according to some probability distribution). 

In equation (\ref{1.2.5}) we have chosen to express $|\psi^{(n)}\rangle$ in terms of the $n$-qubit computation basis, $\{|i_0 i_1 \cdots i_{n-1}\rangle \}$. In general, a different orthonormal basis could have been used which would have resulted in different amplitude coefficients. In fact, it could be the case that for a particular choice of basis all amplitudes become zero except for a single one. Then, $|\psi^{(n)}\rangle$ expressed in this basis is just a tensor product of single-qubit quantum states, i.e., $|\psi^{(n)}\rangle = |\psi_0\rangle \otimes |\psi_1\rangle \otimes \cdots \otimes |\psi_{n-1}\rangle$ for some states $\{ |\psi_i\rangle \}$; such states are called {\em product states}. Quantum states which are not product states---i.e., states for which, independent of the choice of orthonormal basis, at least two amplitudes are non-zero---are called {\em entangled}.

%----------------------------------------------------------------------------------%
\subsection{Open Systems and the Theory of Physical Operations} 
\label{sec:KrausTheorem}

Although as illustrated by the Gottesman-Knill theorem \cite{Gottesman97,Nielsen00} entanglement is not sufficient by itself to guarantee that a quantum computation cannot be efficiently simulated on a classical computer, generating and manipulating highly entangled states is certainly {\em necessary} to giving quantum computers their greater power\footnote{Although, of course, there is no proof but only strong evidence that BQP, i.e., bounded-error  polynomial-time quantum computation, is strictly more powerful than BPP, i.e., bounded-error probabilistic polynomial-time classical computation.}. Unfortunately, it is a fact that entanglement is very fragile and noise or systematic hardware imperfections tend to induce {\em decoherence} by destroying quantum superpositions. 

To obtain a general description of decoherence processes, it is convenient to consider the qubits of our quantum computer as comprising an {\em open} physical system with Hilbert space $\mathcal{H}_S$ which is part of a larger closed system with Hilbert space $\mathcal{H}_{SB}$ including all sources of noise. (Often, we will call $\mathcal{H}_S$ the {\em system} and its ``complement'' $\mathcal{H}_B$ inside $\mathcal{H}_{SB}$ the {\em environment} or {\em bath}; then, $\mathcal{H}_{SB} = \mathcal{H}_S \otimes \mathcal{H}_B$.) The quantum state describing the open system is, in general, no longer a unit vector in $\mathcal{H}_S$. Instead, it is an operator,  $\rho\in \mathcal{B}(\mathcal{H}_S)$, which we call a {\em density matrix}. Density matrices are Hermitian, positive semidefinite and have unit trace. Just as in closed systems, the evolution of the density matrix of an open system due to dynamics that only involve the open system itself is unitary: when the Hamiltonian of the open system is $H_S$, the unitary evolution for time $t$ is given by $U_S=\exp(-iH_S t)$ and the initial density matrix $\rho$ evolves to $U_S\rho U_S^{\dagger}$. Also, the probability of obtaining the $i$th outcome, $x_i$, when the quantity $x$ is measured is ${\rm Prob}(x_i)={\rm Tr}(P_{x_i}\rho )$, and the measurement leaves the open system with the post-measurement density matrix $P_{x_i}\rho P_{x_i}^{\dagger}$ up to normalization. 

The density-matrix formalism generalizes the formalism describing closed systems since the latter can be recovered if one identifies $\rho=|\psi\rangle \langle \psi|$. Moreover, the density-matrix formalism has the advantage of allowing us to describe parts of a closed system without giving a description of the dynamics in the entire closed system. In particular, restricting our attention to $\mathcal{H}_S$, it is possible to obtain a useful characterization of the most general evolution of its state which is consistent with some ``reasonable''  requirements about such an evolution. This characterization is given by the Kraus representation theorem that is stated next. 
\begin{theo}[Kraus Representation Theorem \cite{Schumacher96a,Preskill-notes}]
\label{theo:1}
Let $\rho$ be the initial density matrix of a system with Hilbert space $\mathcal{H}_S$. Let a set of operators $\{ M_k \}$ with $M_k\in \mathcal{B}(\mathcal{H}_S)$ such that $\sum_k M_k^{\dagger} M_k = I_S$ (where $I_S$ is the identity operator in $\mathcal{H}_S$). Then, $\{ M_k \}$ defines the quantum operation, or {\em super-operator},
\begin{equation} 
\label{1.2.6}
\mathcal{E} : \rho \rightarrow \sum\limits_k M_k \rho M_k^\dagger \; .
\end{equation}

\noindent Furthermore, for any linear quantum operation, $\mathcal{E}$, that satisfies the requirements that (i) $\mathcal{E}(\rho)$ is Hermitian, (ii) $\mathcal{E}(\rho)$ has unit trace, and (iii) $ \mathcal{E}(\rho\otimes I_C)$ is positive semidefinite (where $I_C$ is the identity operator in {\em any} other Hilbert space, $\mathcal{H}_C$,  different than $\mathcal{H}_S$), there exists a set, $\{ M_k \}$, of {\em Kraus operators} as defined above such that the operation can be written in the form (\ref{1.2.6}).     
\end{theo}

Requirements (i) and (ii) are consequences of requiring that $\mathcal{E}(\rho)$ be a density matrix. Requirement (iii) is known as the requirement for {\em complete positivity}. If we would forget about $\mathcal{H}_C$, then (iii)  would just be the requirement that $\mathcal{E}(\rho)$ is positive semidefinite as is required for any density matrix. Complete positivity imposes the additional, innocuous looking, requirement that $\mathcal{E}$ evolves the density matrix $\rho \otimes I_C$ for any $\mathcal{H}_C$ decoupled from $\mathcal{H}_S$ also to a density matrix. %Finally, the requirement that $\mathcal{E}$ be linear is associated with the, apparently not entirely justifiable on physical grounds, convention to impose linearity in quantum mechanics.

%------------------------------------------------%
\medskip \medskip \noindent {\bf Superoperators and Stochastic Processes} \medskip

Imagine a quantum operation, $\mathcal{N}$, describing noise acting on our quantum computer which, at some point during the computation, is in the state $\rho$. We often consider a noise model where one of the Kraus operators is proportional to the identity operator $I_S$, $M_0=\sqrt{1- p}\; I_S$ for some real constant $0\leq p\leq 1$. All other nonidentity Kraus operators will satisfy $\sum_{k\not =0} M_k^\dagger M_k = p I_S$. In this situation, we can interpret $\mathcal{N}$ as a {\em stochastic} process which with probability $p$ applies some error to our quantum state $\rho$ and with probability $1-p$ leaves the state unchanged. 

To make the correspondence between $\mathcal{N}$ and a stochastic process more concrete, let us consider an additional system, the ``bath,'' with Hilbert space $\mathcal{H}_B$ such that the dimension of $\mathcal{H}_B$ is at least equal to the number of Kraus operators describing $\mathcal{N}$. Letting $\{|k\rangle_B \}$ be an orthonormal basis in $\mathcal{H}_B$, we can consider the isometry defined by  
\begin{equation} 
\label{1.2.7}
U_{SB} : \rho \otimes |0\rangle \langle 0|_B \rightarrow \sum\limits_{k,l} M_k \rho M_l^\dagger \otimes |k\rangle \langle l|_B \; , \; \forall \rho \in \mathcal{B}(\mathcal{H}_S) \; .
\end{equation}

\noindent $U_{SB}$ can be extended to a unitary in $\mathcal{H}_S \otimes \mathcal{H}_B$ and, furthermore, if we trace over the bath we find
\begin{equation} 
\label{1.2.8}
{\rm Tr}_B \left( U_{SB} \left( \rho \otimes |0\rangle \langle 0|_B \right) U_{SB}^\dagger  \right) = \sum\limits_k M_k \rho M_k^\dagger  \; ,
\end{equation}

\noindent where $M_k \equiv {_B}\langle k| U_{SB} |0\rangle_B $. We can therefore interpret $\mathcal{N}$ as describing a process in which our system interacts with some bath in an initial state $|0\rangle_B$ via the unitary $U_{SB}$, followed by a measurement on the bath along the orthonormal basis $\{|k\rangle_B \}$ whose outcome is {\em unknown}; outcome $k$ occurs with probability $M_k^\dagger M_k$, in which case the post-measurement system state is proportional to $M_k \rho M_k^\dagger$. In particular, if $M_0=\sqrt{1- p}\; I_S$, then the outcome $k=0$ occurs with probability $1-p$ and the post-measurement state in $\mathcal{H}_S$ is the initial state $\rho$, consistent with our interpretation in the previous paragraph. 

%--------------------------------------------------------------%
\subsection{The Quantum Circuit Model}
\label{sec:QCircuitModel}

After this brief introduction to quantum mechanics and the problem of decoherence, it is time to describe the standard circuit model of quantum computation. Since quantum circuits generalize classical circuits, it is natural to start by describing classical computation in terms of circuits.

A classical circuit is a representation of a Boolean function as a composition of other, more elementary Boolean functions. We consider boolean functions---also often called {\em gates}---of the form $f:\mathbb{Z}_2^n \rightarrow \mathbb{Z}_2$; i.e., they take a $n$-bit input to a one-bit output. Let us first fix some set, $\mathcal{G}$, of gates which we will call our {\em gate set} or {\em basis}. Gates in $\mathcal{G}$ play the role of the abstract elementary hardware components of a general-purpose classical computer, and they each have a fixed constant number of input bits. Then, a circuit for computing the Boolean function $g: \mathbb{Z}_2^n \rightarrow \mathbb{Z}_2$ is a finite sequence $g_1, g_2, \dots, g_m$ of gates in $\mathcal{G}$ such that $g(y) = (g_m \circ \cdots \circ g_2 \circ g_1)(y)$, $\forall y\in \mathbb{Z}_2^n$. In order to be well defined, this gate sequence should be represented as an acyclic directed graph, or {\em circuit}: The vertices of the graph correspond to the gates and edges denote bits; the inputs to a gate are a subset of the bits of the input $y$, a subset of the output bits from preceding gates, and possibly some ancillary bits; the outputs of a gate are either input to succeeding gates or are part of the final computation output. A gate set, $\mathcal{G}$, is called {\em universal} if for every Boolean function $g$, there exists a circuit using gates from $\mathcal{G}$ that computes $g$. Examples of universal gate sets are the sets $\{\neg, \wedge  \}$, $\{ \neg, \vee  \}$, and $\{ \Lambda^2(X) \}$; here $\forall a,b,c\in \mathbb{Z}_2$, $\neg(a)=a+1$, $\wedge(a,b) = a\cdot b$, $\vee(a,b) = \neg (\wedge(\neg a, \neg b))$, $\Lambda^2(X)(a,b,c)=(a,b,(a\cdot b)+c)$, and arithmetic is done modulo 2.  

To go from classical circuits to quantum circuits we first need to change from {\em bits} to {\em qubits}:\footnote{The discussion in this and the following chapters could be generalized if $d$-dimensional systems, {\em qudits}, were used instead. But, for simplicity, we will restrict ourselves to discussing qubits.} a quantum computer processes the state of some finite number, $n$, of qubits which are each initialized in some fixed single-qubit state, say $|0\rangle$. Quantum computation proceeds by applying a unitary transformation to these qubits, followed by a final measurement along the computation basis on some subset of them. In analogy to classical circuits, a quantum circuit is a representation of a unitary transformation that acts on $n$ qubits as a finite sequence of elementary unitary transformations or, simply, {\em quantum gates} which operate on a fixed constant number of qubits. These elementary quantum gates play the role of the basic hardware operations of our quantum computer and they can be chosen from a {\em finite} universal gate set which we may again denote by $\mathcal{G}$. 

In a straightforward generalization of this definition of quantum circuits, we may allow initializing each qubit in a different single-qubit state. We may also allow intermediate measurements whose outcome is used to control quantum gates on other qubits, and also measurements along different bases than the computation basis. However, we do {\em not} allow initialization in more complicated multi-qubit states or measurements along the eigenbases of multi-qubit observables. This is important in order not to hide some of the complexity of the algorithm realized by the quantum circuit in the initial preparation or the final measurement. With this definition, the  complexity of a quantum algorithm can be related to the {\em size} of its quantum circuit, i.e., the minimal number of gates from $\mathcal{G}$ required to realize the algorithm. 

The finiteness property of $\mathcal{G}$ is emphasized because unitary transformations form a continuum. It could be possible to allow $\mathcal{G}$ to contain, e.g., all single-qubit rotations, $U_{\vec{n}}(\theta)$, $\forall \vec{n}\in \mathbb{R}^3$, $\forall \theta \in \mathbb{R}$. However, when we consider different physical implementations of quantum computation, it is more natural to only demand that the same finite set of unitary transformations is realized in all of them. More importantly, as we will discuss in the last chapter, fault-tolerant implementations are possible for gates in a finite set but not for gates in a continuous set.

Due to the continuum of unitary operations, the universality property of our gate set, $\mathcal{G}$, is more subtle than in the classical case: The most general gate we can apply to the $n$ initial qubits is a unitary transformation in the unitary group in $2^n$ dimensions, $U(2^n)$. But of course, since our gate set is finite, there will be gates in $U(2^n)$ which we will be unable to implement {\em exactly} using a finite sequence of gates in $\mathcal{G}$. However, as we will discuss in \S \ref{sec:QUniversality}, there exist finite gate sets for which any unitary in $U(2^n)$ can be approximated to any desired accuracy using a quantum circuit with a finite number of gates from $\mathcal{G}$. It is such finite gate sets that we will call {\em universal} for quantum computation.

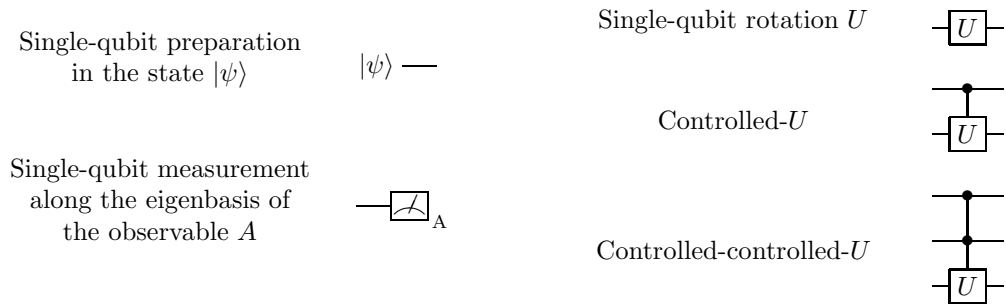
\begin{figure}[tb]
\begin{center} \vspace{0.2cm}
\begin{tabular}{cc}
\parbox{6cm}{
\parbox{4cm}{\begin{center}Single-qubit preparation in the state $|\psi\rangle$ \end{center}} \hskip 0.3cm \Qcircuit @C=1.5ex @R=3ex @!R {
                                         & \push{|\psi\rangle \hspace{0.1cm}} & \qw & \qw
                                        } \vspace{0.2cm} \\  
\parbox{4cm}{\begin{center}Single-qubit measurement along the eigenbasis of the observable $A$\end{center}} \hskip 0.5cm \Qcircuit @C=1.5ex @R=3ex @!R {
                                         & \qw & \meterobs
                                  } \vspace{0.2cm} 
            } & \hskip 1.2cm         
\parbox{6cm}{            
\parbox{4cm}{\begin{center}Single-qubit rotation $U$\end{center}} \hskip 0.3cm \Qcircuit @C=1.5ex @R=3ex @!R {
                                         & & \gate{U} & \qw
                                        } \vspace{0.3cm}  \\    \hskip 0.2cm            
\parbox{4cm}{\begin{center}Controlled-$U$\end{center}} \hskip 0.3cm \parbox{1cm}{ \Qcircuit @C=1.5ex @R=1ex @!R {
                                         & & \ctrl{1} & \qw \\
                                         & & \gate{U} & \qw                     
                                        }} \vspace{0.5cm} \\ \hskip 0.2cm
\parbox{4cm}{\begin{center}Controlled-controlled-$U$\end{center}} \hskip 0.3cm \parbox{1cm}{ \Qcircuit @C=1.5ex @R=1ex @!R {
                                         & & \ctrl{1} & \qw \\
                                         & & \ctrl{1} & \qw \\
                                         & & \gate{U} & \qw
                                        } } } \vspace{0.1cm}                             
\end{tabular}
\caption{\label{fig:1.1.1} Notation for the most common quantum operations. Controlled-$U$ applies $U$ on the second, {\em target}, qubit when the first, {\em control}, qubit is in the state $|1\rangle$, otherwise it acts as the identity. Similarly, controlled-controlled-$U$ applies $U$ on the third, {\em target}, qubit when the first two, {\em control}, qubits are {\em both} in the state $|1\rangle$, otherwise it acts as the identity.}
\end{center}
\end{figure}

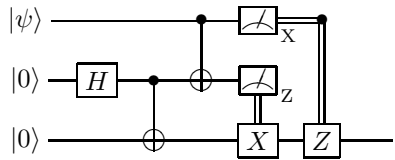
\begin{figure}[t]
\begin{center}
\parbox{1cm}{
\Qcircuit @C=2.3ex @R=2.3ex @!R {
            & \push{|\psi\rangle \hspace{0.1cm}} & \qw      & \qw      & \ctrl{1} & \meterx         & \control \cw \cwx[2] \\
            & \push{|0\rangle \hspace{0.1cm}}    & \gate{H} & \ctrl{1} & \targ    & \meterz \cwx[1] &        \\
            & \push{|0\rangle \hspace{0.1cm}}    & \qw      & \targ    & \qw      & \gate{X}        & \gate{Z} & \qw & \qw
                                        }}
\vspace{0.2cm}                             
\caption{\label{fig:1.1.2} An example of a quantum circuit; the direction of time is from left to right. Here, one qubit in the state $|\psi\rangle$ and two ancillary qubits initialized in the state $|0\rangle$ interact by a sequence of Hadamard and {\sc cnot} gates. Finally, the first qubit is measured in the eigenbasis of the operator $X$ and the second in the computation basis. Conditioned on the measured eigenvalues being $-1$, corrective gates are then applied on the third qubit. It can be verified that the state of the third qubit at the output is $|\psi\rangle$; i.e., the circuit {\em teleports} the state of the first qubit to the third qubit \cite{Bennett93}. %This is a version of {\em quantum teleportation}, which will be discussed in more detail in \S \ref{sec:MeasBased}.
}
\end{center}
\end{figure}

Quantum circuits can be represented by space-time diagrams. In these diagrams, time usually progresses from left to right, lines correspond to the ``world lines'' of qubits and quantum gates are denoted by boxes applied on a subset of these qubits (although, as we will see next, for some gates we use a special notation). The notation for the most common operations in quantum circuit diagrams is shown in figure \ref{fig:1.1.1}, and figure \ref{fig:1.1.2} gives an example of a quantum circuit. Apart from the Pauli operators in equation (\ref{1.2.4}), other common single-qubit rotations are the Hadamard gate,
\begin{equation}
\label{1.2.9}
H \equiv {1\over \sqrt{2}}\left( \begin{array}{cc} 1 & 1 \\ 1 & -1 \end{array}\right) , 
\end{equation} 

\noindent the phase gate,
\begin{equation}
\label{1.2.10}
S \equiv \exp\left({-i {\pi \over 4}Z} \right) = e^{-i {\pi \over 4}} \left( \begin{array}{cc} 1 & 0 \\ 0 & i \end{array}\right) \; , 
\end{equation}

\noindent and the $\pi/8$-gate, $T \equiv S^{1/2}$. We will also often use the two-qubit controlled-{\sc not} gate, {\sc cnot}, which acts in the computation basis as
\begin{equation}
{\rm CNOT}: |a,b\rangle \rightarrow |a,a+b\rangle, \; \forall a,b\in \mathbb{Z}_2 \; ,
\end{equation}
\noindent and whose symbol is \parbox{1cm}{\Qcircuit @C=0.9ex @R=1ex @!R { &  & \ctrl{1} & \qw \\ &  & \targ & \qw }} , 
%
%\begin{equation}
%\label{1.2.11}
%{\rm CNOT}: |a,b\rangle \rightarrow |a,a+b\rangle \; , \; \forall a,b\in \mathbb{Z}_2 \; , \; {\rm whose\; symbol\; is\;\;} \parbox{1cm}{
%\Qcircuit @C=1ex @R=1.7ex @!R {
%   &  & \ctrl{1} & \qw \\
%   &  & \targ & \qw
%                                }} \; \; ,   
%\end{equation}
%
and the controlled-$Z$ gate, {\sc cphase}, which acts in the computation basis as 
\begin{equation}
{\rm CPHASE}: |a,b\rangle \rightarrow (-1)^{a\cdot b}|a,b\rangle, \; \forall a,b\in \mathbb{Z}_2 \;,
\end{equation}
\noindent and whose symbol is \hspace{0.05cm} \parbox{1cm}{\Qcircuit @C=0.8ex @R=3.2ex @!R { & \qw  & \ctrl{1} & \qw & \qw \\ & \qw  & \control \qw & \qw & \qw }} \hspace{0.05cm}.
%
%\begin{equation}
%\label{1.2.11.5}
%\setlength{\unitlength}{1cm}
%{\rm CPHASE}: |a,b\rangle \rightarrow (-1)^{a\cdot b}|a,b\rangle \; , \; \forall a,b\in \mathbb{Z}_2 \; , \; {\rm whose\; symbol\; is\;\;} \hspace{0.2cm}\parbox{1cm}{
%\Qcircuit @C=1ex @R=4.2ex @!R {
%   & \qw  & \ctrl{1} & \qw & \qw\\
%   & \qw  & \control \qw & \qw & \qw
%                                }} \; \; ,   
%\end{equation}
%
Finally, we will use the three-qubit controlled-controlled-{\sc not} or Toffoli gate, $\Lambda^2(X)$, %which, as also mentioned before, acts in the computation basis as $\Lambda^2(X): |a,b,c\rangle \rightarrow |a,b,(a\cdot b)+c\rangle \; , \; \forall a,b,c\in \mathbb{Z}_2$, 
whose action in the computation basis was given previously and whose symbol is \parbox{1cm}{ \Qcircuit @C=0.5ex @R=0.3ex @!R { & & \ctrl{1} & \qw \\ & & \ctrl{1} & \qw \\ & & \targ & \qw }} .
%
%\begin{equation}
%\label{1.2.12}
%\Lambda^2(X) : |a,b,c\rangle \rightarrow |a,b,(a\cdot b)+c\rangle \; , \; \forall a,b,c\in \mathbb{Z}_2 \; , \; {\rm whose\; symbol\; is\;\;} \parbox{1cm}{
%\Qcircuit @C=1.2ex @R=1.7ex @!R {
%   & & \ctrl{1} & \qw \\
%   & & \ctrl{1} & \qw \\
%   & & \targ    & \qw
%                                } } \; \; . 
%\end{equation}

%--------------------------------------------------------------%
\subsection{Accuracy}
\label{sec:Accuracy}

In any physical implementation of a quantum computation, the elementary quantum gates will be executed with some limited accuracy; e.g., the rotation angle in a beam splitter or the length of a pulse can be specified only to some finite precision. In order for the quantum circuit model to be realistic, it is important to show that any desired accuracy for the computation outcome can be achieved if the accuracy of each gate in the quantum circuit scales at most as an inverse polynomial in the size, $L$, of the computation. Equivalently, the number of bits of precision specifying the matrix elements of each quantum gate will need to scale ``slowly'' as at most an inverse polynomial in the {\em logarithm} of $L$ \cite{Bernstein97a}. In particular, quantum computers would resemble analog rather than digital devices if the number of bits of precision scaled as an inverse polynomial in $L$, since then physical parameters such as rotation angles or pulse lengths determining the accuracy of each quantum gate would have to be specified to precision that increased as an exponential in $L$.

Consider an ideal quantum circuit where the sequence of $L$ gates $U_1, U_2, \cdots, U_L$ is applied to the initial $n$-qubit state $|\psi_0\rangle$ where, e.g., $|\psi_0\rangle = |0\rangle^{\otimes n}$. The outcome of the quantum  computation is given by a measurement along the computation basis of a subset of the output qubits. Let $\{q_i^{(\rm ideal)}\}$ denote the ideal probability distribution for the measurement outcomes if all gates in the quantum circuit were executed ideally, and let $\{q_i^{(\rm actual)}\}$ denote the actual probability distribution for the measurement outcomes when gates are executed with some limited accuracy. We can then define the {\em error}, $\delta$, of the actual computation as the $L^1$ distance between these two probability distributions,
\begin{equation}
\label{1.2.13}
\delta = || q^{(\rm actual)}-q^{(\rm ideal)} ||_1 \equiv \sum_i | q_i^{(\rm actual)}-q_i^{(\rm ideal)}| \; ,
\end{equation}

\noindent and let us note the useful property that $|| q^{(1)}-q^{(2)} ||_1 \leq 2$, $\forall q^{(1)}, q^{(2)}$ probability distributions. The {\em accuracy} of the actual computation will then correspond to $1-\delta$. 

%------------------------------------------------%
\medskip \medskip \noindent {\bf A Stochastic Noise Model} \medskip

If noise is stochastic, then we can associate probabilities with the occurrence of faults during the computation. Consider the set $\mathcal{I}_r$ of $r$ specific gates, and let us assume that the probability of faults occurring at all gates in $\mathcal{I}_r$ %(irrespective of what happens at all gates not in $\mathcal{I}_r$) 
is at most $p^r$ for some $0\leq p \leq 1$. For example, this would be the case if faults in each gate are described by an i.i.d. process according to which the gate is implemented ideally with probability at least $1-p$ and it is replaced by a faulty gate with probability at most $p$. If we denote the probability of having at least one faulty gate in the quantum circuit by $P_{\rm fail}\leq L p$,  
\begin{equation}
\label{1.2.14}
q_i^{(\rm actual)} = \left( 1-P_{\rm fail} \right) q_i^{(\rm ideal)} + P_{\rm fail} \; q_i^{(\rm fail)} \; ,
\end{equation}

\noindent for some distribution, $\{q_i^{(\rm fail)}\}$, for the computation outcome when faults have occurred.  Then, we can upper bound 
\begin{equation}
\label{1.2.15}
\delta = P_{\rm fail} \sum\limits_i |q_i^{(\rm fail)} - q_i^{(\rm ideal)}| \leq 2 L p \; .
\end{equation}

\noindent %Therefore, the probabilities of faults in the implementation of the quantum circuit add up at most linearly to determine the final computation error $\delta$. 
In order words, some desired final accuracy, $1-\delta_0$, can be achieved if $p \leq {\delta_0 / 2L}$.
%
%\begin{equation}
%\label{1.2.16}
%p \leq {\delta_0 \over 2L} \;.
%\end{equation}

%------------------------------------------------%
\medskip \medskip \noindent {\bf A Unitary Noise Model} \medskip

In many physical situations of interest, there exist however noise processes which are not stochastic; e.g., systematic imperfections in the implementation of the gate $U_j$ could lead to implementing a different gate $\tilde{U}_j$ instead. In such a noise model we cannot associate a probability with the event that a given gate is faulty. Instead, the error, $\varepsilon$, in the implementation, $\tilde{U}_j$, of the ideal gate, $U_j$, can be defined using a suitable operator norm. Let us assume that for all gates,
\begin{equation}
\label{1.2.17}
|| \tilde{U}_j - U_j ||_{\rm sup} \equiv \sup\limits_{|| |\phi\rangle ||= 1}{|| (\tilde{U}_j - U_j )|\phi\rangle || }  \leq \varepsilon \; ,
\end{equation}

\noindent where the {\em sup} operator norm is unity for unitary operators and has the properties
\begin{equation}
\label{1.2.17.2}
\begin{array}{c}
|| A + B ||_{\rm sup} \leq || A ||_{\rm sup} + || B ||_{\rm sup}  \; ; \; || A \cdot B ||_{\rm sup} \leq || A ||_{\rm sup} \cdot || B ||_{\rm sup} \; ; \; || A \otimes B ||_{\rm sup} = || A ||_{\rm sup} \cdot || B ||_{\rm sup} \; .
%|| A + B ||_{\rm sup} \leq || A ||_{\rm sup} + || B ||_{\rm sup}  \; , \\ 
%|| A \cdot B ||_{\rm sup} \leq || A ||_{\rm sup} \cdot || B ||_{\rm sup} \; ,\\% \;
%|| A \otimes B ||_{\rm sup} = || A ||_{\rm sup} \cdot || B ||_{\rm sup} \; .
\end{array}
\end{equation}

We first note that the ideal probability distribution of the measurement outcomes is given in terms of the ideal final state $|\psi_{\rm final}\rangle = U_L \cdots U_2 U_1|\psi_0\rangle$ as
\begin{equation}
\label{1.2.18}
p_i^{(\rm ideal)} = ||\langle i|\psi_{\rm final}\rangle ||^2 = {\rm Tr}\left(P_i \rho^{(\rm ideal)}  \right)  \; ,
\end{equation}

\noindent where $P_i \equiv |i\rangle \langle i|$ is the projector onto the final state accociated with the outcome $i$, and $\rho^{(\rm ideal)} \equiv |\psi_{\rm final}\rangle \langle \psi_{\rm final} |$. Similarly, if  $|\tilde{\psi}_{\rm final}\rangle = \tilde{U}_L \cdots \tilde{U}_2 \tilde{U}_1|\psi_0\rangle$ is the actual final state, 
\begin{equation}
\label{1.2.19}
p_i^{(\rm actual)} = ||\langle i|\tilde{\psi}_{\rm final}\rangle ||^2 = {\rm Tr}\left(P_i \rho^{(\rm actual)}  \right)   \; ,
\end{equation}

\noindent where $\rho^{(\rm actual)} \equiv |\tilde{\psi}_{\rm final}\rangle \langle \tilde{\psi}_{\rm final} |$. By  substituting in equation (\ref{1.2.13}), 
\begin{equation}
\label{1.2.20}
\delta = \sum_i |{\rm Tr}\left( P_i (\rho^{(\rm actual)} - \rho^{(\rm ideal)})  \right) | \leq \sum_{i,j} |\lambda_j | \cdot \langle \lambda_j |P_i|\lambda_j\rangle \equiv || \rho^{(\rm actual)} - \rho^{(\rm ideal)}||_{\rm tr} \; ,
\end{equation}

\noindent where $\{ \lambda_j \}$ are the eigenvalues of $\rho^{(\rm actual)} - \rho^{(\rm ideal)}$ and $\{|\lambda_j \rangle \}$ are the corresponding eigenvectors. Here, $||\cdot ||_{\rm tr}$ denotes the {\em trace norm}, $||A||_{\rm tr} \equiv {\rm Tr}\left( \sqrt{A^\dagger A} \right)$, $\forall A\in \mathcal{B}(\mathcal{H_S})$ , and in the last step we have used the completeness relation $\sum_i P_i=I$. 
By computing the trace norm in equation (\ref{1.2.20}),  
\begin{equation}
|| \rho^{(\rm actual)} - \rho^{(\rm ideal)}||_{\rm tr} = 2 || |\tilde{\psi}_{\rm final}\rangle - |\psi_{\rm final}\rangle|| \leq  2 || \tilde{U}_L \cdots \tilde{U}_2 \tilde{U}_1 - U_L \cdots U_2 U_1||_{\rm sup} \;,
\end{equation} 
\noindent and finally, by using the triangle inequality of the sup norm and equation (\ref{1.2.17}), 
%
%2 || ( \tilde{U}_L \cdots \tilde{U}_2 \tilde{U}_1 - U_L \cdots U_2 U_1) |\psi_0\rangle|| \leq  2 || \tilde{U}_L \cdots \tilde{U}_2 \tilde{U}_1 - U_L \cdots U_2 U_1||_{\rm sup}$. By using the triangle inequality of the sup norm and equation (\ref{1.2.17}), 

\begin{equation}
\label{1.2.21}
|| \tilde{U}_L \cdots \tilde{U}_2 \tilde{U}_1 - U_L \cdots U_2 U_1||_{\rm sup} \leq L \varepsilon \; ,
\end{equation}

\noindent which implies $\delta \leq 2 L\varepsilon$ \cite{Bernstein97a}. Therefore, we arrive at the same conclusion as for stochastic noise, except that here $\varepsilon$ is a bound on the sup norm distance between the actual and ideal gates instead of a probability.

%------------------------------------------------%
\medskip \medskip \noindent {\bf A Non-Markovian Noise Model} \medskip

The situation we just considered corresponds to the case when the actual noisy gates are unitary operators as, e.g., happens when gate inaccuracies are due to systematic over or under rotations. Let us now generalize this noise model %Nevertheless, by the discussion of physical operations in \S \ref{sec:KrausTheorem}, we can {\em model} any physical 
by considering {\em unitary} faults acting between the computer qubits with Hilbert space $\mathcal{H}_S$ and a common {\em bath} system with Hilbert space $\mathcal{H}_B$; here, the bath is a model for any physically relevant degrees of freedom that have an unwanted coupling to our computer qubits. %Qubits are initialized in the state $|\psi_0\rangle_S$ and the bath is initialized in some standard state, say, $|\phi_0\rangle_B$. 
%We then consider the {\em entire} actual computation as being realized in the larger Hilbert space $\mathcal{H}_S \otimes \mathcal{H}_B$, and the evolution in this larger space is unitary. 

Our ideal gates will now correspond to the unitaries $\{ U_j \otimes I_B \}$, i.e., they implement the desired unitary, $U_j$, in $\mathcal{H}_S$ and act trivially in $\mathcal{H}_B$. Furthermore, let us assume that the actual noisy implementation of the ideal gate $U_j \otimes I_B$ is the unitary $ \Delta_j (U_j \otimes I_B )$ where $\Delta_j$ acts on the support of $U_j$ and on $\mathcal{H}_B$. By expanding in the Pauli basis, we may write $\Delta_j = (I \otimes V_j) + F_j$; here, the {\em fault operator} or, simply, {\em fault} $F_j$ acts nontrivially in the support of $U_j$ and in some way in $\mathcal{H}_B$, and $V_j$, $F_j$ are subject to the constraint that $\Delta_j$ is unitary (e.g., $V_j$ may describe the {\em self-evolution} of the bath). In analogy to assumption (\ref{1.2.17}), let us also assume that for all gates,
\begin{equation}
\label{1.2.22}
|| F_j(U_j \otimes I_B) ||_{\rm sup} = ||F_j ||_{\rm sup}  \leq \varepsilon \; .
\end{equation}
 
\noindent In this noise model, the final state of the computation will be 

\begin{equation}
\label{eq:fault-path-1}
|\tilde{\psi}_{\rm final}\rangle = \Delta_L (U_L \otimes I_B ) \cdots \Delta_2 (U_2 \otimes I_B ) \cdot \Delta_1 (U_1 \otimes I_B ) |\psi_0\rangle_{SB} \; ,
\end{equation} 

\noindent where $|\psi_0\rangle_{SB} = |\psi_0\rangle_S \otimes |\phi_0\rangle_B$ and $|\phi_0\rangle_B$ is arbitrary. Consider next expressing the sequence of unitaries in equation (\ref{eq:fault-path-1}) as a sum of operators by rewriting each $\Delta_j$ as $I \otimes V_j$ plus the fault $F_j$ and opening the parentheses. Let us call each operator in this sum a {\em fault path}, and consider writing this sum as a sum of two terms: The first ``good'' term, $Gd$, contains a single fault path which corresponds to the sequence of the unitaries $\{ U_j \otimes V_j \}$. The second ``bad'' term, $Bd$, is a sum of all remaining fault paths. This allows us to write $|\tilde{\psi}_{\rm final}\rangle = (Gd + Bd)|\psi_0\rangle_{SB}$, where $|\psi_{\rm final}\rangle = Gd |\psi_0\rangle_{SB} $ produces the ideal outcome statistics. Therefore, $|| \, |\tilde{\psi}_{\rm final}\rangle - |\psi_{\rm final}\rangle || = || Bd (|\psi_0\rangle_S \otimes |\phi_0\rangle_B) || \leq || Bd ||_{\rm sup}$. Finally, by regrouping the different fault paths in $Bd$, we rewrite $Bd$ as a sum of terms according to the appearance of the first fault (using some arbitrary time-ordering convention); i.e., in each term, some faulty gate is specified, all gates preceding it are ideal and we sum over fault paths that have or do not have faults in all subsequent gates. There are $L$ terms of this form, each of norm at most $\varepsilon$. Thus, by upper bounding the norm of the sum by the sum of the norms, we obtain $|| Bd ||_{\sup} \leq L \varepsilon$, which implies $\delta \leq 2L\varepsilon$ as before.

\subsection{Quantum Universality}
\label{sec:QUniversality}
%why a finite gate set

As in the case of classical computation, the notion of quantum universality is important for two reasons. First, expressing a quantum algorithm as a quantum circuit that uses gates from a given gate set allows us to classify its complexity by means of the size of the quantum circuit. Second, when we consider noisy quantum computation, it becomes sufficient to construct fault-tolerant implementations for all gates in the universal gate set: when a specific algorithm needs to be realized fault-tolerantly, we first express the algorithm in terms of gates in the universal gate set and, then, we implement each gate in this set fault-tolerantly.

Let us start by giving the formal definition of finite quantum universal gate sets.

\begin{defi}[Finite Quantum Universal Gate Sets]
\label{def:1}
Let a finite set of gates $\mathcal{G}=\{g_i \}$. Then, $\mathcal{G}$ is universal for quantum computation if $\; \forall n\,{>}\,0$, $\;\forall U\,{\in}\,SU(2^n)$, $\forall \varepsilon\,{>}\,0$, $\exists $ a finite sequence $g_1, g_2, \dots , g_L$  such that $||U - g_L \cdots g_2 g_1||_{\rm sup} \leq \varepsilon$.
\end{defi}  

\noindent In order words, the group generated by the gates in $\mathcal{G}$ is {\em dense} in $SU(2^n)$ (we have used $SU(2^n)$ instead of $U(2^n)$ since overall factors are irrelevant for quantum  computation). 

There are several finite universal gate sets which are known to admit a fault-tolerant implementation, and some of them may be preferable to others depending on the particular physical implementation of quantum gates. The following proposition lists the two finite universal gate sets that are most commonly used in fault-tolerant quantum computation.

\begin{propo}
\label{propo:1}
The following gate sets are universal for quantum computation:
\begin{enumerate}
\item Boykin et al. set \cite{Boykin99}: $\mathcal{G} = \{${\sc cnot}$, H, T \}$,
\item Kitaev set \cite{Kitaev97c}: $\mathcal{G}= \{${\sc cnot}$, H, S, {\rm Toffoli} \}$.
\end{enumerate}
\end{propo}

\noindent Another interesting finite universal gate set is the two-element set $\mathcal{G}=\{H, {\rm Toffoli} \}$ \cite{Shi02}, although it is not particularly useful for fault-tolerant quantum computation.

At this point, it is convenient to introduce the following useful classification of quantum gates. 

\begin{defi}[$C_k$ Gate Hierarchy \cite{Gottesman99}]
\label{def:1.2}
Let $C_1$ denote the Pauli group on $n$ qubits which is generated by tensor products of the operators in equation (\ref{1.2.4}) and the identity operator with overall phases ${\pm}1$ or ${\pm}i$. Then, for $k\,{>}\,1$, we recursively define
\[
C_k = \{U|UC_1 U^\dagger \subseteq C_{k-1} \} \; .
\] 
In particular, $C_2$ is called the {\em Clifford group} and contains operators conjugating Pauli operators to Pauli operators. 
\end{defi} 

The Clifford group, $C_2$, is generated by the operators {\sc cnot}, $H$ and $S$---we write $C_2 = \langle${\sc cnot}$, H, S \rangle$ where $\langle \cdot \rangle$ denotes a generating set. The $C_k$ hierarchy has the property that, $\forall k$, $C_k \subseteq  C_{k+1}$ and $C_{k+1}\setminus C_k \not = \emptyset $. In fact, any gate set that generates the Clifford group and contains one additional gate outside the Clifford group is universal for quantum computation \cite{nebe00}. Henceforth, for simplicity, we will call gates not in $C_2$ {\em non-Clifford gates}. The first gate set in proposition \ref{prop:1} contains the non-Clifford gate $T$; the second gate set contains the non-Clifford Toffoli gate. 

An important question is how efficient is the approximation of a given unitary within a desired error $\varepsilon$, i.e., how many gates from a finite universal set, $\mathcal{G}$, are necessary to achieve such an approximation. In addition, it is important to know how efficient it is to {\em find} the gates from $\mathcal{G}$ that realize the desired approximation with a classical computer: if we say that the classical computer {\em compiles} the gate $U\in SU(2^n)$ in terms of gates in $\mathcal{G}$, we need to ask how efficient in time and memory is this compilation as a function of $\varepsilon$. The Solovay-Kitaev theorem answers this question.

\begin{theo}[Solovay-Kitaev Theorem \cite{Kitaev-book,Nielsen05c}]
\label{theo:1.2}
Consider any {\em fixed} number of qubits, $n$, and let $\mathcal{G} = \{ g_i \}\subseteq SU(2^n)$ be a quantum universal gate set that is closed under inversion (i.e., if $g\in \mathcal{G}$ then $g^{-1}\in \mathcal{G}$). Then, $\forall U\,{\in}\,SU(2^n)$, $\forall \varepsilon\,{>}\,0$, $\forall c\,{>}\,0$, there exists a sequence $g_1, g_2, \dots, g_L$ such that \parbox{3.7cm}{$||U - g_L \cdots g_2 g_1 ||_{\rm sup} \leq \varepsilon$}, where $L=${\it O}$\left(\left(log(1/\varepsilon)\right)^{3+c}\right)$ and the classical time for computing the sequence is $T=${\it O}$\left( \left(log(1/\varepsilon)\right)^{3+c}\right)$.
\end{theo}

Thus, the Solovay-Kitaev theorem establishes that gates from $\mathcal{G}$ can efficiently approximate any $U\,{\in}\,SU(2^n)$ within any desired error $\varepsilon$---the efficiency of the approximation is polylogarithmic in $1/\varepsilon$. In addition, this approximation can be found efficiently with a classical computer---the classical running time is also polylogarithmic in $1/\varepsilon$.

Of course, the efficiency of the approximation is in general exponential in the number of qubits, i.e., approximating some $U\,{\in}\,SU(2^n)$ with gates from $\mathcal{G}$ requires quantum circuits of size scaling exponentially in $n$. %\footnote{Recall that each gate in $\mathcal{G}$ acts on a fixed constant number of qubits.} 
But this should not confuse us: The importance of the Solovay-Kitaev theorem is that it implies that any given quantum circuit, $\cal{C}$, which approximates some unitary to some accuracy can be efficiently converted to another quantum circuit, $\cal{C}'$, which achieves an improved accuracy---this conversion is efficient since it is achieved by approximating each gate in $\cal{C}$ which only acts on a fixed constant number of qubits.  

%--------------------------------------------------------------%

%--------------------------------------------------------------%
\section{Quantum Coding}
\label{sec:QCoding}

In this section, I will discuss how quantum information can be encoded in such a way that errors which only affect a sufficiently ``small part'' of it can first be digitized and then corrected. For this discussion, I will assume that the error-correction procedure that diagnoses and corrects errors is realized itself ideally without faults. The more general question of how quantum error correction can be performed with noisy operations is the subject of the theory of fault-tolerant quantum computation and will be discussed in detail in the main body of this thesis.  

In the classical setting, the idea of encoding information in a way that enables error correction is intuitively clear. The most natural classical encoding scheme encodes information by repeating it several times. For example, when we want to transmit the bit value $0$ or $1$ over a noisy transmission line, we may send the message several times as a string of $0$s or $1$s so that the receiver will have a high probability of deducing the intended message by taking a majority vote. The theory of classical error correction includes much more general and efficient schemes than this simple encoding by repetition. However, the principal idea is the same in all these encoding schemes: In order to protect classical information from errors, information is encoded using some form of {\em redundancy} so that if a sufficiently small part of it is affected by errors, the receiver can recover the correct  transmitted information with high probability.  

Protecting quantum information from errors is far more challenging due to the {\em information-disturbance principle}. Put simply, this principle is just the statement that when we gain information about an unknown quantum state we unavoidably disturb it: Since quantum states cannot be copied \cite{Wootters82}, the only possible way to gain information about an unknown quantum state is to perform a measurement on it. But then, the post-measurement state is an eigenstate of the measured operator---as a consequence of the measurement, the initial quantum state {\em collapses} to a post-measurement state which is, in general, different from the initial one. Therefore, quantum error correction procedures must be able to diagnose errors without gaining any information about the encoded quantum states, since if they gain any such information, the encoding will be damaged. 

%A consequence of the {\em no-cloning} theorem which states that there is no physical operation which, for all possible quantum states $|\psi\rangle$, maps $|\psi\rangle$ to $|\psi\rangle \otimes |\psi\rangle$. This implies that, if our information consists of the ensemble of quantum states $\{|\psi_i\rangle \}$, there is no physical copying process that can encode it in the form of the ensemble $\{|\psi_i\rangle \otimes |\psi_i\rangle \otimes \cdots \}$ which would be the analog of the classical encoding by repetition. Thus, the necessary redundancy for protecting quantum information cannot be a result of simply copying the information multiple times in different systems.

Despite this obstacle, quantum codes exist \cite{Shor95b,Steane96}. Their construction is based on the following two general ideas: First, quantum codes  achieve redundancy not by repeating quantum information in different quantum system but by encoding information into quantum states that are {\em entangled} across many different quantum systems. Second, errors in these entangled encoded quantum states can be diagnosed by making multi-qubit measurements which do not reveal information about the encoded information. In essence, quantum codes operate by encoding quantum information into the {\em quantum correlations} between different systems. Intuitively, the idea is that an adversary that can only apply errors on some small fraction of these systems will be unable to gain information about the encoded state since information is hidden in the correlations across a large number of different systems---we may say that information is encoded ``non-locally'' and it is protected as long as errors only act ``locally.''  

The rest of this section is structured as follows: In \S \ref{sec:QECCriteria}, I discuss the basic principles of quantum coding and I state the criteria for quantum error correction. In \S \ref{sec:QStabCodes}, I review the stabilizer formalism that provides a succinct description of a wide class of quantum codes. Next, in \S \ref{sec:QCSSCodes}, I discuss the CSS class of quantum codes which are well-suited for fault-tolerant quantum computation. Finally, in \S \ref{sec:CodeExamples}, I present two examples of quantum codes.

%--------------------------------------------------------------%
\subsection{Quantum Error-Correction Criteria}
\label{sec:QECCriteria}

Consider a {\em system} with a Hilbert space $\mathcal{H}_S$ that includes a natural decomposition into a tensor product of, say, $n$ two-dimensional subsystems which will be the qubits of our quantum computer. According to theorem \ref{theo:1}, any physical noise operation, $\mathcal{N}$, acting on $\mathcal{H}_S$ can be described in terms of its Kraus operators, $\{M_k \}$, which are subject to the constraint $\sum_k M_k^\dagger M_k = I_S$. In particular, each operator $M_k$ can be seen as arising due to a unitary evolution in a larger Hilbert space, $\mathcal{H}_S \otimes \mathcal{H}_B$, that includes a {\em bath} or {\em environment}, B, followed by a measurement in the bath along the orthonormal basis $\{|k\rangle_B \}$ that gave the outcome $k$. Since quantum measurement is an irreversible process, interacting the system and the bath and then performing a measurement on the bath will, in general, result in an irreversible change in the state of the system.  Reversing the action of noise would require reversing the action of the Kraus operators which is only possible if all but one are zero (in which case the non-zero Kraus operator would be unitary); we can state this simply by saying that the only physical operations which are invertible correspond to unitary operations. We cannot therefore hope to be able to invert an arbitrary physical noise operation acting on all the qubits of our quantum computer. In the absence of more information about the structure of noise, the goal of quantum error correction is more modest: the goal is to design a correction procedure for protecting against noise that acts on sufficiently small subsets of qubits. 

Let the initial state in $\mathcal{H}_S$ be the vector $|\psi\rangle_S$. (In the more general case where the state is described by a density matrix, $\rho$, we may always expand $\rho = \sum_i c_i \,|\psi_i\rangle \langle \psi_i|_S\,$, e.g., in terms of the eigenvectors $\{ |\psi_i\rangle_S \}$ of $\rho$, and we can consider the vector $|\psi\rangle = \sum_i \sqrt{c_i} |\psi_i\rangle_S \otimes |i\rangle_R$ where $\{|i\rangle_R \}$ is an orthonormal basis in a virtual {\em reference} system with Hilbert space $\mathcal{H}_R$ which has been added to $\mathcal{H     }_S$. The idea is that if we later trace over the reference system, then we are left with $\rho$, i.e., ${\rm Tr}_R\left(|\psi\rangle \langle \psi| \right) = \rho$. We call this process, by which a density matrix, $\rho$, is seen as arising from a state vector, $|\psi\rangle$, in a larger space, {\em purification} of $\rho$.) By theorem \ref{theo:1}, we can describe $\mathcal{N}$ as performing the operation
\begin{equation}
\label{1.3.1}
|\psi\rangle_S \otimes |0\rangle_B \rightarrow \sum_k M_k |\psi\rangle_S \otimes |k\rangle_B \; ,
\end{equation}

\noindent where each Kraus operator, $M_k$, may be expanded in a fixed basis of operators, $\{E_a \}$, that we will call the {\em error} basis. Then, the action of $\mathcal{N}$ becomes
\begin{equation}
\label{1.3.2}
|\psi\rangle_S \otimes |0\rangle_B \rightarrow \sum_a E_a |\psi\rangle_S \otimes |a\rangle_B \; ,
\end{equation}

\noindent where the states $\{ |a\rangle_B \}$ in $\mathcal{H}_B$ are, in general, not normalized nor orthogonal. If $\{ |a\rangle_B \}$ were in fact orthogonal and normalized then the operators $\{ E_a \}$ could be interpreted as different errors which have occurred on $|\psi\rangle_S$ depending on the state in $\mathcal{H}_B$. Even though this interpretation is not valid in general, it is sufficient to proceed by devising a correction procedure that inverts each $E_a$ separately. Our correction procedure is implemented by a physical recovery operation, $\mathcal{R}$, with Kraus operators $\{R_\mu \}$; then, the composition $\mathcal{R} \circ \mathcal{E}$ acts on $|\psi\rangle_S$ as  
\begin{equation}
\label{1.3.3}
|\psi\rangle_S \otimes |0\rangle_B \otimes |e_0\rangle_A \rightarrow \sum_{a,\mu} R_\mu E_a |\psi\rangle_S \otimes |a\rangle_B \otimes |e_\mu \rangle_A \; ,
\end{equation}

\noindent where $\{|e_\mu\rangle_A \}$ is an orthonormal basis in an ancillary system with Hilbert space $\mathcal{H}_A$ initialized it in the state $|e_0\rangle_A$ which we attach to $\mathcal{H}_S$ in order to implement the recovery. 

Error recovery is successful if the state on the right-hand side of equation (\ref{1.3.3}) is of the form $|\psi\rangle_S \otimes |\phi\rangle_{BA}$, for any state $|\phi\rangle_{BA} \in \mathcal{H}_B\otimes \mathcal{H}_A$.  We have already discussed that this is not always possible for all states, $|\psi\rangle_S$, and all noise operations, $\mathcal{N}$. However, successful error recovery is possible when $|\psi\rangle_S$ is restricted in a {\em subspace} of $\mathcal{H}_S$ and  $\mathcal{N}$ is restricted to act nontrivially on a sufficiently small subset of the $n$ qubits in $\mathcal{H}_S$. The following theorem states the criteria for when quantum error correction is possible.  

\begin{theo}[Quantum Error-Correction Criteria \cite{Knill96,Bennett96a}]
\label{theo:3}
Let an error basis, $\{E_a\}$, and let $P_{L}$ be the projector onto a subspace, $\mathcal{H}_L$, of $\mathcal{H}_S$. Then the following are equivalent:
\begin{enumerate}
\item The noise operation, $\mathcal{N}$, whose Kraus operators are linear combinations of the $\{E_a\}$ with arbitrary complex coefficients is correctable, i.e., there exists an ancillary system with Hilbert space $\mathcal{H}_A$ that is initialized in the state $|e_0\rangle_A$ and a recovery operation, $\mathcal{R}$, acting on $\mathcal{H}_S \otimes \mathcal{H}_A$ such that, $\forall \rho_L \in \mathcal{H}_L$,
  ${\rm Tr}_A \left( \mathcal{R} \left( \mathcal{E} (\rho_L) \otimes |e_0\rangle \langle e_0|_A \right) \right) = \rho_L\;$,

\item $\;\forall E_a, \forall E_b$, $\exists$ a Hermitian complex matrix, $c$, such that $P_{L} E_a^\dagger E_b P_{L} = c_{ab} P_{L}\;$.
\end{enumerate}

%Let an error basis, $\{E_a\}$, and let $P_{L}$ be the projector onto a subspace, $\mathcal{H}_L$, of $\mathcal{H}_S$. Then, for an ancillary system with Hilbert space $\mathcal{H}_A$ that is initialized in the state $|e_0\rangle_A$, a recovery operation, $\mathcal{R}$, acting on $\mathcal{H}_S \otimes \mathcal{H}_A$ exists such that, $\forall \rho_L \in \mathcal{H}_L$, the noise operation, $\mathcal{N}$, whose Kraus operators are linear combinations of the $\{E_a\}$ with arbitrary complex coefficients is correctable in the sense ${\rm Tr}_A \left( \mathcal{R} \left( \mathcal{E} (\rho_L) \otimes |e_0\rangle \langle e_0|_A \right) \right) = \rho_L\;$, if and only if $\;\forall E_a, \forall E_b$, $\exists$ a Hermitian complex matrix $c$ such that $P_{L} E_a^\dagger E_b P_{L} = c_{ab} P_{L}$ .
\end{theo}

Quantum information will be encoded in states $|\psi\rangle_L \in \mathcal{H}_L$; we will call $\mathcal{H}_L$ the {\em code space} and $|\psi\rangle_L$ the {\em logical} $|\psi\rangle$ state. Then, the condition $P_{L} E_a^\dagger E_b P_{L} = c_{ab} P_{L}$ can be understood as saying that (i) two orthogonal logical states must remain orthogonal after errors act on them (so that errors do not make logical states which were initially orthogonal to later overlap), and (ii) different logical states must be transformed by different errors in a way that only depends on the errors and not on the logical states themselves (so that learning about the errors will not give us any information about the logical state). 

We note that due to the linearity of quantum mechanics, if a code can correct all errors in the basis of Pauli operators on up to $t$ qubits, then it can also correct {\em any} operator, $M_k$, supported on up to $t$ qubits (since the Pauli operators form a basis in the operator space). It follows that any noise operation with support on up to $t$ qubits can also be inverted. Therefore, in most cases of interest it is sufficient to design codes that correct Pauli errors such as the stabilizer codes that will be discussed in the next section; henceforth, we will only consider such codes.

%--------------------------------------------------------------%
\subsection{Stabilizer Codes}
\label{sec:QStabCodes}

Let us now briefly review the theory of binary stabilizer codes \cite{Calderbank96a,Gottesman96,Gottesman97} that form a particularly interesting subclass of quantum codes. The code space of a stabilizer code is the {\em simultaneous} eigenspace of a normal abelian subgroup, $\mathcal{S}$, of $C_1$ that does not contain $-I$ or ${\pm}iI$; this subgroup is called the code's {\em stabilizer} since, by convention, $\forall S\in \mathcal{S}$, $S|\psi\rangle_L = |\psi\rangle_L$. 

To understand the error-correction capabilities of a stabilizer code, a useful concept is the code's distance.

\begin{defi}[Distance]
\label{defi:3}
Let the {\em weight} of the Pauli operator $E_a$, ${\rm wgt}(E_a)$, be the number of qubits in its support  (i.e., the number on qubits on which $E_a$ acts different than the identity). Then, with the same notation as in theorem \ref{theo:3}, the distance, $d$, of a code is $d={\rm min} \{ {\rm wgt}(E_a)| P_{L} E_a P_{L} \not \propto P_{L} \}$.
\end{defi}

\noindent In order words, the distance is the smallest number of qubits on which an error, $E_a$, needs to act nontrivially  before the  condition $P_{L} E_a P_{L} \propto P_{L}$ is violated. This implies that $P_{L} E_a^\dagger E_b P_{L} \propto P_{L}$ is satisfied for all $E_a^\dagger E_b$ that act nontrivially on up to $d-1$ qubits, or that the code can correct all errors with support on up to $t=\lfloor {d-1\over 2}\rfloor$ qubits. Assuming the dimension of $\mathcal{H}_L$ is $2^k$, such a code will be denoted as an $[[n,k,d]]$ code---the notation means that $n$ physical qubits are used to encode $k$ logical qubits and the distance of the code is $d$.  

The analysis of stabilizer codes is facilitated by the simple algebra of the Pauli group. Let us define the {\em centralizer} of $\mathcal{S}$ in $C_1$ by $C(\mathcal{S})=\{E \in C_1 | \forall S\in \mathcal{S}, [E,S]=0 \}$; i.e., $C(\mathcal{S})$ is the set of all Pauli operators that commute with every element of the code's stabilizer. Because of the special properties of the Pauli group, the centralizer equals the {\em normalizer}, $N(\mathcal{S})$, of $\mathcal{S}$ in $C_1$, i.e., the set of Pauli operators that leave $\mathcal{S}$ invariant under conjugation. Clearly, $\mathcal{S}$ is a normal subgroup of $N(\mathcal{S})$, and $N(\mathcal{S})/\mathcal{S}$ is a group of order $2^{2k}$ containing the four logical Pauli operators for each of the $k$ logical qubits. Then, $\forall E_a, E_b\in C_1$, either (i) $E_a^\dagger E_b\in \mathcal{S}$, in which case $P_{L} E_a^\dagger E_b P_{L} = P_{L}$, or (ii) $E_a^\dagger E_b\in N(\mathcal{S}) \setminus \mathcal{S}$, in which case $P_{L} E_a^\dagger E_b P_{L} = \mathcal{O}_{ab}P_{L}$ where $\mathcal{O}_{ab}$ is a nontrivial logical Pauli operator, or (iii) $\exists S\in \mathcal{S}$ such that $SE_a^\dagger E_b = - E_a^\dagger E_b S$, which implies $P_{L} E_a^\dagger E_b P_{L} = 0$. Since the error-correction criteria fail only in case (ii), it follows that the distance, $d$, of the code is the minimal weight among all elements in $N(\mathcal{S}) \setminus \mathcal{S}$, i.e., the minimal weight among all logical Pauli operators.

A more succinct description of stabilizer codes is possible if we represent Pauli operators as binary vectors. The subscript $a$ labelling an operator $E_a \in C_1$ can be seen as a length-$2n$ binary vector, where  
\begin{equation}
\label{1.3.4}
     E_a = \bigotimes_{j=1}^n \left( i^{\, a[j]\cdot a[j+n]} \; X^{a[j]} Z^{a[j{+}n]} \right) \, ,
\end{equation}

\noindent and $a[j]$ denotes the $j$th component of $a$. With this notation, two elements $E_a$ and $E_b$ of $C_1$ obey the commutation relation
\begin{equation}
\label{1.3.5}
E_a E_b = (-1)^{a\Lambda b^T} E_b E_a \, ,
\end{equation}
\noindent where `$^T$' denotes matrix transposition, and
\begin{equation}
\Lambda = \left( \begin{array}{cc}
                                      \mathbf{0}_n & \mathbf{I}_n \\
                                      \mathbf{I}_n & \mathbf{0}_n
                     \end{array}
                     \right) \, ;
\end{equation}
\noindent here $\mathbf{0}_n $ and $\mathbf{I}_n$ are the zero and identity $n \times n$ matrices, respectively. 

For a stabilizer code with $k$ logical qubits, the generators $\{G_{a_1},G_{a_2},\dots ,G_{a_{n-k}}\}$ of $\mathcal{S}$ can be represented by a $(n-k)\times 2n$ binary {\em check } matrix,
\begin{equation}
\label{1.3.6}
 G \equiv \left( \begin{array}{c}
                                      a_1\\
                                      a_2\\
                                      \vdots\\
                                      a_{n-k}
                     \end{array}
                     \right) \, .
\end{equation}

\noindent Each generator squares to the identity and has eigenvalues ${\pm}1$ with equal degeneracy. Therefore, the $n-k$ generators specify $2^{n-k}$ mutually orthogonal subspaces, each of dimension $2^n/2^{n-k}=2^k$. In particular, the subspace that is the common $+1$ eigenspace for all generators is the {\em code space}; it is the subspace where, by convention, we encode our $k$ logical qubits. Each subspace can be labeled by a length-$(n{-}k)$ binary vector, $e$, where the eigenvalue of $G_{a_i}$ is $(-1)^{e[i]}$, for $i=1,\dots, n-k$. This vector $e$ is called the {\em syndrome} of the subspace; specifically, the code space has syndrome $e=(0 0 \dots 0)$. Furthermore, equation (\ref{1.3.5}) implies that the action of a Pauli operator $E_a$ on the code space changes the syndrome to the value $e=a \Lambda G^T$.

For a {\em non-degenerate} stabilizer code that corrects $t=\lfloor {d-1\over 2}\rfloor$ errors, all $E_a\in C_1$ of  weight at most $t$ take the code space to mutually orthogonal subspaces with distinct syndromes; thus, under the assumption that no more than $t$ errors occurred, every different value of the syndrome points to a unique Pauli error operator. In implementing error correction, the syndrome, $e$, is first measured, and then $E_a^\dagger$ is applied to invert the error $E_a$, where $E_a$ is the unique Pauli operator with ${\rm wgt}(E_a)\le t$ such that $e=a \Lambda G^T$. If the code is {\em degenerate}, then $E_a$ may not be unique but each $E_a$ of weight up to $t$ satisfying $e=a \Lambda G^T$ is equally effective in correcting the error. Indeed, if $E_a$ and $E_b$ give the same syndrome then $E_a^\dagger E_b$ has syndrome zero (i.e., $E_a^\dagger E_b \in N(\mathcal{S})$) and since ${\rm wgt}(E_a^\dagger E_b)\leq 2t < d$, it is in fact $E_a^\dagger E_b \in \mathcal{S}$; therefore, applying $E_a^\dagger$ to recover from the error $E_b$ results in a combined operator that acts trivially in the encoded information and, so, recovery is successful. The property of a code being degenerate is therefore consequence of the fact that there exist pairs of distinct errors with identical syndromes whose product is in $\mathcal{S}$, and has no analogue in classical coding.

Let us finally comment on the case when the syndrome value does not correspond to any correctable error operator---codes for which this never occurs are called {\em perfect}. If a code is not perfect and a syndrome is obtained that cannot be associated with a correctable error---i.e., if there are two equal-weight errors $E_a$ and $E_b$ that give the same syndrome and $E_a^\dagger E_b\in N(\mathcal{S}) \setminus \mathcal{S}$---successful error recovery is not possible. In that case, error correction simply maps its input to the code space, i.e., it applies some recovery operator which restores the syndrome to zero; for the purposes of our discussion, it will not matter what the convention is for how this recovery operator is chosen. 

%--------------------------------------------------------------%
\subsection{CSS Construction}
\label{sec:QCSSCodes}

A subclass of stabilizer codes is CSS codes \cite{Calderbank95,Steane97} which are especially well-suited for fault-tolerant quantum computation. CSS codes are constructed from two classical binary linear codes whose theory we  will first briefly review.

Letting $\mathbb{F}_2^n$ denote the vector space defined by length-$n$ strings over the binary field $\mathbb{F}_2$, a classical binary linear code, $C=[n,k,d]$, is a subspace of $\mathbb{F}_2^n$, the {\em code space}, which is spanned by the $k$ \parbox{1.35cm}{length-$n$} binary vectors $\{ v_1, v_2, \dots, v_k \}$. The code words are linear combinations of the vectors $\{v_i \}$, i.e., the length-$k$ binary word $w=(w_1, w_2, \dots, w_k)$ is encoded as the code word $w_L \equiv \sum_{i=1}^k w_i v_i$. Code words have the property that $H\cdot w_L^T=0$, where $H$ is the $(n{-}k)\times n$ binary {\em parity check} matrix satisfying $H\cdot v_i^T=0$, $\forall v_i$. Given a code word, $w_L$, which has been affected by an error, $\Delta w_L$, error correction proceeds by computing the length-$(n{-}k)$ binary {\em syndrome} vector, $e=H \cdot (w_L + \Delta w_L)^T = H \cdot (\Delta w_L)^T$, finding the most likely error, $\tilde{\Delta w_L}$, leading to $e$ and applying $\tilde{\Delta w_L}$ to invert the error, $w_L + \Delta w_L + \tilde{\Delta w_L} = w_L$ with high success probability. Let the {\em Hamming weight} of a binary string be the number of $1$s in it. Then, the code's {\em distance}, $d$, is the minimal Hamming weight of a nonzero code word, and all errors of Hamming weight up to $\lfloor {d-1\over 2} \rfloor$ can be corrected. Finally, the {\em dual} code to $C$, $C^\perp$, is the orthogonal complement of $C$ in $\mathbb{F}_2^n$, i.e., all code words of $C^\perp$ are orthogonal to all code words of $C$ (it can be that $C \subseteq C^\perp$ or $C^\perp \subseteq C$). 

CSS quantum codes are constructed from two classical binary linear codes $C=[n,k,d]$ and $C'=[n,k',d']$ such that $(C')^\perp \subseteq C$. The construction proceeds by using $C$ with parity check matrix $H$ to correct $Z$ errors and by using $C'$ with parity check matrix $H'$ to correct $X$ errors, i.e., by creating the quantum check matrix,   
\begin{equation}
\label{1.3.7}
 G \equiv \left( \begin{array}{c|c}
                                      H_{(n-k)\times n)}          & \mathbf{0}_{(n-k)\times n} \\
                                      \mathbf{0}_{(n-k')\times n} & H'_{(n-k')\times n}
                     \end{array}
                     \right) \, ,
\end{equation}

\noindent where subscripts denote the dimensions of the submatrices. The condition $(C')^\perp \subseteq C$ is needed in order for rows of $G$ to commute which requires $H\cdot (H')^T= H' \cdot H^T=0$. Indeed, this requirement implies  that the rows of $H'$ span a code space that is contained in the code space of $C$ (since every code word in $C$ satisfies $H\cdot w_L^T=0$); but the rows of $H'$ span the code space of $(C')^\perp$ since they are orthogonal to every code word in $C'$.

According to equation (\ref{1.3.7}), each row of $H$ becomes a generator of the code's stabilizer that is a tensor product of $X$ operators alone (and the identity on the remaining qubits); such generators we will call {\em $X$-type}. Similarly, each row of $H'$ becomes a generator that is a tensor product of $Z$ operators alone; such generators we will call {\em $Z$-type}. Since there are $2n-k-k'$ independent generators, the quantum code encodes $n-(2n-k-k')=k+k'-n$ logical qubits. Furthermore, by measuring the eigenvalues of the $X$-type (respectively, $Z$-type) generators and computing the syndrome of the classical code $C$ (respectively, $C'$), we can correct $Z$ (respectively, $X$) errors on up to $\lfloor {d-1 \over 2} \rfloor$ (respectively, $\lfloor {d'-1 \over 2} \rfloor$) qubits. And since $Y=iXZ$, a $Y$ error will be corrected as being an $X$ and a $Z$ error on the same qubit. Therefore, the CSS code just constructed is an $[[n,k+k'-n,\min(d,d')]]$ code. 

%--------------------------------------------------------------%
\subsection{Examples}
\label{sec:CodeExamples}

Our first example is a CSS code that is generated from the classical binary $C=[7,4,3]$ code which contains its dual code. The parity check matrix of $C$ is
\begin{equation}
\label{1.3.8}
 H \equiv \left( \begin{array}{ccccccc}
                                      0 & 0 & 0 & 1 & 1 & 1 & 1\\
                                      0 & 1 & 1 & 0 & 0 & 1 & 1\\
                                      1 & 0 & 1 & 0 & 1 & 0 & 1
                     \end{array}
                     \right) \, ,
\end{equation}

\noindent and its code space is spanned by the three rows of $H$ and the additional code word $(1,1,1,0,0,0,0)$. $C^\perp$ is the even subcode of $C$ and its code space is spanned by the rows of $H$. Therefore, $C^\perp \subseteq C$ and we may take $C=C'=[7,4,3]$ in the CSS construction. This results in Steane's $[[7,1,3]]$ code \cite{Steane96}  which encodes one logical qubit in a {\em block} of seven physical qubits and corrects any noise operation whose Kraus operators have support on at most one physical qubit---we will often simply say that the code corrects any single error. 

Our second example is Shor's code \cite{Shor95b}. This code is based on the classical repetition code, $C_Z^{(n)}$, which encodes one logical bit into a string of $n$ bits via the encoding $w\rightarrow w_L=(w,w,\dots,w)$. Its parity check matrix is
\begin{equation}
\label{1.3.9}
 H \equiv \left( \begin{array}{ccccccc}
                                      1 & 1 & 0 & 0 & 0 & \dots \\
                                      0 & 1 & 1 & 0 & 0 & \dots \\
                                      0 & 0 & 1 & 1 & 0 & \dots \\
                                      \vdots & \vdots & \vdots & \vdots & \vdots & 
                     \end{array}
                     \right) \, .
\end{equation}

\noindent Rephrasing $C_Z^{(n)}$ as a quantum stabilizer code, its check matrix is 
\begin{equation}
\label{1.3.10}
 G(C_Z^{(n)}) \equiv \left( \begin{array}{c|c}
                                0 & H      
                     \end{array}
                     \right) \, ,
\end{equation}

\noindent and it follows that it corrects $X$ errors on up to $\lfloor {n-1\over 2}\rfloor$ qubits. Its logical $X$ operator that exchanges $|0\rangle_L$ and $|1\rangle_L$ is $X_L=X\otimes X \otimes X \otimes \cdots$, i.e., an $X$, or {\sc not}, operator acting on all bits. Its logical $Z$ operator that multiplies the phase of $|1\rangle_L$ by $-1$ and leaves $|0\rangle_L$ unchanged is, e.g., $Z_L = Z \otimes I \otimes I \otimes \cdots$, which shows that the code cannot correct any $Z$ errors.

If we interchange the $X$ and $Z$ operators, we can consider the similar code, $C_X^{(n)}$, with parity check matrix 
\begin{equation}
\label{1.3.10.5}
 G(C_X^{(n)}) \equiv \left( \begin{array}{c|c}
                                H & 0      
                     \end{array}
                     \right) \, ,
\end{equation}

\noindent which corrects $Z$ errors on up to $\lfloor {n-1\over 2}\rfloor$ qubits. The logical $Z$ operator of this code is $Z\otimes Z \otimes Z \otimes \cdots$ and the logical $X$ operator is $X \otimes I \otimes I \otimes \cdots$, which implies that $C_X^{(n)}$ corrects no $X$ errors. 

Shor's code is the {\em concatenated} $C_X^{(n)} \circ C_Z^{(n)}$: We partition $n^2$ qubits into $n$ subblocks of $n$ qubits each. Within each subblock we use $C_Z^{(n)}$ to correct $X$ errors. The idea of {\em concatenation} is to further encode the $n$ logical qubits of the $n$ subblocks by using the ``level-2'' code $C_X^{(n)}$; this results in one level-2 logical qubit. Its logical $X$ operator is $X_L \otimes I \otimes \cdots$, where the tensor product structure denotes the $n$ different subblocks. Similarly, its logical $Z$ operator is $Z_L \otimes Z_L \otimes Z_L \otimes \cdots $, i.e., a $Z$ operator applied to the first qubit in each of the $n$ subblocks. Since the weight of these logical $X$ and $Z$ operators is $n$, Shor's code has distance $n$ and it can correct arbitrary errors on up to $\lfloor {n-1\over 2}\rfloor$ qubits; it is an $[[n^2,1,n]]$ code.

%--------------------------------------------------------------%

%--------------------------------------------------------------%

%-----------------------------------------%
\chapter{The Syntax of Fault-Tolerant Quantum Computation}
\label{ch:syntax}
%----------------------------------------------------------%
\section{Introduction}

In this chapter, I will describe quantum circuit designs that are used in implementing quantum computation in a way that is resilient to noise---we will say that such circuits are {\em fault-tolerant}. Similar to the case of fault-tolerant classical computation, the design of fault-tolerant quantum circuits is based on the idea of {\em simulating} each quantum operation in an ideal quantum circuit using operations encoded in an error-correcting code. %Furthermore, intermediate to the simulation of every two successive quantum operations, an error-correction step is inserted which attempts to correct errors that have occurred during the earlier encoded operation before the later encoded operation is executed. Encoded operations and error correction are implemented using composite objects, called {\em gadgets}, that consist of several noisy physical operations. The design of these gadgets must satisfy certain properties that ensure encoding improves the reliability of the fault-tolerant simulation relative to the un-encoded computation. 

The analysis of fault-tolerant quantum circuit simulations has many common features with fault-tolerant simulations of classical circuits as, e.g., analysed by Von Neumann \cite{Neumann55} and by G\'acs \cite{Gacs83,Gacs01}. However, there are two main features in the quantum version of this problem that have no classical analog. First, quantum error correction needs to deal with phase flip ($Z$) in addition to {\em classical} bit flip ($X$) errors, and it must also operate without gaining any information about the encoded information. And second, the fault-tolerant  implementation of a universal set of {\em encoded} (or {\em logical}) operations is significantly more challenging in the quantum than in the classical case and it requires the off-line preparation and verification of special ancillary quantum states known as {\em quantum software}.

For the fault-tolerant methods discussed in this thesis to be effective, %and for the proof of the quantum threshold theorem given in the next chapter, 
there are certain assumptions about the experimental setup and the structure of noise that are essential. The following proposition states them.

\begin{propo}
\label{propo:2}
The following assumptions are necessary for our fault-tolerant methods to be effective and for proving the quantum threshold theorem:
\begin{enumerate}
\item Faults that act collectively on many qubits are highly suppressed in probability or in amplitude depending on the noise model. %---this will be the case for the local noise model I will define in \S \ref{sec:LocalNoise}. 
Furthermore, the noise strength, $\varepsilon$, must be a sufficiently small constant that is {\em independent} of the size of the computation. % i.e., computing with more and more qubits and for longer and longer times must have no effect on the noise strength,
\item There is an inexhaustible supply of fresh ancillary qubits, or the ability to refresh and reuse ancillary qubits an indefinite number of times; otherwise, there will be a limit on the amount of entropy arising from noise that we can flush from the quantum computer \cite{Aharonov96b}.
\item It is possible to apply quantum gates in parallel on disjoint sets of qubits; otherwise we will be unable to fight against noise in memory that afflicts all parts of the computer simultaneously. 
\end{enumerate}
\end{propo}

This chapter is organized as follows: In \S \ref{sec:LocalNoise}, I define the local noise model that will be considered throughout this thesis. Then, in \S \ref{sec:GadgetProperties}, I will state the key properties fault-tolerant quantum circuits need to satisfy. These properties will be stated {\em syntactically} rather than semantically, i.e., as properties of the noisy circuits themselves and independent of the actual quantum state being processed by the quantum computer. This syntactic approach will prove to be extremely useful in proving the basic lemma of this chapter that relates the strength of local noise and the accuracy of the fault-tolerant simulation. We will find that for local noise, fault-tolerant simulations achieve greater accuracy than the unencoded computation provided the noise strength is below a certain critical value. In the next chapter, we will use this result to show that the accuracy of fault-tolerant simulations can be improved to {\em any} desired level. 

%--------------------------------------------------------------%
\section{Local Noise}
\label{sec:LocalNoise}

Consider a quantum circuit computation. Elementary quantum operations in this circuit consist of single-qubit preparations, quantum gates chosen from a finite universal gate set (including identity gates that correspond to storage of a qubit in memory), and single-qubit measurements. Let the term {\em location} denote any of these operations since each operation can be viewed as a point in the space-time quantum circuit diagram; i.e., a location is a label indicating which qubits interact via what quantum operation at what time step of the computation. Due to noise, elementary operations may deviate from their ideal realization resulting in what we may call {\em faults} during the computation.  

%As discussed in \S \ref{sec:QMechanics}, we can in general model noisy operations by unitary operators acting between our computer qubits and a {\em common} bath (where the bath system includes all physically relevant degrees of freedom that interact with the computer qubits and are not under our control). 
Let us start by recalling the non-Markovian noise model we discussed in the introduction. In this noise model, the actual implementation of every ideal gate is a unitary acting on the support of the ideal gate and a common ``bath'' system, where the difference of this unitary from the ideal gate has small sup norm. The noisy implementation of an ideal gate that applies $U_j \otimes I_B$ to the qubits on the support of $U_j$ and the bath is the {\em unitary} $\Delta_j (U_j \otimes I_B)$, where $\Delta_j=(I \otimes V_j) + F_j$ and the {\em fault} $F_j$ acts nontrivially on the support of $U_j$. In this situation, let us say that the noise has {\em strength} $\varepsilon$ if for all gates,
\begin{equation}
\label{2.2.1}
|| F_j ||_{\rm sup} \leq \varepsilon \; .
\end{equation}

A noisy measurement can be modeled as $\Delta_j$ followed by an ideal measurement, and a noisy preparation of a quantum state can be modeled as an ideal preparation followed by $\Delta_j$. %Similarly, a noisy two-qubit gate can be expressed as the ideal gate followed by an operator that can be expanded in the basis of two-qubit Pauli operators; the noise strength $\varepsilon$ is an upper bound on the norm of the sum of the operators in this expansion, excluding the operator $I \otimes I \otimes V_j$. 
For gates that are executed in parallel during the {\em same} time step, a convention is necessary to specify the time ordering of the, possibly non-commuting, operators $\{ \Delta_j \}$ acting on the system and the bath; however, the noise strength does not depend on how the time-ordering ambiguity is resolved.

We also recall the definition of the concept of {\em fault paths}: If some ideal quantum circuit applies the sequence of gates $U_1 \otimes I_B, U_2 \otimes I_B, \cdots, U_L \otimes I_B$, its noisy implementation will correspond to the actual unitary $\Delta_L(U_L \otimes I_B) \cdots \Delta_1(U_1 \otimes I_B)$ \footnote{Our operator-ordering convention is that operators on the right act first on the input state.}. By expanding the $\{ \Delta_j \}$, we obtain a sum of terms where, for each term, either $I  \otimes V_j$ or the fault operator, $F_j$, has been applied at the noisy location $j$. Each term in this sum is a fault path and it indicates which locations in the quantum circuit are implemented ideally and which locations are faulty. 

For each fault path, we may consider the corresponding operator acting on the bath. If different fault paths are labeled by perfectly distinguishable states of the bath, we expect that probabilities can be associated with faults. More specifically, if for all locations the operators $I\otimes V_j$ and $F_j$ map the bath to orthogonal states, then we expect that different fault paths will not interfere so that our noise model describes faults that are independent and stochastic (in essence, we assume that there exists a mechanism in the bath that effectively projects onto one of these perfectly distinguishable states, thus preventing the interference between different fault paths). In this case, the probability of a fault is at most $p=\varepsilon^2$---we may regard $F_j$ as the {\em amplitude} weighting a fault, whose norm square is a probability. If instead $\{ F_j\}$ decompose into a tensor-product form between the system and the bath, then our  noise model describes unitary faults which do not involve the bath. In general, even if faults involve the bath, it may not be the case that different faults are associated with perfectly distinguishable states of the bath. In that case we cannot associate a well-defined probability with faults at different locations and different fault paths will {\em interfere} as they can add coherently. Nevertheless, using the properties (\ref{1.2.17.2}), the sup norm of the {\em sum} of all fault paths with faults at some $r$ {\em specific} locations is at most $\varepsilon^r$ %(where, at all other locations, we have considered the full unitaries $\{ \Delta_j(U_j \otimes I_B) \, | \, j\not \in \mathcal{I}_r \}$) 
\cite{Terhal04}. This fact motivates the following definition. 

\begin{defi}[Local Noise] 
\label{def:4}
Consider a noisy quantum circuit realized as a unitary transformation acting on a system and a bath. We may express    this unitary transformation as a sum of fault paths where, for each fault path, each location is either faulty or ideal. Let ${\cal I}_r$ denote a set of $r$ specific locations, and let $F({\cal I}_r)$ denote the sum over all fault paths for which all locations in ${\cal I}_r$ are faulty (with the fault paths unrestricted outside of ${\cal I}_r$).  Then, noise is {\em local} with strength $\varepsilon$ if, for all $\mathcal{I}_r$, $|| F({\cal I}_r) ||_{\rm sup} \leq \varepsilon^r$. (If faults are stochastic, then noise is local with strength $p$ if, for all ${\cal I}_r$, the probability of $F({\cal I}_r)$ is at most $p^r$.)
\end{defi}

It should be emphasized that the noise model described in the beginning of this section is a particular example of  local noise---there exist other noise models that also satisfy the condition of definition \ref{def:4} as, e.g., the noise model in \cite{Aharonov05}. The defining characteristic of local noise is that there is an exponential penalty for faults that occur collectively at once on many qubits that do not interact in the ideal computation.\footnote{For example, if $\mathcal{I}_2$ labels two qubits that do not interact during some time step in the quantum computation, then $F(\mathcal{I}_2)\leq \varepsilon^2$. In contrast, if the same two qubits interact via a {\sc cnot} at the same time step, then they are both considered to be part of {\em one} location, $\mathcal{I}_1$, and $F(\mathcal{I}_1)\leq \varepsilon$. } %For instance, in the noise model described in the beginning of this section, at each time step, faults can act collectively only on those sets of qubits that interact during the execution of a quantum gate. 
However, faults at different locations are allowed to act on a common bath that can store information about faults for an indefinitely long time. Hence, local noise is in general {\em non-Markovian} and the presence of a common bath allows faults to be correlated both in time and in space. We will see that these correlations do not prevent fault-tolerant error correction from being effective, provided the noise strength is sufficiently weak.

From a microscopic perspective, the physics of local noise is most naturally formulated in terms of a Hamiltonian, $H$, that describes the joint evolution of the system and the bath. We may generally express $H$ as
\begin{equation}
\label{2.2.2}
H = H_S + H_B +H _{SB} \;,
\end{equation}

\noindent where $H_S$ is the time-dependent Hamiltonian of the system that generates the evolution of the ideal quantum computation, $H_B$ is an arbitrary Hamiltonian of the bath, and the time-dependent Hamiltonian $H_{SB}$ couples the system to the bath. In particular, the local noise model described in the beginning of this section can be generated by a coupling of the form
\begin{equation}
\label{2.2.3}
H_{SB} = \sum_j H_{SB,j}  \;,
\end{equation} 

\noindent where each term $H_{SB,j}$ describes how the set of qubits on the support of location $j$ interact with the bath; i.e., at any given time step, different qubits are {\em directly} coupled by $H_{SB}$ only if the ideal computation Hamiltonian, $H_S$, also couples them in the same time step. We emphasize that because the qubits at every location $j$ are coupled via $H_{SB,j}$ to a {\em common} bath, faults may be correlated both spatially and temporally. In particular, these correlations may be chosen adversarially since we have not imposed any locality condition on the Hamiltonian of the bath, $H_B$, that is completely arbitrary. %Also, we note that we can shift $H_{SB}$ by a constant in order to optimize $\varepsilon$.

We can express the time evolution governed by the Hamiltonian $H$ in terms of {\em time-resolved} fault paths \cite{Terhal04}: In order to study the evolution for total time $T$, we divide $T$ into $N$ time intervals each of width $\Delta t=T/N$, and take the limit $N\rightarrow \infty$ so that terms of order $(\Delta t)^2$ can be neglected. Then, the  evolution operator for the total time interval $(0,T)$ can be Lie-Trotter expanded as
\begin{equation}
\label{2.2.4}
U(0,T) = \lim_{N\rightarrow \infty} \prod_{i=1}^N U_S(\Delta t_i) U_{SB}(\Delta t_i) U_B(\Delta t_i) \;,
\end{equation}

\noindent where $U_S(\Delta t_i)$, $U_B(\Delta t_i)$ describe the time evolution during the $i$th interval in the system and the bath, respectively\footnote{There is an implicit time ordering in equation (\ref{2.2.4}) since the operators acting at different time intervals may not commute.}. By taking the coupling Hamiltonian, $H_{SB}$, as in equation (\ref{2.2.3}), $U_{SB}(\Delta t_i)$ for the $i$th interval can be expanded as
\begin{equation}
\label{2.2.5}
U_{SB}(\Delta t_i) \approx \prod_j \left( I_{SB} -i H_{SB,j} \Delta t \right) \; ,
\end{equation}

\noindent where $j$ runs over all locations in the time interval $(0,T)$ (i.e., if $(0,T)$ corresponds to one time step in the computation, $j$ runs over all gates that are executed during that time step). By opening the parentheses in equation (\ref{2.2.5}), we obtain a sum of fault paths: at each time interval labelled by $i$, either the identity, $I_{SB}$, or the {\em micro-fault}, $-i H_{SB,j} \Delta t$, is inserted at each location $j$. 

Suppose that the maximum time required for implementing an operation in the quantum circuit is $\tau$, so that each location can be viewed as a {\em coarse graining} of $N$ consecutive time intervals of width at most $\Delta t=\tau/N$. We also observe that a specific coarse-grained location, $j$, is faulty if there is a micro-fault at any one of the $N$ time intervals it contains. %In order words, for a coarse-grained location $j$ to be ideal, $U_{SB}$ for fixed $j$ in eq.$\,$(\ref{2.2.5}) has been replaced by the identity $I_{SB}$ in all $N$ time intervals in eq.$\,$(\ref{2.2.4}). For each time interval leballed by $i$, let $Gd_i = U_S(\Delta t_i) U_B(\Delta t_i)$ and $Bd_i = -i \Delta t_i U _S(\Delta t_i) H_{SB,a} U_B(\Delta t_i)$. 
Then, in the time-resolved fault path expansion at location $j$, the term $\prod_{i=1}^N U_S(\Delta t_i) U_B(\Delta t_i)$ corresponds to the ideal implementation, whereas the sum, $F_j$, of all other terms corresponds to a faulty implementation. 

Let us now calculate an upper bound on $||F_j ||_{\rm sup}$. In all terms in the sum $F_j$, a micro-fault is inserted in at least one of the $N$ time intervals. Therefore, we can rewrite $F_j$ as a sum of $N$ terms where a micro-fault is inserted in some time interval, each preceding time interval has been replaced by $I_{SB}$ and each succeeding time interval has been replaced by the unitary $U_{SB}(\Delta t_i)$. Thus, $||F_j ||_{\rm sup} \leq \lim_{N\rightarrow \infty} N ||  -i H_{SB,j} \Delta t ||_{\rm sup} = \tau || H_{SB,j} ||_{\rm sup}$. If we now suppose that at all noisy locations the system-bath coupling Hamiltonian satisfies
\begin{equation}
\label{2.2.6}
|| H_{SB,j} ||_{\rm sup} \leq \lambda_0 \; ,
\end{equation}

\noindent we can conclude that the norm of a fault acting at a specific coarse-grained location, $j$, satisfies $||F_j ||_{\rm sup} \leq \lambda_0 \tau$. This provides a justification in terms of a microscopic Hamiltonian model for the more phenomenological non-Markovian noise model we described earlier and the assumption (\ref{2.2.1}). Furthermore, if we specify $r$ coarse-grained locations, $\mathcal{I}_r$, and by letting $F(\mathcal{I}_r)$ denote the sum over all time-resolved fault paths with faults at those $r$ coarse-grained locations (with the time-resolved fault paths  unrestricted elsewhere), then we similarly obtain $|| F(\mathcal{I}_r) ||_{\rm sup} \leq \prod_{j\in \mathcal{I}_r} \tau || H_{SB,j} ||_{\rm sup} \leq  (\lambda_0 \tau)^r$ \cite{Terhal04}. %by organizing the sum in terms of the first of the $N$ intervals where a fault occurs within each coarse-grained location and by using eq.$\,$(\ref{1.2.17.2}),
Hence, the Hamiltonian noise model described by equations (\ref{2.2.2}) and (\ref{2.2.3}) and the condition (\ref{2.2.6}) is local with strength $\varepsilon \equiv \lambda_0 \tau$. 

%----------------------------------------------------------%
\section{Fault-Tolerant Quantum Circuit Simulations}
\label{sec:GadgetProperties}

In the fault-tolerant implementation of some ideal quantum computation, each operation will be {\em simulated} by an encoded operation that acts on the logical qubits of a quantum error-correcting code. Encoded operations are implemented by simulation procedures or, simply, {\em gadgets} that are composite objects and consist of elementary noisy physical operations. Apart from gadgets that implement encoded operations, there also exist gadgets that implement quantum error correction. In describing the operation of gadgets, we will speak of the number of {\em faults} that are {\em contained} in them; as defined previously, we say that a fault has occurred in a location inside a gadget when the actual noisy operation at that location deviates from the ideal operation. On the other hand, the result of faults is to cause {\em errors} in the logical state of the code blocks that are processed by the gadgets. 

Roughly speaking, gadgets are designed to prevent errors introduced by faults inside them from propagating to become many more errors within the same code block. To make this intuitive description precise, I will first define what is meant by {\em simulation} of some ideal operation and also what property makes a simulation {\em fault-tolerant}.  Then, I will describe certain key properties that gadgets realizing such simulations need to satisfy. However, I will postpone the discussion about how gadgets with these properties are constructed until chapter \ref{ch:lower-bounds}. 

For the discussion that follows some notation will be helpful: Elementary physical operations will be called {\em level-0 gadgets} or {\em 0-Ga}s. Our 0-Gas include the quantum gates of a finite quantum universal gate set, the  preparation of a single qubit in the $|0\rangle$ state or {\em 0-preparation}, and the measurement of a single qubit in the computation basis that records the classical outcome or {\em 0-measurement}. Although this set of 0-Gas is sufficient,  it will be convenient for our gadget constructions in chapter \ref{ch:lower-bounds} to add a 0-Ga that prepares a single qubit in the ${+}1$ eigenstate of $X$, $|+\rangle$, and another 0-Ga that measures a single qubit along the eigenbasis of $X$ and records the classical outcome. 

For each 0-Ga, there is a gadget implementing the corresponding {\em encoded} (or {\em logical}) operation; such gadgets will be called {\em level-1 gadgets} or {\em 1-Ga}s and they act on {\em level-1 code blocks} or {\em 1-blocks}. Gadgets that implement quantum error correction will be called {\em level-1 error-correction gadgets} or {\em 1-EC}s. Here, the term {\em level} is used to indicate whether coding has been used: physical operations are not encoded, whereas gadgets use the encoding of some quantum code.\footnote{In the next chapter, we will also define {\em level-2 gadgets}, etc., and the notation anticipates this fact.}

\begin{figure}[tbh]
\begin{center}
\setlength{\unitlength}{1pt}
\vspace{0.5cm}
\begin{picture}(160,54)
\put(0,19){\line(1,0){5}}
\put(0,35){\line(1,0){5}}
\put(5,13){\framebox(20,27){\footnotesize 0-Ga}}
\put(25,19){\line(1,0){5}}
\put(25,35){\line(1,0){5}}
\put(40,21){\makebox(30,12){$\Longrightarrow$}}
\put(78,12){\line(1,0){10}}
\put(78,42){\line(1,0){10}}
\put(88,0){\framebox(28,54){1-Ga}}
\put(116,12){\line(1,0){10}}
\put(116,42){\line(1,0){10}}
\put(126,0){\framebox(24,24){1-EC}}
\put(150,12){\line(1,0){10}}
\put(126,30){\framebox(24,24){1-EC}}
\put(150,42){\line(1,0){10}}
\end{picture}
\end{center}
\caption{\label{fig:2.1} A schematic of the replacement rules generating the fault-tolerant simulation applied to a two-qubit gate. Each 0-Ga in the ideal circuit is replaced by a 1-Rec which consists of the 1-Ga corresponding to the 0-Ga followed by a 1-EC acting on each 1-block output from the 1-Ga.}
\end{figure}
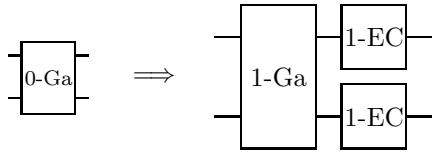

To obtain the fault-tolerant simulation, we replace each 0-Ga in the ideal circuit by the corresponding 1-Ga {\em followed} by a 1-EC in all output 1-blocks. The exception is measurement 0-Gas that have no quantum output and which are replaced by the corresponding 1-Gas alone. We call each 1-Ga grouped together with the 1-ECs succeeding it a {\em level-1 rectangle} or {\em 1-Rec} (measurement 1-Gas form 1-Recs by themselves). %the term {\em rectangle} is motivated by the fact that in the fault-tolerant simulation laid out as a space-time circuit diagram, 1-Gas grouped together with their succeeding 1-ECs cover a rectangular area. 
Figure \ref{fig:2.1} shows the replacement rule for a two-qubit 0-Ga.  

These replacement rules generate the fault-tolerant simulation of any ideal quantum circuit. To make precise the notion of simulation, it is helpful to introduce the notion of an {\em ideal decoder} or {\em i-decoder}. This device performs {\em ideal} (i.e., faultless) error correction on its input 1-block and subsequently decodes  the 1-block to a single qubit tracing over (i.e., discarding) all other output ancillary qubits. It should be emphasized that i-decoders are useful as mathematical {\em tools} in our analysis but do {\em not} correspond to any physical operation applied during the simulation\footnote{Although, certainly, noisy decoders can be constructed as physical devices.}. Let the 1-Rec that simulates some ideal 0-Ga be called {\em correct}; the following definition explains what is meant by such a simulation.    
\begin{defi}[Correctness]
\label{def:5} 
A 1-Rec is {\em correct} if the 1-Rec followed by i-decoders is equivalent to the i-decoders followed by the ideal 0-Ga that the 1-Rec simulates. For a single-qubit gate 0-Ga, correctness means schematically:
\end{defi}
\vspace{0.2cm}
\setlength{\unitlength}{1pt}
\begin{picture}(292,24)
\put(0,12){\line(1,0){10}}
\put(10,0){\framebox(48,24){\shortstack{correct\\1-Rec}}}
\put(58,12){\line(1,0){10}}
\put(68,0){\framebox(48,24){\shortstack{i-decoder}}}
\put(116,12){\line(1,0){10}}
\put(126,6){\makebox(20,12){=}}
\put(146,12){\line(1,0){10}}
\put(156,0){\framebox(48,24){\shortstack{i-decoder}}}
\put(204,12){\line(1,0){10}}
\put(214,0){\framebox(48,24){\shortstack{ideal\\$0$-Ga}}}
\put(262,12){\line(1,0){10}}
\put(272,6){\makebox(20,12){.}}
\end{picture}
\vspace{0.5cm}

\noindent In order words, suppose that an ideal 0-Ga applies the unitary transformation $U$ to $r$ qubits. If the input to the corresponding 1-Rec is $r$ 1-blocks in, say, the state $|\psi\rangle_L$, applying i-decoders to each of the $r$ 1-blocks maps $|\psi\rangle_L$ to $|\psi\rangle$. Then, if the 1-Rec is correct, the i-decoders following it map the output of the 1-Rec to $U|\psi\rangle$. In this sense, correctness captures the idea of what it means for the 1-Rec to simulate the corresponding ideal 0-Ga. 

Definition \ref{def:5} applies to 1-Recs simulating gate 0-Gas and requires a minor adaptation for single-qubit preparation or measurement 1-Recs. For preparation 1-Recs, correctness is meant as the property that a preparation 1-Rec followed by an i-decoder is equivalent to the ideal preparation 0-Ga---a correct preparation 1-Rec {\em annihilates} an i-decoder, or schematically:

\vspace{0.5cm}
\setlength{\unitlength}{1pt}
\begin{picture}(272,24)
\put(0,0){\framebox(58,24){\shortstack{correct\\prep. 1-Rec}}}
\put(58,12){\line(1,0){10}}
\put(68,0){\framebox(48,24){\shortstack{i-decoder}}}
\put(116,12){\line(1,0){10}}
\put(126,6){\makebox(20,12){=}}

\put(146,0){\framebox(48,24){\shortstack{ideal\\$0$-prep.}}}
\put(194,12){\line(1,0){10}}
\put(204,6){\makebox(20,12){.}}
\end{picture}
\vspace{0.3cm}

\noindent For measurement 1-Recs (which consist of the measurement 1-Gas alone), correctness is meant as the property that a measurement 1-Rec is equivalent to an i-decoder followed by the ideal measurement 0-Ga---a correct measurement 1-Ga {\em creates} an i-decoder, or schematically:

\vspace{0.5cm}
\setlength{\unitlength}{1pt}
\begin{picture}(272,24)
\put(0,12){\line(1,0){10}}
\put(10,0){\framebox(58,24){\shortstack{correct\\meas. 1-Rec}}}
\put(68,6){\makebox(20,12){=}}
\put(88,12){\line(1,0){10}}
\put(98,0){\framebox(48,24){\shortstack{i-decoder}}}
\put(146,12){\line(1,0){10}}
\put(156,0){\framebox(48,24){\shortstack{ideal\\$0$-meas.}}}
\put(202,6){\makebox(20,12){.}}
\end{picture}
\vspace{0.3cm}

Now that we have translated the idea of gate simulation to the notion of correctness, we are ready to discuss what it means for a simulation to be fault-tolerant. Intuitively, a simulation is fault-tolerant if it is robust against faults. More concretely, if we use a quantum error-correcting code that corrects arbitrary Pauli errors on up to $t$ qubits in a 1-block, we expect that a noisy simulation will be robust against faults that act nontrivially on no more than $t$ locations. 

\begin{figure}[tbh]
\begin{center}
\setlength{\unitlength}{1pt}
\begin{picture}(201,96)
\put(5,12){\line(1,0){5}}
\put(10,0){\framebox(24,24){1-EC}}
\put(34,12){\line(1,0){14}}
\put(5,44){\line(1,0){5}}
\put(10,32){\framebox(24,24){1-EC}}
\put(34,42){\line(1,0){14}}
\put(48,0){\framebox(28,56){1-Ga}}
\put(76,12){\line(1,0){14}}
\put(76,44){\line(1,0){5}}
\put(85,44){\line(1,0){5}}
\put(90,0){\framebox(24,24){1-EC}}
\put(114,12){\line(1,0){5}}
\put(90,32){\framebox(24,24){1-EC}}
\put(114,44){\line(1,0){5}}
\put(123,44){\line(1,0){5}}
\put(85,76){\line(1,0){5}}
\put(90,64){\framebox(24,24){1-EC}}
\put(114,76){\line(1,0){14}}
\put(128,32){\framebox(28,56){1-Ga}}
\put(156,44){\line(1,0){14}}
\put(156,76){\line(1,0){14}}
\put(170,32){\framebox(24,24){1-EC}}
\put(194,44){\line(1,0){5}}
\put(170,64){\framebox(24,24){1-EC}}
\put(194,76){\line(1,0){5}}
\put(3,-4){\dashbox(118,64){}}
\put(83,28){\dashbox(118,64){}}

\end{picture}
\end{center}
\caption{\label{fig:2.2} The two 1-exRecs, indicated by dashed lines, that simulate two succeeding gate 0-Gas are overlapping since they share a 1-EC which is a trailing 1-EC of the earlier 1-exRec and a leading 1-EC of the later 1-exRec.}
\end{figure}
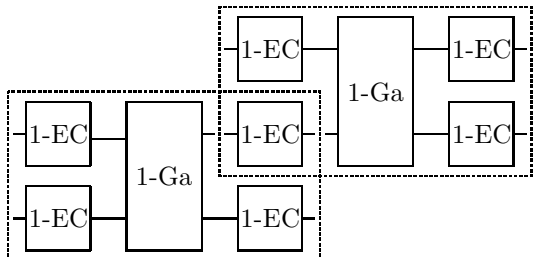

To make this requirement even more precise, we will need to consider {\em level-1 extended rectangles} or {\em 1-exRec}s that are formed by combining each gate 1-Ga with the 1-ECs {\em both} preceding and following it. In the case of preparation 1-Gas, the 1-exRec is the 1-Ga grouped with the 1-ECs following it since there is no quantum input; for measurement 1-Gas, the 1-exRec is the 1-Ga grouped with the 1-ECs preceding it since there is no quantum output. According to this definition, we observe that the 1-exRecs that simulate two successive 0-Gas have at least one 1-EC in common; we will say that these 1-exRecs {\em overlap}. For instance, figure \ref{fig:2.2} shows the 1-exRecs corresponding to the fault-tolerant simulation of two successive two-qubit gate 0-Gas.

The intuitive reason why we need to consider 1-exRecs instead of 1-Recs when considering fault tolerance is the following: Whether a 1-Rec is correct or not depends not only on the faults that it contains but also on the state of its input 1-blocks. In turn, the state of these input 1-blocks depends on the faults inside the 1-ECs preceding the 1-Rec. Although one could imagine that other faults in the past of the computation are also important, we will show that this is not the case: The correctness of a 1-Rec only depends on how many faults have occurred inside the 1-exRec that contains it. We will say that a 1-exRec which does not contain ``too many'' faults is {\em good}; the following definition explains how many is too many.
\begin{defi}[Goodness]
\label{defi:6}
Consider a fault-tolerant quantum circuit simulation based on a distance-$d$ quantum error-correcting code, and let $t=\lfloor {d-1\over 2}\rfloor$. A 1-exRec is {\em good} if it contains at most $t$ faults; if it is not good it is {\em bad}.
\end{defi}

The main result in this chapter is that the goodness of a 1-exRec implies the correctness of the 1-Rec it contains. The proof of this result relies on the existence of explicit constructions of gadgets (i.e., 1-Gas and 1-ECs) with suitable properties. To state these properties, let us introduce the notion of an {\em ideal $s$-filter} or simply {\em  $s$-filter}, where $s$ is an integer between zero and $n$, the number of physical qubits in the 1-block. An $s$-filter is a device that performs an {\em ideal} (i.e., faultless) orthogonal projection onto the subspace spanned by syndromes corresponding to Pauli errors acting on at most $s$ qubits in the 1-block; i.e., it leaves the 1-block {\em unchanged} if a {\em hypothetical} ideal syndrome measurement would {\em always} indicate a recovery Pauli operator of weight no more than $s$---it does {\em not} perform syndrome {\em measurement} and does {\em not} apply the recovery operator---and rejects the 1-block (and the whole computation) otherwise. If a 1-block is not rejected by an $s$-filter, we say that the 1-block {\em passes through} the $s$-filter. Just as i-decoders, $s$-filters are {\em tools} in our analysis and do not correspond to any physical operation applied during the fault-tolerant simulation. We will use $s$-filters as {\em tags} we can insert in our quantum circuit diagrams to indicate that the syndrome at some given point would indicate no more than $s$ Pauli errors {\em assuming} faultless syndrome measurement were performed at that point; we will be careful to insert these tags so that they act as the identity on the actual computation.  

For brevity, let a 1-Ga or 1-EC that contains no more than $r$ faults be called {\em $r$-good}. The key gadget properties starting with the properties for 1-EC gadgets are as follows. 
\begin{props}[1-EC Properties]
\label{prop:1} Let $t=\lfloor {d-1\over 2}\rfloor$. 1-EC gadgets for a distance-$d$ quantum error-correcting code satisfy:
\end{props}
\vspace{-0.2cm}
\noindent {\it (a) $\forall r\leq t$, the output of an {\em $r$-good} 1-EC passes through an $r$-filter, or}

\vspace{0.5cm}
\setlength{\unitlength}{0.91pt}
\begin{picture}(368,24)
\put(0,12){\line(1,0){10}}
\put(10,0){\framebox(48,26){\shortstack{$r$-good\\1-EC}}}
\put(58,12){\line(1,0){10}}
\put(68,6){\makebox(20,12){=}}
\put(88,12){\line(1,0){10}}
\put(98,0){\framebox(48,26){\shortstack{$r$-good\\1-EC}}}
\put(146,12){\line(1,0){10}}
\put(156,0){\framebox(48,26){$r$-filter}}
\put(204,12){\line(1,0){10}}
%\put(205,0){\makebox(68,26){$(r \le t)$}}
\put(223,12){,}
\end{picture} 
\vspace{0.2cm}

\noindent {\it (b) $\forall s,r$ such that $s+r\leq t$, if a 1-block which passes through an $s$-filter is input to an $r$-good 1-EC followed by an i-decoder, the output is the same as with the $r$-good 1-EC omitted, or }

\vspace{0.5cm}
\setlength{\unitlength}{0.91pt} 
\begin{picture}(436,24)
\put(0,12){\line(1,0){10}}
\put(10,0){\framebox(48,26){$s$-filter}}
\put(58,12){\line(1,0){10}}
\put(68,0){\framebox(48,26){\shortstack{$r$-good\\1-EC}}}
\put(116,12){\line(1,0){10}}
\put(126,0){\framebox(48,26){\shortstack{i-decoder}}}
\put(174,12){\line(1,0){10}}

\put(184,6){\makebox(20,12){=}}
\put(204,12){\line(1,0){10}}
\put(214,0){\framebox(48,26){$s$-filter}}
\put(262,12){\line(1,0){10}}
\put(272,0){\framebox(48,26){\shortstack{i-decoder}}}
\put(320,12){\line(1,0){10}}
%
%\put(330,0){\makebox(68,26){$(r+s\le t)$}}
\put(340,12){.}
\end{picture}
\vspace{0.5cm}

Property \ref{prop:1}(a) is the statement that on an {\em arbitrary} input, a 1-EC which contains at most $r$ faults produces an output on which ideal syndrome measurement would find at most $r$ Pauli errors. Property \ref{prop:1}(b) states that if the sum of the Pauli errors in the input 1-block and the faults in the 1-EC is no greater than $t$, then ideal decoding of the output gives the same output as if the 1-EC were not present (or, since the i-decoder performs ideal error correction itself, if the 1-EC were present but contained no faults). Both properties reflect the requirement that $r\leq t$ faults inside a 1-EC gadget must not lead to errors which propagate to more than $r$ qubits in the 1-block. 

The properties for 1-Ga gadgets are:
\begin{props}[1-Ga Properties]
\label{prop:2} Let $t=\lfloor {d-1\over 2}\rfloor$. Consider $k$ 1-blocks which respectively pass through $\{ s_1, s_2, \cdots, s_k\}$-filters and let them be input to an $r$-good 1-Ga gadget for a distance-$d$ quantum  error-correcting code simulating a $k$-qubit gate, where $s\equiv \sum_{i=1}^k s_i + r \leq t$. Then,
\end{props}
\vspace{-0.2cm}
\noindent {\it (a) all output 1-blocks pass through $s$-filters, or }

\vspace{0.5cm}
\setlength{\unitlength}{0.91pt}
\begin{picture}(456,24)
\put(0,12){\line(1,0){10}}
\put(10,0){\framebox(58,26){$\{s_i\}$-filters}}
\put(68,12){\line(1,0){10}}
\put(78,0){\framebox(48,26){\shortstack{$r$-good\\1-Ga}}}
\put(126,12){\line(1,0){10}}
\put(136,6){\makebox(20,12){=}}
\put(156,12){\line(1,0){10}}
\put(166,0){\framebox(58,26){$\{s_i\}$-filters}}
\put(224,12){\line(1,0){10}}
\put(234,0){\framebox(48,26){\shortstack{$r$-good\\1-Ga}}}
\put(282,12){\line(1,0){10}}
\put(292,0){\framebox(48,26){$s$-filters}}
\put(340,12){\line(1,0){10}}
%
%\put(340,0){\makebox(90,26){$(s=r+\sum_i s_i \le t)$}}
\put(360,12){,}
\end{picture}
\vspace{0.2cm}

\noindent {\it (b) the $r$-good 1-Ga followed by i-decoders is equivalent to applying the i-decoders first followed by the {\em ideal} $k$-qubit 0-Ga, or }

\vspace{0.5cm}
\setlength{\unitlength}{0.91pt} 
\begin{picture}(432,24)
\put(0,12){\line(1,0){10}}
\put(10,0){\framebox(58,26){$\{s_i\}$-filters}}
\put(68,12){\line(1,0){10}}
\put(78,0){\framebox(48,26){\shortstack{$r$-good\\1-Ga}}}
\put(126,12){\line(1,0){10}}
\put(136,0){\framebox(58,26){\shortstack{i-decoders}}}
\put(194,12){\line(1,0){10}}

\put(204,6){\makebox(20,12){=}}
\put(224,12){\line(1,0){10}}
\put(234,0){\framebox(58,26){$\{s_i\}$-filters}}
\put(292,12){\line(1,0){10}}
\put(302,0){\framebox(58,26){\shortstack{i-decoders}}}
\put(360,12){\line(1,0){10}}

\put(370,0){\framebox(48,26){\shortstack{ideal\\$0$-Ga}}}
\put(418,12){\line(1,0){10}}
%
%\put(398,0){\makebox(68,26){$(r+\sum_i s_i\le t)$}}
\put(438,12){.}
\end{picture}
\vspace{0.5cm}

For the properties above, the $s_i$-filter is applied to the $i$th input 1-block and, for property \ref{prop:2}(a),  each one of the $k$ $s$-filters on the right is applied to one of the $k$ output 1-blocks. 

Property \ref{prop:2} requires a minor adaptation for preparation and measurement 1-Gas. For 1-preparations and $\forall r\leq t$, property \ref{prop:2}(a) becomes:

\vspace{0.5cm}
\setlength{\unitlength}{0.91pt}
\begin{picture}(194,24)

\put(0,0){\framebox(48,24){\shortstack{$r$-good\\1-prep.}}}
\put(48,12){\line(1,0){10}}
\put(58,6){\makebox(20,12){=}}

\put(78,0){\framebox(48,24){\shortstack{$r$-good\\1-prep.}}}
\put(126,12){\line(1,0){10}}
\put(136,0){\framebox(48,24){$r$-filter}}
\put(184,12){\line(1,0){10}}
\put(205,12){,}
\end{picture}
\vspace{0.3cm}

\noindent
and property \ref{prop:2}(b) becomes:

\vspace{0.5cm}
\setlength{\unitlength}{0.91pt}
\begin{picture}(268,24)

\put(0,0){\framebox(48,24){\shortstack{$r$-good\\1-prep.}}}
\put(48,12){\line(1,0){10}}
\put(58,0){\framebox(48,24){\shortstack{i-decoder}}}
\put(106,12){\line(1,0){10}}

\put(116,6){\makebox(20,12){=}}
\put(136,0){\framebox(48,24){\shortstack{ideal\\$0$-prep.}}}
\put(184,12){\line(1,0){10}}
\put(205,12){.}
\end{picture}
\vspace{0.3cm}

\noindent
For 1-measurements, which have classical bits as outputs, there is no analog of property \ref{prop:2}(a) and, $\forall r,s$ such that $r+s\le t$, property \ref{prop:2}(b) becomes:

\vspace{0.5cm}
\setlength{\unitlength}{0.91pt}
\begin{picture}(378,24)
\put(0,12){\line(1,0){10}}
\put(10,0){\framebox(48,24){$s$-filter}}
\put(58,12){\line(1,0){10}}
\put(68,0){\framebox(48,24){\shortstack{$r$-good\\1-meas.}}}

\put(116,6){\makebox(20,12){=}}
\put(136,12){\line(1,0){10}}
\put(146,0){\framebox(48,24){$s$-filter}}
\put(194,12){\line(1,0){10}}
\put(204,0){\framebox(48,24){\shortstack{i-decoder}}}
\put(252,12){\line(1,0){10}}

\put(262,0){\framebox(48,24){\shortstack{ideal\\$0$-meas.}}}

\put(320,12){,}
\end{picture}
\vspace{0.3cm}

\noindent where the classical processing of the measurement outcomes is either performed in encoded form using a classical error-correcting code or, if the outcomes are decoded to the level of single classical bits, the classical gates that perform the decoding are faultless. Otherwise, a single fault in the final classical decoding step (e.g., a fault in our {\em reading} of the final outcome from the display of a detector) could cause an error in the outcome and the property is not satisfied.

It is now time to state and prove the basic lemma of this chapter.

\begin{lem}[Goodness Implies Correctness] 
\label{lem:1}
Consider a fault-tolerant quantum circuit simulation based on a distance-$d$ quantum error-correcting code that is executed using gadgets satisfying properties \ref{prop:1} and \ref{prop:2}. Then, the 1-Rec contained in a good 1-exRec is correct. For a good 1-exRec corresponding to a single-qubit gate 0-Ga, the lemma means schematically: 
\end{lem}
\vspace{0.2cm}
\setlength{\unitlength}{0.91pt} \hspace{-0.4cm}
\begin{picture}(446,24)
\put(0,12){\line(1,0){10}}
\put(10,0){\framebox(48,24){1-EC}}
\put(58,12){\line(1,0){10}}
\put(68,0){\framebox(48,24){1-Ga}}
\put(116,12){\line(1,0){10}}
\put(126,0){\framebox(48,24){1-EC}}
\put(174,12){\line(1,0){10}}
\put(184,0){\framebox(48,24){\shortstack{i-decoder}}}
\put(232,12){\line(1,0){10}}

\put(242,6){\makebox(20,12){=}}
\put(262,12){\line(1,0){10}}
\put(272,0){\framebox(48,24){1-EC}}
\put(320,12){\line(1,0){10}}
\put(330,0){\framebox(48,24){\shortstack{i-decoder}}}
\put(378,12){\line(1,0){10}}
\put(388,0){\framebox(48,24){\shortstack{ideal\\$0$-Ga}}}
\put(436,12){\line(1,0){10}}
\put(446,6){\makebox(20,12){.}}
\end{picture}
\vspace{0.5cm}

\noindent {\bf Proof}. We will only give the proof for single-qubit gate 0-Gas---the argument is essentially the same for multi-qubit gate, preparation or measurement 0-Gas. By definition, a 1-exRec is good if it contains no more than $t=\lfloor {d-1\over 2} \rfloor$ faults. Suppose there are $s$ faults in the leading 1-EC, $r$ faults in the 1-Ga, and $s'$ faults in the trailing 1-EC, with $s + r + s'\leq t$. Then, we can write the left-hand side of our equation schematically as:

\vspace{0.5cm}
\setlength{\unitlength}{0.91pt}
\begin{picture}(242,24)
\put(0,12){\line(1,0){10}}
\put(10,0){\framebox(48,24){\shortstack{$s$-good\\$1$-EC}}}
\put(58,12){\line(1,0){10}}
\put(68,0){\framebox(48,24){\shortstack{$r$-good\\$1$-Ga}}}
\put(116,12){\line(1,0){10}}
\put(126,0){\framebox(48,24){\shortstack{$s'$-good\\$1$-EC}}}
\put(174,12){\line(1,0){10}}
\put(184,0){\framebox(48,24){\shortstack{i-decoder}}}
\put(232,12){\line(1,0){10}}
\put(242,6){\makebox(20,12){.}}
\end{picture}
\vspace{0.3cm}

\noindent Using property \ref{prop:1}(a), we can insert an $s$-filter after the leading 1-EC and then, using property \ref{prop:2}(a), we can insert an $(s{+}r)$-filter after the 1-Ga to obtain the equivalent circuit:

\vspace{0.5cm}
\setlength{\unitlength}{0.91pt}
\begin{picture}(378,24)
\put(0,6){\makebox(20,12){=}}
\put(20,12){\line(1,0){10}}
\put(30,0){\framebox(48,24){\shortstack{$s$-good\\$1$-EC}}}
\put(78,12){\line(1,0){10}}
\put(88,0){\framebox(48,24){$s$-filter}}
\put(136,12){\line(1,0){10}}
\put(146,0){\framebox(48,24){\shortstack{$r$-good\\$1$-Ga}}}
\put(194,12){\line(1,0){10}}
\put(204,0){\framebox(48,24){\shortstack{$(s{+}r)$-\\filter}}}
\put(252,12){\line(1,0){10}}
\put(262,0){\framebox(48,24){\shortstack{$s'$-good\\$1$-EC}}}
\put(310,12){\line(1,0){10}}
\put(320,0){\framebox(48,24){\shortstack{i-decoder}}}
\put(368,12){\line(1,0){10}}
\put(378,6){\makebox(20,12){.}}
\end{picture}
\vspace{0.3cm}

\noindent
Using property \ref{prop:1}(b), we can now omit the trailing 1-EC and then, using property \ref{prop:2}(a) backwards, we can omit the $(s{+}r)$-filter that follows the 1-Ga to obtain the equivalent circuit:

\vspace{0.5cm}
\setlength{\unitlength}{0.91pt}
\begin{picture}(362,24)
\put(0,6){\makebox(20,12){=}}
\put(20,12){\line(1,0){10}}
\put(30,0){\framebox(48,24){\shortstack{$s$-good\\$1$-EC}}}
\put(78,12){\line(1,0){10}}
\put(88,0){\framebox(48,24){$s$-filter}}
\put(136,12){\line(1,0){10}}
\put(146,0){\framebox(48,24){\shortstack{$r$-good\\$1$-Ga}}}
\put(194,12){\line(1,0){10}}
\put(204,0){\framebox(48,24){\shortstack{i-decoder}}}
\put(252,12){\line(1,0){10}}
\put(262,6){\makebox(20,12){.}}
\end{picture}
\vspace{0.3cm}

\noindent Finally, using property \ref{prop:2}(b), we can move the i-decoder to the left thereby converting the 1-Ga to the ideal 0-Ga it simulates and then, using property \ref{prop:1}(a) backwards, we can remove the $s$-filter that follows the leading 1-EC. We thus obtain the equivalent circuit:

\vspace{0.5cm}
\setlength{\unitlength}{0.91pt}
\begin{picture}(204,24)
\put(0,6){\makebox(20,12){=}}
\put(20,12){\line(1,0){10}}
\put(30,0){\framebox(48,24){\shortstack{$s$-good\\$1$-EC}}}
\put(78,12){\line(1,0){10}}
\put(88,0){\framebox(48,24){\shortstack{i-decoder}}}
\put(136,12){\line(1,0){10}}
\put(146,0){\framebox(48,24){\shortstack{ideal\\$0$-Ga}}}
\put(194,12){\line(1,0){10}}
\put(204,6){\makebox(20,12){.}}
\end{picture}
\vspace{0.3cm}

\noindent This proves the lemma. 

\rightline{$\square$}

%----------------------------------------------------------%
\section{Accuracy}

Lemma \ref{lem:1} implies that if all 1-exRecs are good, the fault-tolerant simulation will produce exactly the same outcome probability distribution as the ideal simulated circuit: If all measurement 1-exRecs are good, then we can replace the 1-Recs contained in them by equivalent circuits of i-decoders followed by the ideal simulated measurement 0-Gas. Next, if all gate 1-exRecs are good, we can propagate these i-decoders to the left thereby transforming all gate 1-Recs to equivalent circuits applying the ideal simulated gate 0-Gas. Finally, if all preparation 1-exRecs are good, we can annihilate the i-decoders thereby obtaining an equivalent circuit with ideal preparation 0-Gas. Overall, if all 1-exRecs are good, the fault-tolerant simulation is equivalent to the ideal simulated circuit. Put simply, what happens is that the goodness of all 1-exRecs constrains the faults inside gadgets to be such that they never cause errors on more than $t$ qubits in any 1-block; hence, errors are always successfully corrected and the fault-tolerant simulation as a whole is successful. 

%------------------------------------------------%
\medskip \medskip \noindent {\bf Stochastic Local Noise} \medskip

Suppose that we simulate an ideal circuit of size $L$, and suppose noise is local and stochastic with strength $p$. For a  fault-tolerant simulation based on a distance-$d$ error-correcting code to be unsuccessful, at least $t+1$ faults must have occurred inside one of the $L$ 1-exRecs where $t=\lfloor {d-1\over 2 }\rfloor$. Since the probability of all fault paths with at least $t+1$ faults in {\em specified} locations is at most $p^{t+1}$, if we let $C$ denote the number of locations in the largest 1-exRec, the probability that a specified 1-exRec contains at least $t+1$ faults is at most ${C\choose t+1} p^{t+1}$. This implies that the error, $\delta$, of the fault-tolerant simulation can be upper bounded as in equation (\ref{1.2.15}) by substituting $p \rightarrow {C\choose t+1} p^{t+1}$,   
\begin{equation}
\label{2.2.7}
\delta \leq 2 L {C\choose t+1} p^{t+1} \;.
\end{equation}

\noindent Hence, the fault-tolerant simulation has an improved accuracy compared to the unencoded computation  provided ${C\choose t+1} p^{t+1} < p$, or $p < p_{\rm crit}$ where 
\begin{equation}
\label{2.2.8}
p_{\rm crit} \equiv {C\choose t+1}^{-1/t} \;.
\end{equation}

%------------------------------------------------%
\medskip \medskip \noindent {\bf Local Noise} \medskip

We can arrive at a similar conclusion if we consider local noise that is not stochastic. In this case, we consider the fault-tolerant simulation as a noisy quantum circuit represented by a unitary acting between the system qubits and a common bath. By performing a Lie-Trotter expansion of the Hamiltonian, $H_{SB}$, that couples the system and the bath, we can express the evolution at some specific coarse-grained location and the bath as a sum of an operator that acts as the identity on the coarse-grained location {\em plus} a fault operator acting nontrivially on the coarse-grained location. Repeating for all coarse-grained locations, we obtain a fault-path expansion of the entire fault-tolerant simulation. 

We can now group fault paths in two terms: The first term, $Gd^{(1)}$, is a sum over all {\em good} fault paths; a fault path is good if it applies a fault operator in no more than $t$ coarse-grained locations within each of the $L$ 1-exRecs. Lemma \ref{lem:1} implies that all good fault paths will give identical outcome statistics as the ideal computation; hence, their sum $Gd^{(1)}$ will also give the ideal outcome statistics. The second term, $Bd^{(1)}$, is a sum over all remaining fault paths which we will call {\em bad}. A fault path is bad if it applies a fault in at least $t+1$ coarse-grained locations inside at least one of the $L$ 1-exRecs. The two terms $Gd^{(1)}$ and $Bd^{(1)}$ play an analogous role to the $Gd$ and $Bd$ operators in \S \ref{sec:Accuracy}; specifically, the error, $\delta$, of the fault-tolerant simulation can be upper bounded by the sup norm of the sum over all bad fault paths, $\delta \leq 2||Bd^{(1)}||_{\rm sup}$.

In the special case when local noise is due to a noise process that satisfies condition (\ref{2.2.1}), upper bounding $||Bd^{(1)}||_{\rm sup}$ is straightforward: We order the $L$ 1-exRecs in some arbitrary way and we also consider some ordering of the coarse-grained locations inside each 1-exRec. We now organize $Bd^{(1)}$ as a sum of $L$ terms where for the $i$th term the $i$th 1-exRec is bad, all 1-exRecs preceding it are good and we put no restriction on the 1-exRecs following it. Then, we express each of these $L$ terms as a sum of terms where for each term a different set, $\mathcal{I}_{t+1}$, of $t+1$ faulty coarse-grained locations is identified inside the bad 1-exRec, the ideal gates are applied in all coarse-grained locations that precede the latest of the $t+1$ faults and are not in $\mathcal{I}_{t+1}$ and the {\em unitary} coupling described by $H_{SB}$ is applied between the bath and all coarse-grained locations that are not directly coupled by $H_{SB}$ to any of the $t+1$ faults and are subsequent to the latest of them. There are at most ${C\choose t+1}$ terms of this form and each has sup norm at most $\varepsilon^{t+1}$. Thus, we conclude that  
\begin{equation}
\label{2.2.9}
||Bd^{(1)}||_{\rm sup} \leq L {C\choose t+1} \varepsilon^{t+1} \;.
\end{equation}

\noindent Therefore, the error of the fault-tolerant simulation is smaller than the error, $2L\varepsilon$, of the unencoded computation provided $\varepsilon < \varepsilon_{\rm crit}$ where

\begin{equation}
\label{2.2.9.2}
\varepsilon_{\rm crit} \equiv {C\choose t+1}^{-1/t} \;.
\end{equation}

To obtain an upper bound on $Bd^{(1)}$ for general local noise is more challenging. The essential difference is that now condition (\ref{2.2.1}) does not necessarily hold for every coarse-grained location irrespective of whether other coarse-grained locations are faulty or ideal. Instead, definition \ref{def:4} only allows us to upper bound the sup norm of the {\em sum} of all fault paths with faults at some $r$ specific coarse-grained locations\footnote{Note that this summation corresponds to considering the unitary coupling described by $H_{SB}$ between the bath and all other coarse-grained locations not coupled directly with the $r$ specified faulty ones.}. In particular, the method we used in the previous paragraph to organize $Bd^{(1)}$ as a sum of at most $L{C\choose t+1}$ terms does not immediately lead to the desired bound: The problem is that in each such term we have imposed a restriction on some locations outside the set $\mathcal{I}_{t+1}$; namely that the {\em ideal} gates are applied in all coarse-grained locations that precede the latest of the $t+1$ faults. Hence, each such term is {\em not} a sum over {\em all} fault paths with faults at the locations in $\mathcal{I}_{t+1}$ (i.e., some fault paths are excluded) and, so, we cannot directly upper bound its sup norm by $\varepsilon^{t+1}$. 

The solution to this problem is based on the idea that we can perform our Lie-Trotter expansion ``in reverse,'' i.e., we can reexpress the ideal realization of a given coarse-grained location as the unitary coupling described by $H_{SB}$ {\em minus} a fault operator that acts on this coarse-grained location. Our motivation for doing this is that in this way we can eliminate the restriction that we imposed above on some locations outside the set $\mathcal{I}_{t+1}$ to be ideal. In the following lemma we develop this idea and perform the combinatorics.
 
\begin{lem}[Fault Path Inclusion-Exclusion] 
\label{lem:2}

Consider a noisy quantum circuit with $C$ locations which is subject to {\em local} noise with strength $\varepsilon$. We express the unitary transformation acting on the system and the bath as a sum of fault paths where, for each fault path, each location is either faulty or ideal, and let $F(\mathcal{I}_{r\geq s})$ denote the sum over all fault paths with at least $s$ faults. Then, 
\begin{equation}
\label{2.2.12}
||F(\mathcal{I}_{r\geq s})||_{\rm sup} \leq {C \choose s}\varepsilon^s e^{(C-s)\varepsilon} \;.
\end{equation}
\end{lem}

\noindent {\bf Proof}. Let us choose some arbitrary ordering of the $C$ coarse-grained locations that does not need to correspond to the actual time ordering of the operations in the quantum circuit---speaking of {\em earlier} or {\em later} locations should be understood as referring to this ordering. Then, $F(\mathcal{I}_{r\geq s})$ can be expressed as a sum of at most ${C \choose s}$ terms, where each term corresponds to a sum over all fault paths such that a particular set, $\mathcal{I}_s$, of $s$ faulty coarse-grained locations is identified and the ideal gates are applied in all coarse-grained locations that precede the latest of the $s$ specified faults and are not in $\mathcal{I}_s$ (i.e., the different fault paths in the sum act identically in the $s$ specified locations and the locations preceding them and may differ on the coarse-grained locations that succeed the latest of the $s$ specified faults). Let us say that this is the  {\em original expansion} of $F(\mathcal{I}_{r\geq s})$. %We will also say that each term in the original expansion is {\em labelled} by the particular configuration of $s$ faults to which it corresponds.

Next, in each term of the original expansion, we consider the ideal gates that have been applied to locations preceding the latest of the $s$ faults. We proceed to reexpress all of them as the unitary transformation that couples those locations and the bath {\em minus} a fault operator. We thus obtain the {\em derived expansion} which can be expressed as a sum of terms, where each term corresponds to a sum over all fault paths such that $m$ specific  coarse-grained locations are identified as faulty with $s\leq m\leq C$ and without other restrictions. %We will also say that each term in the derived expansion is {\rm labelled} by the particular configuration of $m$ faults to which it corresponds. 
We now need to count how many times a fault path where $m$ faulty coarse-grained locations have been specified appears in this derived expansion. 

Let $F(\mathcal{I}_{r=m})$ denote the sum over all the ${C\choose m}$ different terms in the derived expansion that identify exactly $m$ faulty coarse-grained locations. Each of these terms may arise from at most ${m-1 \choose s-1}$ terms in the original expansion, since we need to count the number of ways of choosing which $s-1$ of the first $m-1$ faulty coarse-grained locations belonged in the original expansion and the maximum, ${m-1 \choose s-1}$, corresponds to the case when each faulty coarse-grained location is coupled independently with the bath. Therefore, the sup norm of each of these terms is at most ${m-1\choose s-1}\varepsilon^{m}$, and so the sup norm of $F(\mathcal{I}_{r=m})$ can be upper bounded by ${C \choose m} {m-1 \choose s-1} \varepsilon^{m}$ %(again, the sign $(-1)^{m-s}$ has no consequence). 
(the sign of each term in the derived expansion may be $+1$ or $-1$, but these signs have no consequence for our bound). 

Since our derived expansion expresses $F(\mathcal{I}_{r\geq s})$ as $\sum_{m=s}^{C} F(\mathcal{I}_{r= m})$, we conclude that

\[
|| F(\mathcal{I}_{r\geq s}) ||_{\rm sup}  \leq  \sum_{m=s}^C {C \choose m} {{m-1}\choose{s-1}} \varepsilon^m = {C\choose s}\varepsilon^s \sum_{m=s}^C {s \over m} {C-s \choose m-s} \varepsilon^{m-s} 
\]
\begin{equation}  
\label{2.2.13}
                   \hspace{0.5cm}         \leq  {C\choose s}\varepsilon^s \sum_{m=0}^\infty {(C-s)^{m} \over m!} \varepsilon^{m} = {C \choose s}\varepsilon^s e^{(C-s)\varepsilon} \;.
\end{equation}

This proves the lemma.

\rightline{$\square$}

We can now use lemma \ref{lem:2} twice to upper bound $||Bd^{(1)}||_{\rm sup}$. Recall that $Bd^{(1)}$ is the sum of all fault paths with at least $t+1$ faults in at least one of the $L$ 1-exRecs. By choosing an ordering of the $L$ 1-exRecs, our first step is the same as before: we write $Bd^{(1)}$ as a sum of terms where for the $i$th term the $i$th 1-exRec is bad, all 1-exRecs preceding it are good and we put no restriction on the succeeding 1-exRecs; this is now our {\em original expansion}. We next reexpress every good 1-exRec as a sum of one term where we apply the unitary coupling between the bath and the coarse-grained locations inside the 1-exRec {\em minus} a second term which contains all fault paths that make the 1-exRec to be {\em bad}; this leads to our {\em derived expansion}. By using the same argument as in the proof of lemma \ref{lem:2}, we can express $Bd^{(1)}$ as a sum of terms where for each term $m$ specific 1-exRecs are bad with $1\leq m \leq L$ and we put no restriction on the other 1-exRecs. It follows that we can upper bound the sup norm of $Bd^{(1)}$ as in equation (\ref{2.2.12}) with $C\rightarrow L$, $s\rightarrow 1$, and $\varepsilon \rightarrow || F(\mathcal{I}_{r\geq s}) ||_{\rm sup}$;
\begin{equation}  
\label{2.2.14}
|| Bd^{(1)} ||_{\rm sup} \leq L || F(\mathcal{I}_{r\geq t+1}) ||_{\rm sup} \cdot e^{(L-1) \cdot || F(\mathcal{I}_{r\geq t+1}) ||_{\rm sup}} \;,
\end{equation}

\noindent where $F(\mathcal{I}_{r\geq t+1})$ denotes the sum of all fault paths that make a specific 1-exRec bad since all these fault paths must contain at least $t+1$ faults. Assuming $\varepsilon$ is sufficiently small so that $(L-1) \cdot || F(\mathcal{I}_{r\geq t+1}) ||_{\rm sup} \leq 1$ and by using lemma \ref{lem:2} again to upper bound $|| F(\mathcal{I}_{r\geq t+1}) ||_{\rm sup}$,
\begin{equation}  
\label{2.2.14.5}
|| Bd^{(1)} ||_{\rm sup} \leq e L \kappa {C \choose t+1}\varepsilon^{t+1}    \;,
\end{equation}

\noindent where $\kappa \geq e^{(C-t-1)\varepsilon} $. The error, $\delta$, of the fault-tolerant simulation can then be upper bounded by 
\begin{equation}  
\label{2.2.15}
\delta \leq 2 || Bd^{(1)} ||_{\rm sup} \leq 2 e L \kappa {C \choose t+1}\varepsilon^{t+1}   \;,
\end{equation}

\noindent which is smaller than the error, $2L\varepsilon \cdot e^{(L-1)\varepsilon} \leq 2eL \varepsilon$ assuming $(L-1) \, \varepsilon \leq 1$, of the unencoded computation provided $\kappa {C \choose t+1}\varepsilon^{t+1} < \varepsilon$, or $\varepsilon < \varepsilon_{\rm crit}$ where 
\begin{equation}  
\label{2.2.16}
\varepsilon_{\rm crit} \equiv \left( \kappa {C \choose t+1} \right)^{-1/t}   \;,
\end{equation}

\noindent for any $\kappa \geq e^{(C-t-1)\varepsilon_{\rm crit}}$. %(Our assumptions are that $(L-1) {C \choose t+1}\varepsilon_{\rm crit}^{t+1} e^{(C-t-1)\varepsilon_{\rm crit}} \leq 1$ and also $(L-1)\, \varepsilon_{\rm crit} \leq 1$.)

In obtaining the derived expansion of $Bd^{(1)}$ above, there is a detail that we have ignored: since two successive bad 1-exRecs overlap, their badness can be due to sets of faults that are not disjoint. Although this seems to complicate the analysis, we can in fact modify our definition of goodness in such a way that any two successive bad 1-exRecs will be designated to be simultaneously bad only due to sets of faults that are disjoint. With this modified definition which we will give in \S \ref{sec:Recursive}, our calculation above is valid. 

%----------------------------------------------------------%

\section{History and Acknowledgements}

The basic principles of fault-tolerant quantum circuit simulations and explicit gadget constructions satisfying properties \ref{prop:1} and \ref{prop:2} (to be discussed in chapter \ref{ch:lower-bounds}) were first given by Shor \cite{Shor96}. In this and in all subsequent work, the properties of fault-tolerant gadgets were discussed {\em semantically}, i.e., in terms of an explicit representation of the errors afflicting the quantum state being processed by these gadgets. 

After Gottesman, Preskill and I began thinking about the problem of proving the quantum threshold theorem for distance-3 codes, it quickly became clear that the criterion for fault tolerance would involve extended rectangles  whose analysis we found to be difficult semantically. In fact, our definition of goodness  had previously been used by Knill, Laflamme and Zurek \cite{Knill96b}, but their argument was informal and it was not at all clear how to apply it to recursive fault-tolerant simulations (to be discussed in chapter \ref{ch:threshold}). During and shortly after the QIP 2005 workshop, Gottesman suggested that we use an insight from \cite{Crepeau02} to phrase our definition of correctness in terms of ideal decoding circuits; this {\em syntactic} definition led to the formulation of lemma \ref{lem:1}.   

An analysis of fault-tolerant quantum computation for a coherent noise model was first formulated in an insightful paper by Terhal and Burkard \cite{Terhal04}. Their analysis considered a noise model where a locality condition is imposed not only on the interaction between the system and the bath, but also on interactions among the degrees of freedom within the bath. The essential ingredient for generalizing the proof in \cite{Terhal04} was the formulation of lemma \ref{lem:2} for the Hamiltonian noise model described in \S \ref{sec:LocalNoise}. Lemma \ref{lem:2} for this noise model was proved by Preskill and it was included in \cite{Aliferis05b} with a simplified proof due to Gottesman. The version of this lemma given here applies to general local noise; the proof is essentially the same as in \cite{Aliferis05b}.  

%----------------------------------------------------------%

%-----------------------------------------%
\chapter{Recursive Simulations and the Quantum Threshold Theorem}
\label{ch:threshold}
%----------------------------------------------------------%
%\hskip 7.8cm \parbox{6.7cm}{\small {\it The most valuable intuitions are the last to be attained; the most valuable of all are those which determine methods. ---Nietzsche, Antichrist {\rm 13}.}} \vskip 1cm
%---------------------------------------------------------------------%
\section{Introduction}

%In the previous chapter, I discussed how fault-tolerant simulations based on a code correcting errors on up to $t$ qubits achieve an accuracy of {\it O}$(\varepsilon^{t+1})$, where $\varepsilon$ is the strength of local noise. 

In this chapter, I will discuss recursive fault-tolerant quantum circuit simulations and prove the quantum threshold theorem for local noise. The idea of recursive simulations is intuitive: In the previous chapter, I described how any ideal quantum circuit can be simulated using gadgets that perform the computation in an encoded form using some quantum error-correcting code. Furthermore, these simulations are fault-tolerant in the sense that the noisy simulation is more accurate than the noisy unencoded computation provided the strength of local noise is below some critical value. But, if for some noise strength the fault-tolerant simulation improves the accuracy of the unencoded computation, we expect that a fault-tolerant simulation of the fault-tolerant simulation will improve it even more! And we can continue applying this idea again and again building a hierarchy of simulations that achieves any desired accuracy. We will then identify the critical noise strength as the {\em accuracy threshold} for quantum computation.  

By construction, at any level, $k$, of a recursive fault-tolerant simulation, the simulation will be mapped to the corresponding simulation at level $k{-}1$ if we replace every 1-Rec inside it by a 0-Ga. If we replace every 1-Rec inside a simulation at level $k{-}1$ by a 0-Ga, the simulaton at level $k{-}1$ will be mapped to the corresponding simulation at level $k{-}2$, etc. Our analysis of the accuracy of recursive fault-tolerant simulations will exploit exactly this property: We will show that the fault-tolerant simulation at level $k$ with noise strength $\varepsilon$ is equivalent to another fault-tolerant simulation at level $k{-}1$ with a {\em transformed} noise strength,  $\varepsilon^{(1)}$. Then, by applying the same argument a second time, the fault-tolerant simulation at level $k{-}1$ with noise strength $\varepsilon^{(1)}$ is equivalent to another fault-tolerant simulation at level $k{-}2$ with a {\em transformed} noise strength, $\varepsilon^{(2)}$. After we apply this argument $k$ times, we will obtain an equivalent simulation at level 0 with noise strength $\varepsilon^{(k)}$; but, a level-0 simulation is an execution of the ideal circuit with noisy physical operations! We conclude that the initial fault-tolerant simulation at level $k$ is equivalent to an execution of the ideal quantum circuit with operations which have noise strength $\varepsilon^{(k)}$. Provided the physical noise strength, $\varepsilon$, is below the accuracy threshold value, $\varepsilon^{(k)}$ can be made arbitrarily small, thereby showing that the level-$k$ simulation can achieve any desired accuracy. This is the outline of our proof of the quantum threshold theorem. 

The remaining of this chapter is organized as follows: In \S \ref{sec:Recursive}, I discuss the replacement rules that give rise to recursive fault-tolerant simulations. Then, in \S \ref{sec:LevelReduction}, I prove the {\em level-reduction} lemma which shows that recursive fault-tolerant simulations subject to local noise can be analysed ``one level at a time.'' Finally, in \S \ref{sec:Threshold}, I use this lemma to prove the quantum threshold theorem for local noise. I also discuss a method for obtaining improved rigorous lower bounds on the quantum accuracy threshold by means of a more detailed combinatorial analysis; such an analysis is performed explicitly in chapter \ref{ch:lower-bounds} for a particular code.

%----------------------------------------------------------%
\section{Recursive Fault-Tolerant Simulations}
\label{sec:Recursive}

\begin{figure}[tb]
\begin{center}
%\leavevmode
%\epsfysize=4.5cm
%\epsfbox{recursive.eps}
\includegraphics[width=6cm]{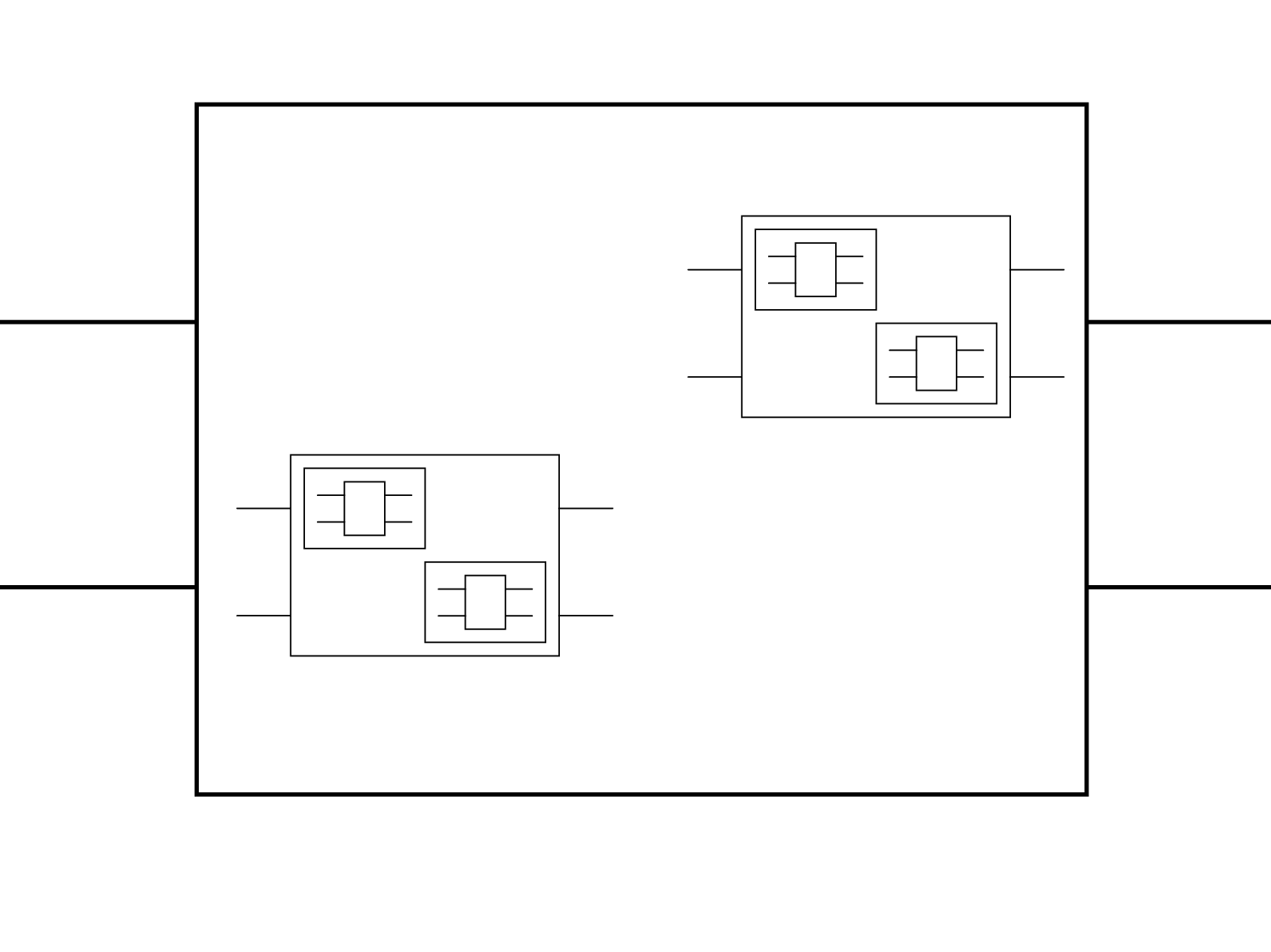}
\vspace{-0.8cm}     
\end{center}
\caption{\label{fig:recursive} A recursive simulation. A level-$k$ gadget is built from level-$(k{-}1)$ gadgets, which are built from level-$(k{-}2)$ gadgets, etc.}
\end{figure}

The quantum computation performed by a recursive fault-tolerant simulation is protected by a {\em concatenated} quantum code \cite{Knill96c}. The code block of a concatenated code is constructed as a hierarchy of codes within codes---the code block at any level, $k$, of this hierarchy is built from logical qubits encoded at level $k-1$ of the hierarchy. And there exists a corresponding hierarchy of gadgets: At the physical level, which we may call {\em level 0},  gadgets are just the noisy physical operations that we have called 0-Gas. At the next level of this recursion, {\em level 1,} each ideal gate is simulated by a 1-Rec that acts on 1-blocks of some quantum error-correcting code, $\mathcal{C}$, as was discussed in the previous chapter. One level higher in the recursion, {\em level 2}, an ideal gate is simulated by a {\em 2-Rec} that acts on the code blocks of the concatenated quantum code $\mathcal{C}\circ \mathcal{C}$; we will call such code blocks {\em 2-blocks}. A 2-Rec is constructed by replacing each 0-Ga inside a 1-Rec by the 1-Rec corresponding to the 0-Ga. At level $k$ of the recursion, ideal gates are simulated by {\em $k$-Recs} that are constructed by replacing each 0-Ga inside a $(k{-}1)$-Rec by the corresponding 1-Rec; {\em $k$-Rec}s act on $k$-blocks, i.e., code blocks of a quantum code that is concatenated $k$ times. If $k$ levels of recursion are used in total, then there will be one $k$-Rec for each location in the ideal quantum circuit we simulate. We will often call the computation realized by $k$-Recs a {\em level-$k$ simulation}; in particular, a level-$0$ simulation is a direct implementation of some quantum circuit with noisy 0-Gas. Figure \ref{fig:recursive} shows a schematic of a quantum circuit executing a recursive simulation.

Lemma \ref{lem:1} in the previous chapter established that the 0-Recs inside good 1-exRecs (i.e., 1-exRecs that contain no more than $t=\lfloor {d-1\over 2}\rfloor$ faults, where $d$ is the distance of the quantum code used) are correct (i.e., they perform an accurate simulation of the corresponding 0-Ga). This allowed us to conclude that if all 1-exRecs are good, the level-$1$ simulation realized by 1-Recs can be replaced by an equivalent level-$0$ simulation where all 1-Recs are converted to the ideal 0-Gas they simulate. 

Now, we would like to generalize this conclusion. We would like to prove that no matter how many faults have occurred, the level-$1$ simulation can always be replaced by an equivalent level-0 simulation where the 1-Rec inside every good 1-exRec has been converted to the ideal simulated 0-Ga, and the 1-Rec inside every bad 1-exRec has been converted to some faulty implementation of the simulated 0-Ga. But, from the onset, we encounter a problem in identifying bad 1-exRecs with faulty 0-Gas: Two successive 1-exRecs overlap (i.e., they contain at least one 1-EC gadget in common) and, for this reason, they can both be bad due to sets of faults that are {\em not} disjoint. 

\begin{figure}[tbh]
\begin{center}
\vspace{0.5cm}
\begin{picture}(250,48)
\put(2,22){\line(1,0){5}}
\put(7,4){\framebox(36,36){1-EC}}
\put(43,22){\line(1,0){14}}
\put(57,4){\framebox(36,36){1-Ga}}
\put(85,26){$\times$}
\put(93,22){\line(1,0){5}}
\put(102,22){\line(1,0){5}}
\put(107,4){\framebox(36,36){1-EC}}
\put(113,6){$\times$}
\put(143,22){\line(1,0){5}}
\put(152,22){\line(1,0){5}}
\put(157,4){\framebox(36,36){1-Ga}}
\put(193,22){\line(1,0){14}}
\put(207,4){\framebox(36,36){1-EC}}
\put(231,32){$\times$}
\put(243,22){\line(1,0){5}}
\put(0,0){\dashbox(150,44){}}
\put(100,-4){\dashbox(150,52){}}
\end{picture}
\end{center}
\caption{\label{fig:bad-overlapping-exRecs} Two non-independent successive bad 1-exRecs, indicated by dashed lines, with fault locations indicated by $\times$. Because one of the three faults is contained in the shared 1-EC, the bad 1-exRecs are not independent events. In this situation, we may regard the earlier 1-Ga as a gadget that simulates the corresponding ideal 0-Ga accurately, and the later 1-exRec rather than the later 1-Rec as a gadget that simulates the corresponding ideal 0-Ga inaccurately.}
\end{figure}
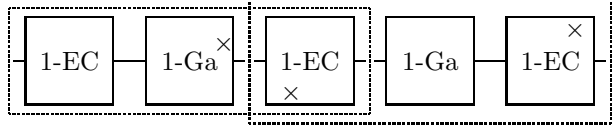

Suppose, e.g., that we use a distance-3 code and that two successive 1-exRecs are both bad due to one fault in the shared 1-EC, one fault in the 1-Ga of the earlier 1-exRec and one fault in the trailing 1-EC of the later 1-exRec (Fig.~\ref{fig:bad-overlapping-exRecs}). These two overlapping 1-exRecs simulate two successive 0-Gas. Ordinarily, we would think of the earlier of the two 1-Recs as the gadget that simulates the earlier of the two 0-Gas. But, in this case, we may instead {\em remove} the shared 1-EC from the earlier 1-Rec and regard the 1-Ga contained in the earlier 1-Rec as the gadget that simulates the earlier 0-Ga; we will say that the earlier 1-exRec has been {\em truncated}. Because the truncated earlier 1-exRec contains only one fault, we expect that it simulates the earlier 0-Ga accurately. The shared 1-EC which we removed from the earlier 1-exRec will be added to the later 1-Rec; so, we will regard the entire later 1-exRec---rather than the 1-Rec it contains---as the gadget that simulates the later 0-Ga; and it is only this later simulation that is inaccurate. Thus, our intuition is that the two overlapping bad 1-exRecs in figure \ref{fig:bad-overlapping-exRecs} are really no worse than a single bad 1-exRec.

More generally, we wish to regard the badness of different 1-exRecs as arising due to disjoint sets of faults. We will say that two bad 1-exRecs are {\em independent} if they are bad due to a disjoint set of faults; when this is not the case, we will say that they are {\em non-independent}. As indicated by the example above, our criterion for when two overlapping bad 1-exRecs are independent is that the earlier bad 1-exRec must {\em remain bad} when all shared 1-EC gadgets are removed. To reflect this requirement, we revise definition \ref{defi:6} as follows.
\begin{defi}[Goodness (revised)]
\label{defi:7}
Consider a fault-tolerant quantum circuit simulation based on a distance-$d$ quantum error-correcting code, and let $t=\lfloor {d-1\over 2}\rfloor$. A 1-exRec is {\em good} if it contains at most $t$ faults; if it is not good it is {\em bad}. Two bad 1-exRecs are {\em independent} if they are nonoverlapping or if they overlap and the earlier 1-exRec is still bad when the shared 1-ECs are removed. 
\end{defi}

The fact that two non-independent bad 1-exRecs can be viewed as simulating just one rather than two faulty 0-Gas will be proven rigorously in \S \ref{sec:BadRecs}. But before giving more details about treating non-independent bad 1-exRecs, let us first use this fact to prove the basic level-reduction lemma.

%--------------------------------------------------------------%
\section{The Level-Reduction Lemma}
\label{sec:LevelReduction}

Our goal is to assess the error, $\delta$, of the level-$k$ simulation as a function of the level $k$ of recursion. We will imagine that local noise of strength $\varepsilon$ is acting on all elementary physical operations or, simply, {\em locations} realizing this level-$k$ simulation; then the question is how $\delta$ is related to $\varepsilon$ and $k$. We will answer this question indirectly: We will first show that the level-$k$ simulation is equivalent to a level-$(k{-}1)$ simulation that is subject to local noise of {\em transformed} strength, $\varepsilon^{(1)}$. %Using the same argument again, $\delta$ can be upper bounded by the accuracy of an equivalent level-$(k{-}2)$ simulation which is subject to local noise of strength $\varepsilon^{(2)}$, etc. 
The following lemma makes this statement precise.

\begin{lem}[Level Reduction]
\label{lem:3}
Consider a recursive fault-tolerant simulation that is generated by the self-similar replacement rules described above on gadgets satisfying properties \ref{prop:1} and \ref{prop:2} for a distance-$d$ quantum error-correcting code, and let $t=\lfloor{d-1\over 2}\rfloor$ and $C$ be the number of locations in the largest 1-exRec. Let the level-$k$ simulation ($k\geq 1$) be subject to local noise of strength $\varepsilon$ (or $p$ if noise is stochastic). Then, there exists a level-$(k{-}1)$ simulation that produces the same output probability distribution and is subject to local noise of strength
\[
\varepsilon^{(1)} \leq \kappa {C\choose t+1} \varepsilon^{t+1} \; ,
\]

\noindent where $\kappa \geq e^{(C-t-1)\varepsilon}$ (or $p^{(1)} \leq {C\choose t+1} p^{t+1}$ if noise is stochastic).
\end{lem}

\noindent {\bf Proof}. We start by considering all measurement 1-exRecs. Using lemma \ref{lem:1}, we replace the 1-Rec inside every good measurement 1-exRec by an i-decoder followed by the ideal simulated measurement 0-Ga. We then continue by moving the i-decoders to the left of the 1-Rec inside every good gate 1-exRec, thereby replacing the gate  1-Recs by the ideal simulated gate 0-Gas. Finally, we annihilate all i-decoders inside good preparation 1-exRecs, thereby replacing the preparation 1-Recs by the ideal simulated preparation 0-Gas. In addition, every time i-decoders are stuck to the right of a bad gate 1-exRec, the i-decoders can be moved to the left of the {\em entire} bad 1-exRec which is replaced by some {\em faulty} implementation of the simulated gate 0-Ga; and similarly for bad measurement or preparation 1-exRecs (see \S \ref{sec:BadRecs}). Overall, by means of this ``wave'' of i-decoders moving from the right to the left of all level-$1$ gadgets, we convert every 1-Rec to either an ideal or a faulty implementation of the 0-Ga it simulates. Hence, one level of coding has been removed and the initial level-$k$ simulation has been converted to a level-$(k{-}1)$ simulation that produces exactly the same output.

It remains to consider the {\em effective} noise afflicting this level-$(k{-}1)$ simulation. Consider a set, $\mathcal{I}^{(1)}_r$, of $r$ locations in the level-$(k{-}1)$ simulation and let $F(\mathcal{I}^{(1)}_r)$ denote the sum of all fault paths with faults at all locations in $\mathcal{I}^{(1)}_r$. Each location in the level-$(k{-}1)$ simulation arises from a 1-Rec in the initial level-$k$ simulation and, furthermore, $r$ specific locations are faulty only if the corresponding 1-exRecs in the initial level-$k$ simulation were all {\em independently} bad. Recall that each independent bad 1-exRec contains at least $t+1$ faults. For local stochastic noise, the sum of all fault paths with $t+1$ faults at specified locations in each of the $r$ bad 1-exRecs has probability at most $\left( p^{t+1} \right)^r$ and, so, the probability of $F(\mathcal{I}^{(1)}_r)$ is at most $\left(p^{(1)}\right)^r$ where 
\begin{equation}
\label{3.1.1}
p^{(1)} \leq {C\choose t+1} p^{t+1} \; .
\end{equation}  

\noindent Hence, we conclude that the equivalent level-$(k{-}1)$ simulation that results after level reduction is afflicted by local stochastic noise of strength $p^{(1)}$.

For local noise that is not stochastic, we can follow the argument in proving lemma \ref{lem:2} to obtain the derived expansion of $F(\mathcal{I}^{(1)}_r)$. Let $F_i(\mathcal{I}_{r= m_i})$ denote the sum of all fault paths such that the $i$th of the $r$ specified bad 1-exRecs contains $m_i$ faults with $t+1 \leq m_i \leq C$ (see \S \ref{sec:Overlap} for dealing with overlapping bad 1-exRecs), and also let $m=\sum_i m_i$. Then, $F(\mathcal{I}^{(1)}_r)$ can be expanded as 
\begin{equation}
\label{3.1.1.5}
F(\mathcal{I}^{(1)}_r) = \sum_{m_1=t+1}^{C} \cdots \sum_{m_r=t+1}^{C} F_1(\mathcal{I}_{r= m_1}) \dots F_r(\mathcal{I}_{r= m_r}) \; ,
\end{equation}

\noindent so that, by performing a similar combinatorial analysis as in the proof of lemma \ref{lem:2}, 

\begin{equation}
\label{3.1.2}
||F(\mathcal{I}^{(1)}_r)||_{\rm sup} \leq \sum_{m_1=t+1}^{C} {m_1 - 1\choose t} {C\choose m_1} \cdots \sum_{m_r=t+1}^{C} {m_r - 1\choose t} {C\choose m_r} \varepsilon^{m} \; .
\end{equation}  

\noindent Similar to equation (\ref{2.2.13}), we conclude that $||F(\mathcal{I}^{(1)}_r)||_{\rm sup} \leq \left(\varepsilon^{(1)} \right)^r$ where
\begin{equation}
\label{3.1.3}
\varepsilon^{(1)} \leq  {C \choose t+1} \varepsilon^{t+1} e^{(C-t-1)\varepsilon} \; .
\end{equation}

\noindent Therefore, the level-$(k{-}1)$ simulation that results after level reduction is afflicted by local noise of strength $\varepsilon^{(1)}$, as desired.

\noindent This proves the lemma. 

\rightline{$\square$}

%--------------------------------------------------------------%
\subsection{Independent Bad Rectangles as Simulated Faults}
\label{sec:BadRecs}

If all 1-exRecs are good, then we can apply lemma \ref{lem:1} in succession to convert all of them to the ideal 0-Gas they simulate. But in the proof of the level-reduction lemma, we also need to consider the case of bad 1-exRecs and, in this case, lemma \ref{lem:1} does not apply. What we would like to achieve is an equivalence allowing us to convert the 1-Rec inside a bad 1-exRec that is followed by i-decoders to a circuit where i-decoders appear first and they are followed by some faulty implementation of the corresponding simulated 0-Ga. However, this is in general not possible: the 1-Rec inside a bad 1-exRec is not simulating a well-defined faulty 0-Ga independent of the input to the 1-Rec. Indeed, if we attempt to move the i-decoders from the right of the 1-Rec inside a bad 1-exRec to in front of it and thereby transform the 1-Rec to a faulty 0-Ga, the particular faulty 0-Ga we obtain may depend on the syndrome that the i-decoders find. For instance, faults in the 1-Rec might combine differently with a  fault which has previously acted on the first qubit in an input 1-block than with a fault which has previously acted  on the second qubit in the same input 1-block; then, even though applying an i-decoder to the input 1-block yields the same state in both cases,  applying i-decoders to the output of the 1-Rec in these two cases might yield different output states.

This observation tells us what direction we should take: we should look for an equivalence to faulty 0-Gas that {\em do} depend on the syndrome i-decoders find. To make the dependence on the syndrome information explicit, we will introduce a new version of i-decoders that we will call {\em coherent i-decoders}, or ${\cal D}$'s. Whereas i-decoders trace over all output ancillary qubits, coherent i-decoders do not; they maintain the {\em coherence} between the syndrome that ideal error-correction finds and the state of the decoded qubit. More specifically, let $|\psi\rangle_L$ denote some logical single-qubit state, and let $\{E_i\}$ denote the set of correctable Pauli errors of an $[[n,k,d]]$ stabilizer code; then the action of the coherent i-decoder is
\begin{equation}
\label{3.1.4}
 {\cal D}: E_i |\psi\rangle_L \mapsto |\psi\rangle \otimes |e_i\rangle \; ,
\end{equation}

\noindent where $|\psi\rangle$ is the decoded single-qubit state carried, e.g., by the first output qubit, and $|e_i\rangle$ denotes the state of remaining $n{-}1$ qubits that records the syndrome. 

If the states $\{E_i|\psi\rangle_L \}$ are a complete basis (i.e., if the code is perfect), then equation (\ref{3.1.4}) completely characterizes the action of the coherent i-decoder. Otherwise, since the action in equation (\ref{3.1.4}) is inner-product preserving, we may always choose some unitary extension of ${\cal D}$. Then, its inverse,  ${\cal D}^{-1}$, is a {\em coherent i-encoder} that takes a qubit in the input state $|\psi\rangle$ and $n{-}1$ qubits in the state $|e_i\rangle$ and encodes them in the state $|\psi\rangle_L$ with Pauli error $E_i$. Of course, an i-decoder is just a coherent i-decoder that discards the $n{-}1$ output qubits carrying the syndrome information, or schematically:

\vspace{0.5cm}
\begin{picture}(152,48)
\put(0,36){\line(1,0){10}}
\put(10,24){\framebox(48,24){\shortstack{i-decoder}}}
\put(58,36){\line(1,0){10}}
\put(68,30){\makebox(20,12){=}}
\put(88,36){\line(1,0){10}}
\put(98,24){\framebox(24,24){${\cal D}$}}
\put(122,36){\line(1,0){10}}
\put(110,24){\line(0,-1){15}}
\put(110,9){\line(1,0){10}}
\put(114,3){\makebox(20,12){$\times$}}
\put(132,30){\makebox(20,12){.}}
\end{picture}

\noindent This implies that lemma \ref{lem:1} in the previous chapter holds if i-decoders are replaced by coherent i-decoders: Indeed, lemma \ref{lem:1} is precisely the statement that the 1-Rec inside a good 1-exRec is correct {\em independent} of what syndrome the i-decoders find (since, otherwise, if there was a dependence on the syndrome, lemma \ref{lem:1} would not always hold independent of the syndrome that the i-decoders discard). For a 1-Rec simulating a single-qubit 0-Ga, lemma \ref{lem:1} now takes the form of the equality 

%%% exRec-Cor with syndrome propagation
\vspace{0.5cm}
\begin{picture}(324,48)
\put(0,36){\line(1,0){10}}
\put(10,24){\framebox(30,24){1-EC}}
\put(40,36){\line(1,0){10}}
\put(50,24){\framebox(48,24){1-Rec}}
\put(98,36){\line(1,0){10}}
\put(108,24){\framebox(24,24){${\cal D}$}}
\put(132,36){\line(1,0){10}}
\put(120,24){\line(0,-1){15}}
\put(120,9){\line(1,0){22}}
\put(142,30){\makebox(20,12){=}}
\put(162,36){\line(1,0){10}}
\put(172,24){\framebox(30,24){1-EC}}
\put(202,36){\line(1,0){10}}
\put(212,24){\framebox(24,24){${\cal D}$}}
\put(236,36){\line(1,0){10}}
\put(246,24){\framebox(48,24){\shortstack{ideal\\0-Ga}}}
\put(294,36){\line(1,0){10}}
\put(224,24){\line(0,-1){15}}
\put(224,9){\line(1,0){22}}
\put(246,0){\framebox(48,18){${\cal O}_{\rm syndr}$}}
\put(294,9){\line(1,0){10}}
\put(304,30){\makebox(20,12){,}}
\end{picture}
%\vspace{0.1cm}

\noindent where ${\cal O}_{\rm syndr}$ is some operation acting on the syndrome.

Since coherent i-decoders are unitary, we are free to create pairs of coherent i-decoders and i-encoders (or simply, {\em decoder-encoder pairs}) via the identity $I= {\cal D} \cdot {\cal D}^{-1}$ which we can insert at any point during the quantum computation. In particular, we can insert decoder-encoder pairs preceding a bad 1-exRec in order to convert it to a faulty 0-Ga that depends on the syndrome that is processed by the coherent i-decoders. For a 1-exRec corresponding to a single-qubit 0-Ga, we can illustrate this as 

%%% bad exRec identity
\vspace{0.5cm}
\begin{picture}(312,48)
\put(0,36){\line(1,0){10}}
\put(10,24){\framebox(48,24){\shortstack{bad\\1-exRec}}}
\put(58,36){\line(1,0){10}}
\put(68,24){\framebox(24,24){${\cal D}$}}
\put(92,36){\line(1,0){10}}
\put(80,24){\line(0,-1){15}}
\put(80,9){\line(1,0){22}}
\put(102,30){\makebox(20,12){=}}
\put(122,36){\line(1,0){10}}
\put(132,24){\framebox(24,24){${\cal D}$}}
\put(156,36){\line(1,0){10}}
\put(144,24){\line(0,-1){15}}
\put(144,9){\line(1,0){34}}
\put(178,24){\line(0,-1){15}}
\put(166,24){\framebox(48,24){\shortstack{faulty\\0-Ga}}}
\put(214,36){\line(1,0){10}}
\put(202,24){\line(0,-1){15}}
\put(202,9){\line(1,0){22}}
\put(224,30){\makebox(20,12){,}}
\end{picture}

\noindent where the faulty 0-Ga is defined as the operation

%%% definition of faulty 0-Ga
\vspace{0.5cm}
\begin{picture}(244,48)
\put(0,36){\line(1,0){10}}
\put(10,24){\framebox(48,24){\shortstack{faulty\\0-Ga}}}
\put(58,36){\line(1,0){10}}
\put(0,9){\line(1,0){22}}
\put(22,24){\line(0,-1){15}}
\put(46,24){\line(0,-1){15}}
\put(46,9){\line(1,0){22}}
\put(68,30){\makebox(20,12){=}}
%
%\put(122,36){\line(1,0){10}}
\put(88,36){\line(1,0){10}}
\put(98,24){\framebox(24,24){${\cal D}^{-1}$}}
\put(88,9){\line(1,0){22}}
\put(110,24){\line(0,-1){15}}
\put(122,36){\line(1,0){10}}
\put(132,24){\framebox(48,24){\shortstack{bad\\1-exRec}}}
\put(180,36){\line(1,0){10}}
\put(190,24){\framebox(24,24){${\cal D}$}}
\put(214,36){\line(1,0){10}}
\put(202,24){\line(0,-1){15}}
\put(202,9){\line(1,0){22}}
\put(224,30){\makebox(20,12){.}}
\end{picture}

It is important that decoder-encoder pairs are inserted on every 1-block preceding the {\em entire} bad 1-exRec because, in the next step, the coherent i-decoders will move to the left of the preceding 1-exRecs. We only want to consider these preceding 1-exRecs as bad if they are bad {\em independently} of the current 1-exRec. Thus, we  insert our decoder-encoder pairs so that we {\em truncate} these preceding 1-exRecs from all 1-EC gadgets they share with the current bad 1-exRec. If the preceding truncated 1-exRecs are indeed bad independently of the current 1-exRec then, in the next step, we insert decoder-encoder pairs preceding them as well. And we repeat the same trick every time coherent i-decoders encounter a bad 1-exRec, thereby converting every independent bad 1-exRec to a faulty 0-Ga that is correlated with the syndrome.  

On the other hand, every non-independent bad 1-exRec will, after truncation of the 1-EC gadgets it shares with its succeeding 1-exRecs, become good. Good truncated 1-exRecs are correct as can be easily seen as follows: First, assume i-decoders (instead of coherent i-decoders) are placed to the right of a truncated good 1-exRec. We can restore all truncated 1-ECs by pulling out ideal (i.e., faultless) 1-ECs from the i-decoders, since clearly

\vspace{0.5cm}
\begin{picture}(292,24)
\put(0,12){\line(1,0){10}}
\put(10,0){\framebox(48,24){\shortstack{i-decoder}}}
\put(58,12){\line(1,0){10}}
\put(68,6){\makebox(20,12){=}}
\put(88,12){\line(1,0){10}}
\put(98,0){\framebox(48,24){\shortstack{ideal\\1-EC}}}
\put(146,12){\line(1,0){10}}
\put(156,0){\framebox(48,24){\shortstack{i-decoder}}}
\put(204,12){\line(1,0){10}}
\put(214,6){\makebox(20,12){.}}
\end{picture}
\vspace{0.3cm}

\noindent Now, the good truncated 1-exRecs have become equivalent to full good 1-exRecs and lemma \ref{lem:1} applies. But, lemma \ref{lem:1} being true when phrased in terms of i-decoders implies that the correctness of good truncated 1-exRecs does not depend on the syndrome that the i-decoders find; hence, the lemma will also apply if coherent i-decoders are used instead. This shows that, in general, coherent i-decoders can be propagated to the left of the 1-Rec inside every good 1-exRec (truncated or not), thereby converting the 1-Rec to the ideal simulated 0-Ga applied on the decoded qubit(s) and some operation applied on the syndrome. 

In effect, the syndrome acts as a common {\em bath} with which qubits at all faulty 0-Gas interact after level reduction: Consider a sequence of 1-exRecs that contains $r$ independent bad $1$-exRecs. By creating i-decoders, moving them to the left and finally annihilating them transforms the 1-exRecs to an equivalent circuit of 0-Gas with $r$ faults; and these faults can be correlated since they correspond to the operation simulated by the bad 1-exRecs that depends on the syndrome. For instance, two bad 1-exRecs separated by a sequence of good intermediate 1-exRecs can be converted to two correlated faulty 0-Gas that are separated by ideal 0-Gas, or schematically: 

%%% good between two bad circuits
\vspace{0.5cm}
\begin{picture}(444,48)
\put(0,36){\line(1,0){10}}
\put(10,24){\framebox(40,24){\shortstack{bad\\1-exRec}}}
\put(50,36){\line(1,0){10}}
\put(60,24){\framebox(52,24){\shortstack{good\\1-exRecs}}}
\put(112,36){\line(1,0){10}}
\put(122,24){\framebox(40,24){\shortstack{bad\\1-exRec}}}
\put(162,36){\line(1,0){10}}
\put(172,24){\framebox(24,24){${\cal D}$}}
\put(196,36){\line(1,0){10}}
\put(184,24){\line(0,-1){15}}
\put(184,9){\line(1,0){22}}
\put(206,30){\makebox(20,12){=}}
\put(226,36){\line(1,0){10}}
\put(236,24){\framebox(24,24){${\cal D}$}}
\put(260,36){\line(1,0){10}}
\put(248,24){\line(0,-1){15}}
\put(248,9){\line(1,0){34}}
\put(282,24){\line(0,-1){15}}
\put(270,24){\framebox(36,24){\shortstack{faulty\\0-Ga}}}
\put(306,36){\line(1,0){10}}
\put(294,24){\line(0,-1){15}}
\put(294,9){\line(1,0){22}}
\put(316,0){\framebox(52,18){${\cal O}_{\rm syndr}$}}
\put(368,36){\line(1,0){10}}
\put(316,24){\framebox(52,24){\shortstack{ideal\\0-Gas}}}
\put(368,9){\line(1,0){18}}
\put(386,24){\line(0,-1){15}}
\put(378,24){\framebox(36,24){\shortstack{faulty\\0-Ga}}}
\put(414,36){\line(1,0){10}}
\put(402,24){\line(0,-1){15}}
\put(402,9){\line(1,0){22}}
\put(424,30){\makebox(20,12){.}}
\end{picture}
\vspace{0.3cm}

It follows that even if noise is uncorrelated at the physical level, effective correlations between faults will arise as one considers the different coding levels in a recursive fault-tolerant simulation. Local noise has the property that it includes the possibility of such correlations and, for this reason, local noise acting in the level-$k$ simulation remains local in the equivalent level-$(k{-}1)$ simulation after level reduction and only the noise strength is transformed. In this sense, we may say that local noise is ``stable'' under level reduction.

%--------------------------------------------------------------%
\subsection{The Combinatorics for Overlapping Rectangles}
\label{sec:Overlap}

Finally, we need to justify why the bound (\ref{3.1.3}) is valid even though some of the $r$ specified bad 1-exRecs in the level-$k$ simulation may overlap. We wish to express $F(\mathcal{I}^{(1)}_r)$ as a sum of terms where  for each term, each one of the $r$ specified bad 1-exRecs contains from $t+1$ up to some maximum number of faults and it is bad {\em independently} from all other bad 1-exRecs. Because of the requirement for independence, the maximum number of faults in each of the $r$ specified bad 1-exRecs can be either the number of locations in the full 1-exRec (if none of the 1-exRecs immediately following are bad), or the number of locations in the truncated 1-exRec (if one of the 1-exRecs immediately following is one of the other $r$ specified bad 1-exRecs). 

Nevertheless, for any specified fault path, whether each bad 1-exRec is truncated or not is unambiguously determined and, in order for a 1-exRec to be bad independently, it must contain at least $t+1$ faults. Therefore, we can still apply the bound (\ref{3.1.2}) with the only change that for each fault path in the sum some 1-exRecs may be truncated, and which 1-exRecs are truncated varies from fault path to fault path so that the upper limits of each summation may vary. Since a truncated 1-exRec contains fewer locations than the full 1-exRec, the bound (\ref{3.1.2}) applies in general independent of whether some of the $r$ specified bad 1-exRecs overlap.

%--------------------------------------------------------------%
\section{The Quantum Threshold Theorem for Local Noise}
\label{sec:Threshold}

The virtue of the level-reduction lemma is that it is {\em composable}. We can apply it $k$ times in succession to convert a level-$k$ simulation to a level-$0$ simulation that produces the same output statistics; the level-$0$ simulation is directly implementing the desired quantum circuit but with a transformed noise strength. Consider first  local stochastic noise of strength $p$ acting on the initial level-$k$ simulation. Then, by iterating $k$ times the recursion equation $p^{(j)} \leq {C\choose t+1} \left(p^{(j-1)}\right)^{t+1}$ with $p^{(0)}\equiv p$, the noise strength in the equivalent level-$0$ simulation is 
\begin{equation}
\label{3.1.5}
 p^{(k)} \leq p_{\rm thr} \left( {p \over p_{\rm thr}} \right)^{(t+1)^k} \; , \; {\rm where} \; \; \;  p_{\rm thr} \equiv {C\choose t+1}^{-1/t} \; .
\end{equation}

%\noindent where
%
%\begin{equation}
%\label{3.1.6}
% p_{\rm thr} \equiv {C\choose t+1}^{-1/t} \; .
%\end{equation}

\noindent Hence, $p_{\rm thr}$ is the {\em threshold} noise strength, i.e., for $p<p_{\rm thr}$ the {\em effective}  noise that acts on the level-$k$ simulation has strength that decreases doubly-exponentially with $k$ as shown by equation (\ref{3.1.5}). Using equation (\ref{1.2.15}) with $p \rightarrow p^{(k)}$, we conclude that the error, $\delta$, of the level-$k$ simulation is 
\begin{equation}
\label{3.1.7}
\delta \leq 2 L\, p_{\rm thr} \left( {p \over p_{\rm thr}} \right)^{(t+1)^k} \; .
\end{equation}

\noindent Therefore, any desired accuracy, $1-\delta_0$, can be achieved by choosing $k$ such that
\begin{equation}
\label{3.1.8}
(t+1)^k \geq  {\log \left( 2L\, p_{\rm thr} / \delta_0 \right) \over \log \left( p_{\rm thr}/ p \right) } \; .
\end{equation}

\noindent If $c$ is the size (i.e., number of locations) of the largest 1-Rec and if $d$ if the largest depth (i.e., number of time steps) in any 1-Rec then, because of the self-similarity of recursive simulations, all $k$-Recs will have size at most $c^k=\left((t+1)^k\right)^{\log_{t+1} c}$ and depth at most $d^k=\left((t+1)^k\right)^{\log_{t+1} d}$.

Similarly, for local noise that is not stochastic, if noise of strength $\varepsilon$ acts on the initial level-$k$ simulation, then the equivalent level-$0$ simulation is afflicted by local noise of strength
\begin{equation}
\label{3.1.9}
 \varepsilon^{(k)} \leq \varepsilon_{\rm thr} \left( {\varepsilon \over \varepsilon_{\rm thr}} \right)^{(t+1)^k} \; , \; {\rm where} \; \; \; \varepsilon_{\rm thr} \equiv \left( \kappa {C\choose t+1} \right)^{-1/t} \; ,
\end{equation}
%
%\noindent where
%
%\begin{equation}
%\label{3.1.10}
% \varepsilon_{\rm thr} \equiv \left( \kappa {C\choose t+1} \right)^{-1/t} \; ,
%\end{equation}

\noindent for $\kappa \geq e^{(C-t-1)\varepsilon_{\rm thr}}$ provided $\varepsilon<\varepsilon_{\rm thr}$; thus, $\varepsilon_{\rm thr}$ is the {\em threshold} noise strength. Similar to equation (\ref{2.2.15}), the error, $\delta$, of the level-$k$ simulation can then be upper bounded by 
\begin{equation}
\label{3.1.11}
 \delta \leq 2 e L \varepsilon^{(k)} \leq 2 e L \varepsilon_{\rm thr} \left( {\varepsilon \over \varepsilon_{\rm thr}} \right)^{(t+1)^k} \; .
\end{equation}

\noindent Therefore, we arrive at a similar conclusion as for local stochastic noise, i.e., any desired accuracy,  $1-\delta_0$, can be achieved provided $k$ is chosen so that
\begin{equation}
\label{3.1.12}
(t+1)^k \geq  {\log \left( 2 e L\, \varepsilon_{\rm thr} / \delta_0 \right) \over \log \left( \varepsilon_{\rm thr}/ \varepsilon \right) } \; .
\end{equation}

With the help of the level-reduction lemma, we have thus proved the quantum threshold theorem for local noise that is stated next.

\begin{theo}[Quantum Accuracy Threshold for Local Noise]
\label{theo:4}
Suppose gadgets can be constructed such that they satisfy properties \ref{prop:1} and \ref{prop:2} for a distance-$d$ quantum error-correcting code, and let $t=\lfloor{d-1\over 2}\rfloor$. Let $c$ be the largest size and $d$ be the largest depth in any 1-Rec, and let $C$ be the size of the largest 1-exRec. Suppose that noisy quantum circuits are afflicted by local noise of strength
\[
\varepsilon < \varepsilon_{\rm thr} \equiv \left( \kappa {C\choose t+1} \right)^{-1/t} \;,
\]

\noindent where $\kappa \geq e^{(C-t-1)\varepsilon_{\rm thr}}$ (or $p< p_{\rm thr} \equiv {C\choose t+1}^{-1/t}$ if noise is stochastic). Then, for any fixed $\delta_0>0$, any ideal quantum circuit of size $L$ and depth $D$ can be simulated with error $\delta_0$ by a noisy circuit of size $O\left(L(\log L)^{{\log_{t+1} c}}\right)$ and depth $O\left(D(\log L)^{{\log_{t+1} d}}\right)$.
\end{theo}

As was mentioned in the statement of properties \ref{prop:2}, an assumption in our construction of gadgets simulating measurements is that either classical processing is performed in encoded form using a classical error-correcting code or, if classical information is decoded to single classical bits, the classical gates that manipulate these bits are faultless. There is still a quantum accuracy threshold if the classical processing is noisy with noise strength $p_{\rm cl}$, but the error, $\delta_0$, is limited to be of order $p_{\rm cl}$ in this case: one can perform fault-tolerant quantum error correction either by measuring the syndrome and doing the classical processing using a classical error-correcting code, or by performing the error recovery coherently using quantum operations and {\em without} quantum measurement. Then, the only obstacle is decoding the final outcome of the computation to the level of single classical bits, and the error of this decoding will be limited to be of order $p_{\rm cl}$ due to the noisy classical gates.

%We can optimize our estimate of $\eta_0$ by setting $C= e^{(A-s)\eta_0}$ and solving eq.~(\ref{solve-for-eta}) for $\eta_0$, or more simply we may choose $C=e$, since the inequality $(A-s)^{s-1}\le \eta_0^{-(s-1)}=e {A\choose s}$ is satisfied for typical level-1 simulations of interest (e.g., for $s=2$).

%--------------------------------------------------------------%
\subsection{Improving the Threshold Estimate}
\label{sec:Malignant}

In theorem \ref{theo:4}, we estimated the quantum accuracy threshold by considering all possible ways, ${C\choose t+1}$, of placing $t+1$ faults in the largest 1-exRec. This estimate is however pessimistic since for some of these  configurations of faults, the 1-Rec contained inside the 1-exRec will still be correct. Let us now discuss how this observation can help us obtain improved rigorous lower bounds on the quantum accuracy threshold.

Let us say that a set of locations inside a 1-exRec is {\em benign} if the 1-Rec contained in the 1-exRec is correct for arbitrary fault operators acting on all locations in the set. Naturally, if a set of locations is not benign, we will say it is {\em malignant}. Then, in this language, lemma \ref{lem:1} says that all sets of locations with cardinality at most $t$ are benign; but we expect that many sets of locations with more than $t$ faults will also be benign. We can generalize definition \ref{defi:7} accordingly.
\begin{defi}[Goodness (revised and generalized)]
\label{defi:8}
Consider a fault-tolerant quantum circuit simulation based on a distance-$d$ quantum error-correcting code, and let $t=\lfloor {d-1\over 2}\rfloor$. A 1-exRec is {\em bad} if it contains faults at a malignant set of locations; if it is not bad  it is {\em good}. Two bad 1-exRecs are {\em independent} if they are nonoverlapping or if they overlap and the earlier 1-exRec is still bad when the shared 1-ECs are removed. 
\end{defi}

\noindent Then, lemma \ref{lem:1} is true by definition. Furthermore, with this generalized definition of goodness, lemma \ref{lem:3} will yield an improved estimate on the strength of noise that acts on the equivalent simulation that results after level reduction: Let us denote by $A$ the number of malignant sets of $t+1$ locations and let us pessimistically consider that all sets of at least $t+2$ locations are malignant. Then, for local stochastic noise, the bound (\ref{3.1.1}) can be replaced by the improved estimate
\begin{equation}
\label{3.1.13}
p^{(1)} \leq A p^{t+1} + B p^{t+2} \; ,
\end{equation}  

\noindent where $B\equiv {C\choose t+2}$ is the number of sets of $t+2$ locations in the largest 1-exRec. Equivalently, 
\begin{equation}
\label{3.1.14}
p^{(1)} \leq A' p^{t+1} \; ,
\end{equation}

\noindent which indicates an accuracy threshold, $p_{\rm thr} \equiv (A')^{-1/t}$. Here, $A'$ is obtained by solving the equation $A + B p_{\rm thr} = A'$, or
\begin{equation}
\label{3.1.15}
 A +B (A')^{-1/t} = A' \;.
\end{equation}

\noindent For instance, for $t=1$, we obtain $A'=\frac{1}{2}A\left(1 +\sqrt{1+4B/A^2}\right)$.

For local noise that is not stochastic, the sum of all fault paths with at least $t+2$ faults can be expressed as in lemma \ref{lem:2} with $s=t+2$. We need to add to this an additional contribution due to the sum of all fault paths  with faults at {\em exactly} $t+1$ malignant coarse-grained locations; let us denote this term as  $F(\mathcal{M}_{t+1})$. For each specified set of $t+1$ malignant coarse-grained locations, we place a fault operator at each of the locations in the set, and we place the ideal operations at all other locations inside the 1-exRec. Then, as in the proof of lemma \ref{lem:2}, we expand each ideal operation as the unitary coupling with the bath {\em minus} a fault operator. In this derived expansion, each term with faults at $m$ specific coarse-grained locations appears at most ${m\choose t+1}$ times (if every subset of $t+1$ out of the $m$ coarse-grained locations is malignant) and may appear as few as zero times (if no subset of $t+1$ out of the $m$ coarse-grained locations is malignant). On the other hand, each set of $m$ faulty coarse-grained locations occurs at most ${m-1 \choose t+1}$ times in the derived expansion arising from the original expansion of all fault paths with at least $t+2$ faults. Furthermore, these terms are opposite in sign to the terms that arise from the malignant sets of $t+1$ coarse-grained locations (the signs are $(-1)^{m-t-1}$ and $(-1)^{m-1-t-1}$ in the two cases). Therefore, after  combining the two expansions, each fault path with $m$ specific faults appears a number of times which is at most the larger of ${m-1\choose t+1}$ and ${m\choose t+1}- {m-1\choose t+1}={m-1\choose t}$. Since, $\forall m\geq t+2$, ${m-1\choose t} \leq (t+1) {m-1\choose t+1}$, we can upper bound the sum over all fault paths with faults in all malignant sets of $t+1$ coarse-grained locations or faults in at least $t+2$ coarse-grained locations similar to equation (\ref{2.2.13}):
\begin{equation}  
\label{3.1.16}
\varepsilon^{(1)} \leq || F(\mathcal{M}_{t+1}) + F(\mathcal{I}_{r\geq t+2}) ||_{\rm sup} \leq A \varepsilon^{t+1} + B \varepsilon^{t+2} \;,
\end{equation}

\noindent where $A$ is again the number of malignant sets of $t+1$ coarse-grained locations, and $B\equiv( t+1)\kappa {C\choose t+2}$ with $\kappa \geq e^{(C-t-2)\varepsilon}$. As before, we can write
\begin{equation}  
\label{3.1.17}
\varepsilon^{(1)} \leq A' \varepsilon^{t+1} \;,
\end{equation}

\noindent which implies $\varepsilon_{\rm thr} \equiv (A')^{-1/t}$ ($A'$ is again defined by equation (\ref{3.1.15})).

Overall, using the concept of malignant sets of locations, we have proved the following theorem.

\begin{theo}[Quantum Accuracy Threshold for Local Noise (revised)]
\label{theo:5}
Suppose gadgets can be constructed such that they satisfy properties \ref{prop:1} and \ref{prop:2} for a distance-$d$ quantum error-correcting code, and let $t=\lfloor{d-1\over 2}\rfloor$. Let $c$ be the largest size and $d$ be the largest depth in any 1-Rec, and let $C$ be the size of the largest 1-exRec. Suppose that noisy quantum circuits are afflicted by local noise of strength
\[
\varepsilon < \varepsilon_{\rm thr} \equiv \left( A' \right)^{-1/t} \;,
\]

\noindent (or $p< p_{\rm thr} \equiv (A')^{-1/t}$ if noise is stochastic) with $A'$ as in equation (\ref{3.1.15}) where $A$ is the largest number of malignant sets of $t+1$ locations in any 1-exRec and $B\geq (t+1) e^{(C-t-2)\varepsilon_{\rm thr}} {C\choose t+2}$ (or $B={C\choose t+2}$ if noise is stochastic). Then, for any fixed $\delta_0>0$, any ideal quantum circuit of size $L$ and depth $D$ can be simulated with error $\delta_0$ by a noisy circuit of size $O\left(L(\log L)^{{\log_{t+1} c}}\right)$ and depth $O\left(D(\log L)^{{\log_{t+1} d}}\right)$.
\end{theo}

It is clear that we can obtain a further improvement in our estimate of the quantum accuracy threshold by resolving the sets of $t+2$ or more coarse-grained locations into malignant and benign; but we will not discuss such extensions here.

%----------------------------------------------------------%
\section{History and Acknowledgements}

The analysis of recursive simulations and proofs of the quantum threshold theorem for independent stochastic noise  were first given by Aharonov and Ben-Or \cite{Aharonov99comb}, by Kitaev \cite{Kitaev97c}, and by Knill, Laflamme, and Zurek \cite{Knill96b}; similar discussions were also given in \cite{Preskill98} and \cite{Gottesman97}. All these proofs, with the exception of the discussion in \cite{Knill96b} which is given without proof, applied to computation that is encoded using a code of distance at least 5. At the same time, most estimates for the quantum accuracy threshold were derived using the distance-3 Steane code \cite{Steane96}; this created an unsatisfactory situation since the formal proofs did not apply to this code.

Knill, Laflamme and Zurek in \cite{Knill96b} sketched a method of analysis of fault-tolerant quantum computation based on a distance-3 code, but their argument only applied to level-1 simulations; in particular, it was not clear how to extend that argument to higher levels of a recursive simulation as is necessary for deriving an accuracy threshold condition. Starting in the spring of 2004, Gottesman, Preskill and I considered extending the argument in \cite{Knill96b} to obtain a rigorous proof. Thinking in terms of the notions of goodness and correctness proved to be extremely beneficial in deriving the proof in \cite{Aliferis05b} which is simple and, I think, elegant. The proof in \cite{Aliferis05b} follows the tradition set by \cite{Aharonov99comb} and analyzes the accuracy of level-$k$ gadgets using induction on $k$; in \cite{Aliferis05b} the inductive step was taken via a procedure we called {\em threshold dance} and which was suggested by Preskill. Shortly after \cite{Aliferis05b}, Reichardt also proved a quantum threshold theorem for distance-3 codes for the case of independent stochastic Pauli noise \cite{Reichardt05}.

Encouraged by our success with distance-3 codes, Gottesman, Preskill and I subsequently started thinking about distance-2 codes motivated by the scheme of postselected quantum computation and the exciting numerical results of Knill \cite{Knill04,Knill05}. It was by thinking about the proof of the threshold theorem for postselected quantum computation that I understood how to obtain the proof of the quantum threshold theorem using a level-reduction procedure. In retrospect, Knill, Laflamme, and Zurek had the right intuition in \cite{Knill96b} to considered level-1 simulations only: level reduction shows that recursive simulation can really be analysed one level at a time and so, in a certain sense, level 1 is all that matters! 

%----------------------------------------------------------%

%-----------------------------------------%
\chapter{Further Applications of Level Reduction}
\label{ch:extensions}
%----------------------------------------------------------%
\section{Introduction}

The proof of the quantum threshold theorem in the previous chapter was based on the level-reduction lemma. In essence, this lemma told us that recursive simulations can be analysed by considering how physical noise translates into an effective noise that acts on the encoded operations. Put differently, the concept of level reduction tells us that there are two questions to be addressed in order to prove the quantum threshold theorem: (i) Is our noise model general enough so that its structure is  preserved under level reduction? And, if yes (as for local noise), (ii) how are the parameters characterizing this noise model (i.e., the noise strength for local noise) transformed under level reduction? The answer to the first question will tell us whether the quantum threshold theorem can be proved for a given noise model; the answer to the second question will tell us what the accuracy threshold condition is.

In this chapter, I will discuss how this intuition for the analysis of recursive simulations can be used to prove the quantum threshold theorem in two new settings. First, in \S \ref{sec:Leakage}, I will extend the proof to apply to local leakage noise; i.e., our noise model will include the possibility that a qubit may become inaccessible to our control operations by {\em leaking} to additional degrees of freedom beyond the two-dimensional Hilbert space ideally characterizing it. The proof will proceed by constructing gadgets that under a {\em leakage-reduction} procedure, transform local leakage noise to regular local noise for which the level-reduction lemma applies. Second, in \S \ref{sec:MeasBased}, I will extend the proof to measurement-based models of quantum computation where elementary quantum gates are themselves simulated by gadgets. In this case again, the proof will proceed by showing that under a suitable {\em model-conversion} procedure, noisy measurement-based quantum computation is equivalent to noisy quantum circuit computation for which the level-reduction lemma applies. 

%----------------------------------------------------------%

%--------------------------------------------------------------%
\section{Leakage Noise}
\label{sec:Leakage}

%--------------------------------------------------------------%
%\subsection{Introduction: The Problem with Leakage}

In many physical implementations, a qubit is a two-dimensional {\em subspace} of a higher-dimensional Hilbert space.  %This can happen when we use encoded physical states as elementary qubits as in the decoherence-free-subspace (DFS) formalism.
Most usually this happens when a qubit is simply two low-lying levels in a many-level physical system; examples include an ion-trap qubit, an optical qubit which is the dual-rail subspace of two photonic modes \cite{Knill00}, a Josephson-junction qubit which is formed by the lowest energy states in a double well potential \cite{Orlando99}, and a quantum-dot qubit which is encoded in electron-spin states \cite{DiVincenzo00, Petta05}. For such qubits, {\em leakage faults} transfer amplitude in and out of the two-dimensional subspace defining the qubit and are likely to be an important source of errors. In this section, I will refer to the two-dimensional qubit space as the {\em system space}, and to its complement---i.e., the space occupied by qubits that have leaked out of the system space---as the {\em leakage space}.

The first protection against leakage faults is the use of low-decoherence qubit encodings and dynamical decoupling methods (as analyzed, e.g., in \cite{Tian00} and \cite{Byrd05}). After these methods are applied to reduce the occurrence of leakage, quantum error correction will need to be used to protect against the remaining rare leakage faults. However, the problem is that standard quantum error-correcting codes are not designed to correct errors due to leakage faults directly since the correctable error operators are typically only Pauli operators with support on the system space. Thus, before quantum error correction is applied, we first need to {\em convert} errors due to leakage faults to regular errors that occur in the system space. 

\begin{figure}[tbh]
\begin{center}
\setlength{\unitlength}{1cm}
\vspace{0.2cm} 
\parbox{1cm}{
\Qcircuit @C=2.3ex @R=2.3ex @!R {
            & \push{|\psi\rangle \hspace{0.1cm}}   & \ctrl{1} & \gate{X} & \ctrl{1} & \gate{X} & \qw & \push{\hspace{-0.2cm} |\psi\rangle }  \\
            & \push{|0\rangle \hspace{0.1cm}}      & \targ    & \qw      & \targ    & \meterz 
                                        }}
\vspace{0.2cm}                             
\caption{\label{fig:4.1} A circuit for detecting leakage in implementations where the {\sc cnot} gate acts trivially  if its control qubit is in the leakage space. To test a qubit for leakage, we use an ancillary qubit initialized in the state $|0\rangle$ and a sequence of two {\sc cnot} and two bit-flip, $X$, gates. If the qubit is in the system space, then its state is preserved but the state of the ancillary qubit is flipped to the state $|1\rangle$; hence, the measurement gives outcome $-1$. Otherwise, the ancillary qubit remains in the state $|0\rangle$ if leakage has occurred and the measurement gives outcome $+1$.  }
\end{center}
\end{figure}
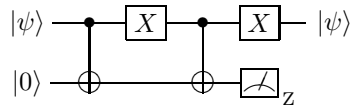

Converting qubits that have leaked out of the system space to erroneous qubits {\em in} the system space can be achieved in two ways: One way is to detect leakage and replace any leaked qubits by {\em new} qubits in the system space. For instance, in optical schemes for quantum computation, parity measurements of the photon occupation number in different modes can be performed that will indicate whether photon loss has occurred \cite{Knill00}. Another example is the indirect detection of leakage for ion-trap qubits by observing the emittion of photons from a cavity \cite{Pellizzari95}. Some leakage detection is also possible in solid-state schemes where qubits are encoded in two electron-spin states of a double quantum dot \cite{Petta05}. More generally, leakage can be detected by running each qubit through an appropriate leakage-detection circuit whose construction depends on the nature of the leakage space and the implementation of quantum gates. For instance, in implementations where the {\sc cnot} gate acts trivially if its control qubit is {\em not} in the system space, leakage detection can be achieved by the circuit in figure \ref{fig:4.1} \cite{Gottesman97,Preskill97}.

Let us refer to any procedure for detecting leakage as a {\em leakage-detection unit}, or {\em LDU}. If leakage detection specifically points to one qubit, we can simply replace the qubit by a new qubit initialized in some fixed state. But, in general, LDUs may act on multiple qubits and may not give us specific information about which qubit has leaked (e.g., in ion-trap schemes, detecting a photon escaping from the cavity indicates that one of the qubits inside the trap leaked but not which one). If leakage is detected on ancillary qubits, we can throw away all qubits and repeat the ancilla preparation circuit anew some fixed number of times. Otherwise, if leakage is detected on data qubits, we can use quantum teleportation to transfer the state of each qubit to a new qubit in the system space. 

Alternatively to using leakage detection, we can constantly refresh qubits so that any leaked ones will be quickly replaced by erroneous qubits in the system space without us detecting  whether leakage occurred; let us  refer to such a procedure as a {\em leakage-reduction unit}, or {\em LRU}. The quantum teleportation circuit in figure \ref{fig:1.1.2} is a natural implementation of an LRU \cite{Mochon04}. Indeed, if no leakage fault occurs during the execution of the teleportation circuit, the state of the third output qubit will always be in the system space independent of whether the first input qubit had leaked out of its system space or not. Hence, quantum teleportation offers a general solution to leakage reduction for cases when leakage detection, which depends on the specific implementation and the nature of the leakage space, is either limited or difficult to implement. 

The remaining of this section is organized as follows: First, in \S \ref{sec:LeakageModel}, I describe a Hamiltonian  local leakage noise model that generalizes the Hamiltonian noise model discussed in \S \ref{sec:LocalNoise}. In \S \ref{sec:LeakageReduction}, I describe how quantum circuits need to be modified to protect against leakage faults %by inserting an LRU before every elementary gate (or LDUs if leakage detection is possible); of course, LRUs (or LDUs) are subject to regular and leakage faults themselves. Then, 
and I prove the central result of this analysis, the {\em leakage-reduction} lemma. %which shows that recursive simulations that use LRUs and are afflicted by local leakage noise are equivalent to recursive simulations without LRUs which are afflicted by regular local noise. 
In \S \ref{sec:LeakageThreshold}, I use the leakage-reduction lemma together with the level-reduction lemma from the previous chapter to prove the quantum threshold theorem for local leakage noise. Finally, in \S \ref{sec:Stretching}, I discuss conditions for where LRUs must be placed inside a quantum circuit and where they can be omitted. %In chapter \ref{sec:xxx} we illustrate these concepts with circuits for the Steane [[7,1,3]] code.

%--------------------------------------------------------------%
\subsection{The Local Leakage Noise Model}
\label{sec:LeakageModel}

Let ${\cal H}_S$ be the system Hilbert space of all qubits, with ${\cal H}_{S[i]}$ the Hilbert space of qubit $i$. There is an extension of each ${\cal H}_{S[i]}$, ${\cal H}_{S_{ext}[i]}={\cal H}_{S[i]} \oplus {\cal H}_{L[i]}$, so that ${\cal H}_{L[i]}$ is the leakage space of qubit $i$. Furthermore, we will assume that ${\cal H}_{L}=\otimes_i {\cal H}_{L[i]}$, i.e., the leakage spaces of different qubits are disjoint; this condition is fulfilled in most physical systems. Similar to \S \ref{sec:LocalNoise}, we will assume the following Hamiltonian
\begin{equation} 
\label{4.2.1}
H = H_S + H_B + H _{SB} + H_{\rm leak} \;, 
\end{equation}

\noindent where the additional leakage part of the Hamiltonian is of the form
\begin{equation} 
\label{4.2.2}
H_{\rm leak}(t)=H_{SL(B)}(t)+H_L(t) \;.
\end{equation}

\noindent Here, $H_L$ describes the arbitrary evolution in the leakage spaces and any coupling of the leakage spaces to the bath, and the time-dependent $H_{SL(B)}$ is a linear combination of operators coupling the leakage and system spaces and, potentially, the bath, $B$. An example of leakage that includes a coupling to a bath is a trapped ion in a cavity where transitions into or out of the two-level subspace can create or annihilate photons in the cavity. We will take $H_{SL(B)}$ to be of the form
\begin{equation} 
\label{4.2.3}
H_{SL(B)}(t)=\sum_{j} H_{SL(B),j} \;,
\end{equation}

\noindent where $H_{SL(B),j}$ only couples the system and leakage spaces of the set of qubits on the support of  location $j$ in the ideal quantum computation.

This noise model can be analysed using time-resolved fault paths as in \S \ref{sec:LocalNoise}. Let us assume that for all noisy locations the system-bath coupling for regular faults has the form of equation (\ref{2.2.3}) and satisfies condition (\ref{2.2.6}), and let us also make a similar assumption for the coupling between the system and leakage spaces, $|| H_{SL(B),j} ||_{\rm sup} \leq \lambda_1$. Then, we can conclude that the sum of all fault paths with faults at $r$ specific coarse-grained locations, $\mathcal{I}_r$, has sup norm $|| F(\mathcal{I}_r) ||_{\rm sup} \leq  ((\lambda_0 + \lambda_1) \tau)^r$. Hence, our leakage noise model is local with strength $\varepsilon \equiv (\lambda_0 + \lambda_1) \tau$.

\subsection{The Leakage-Reduction Lemma}
\label{sec:LeakageReduction}

Before proceeding further, at this point it will be helpful to pause and discuss in more detail why leakage faults require a separate fault tolerance analysis. What distinguishes leakage faults from regular qubit faults is that once a qubit has leaked, its subsequent self-evolution or interaction with other qubits will depend on the details of the leakage space which vary in different implementations. In particular, if one of the input qubits to a multi-qubit quantum gate has leaked, the gate will in principle operate on the extended space of {\em all} its input qubits even if it is executed ideally. Therefore, leakage errors cannot be propagated in a simple way through ideal gates since this propagation depends on the particular gate implementation, i.e., the specification of how gates operate on the extended space. Since the design of fault-tolerant quantum circuits is guided by the particular ways in which errors propagate through ideal gates, a circuit designed to be robust again regular qubit faults will not, in general, be effective in maintaining fault tolerance against leakage.

%Consider a quantum circuit realizing a recursive simulation designed to protect against regular qubit faults. To ensure that all ideal 0-Ga's in this circuit act in the system space of their inputs, we insert an LRU before every \emph{gate} 0-Ga. 
To ensure that all faultless 0-Gas in a quantum circuit act in the system space of their inputs, we will consider  inserting an LRU before every \emph{gate} 0-Ga. 
%After inserting LRUs before every gate 0-Ga, we can think of 1-exRecs as being transformed to what we can call LRU-1-exRecs. Similarly, $k$-exRecs are transformed to LRU-$k$-exRecs, etc.
We need not place LRUs preceding measurement 0-Gas which have only classical output or preceding qubit-preparation 0-Gas which have no quantum input. When an LRU is executed without a fault, it guarantees that (i) if its input is in the system space then the identity operation is applied, and (ii) if its input is in the leakage space then the output qubit state is in the system space. As discussed above, the quantum teleportation circuit in figure \ref{fig:1.1.2} is a natural way to satisfy these two requirements. If leakage detection per qubit is possible, we can alternatively place an LDU before every gate 0-Ga. An LDU outputs a classical bit indicating whether leakage has occurred, in which case the leaked qubit can be replaced by a new qubit in some fixed state in the system space. In the discussion that follows, we will assume LRUs precede gate 0-Gas, but the idea is essentially the same if LDUs are used instead.  

Given a quantum circuit with LRUs before every gate 0-Ga, it is clear that {\em if} the LRUs are faultless, this circuit subjected to leakage faults is equivalent to the circuit without LRUs subjected to regular qubit faults. This is because LRUs replace any leaked input with a state in the system space before the next 0-Ga is applied. However an LRU can fail. It can fail due to a regular qubit fault, but it can also fail due to a leakage fault; e.g., if LRUs are implemented by quantum teleportation, there can be a leakage fault acting on the outgoing qubit. %Hence, a new fault-tolerance analysis is required to show that fault paths with sufficiently few leakage or regular faults will give the ideal outcome probability distribution; but, fortunately, this analysis will only require a minor extension of the results in Chapters \ref{ch:syntax} and \ref{ch:threshold}. 
The essential insight for analyzing leakage is that an LRU implements leakage correction at the physical level, {\em level $0$}, just as a $k$-EC gadget performs regular error correction at level $k$ of a recursive simulation. The leakage-reduction lemma that follows formalizes this intuition: This lemma allows us to transform a quantum circuit where level-$0$ leakage correction is performed by LRUs to a quantum circuit where LRUs are removed and the effective noise only includes regular qubit faults. %After proving this lemma, we can apply the threshold theorem for regular local noise to this equivalent simulation thus proving the threshold theorem for our local leakage noise model.

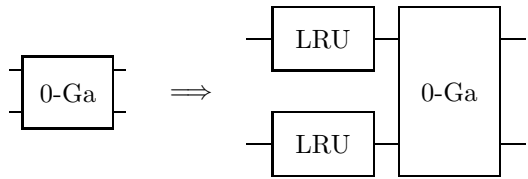
\begin{figure}[tbh]
\begin{center}
\setlength{\unitlength}{0.035cm}
\vspace{0.5cm}
%\leavevmode
%\epsfysize=2.5cm
%\epsfbox{leak-18.eps}
%\includegraphics[width=6.5cm]{leak-18}
%\epsfig{file=leak-18.eps,width=6.5cm}
%
\begin{picture}(172,14)
\put(-10,-16){\line(1,0){5}}
\put(-10,0){\line(1,0){5}}
\put(-5,-22){\framebox(34,27){\shortstack{$0$-Ga}}}
\put(29,-16){\line(1,0){5}}
\put(29,0){\line(1,0){5}}
\put(44,-15){\makebox(30,12){$\Longrightarrow$}}
\put(80,12){\line(1,0){10}}
\put(90,0){\framebox(38,24){\shortstack{LRU}}}
\put(80,-28){\line(1,0){10}}
\put(90,-40){\framebox(38,24){\shortstack{LRU}}}
\put(128,12){\line(1,0){10}}
\put(128,-28){\line(1,0){10}}
\put(138,-40){\framebox(38,64){\shortstack{$0$-Ga}}}
\put(176,12){\line(1,0){10}}
%\put(106,0){\framebox(38,24){\shortstack{LRU}}}
%
\put(176,-28){\line(1,0){10}}
%\put(106,-40){\framebox(38,24){\shortstack{LRU}}}
%
%\put(144,12){\line(1,0){10}}
%\put(144,-28){\line(1,0){10}}
%
%\put(190,-15){\makebox(20,12){.}}
\end{picture}
\setlength{\unitlength}{1cm}
\vspace{1.5cm}
\caption{\label{fig:4.2} To combat leakage faults, we insert LRUs on the inputs of every gate 0-Ga. The combination of a 0-Ga with its preceding LRUs we call a {\em 0-Rec}.}
\end{center}
\end{figure}

But we first need some additional definitions; let us start by defining {\em level-0 rectangles}, or {\em 0-Rec}s. For quantum circuits without LRUs this would be trivial: a 0-Rec is just a 0-Ga and it is bad (and incorrect) when it is faulty. When LRUs are inserted in a quantum circuit, for every gate 0-Ga we define a {\rm 0-Rec} as the 0-Ga preceded by LRUs on all its inputs (Fig.~\ref{fig:4.2}); for a qubit-preparation or measurement 0-Ga, the {\rm 0-Rec} coincides with the 0-Ga. We will also say that a 0-Rec {\em simulates} the 0-Ga that remains after the LRUs are removed.

Similar to definition \ref{def:5}, we would like to define what it means for 0-Recs to be correct. In this new definition, the role of an ideal decoder will now be played by an {\em ideal LRU,} or {\em i-LRU}. This device can be thought of as an ideal (i.e., faultless) execution of quantum teleportation where the classical bits are discarded after any Pauli correction operators are applied; just as i-decoders, i-LRUs are tools for our analysis and do not represent physical operations applied during the computation. The definition of correctness and goodness for 0-Recs is as follows.

\begin{defi}[Correctness for 0-Recs]
\label{def:9}
A 0-Rec is {\em correct} if the 0-Rec followed by i-LRUs is equivalent to i-LRUs followed by the ideal 0-Ga that the 0-Rec simulates.
\end{defi}

\begin{defi}[Goodness for 0-Recs]
\label{def:10}
A 0-Rec is {\em good} if it contains {\em no} leakage or regular faults, otherwise it is bad.
\end{defi}

Since a good 0-Rec contains no faults, its {\rm LRU}s operate ideally; hence, the following lemma holds.

\begin{lem}
\label{lem:3.5}
A good 0-Rec is correct.
\end{lem}

Definition \ref{def:10} also applies to preparation and measurement 0-Recs which do not contain any LRUs: these 0-Recs are also good when they are faultless. Moreover, the correctness of a good measurement 0-Rec implies that it can be replaced by an i-LRU followed by the corresponding ideal measurement 0-Ga providing a means to create i-LRUs (see \S \ref{sec:badtofaulty}). Similarly, correctness implies that a good preparation 0-Rec followed by an i-LRU can be replaced by the corresponding ideal preparation 0-Ga alone, thus annihilating the i-LRU.

We can now state our basic lemma.

\begin{lem}[Leakage Reduction]
\label{lem:4} 
Consider any quantum circuit where {\rm LRU}s have been inserted preceding every gate {\rm 0-Ga}, and let $C_{\rm leak}$ be the maximum number of locations in any 0-Rec. Then, if this circuit is subject to local leakage noise of strength $\varepsilon$, there is a quantum circuit with the LRUs removed that produces the same output probability distribution and is subject to regular local noise with strength
\[
\tilde{\varepsilon} \leq  C_{\rm leak} \, \varepsilon \; .
\]

%\noindent where $\kappa_0 \geq e^{(C_{\rm leak} - 1)\varepsilon}$.
\end{lem}

\noindent {\bf Proof}. The proof is similar to the proof of the level-reduction lemma. We imagine first creating i-LRUs out of measurement 0-Recs thereby transforming them to measurement 0-Gas. Next, we imagine propagating these i-LRUs to the left through gate 0-Recs thereby transforming them to gate 0-Gas. Finally, we imagine annihilating the i-LRUs inside preparation 0-Recs thereby transforming them to preparation 0-Gas. By lemma \ref{lem:3.5}, every good 0-Rec will be thus replaced by the ideal simulated 0-Ga; in \S \ref{sec:badtofaulty}, we will also show that every bad 0-Rec can be transformed to a faulty 0-Ga that operates on the system space. Hence, with this maneuver which we can visualize as an ``i-LRU wave'' propagating from the right to the left of the quantum circuit which includes LRUs, we will obtain a circuit with the LRUs removed that realizes exactly the same computation, and this circuit is subject to regular qubit faults alone.

Consider some set, $\mathcal{I}^{(0)}_r$, of $r$ specific locations in this equivalent circuit. In order for these locations to be faulty, the corresponding 0-Recs in the initial circuit must all be bad, i.e., they must each contain at least one fault. Then, by using a similar counting argument as in the proof of the level-reduction lemma, we can upper bound the sup norm of the sum of all fault paths with faults at all 0-Recs corresponding to the locations in $\mathcal{I}^{(0)}_r$ by
\begin{equation} 
\label{4.2.4}
|| F(\mathcal{I}^{(0)}_{r}) ||_{\rm sup} \leq \left( \kappa_0 C_{\rm leak} \varepsilon \right)^r \;,
\end{equation}

\noindent where $\kappa_0 \geq e^{(C_{\rm leak} -1 )\varepsilon}$. In fact, it is sufficient to take $\kappa_0=1$ as can be seen by repeating the analysis of time-resolved fault paths in \S \ref{sec:LocalNoise} and considering each 0-Rec as {\em one} operation that takes time $C_{\rm leak} \, \tau$. We conclude that the noise acting on the equivalent circuit is local with strength $\tilde{\varepsilon} \leq C_{\rm leak} \, \varepsilon$.

\noindent This proves the lemma. 

\rightline{$\square$}

There is a corresponding threshold condition for the case when local leakage noise is stochastic. However, it is important to emphasize that in most physical settings, leakage is an inherently non-Markovian process, in particular when it does not involve interactions with a bath which can induce decoherence between different fault paths. In cases when leakage faults can be assigned a probability, $p$, the same analysis holds where now the noise strength afflicting the equivalent circuit will be $\tilde{p} \leq C_{\rm leak} \, p$.

%----------------------------------------------------------------------------------------------------%
\subsubsection{Bad Level-0 Rectangles As Simulated Faults}
\label{sec:badtofaulty}

It remains to explain how i-LRUs can be moved to the left of bad 0-Recs thereby transforming them to faulty 0-Gas that act on the system space. The idea for accomplishing this is essentially the same as for converting bad 1-exRecs to simulated faulty 0-Gas as discussed in \S \ref{sec:BadRecs}. 

\begin{figure}[tb]
\begin{center}
\vspace{0.2cm}  
\parbox{1cm}{
\Qcircuit @C=2.3ex @R=2.3ex @!R {
            & \push{|\psi\rangle \hspace{0.1cm}} & \qw      & \qw      & \ctrl{1} & \gate{H}     & \qw      & \ctrl{2} & \qw & \qw \\
            & \push{|0\rangle \hspace{0.1cm}}    & \gate{H} & \ctrl{1} & \targ    & \qw          & \ctrl{1} & \qw  & \qw & \qw \\
            & \push{|0\rangle \hspace{0.1cm}}    & \qw      & \targ    & \qw      & \qw          & \targ    & \control \qw & \qw & \qw
                                        }}
\vspace{0.2cm}                             
\caption{\label{fig:4.3} A coherent version of quantum teleportation that can, e.g., by used to construct our coherent i-LRUs. Instead of measuring the first two qubits, we apply the Pauli corrections coherently via a Hadamard rotation followed by a {\sc cnot} and a {\sc cphase} gate. If we define the {\sc cnot} and {\sc cphase} gates in this circuit  to act trivially if their control qubits are in the leakage space, then independent of the state of the input qubit, the third qubit is always in some state in the system space. }
\end{center}
\end{figure}
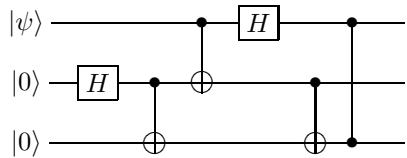

The first step is to consider the coherent version of an i-LRU. Like i-decoders, i-LRUs generate a syndrome; taking i-LRUs to be ideal quantum teleportation circuits, the syndrome consists of the measurement bits. We can then define a coherent invertible teleportation in which one, instead of measuring, performs {\sc cnot} and  {\sc cphase} gates on the output qubit (Fig.~\ref{fig:4.3}). Furthermore, it is possible to {\em define} the {\sc cnot} and {\sc cphase} gates in this i-LRU circuit to act trivially on their target qubits when their control qubits are in the leakage space. Then, the output qubit will always be in the system space independent of whether the input qubit has leaked or not; leakage will be confined to the {\em syndrome qubits}, i.e., the first two output qubits in figure \ref{fig:4.3}. Let us denote a  coherent i-LRU as $\mathcal{LRU}$, so that its inverse is denoted as $\mathcal{LRU}^{-1}$. Hence, for our analysis, we can create {\em LRU pairs}, $\mathcal{LRU}^{-1} \cdot \mathcal{LRU}=I$, and insert them at any point during the quantum computation.  

Consider first those qubits in a quantum circuit that are measured. A measurement can be viewed as an ${\cal LRU}$ followed by another measurement whose input is always in the system space: We can insert an LRU pair before every measurement 0-Ga, ${\cal M}$, and identify ${\cal M}'= {\cal M} \cdot {\cal LRU}^{-1}$ as a measurement 0-Ga that  always acts on the system space (and depends on the syndrome qubits input to the ${\cal LRU}^{-1}$). We can next propagate the ${\cal LRU}$s of each LRU pair to the left through good 0-Recs transforming them to the ideal simulated 0-Gas. Indeed, since lemma \ref{lem:3.5} applies to i-LRUs that measure the syndrome qubits and finally discard the syndrome information, the same is also true for coherent i-LRUs and do not measure the syndrome qubits. When we encounter a bad 0-Rec, we insert an LRU pair preceding it in all its inputs. The leading ${\cal LRU}$ in each LRU pair can then be moved further to the left through the remaining 0-Recs until we encounter another bad 0-Rec and repeat by inserting new LRU pairs preceding it.

Furthermore, a bad 0-Rec grouped together with the ${\cal LRU}^{-1}$ or ${\cal LRU}^{-1}$s preceding it and the ${\cal LRU}$ or ${\cal LRU}$s  succeeding it can be interpreted as a faulty 0-Ga that acts in the system space. For example, consider a 0-Rec simulating a single-qubit 0-Ga. The input to the ${\cal LRU}^{-1}$ is a qubit always in the system space and some input syndrome qubits. Similarly, the ${\cal LRU}$ outputs a qubit always in the system space and some syndrome qubits. Hence, the combined operation ${\cal LRU} \cdot\,$0-Rec$\,\cdot {\cal LRU}^{-1}$ can be viewed as some 0-Ga supported in the system space alone. For a good 0-Rec, this 0-Ga is exactly the 0-Ga simulated by the 0-Rec---this is our correctness property. What we have shown is that even bad 0-Recs can be viewed as simulating some faulty 0-Ga that is supported on the system space; and different faulty 0-Gas may be {\em correlated} as they depend on the syndrome processed by the LRU pairs.   

%-------------------------------------------------------------------------%
\subsection{The Quantum Threshold Theorem for Local Leakage Noise}
\label{sec:LeakageThreshold}

The leakage-reduction lemma states that local leakage noise is effectively transformed to regular local noise if we insert LRUs preceding every gate 0-Ga in a quantum circuit. We will now consider inserting LRUs preceding every gate 0-Ga in a quantum circuit, $\mathcal{C}_{\rm rec}$, that realizes a recursive fault-tolerant simulation designed to be robust against regular local noise as described in chapter \ref{ch:threshold}; let the resulting quantum circuit be denoted as $\mathcal{C}_{\rm rec}^{\rm LRU}$. Then, the leakage-reduction lemma tells us that local leakage noise of strength $\varepsilon$ acting on $\mathcal{C}_{\rm rec}^{\rm LRU}$ is equivalent to regular local noise of strength at most $C_{\rm leak} \varepsilon$ acting on $\mathcal{C}_{\rm rec}$. Applying theorem \ref{theo:5} to $\mathcal{C}_{\rm rec}$ completes the proof of the quantum threshold theorem for local leakage noise which is stated next.

\begin{theo}[Quantum Accuracy Threshold for Local Leakage Noise]
\label{theo:6}
Suppose gadgets can be constructed such that they satisfy properties \ref{prop:1} and \ref{prop:2} for a distance-$d$ quantum error-correcting code, and let $t=\lfloor{d-1\over 2}\rfloor$. Let $c$ be the largest size and $d$ be the largest depth in any 1-Rec. Let $C$ be the size of the largest 1-exRec, and $C_{\rm leak}$ be the size of the largest 0-Rec. Suppose that noisy quantum circuits are afflicted by local leakage noise of strength
\[
\varepsilon < \varepsilon_{\rm thr} \equiv \left( A' \right)^{-1/t} / C_{\rm leak} \;,
\]

\noindent (or $p< p_{\rm thr} \equiv (A')^{-1/t} / C_{\rm leak}$ if noise is stochastic) with $A'$ as in equation (\ref{3.1.15}) where $A$ is the largest number of malignant sets of $t+1$ faults in any 1-exRec and $B\geq (t+1) e^{(C-t-2)\varepsilon_{\rm thr}} {C\choose t+2}$ (or $B={C\choose t+2}$ if noise is stochastic). Then, for any fixed $\delta_0>0$, any ideal quantum circuit of size $L$ and depth $D$ can be simulated with error $\delta_0$ by a noisy circuit of size $O\left(L(\log L)^{{\log_{t+1} c}}\right)$ and depth $O\left(D(\log L)^{{\log_{t+1} d}}\right)$.
\end{theo}

The value of $C_{\rm leak}$ depends on our method for realizing LRUs. If $r$ is the number of elementary operations in an LRU, then $C_{\rm leak} \leq 2r+1$ since we can restrict to 0-Recs simulating two-qubit 0-Gas and such 0-Recs contain two LRUs. In particular, if LRUs are implemented by the quantum teleportation circuit in figure \ref{fig:1.1.2}, then $r=7$ and the accuracy threshold for local leakage noise is diminished by at most an order of magnitude compared to the accuracy threshold for regular local noise. However, this conclusion is overly pessimistic since it is based on our assumption that LRUs are inserted preceding every gate 0-Ga. Although this assumption is helpful for proving theorem \ref{theo:6}, it is not necessary. In the next section, we will discuss conditions for when LRUs can be omitted while still maintaining protection against leakage. %As I will discuss in the next section, protecting against  leakage requires LRUs to be inserted preceding only a typically small fraction of all gate 0-Gas.

%It is clear that leakage reduction implemented by teleportation may require frequent measurement. This could potentially be problematic in situations where measurements are slow compared to fundamental gate-times, as e.g. in many (solid-state) implementations of quantum computation. However, data qubits do not always need to wait for the measurements to finish. As with regular error-correction where measurements are used, the Pauli corrections that result from teleportation can in many cases be kept in a classical memory and they need only be used to adapt the non-Clifford parts of the computation. In fact, since non-Clifford gates are used in logical parts of the computation and not during error-correction, the measurement outcomes of teleportations can be combined with the results of error-correction to give a joint Pauli correction operator. Therefore, assuming classical computation is fast and robust, leakage reduction via teleportation will not suffer any additional time overheads than the overheads already associated with regular error-correction that uses measurements.

%-----------------------------------------------------------------------------------------------------------%
\subsection{Contracting Level-0 Rectangles}
\label{sec:Stretching}

Our motivation for omitting LRUs between successive 0-Gas is twofold: First, eliminating unnecessary LRUs will make the fault-tolerant circuits more efficient. And, second, we expect that this will help increase the accuracy threshold for leakage noise bringing it closer to the threshold for regular noise. The idea of omitting LRUs also inspires a similar modification in fault-tolerant circuits designed to protect against regular qubit errors. In that case, it may be similarly advantageous to omit error-correction gadgets; we will return to this point in the next chapter. % between the application of successive gate gadgets. For instance, if the strength of noise in memory is low enough, it may be advantageous to use a replacement rule that encodes one memory location by parallel memory locations at the next level of recursion {\em without} a following error-correction gadget (see figure \ref{fig:2.1}).

\begin{figure}[h]
\begin{center}
\setlength{\unitlength}{0.03cm}
\vspace{0.7cm} 
%\leavevmode
%\epsfysize=3.2cm
%\epsfbox{leak-20.eps}
%\includegraphics[width=7.5cm]{leak-20}
%\epsfig{file=leak-20.eps,width=7.5cm}
%
\begin{picture}(172,90)
\put(-50,92){\line(1,0){10}}
\put(-40,80){\framebox(38,24){\shortstack{LRU}}}
\put(-50,52){\line(1,0){10}}
\put(-40,40){\framebox(38,24){\shortstack{LRU}}}
\put(-2,92){\line(1,0){10}}
\put(-2,52){\line(1,0){10}}
\put(8,40){\framebox(38,64){\shortstack{$0$-Ga}}}
\put(46,92){\line(1,0){10}}
%%\put(56,80){\framebox(38,24){\shortstack{LRU}}}
%
\put(46,52){\line(1,0){50}}
%
%%\put(94,92){\line(1,0){10}}
\put(94,52){\line(1,0){10}}
\put(46,12){\line(1,0){10}}
\put(56,0){\framebox(38,24){\shortstack{LRU}}}
\put(94,12){\line(1,0){10}}
\put(104,0){\framebox(38,64){\shortstack{$0$-Ga}}}
\put(142,52){\line(1,0){10}}
\put(142,12){\line(1,0){10}}
\put(152,40){\framebox(38,24){\shortstack{LRU}}}
\put(190,52){\line(1,0){10}}
\put(152,0){\framebox(38,24){\shortstack{LRU}}}
\put(190,12){\line(1,0){10}}
\end{picture}
\setlength{\unitlength}{0.035cm}
\caption{\label{StRec-example}  An example of a 0-conRec obtained by contracting two successive 0-Recs that  simulate two two-qubit 0-Gas.} \vspace{-0.3cm}
\end{center}
\end{figure}
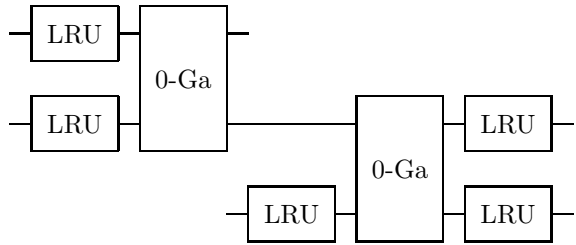

Consider the union of two or more consecutive 0-Recs with any intermediate LRUs omitted. We will call such an object a {\em contracted level-0 rectangle} or {\em 0-conRec}; an example is shown in figure \ref{StRec-example}. If this   0-conRec contains no faults, then the initial LRUs will operate ideally and the 0-conRec will be correct. In this case, the omission of the intermediate LRU has no consequence. However, assume instead that a leakage fault occurred in the first 0-Ga inside the 0-conRec only. Then we cannot interpret the second 0-Ga as an ideal 0-Ga even though it is executed without fault. This is because if the input to a 0-Ga is not in the system space, then the action of the 0-Ga can in principle be arbitrary in the extended space of {\em all} its inputs. This shows that in general, if the first 0-Ga inside a 0-conRec fails, all subsequent 0-Gas inside the 0-conRec will fail as well. 

In general, lemma$\,$\ref{lem:3.5} applies also to 0-conRecs; i.e., a 0-conRec without faults followed by i-LRUs is equivalent to i-LRUs followed by the ideal sequence of 0-Gas the 0-conRec simulates. On the other hand, if a 0-conRec is bad, it can be viewed as implementing the same sequence of 0-Gas but with all 0-Gas potentially faulty\footnote{This can be achieved by inserting LRU pairs inside the bad 0-conRec as in \S \ref{sec:badtofaulty}.}.

Consider finally a recursive simulation where LRUs have been inserted preceding every gate 0-Ga to protect against leakage. Let us imagine we have contracted $r$ successive 0-Recs to form a 0-conRec, and also let $\mathcal{I}_r^{(0)}$ be the set of locations that are simulated by this 0-conRec. Then, this 0-Rec contraction is allowed if no subset of  $\mathcal{I}_r^{(0)}$ forms a malignant set inside any 1-exRec. For instance, a case where contracting 0-Recs is allowed is in a sequence of 0-Recs simulating single-qubit gate or memory 0-Gas: in this case, it suffices to keep the LRU at the beginning and omit all others, since any sequence of single-qubit faulty 0-Gas is no worse than just one fault in any of these 0-Gas. Another example is a 0-Rec simulating a preparation 0-Ga followed by a 0-Rec simulating a two-qubit gate 0-Ga, where again the intermediate LRU can be omitted\footnote{Then, note that the resulting 0-conRec will contain no LRUs.}.  

%------------------------------------------------------------------%

%----------------------------------------------------------------------%
\section{Measurement-Based Quantum Computation}
\label{sec:MeasBased}

In this section, I will discuss fault-tolerant quantum computation in cases where every gate in an ideal quantum circuit is {\em simulated} by a sequence of quantum measurements on an larger Hilbert space that includes ancillary qubits. I will refer to such schemes where the basic primitive is quantum measurement as models of {\em measurement-based} quantum computation. Although several such models were initially proposed \cite{Raussen01,Nielsen01,Leung01,Leung03}, it has become clear that they are all related to one another \cite{Verstraete03, Aliferis04, Childs04, Perdrix04} and, moreover, they can all be understood in terms of a simple computation primitive which is known as {\em gate teleportation}.

The concept of gate teleportation was developed to address the problem of constructing a quantum universal set of gate gadgets satisfying properties \ref{prop:2} \cite{Shor96,Gottesman97,Gottesman99,Zhou00}. On the other hand, the interest in measurement-based quantum computation stems primarily from questions relating to physical implementations. In particular, computation in measurement-based models proceeds by preparing suitable ancillary states {\em off-line}, i.e., separate to the main computation and, after preparation is complete, performing a joint quantum measurement on the ancillary qubits and some data qubits that is equivalent to applying a quantum gate on the latter. Thus, since some of the complexity of performing quantum gates is moved to the off-line preparation of ancillary states, measurement-based models are beneficial in those experimental situations where the direct implementation of quantum gates is problematic (see, e.g., \cite{Nielsen04}).  
    
The rest of this section is structured as follows: In \S \ref{sec:GateTel}, I discuss the concept of gate teleportation and how it relates to measurement-based quantum computation. Then, in \S \ref{sec:ModelReduction}, I will prove the {\em model-conversion} lemma which shows that noisy measurement-based quantum computation can be analyzed in terms of an equivalent noisy quantum-circuit computation. This will then naturally lead to a proof of the threshold theorem for local noise for measurement-based quantum computation which I give in \S \ref{sec:MBThreshold}. %Finally, in \S \ref{sec:LeakInMBC}, I discuss the problem of leakage in  measurement-based computation and show that, because of the frequent use of quantum measurement, measurement-based computation schemes can be made robust against local leakage noise without the use of explicit LRUs. 

%----------------------------------------------------------------------%
\subsection{Gate Teleportation and Quantum Computation by Measurements}
\label{sec:GateTel}

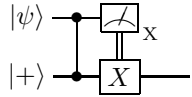
\begin{figure}[t]
\begin{center}
\vspace{0.2cm}
\setlength{\unitlength}{1cm}
\parbox{1cm}{
\Qcircuit @C=2.2ex @R=2.2ex @!R { 
            & \push{|\psi\rangle \hspace{0.1cm}} & \ctrl{1} & \meterx \cwx[1] \\
            & \push{|+\rangle \hspace{0.1cm}}    & \control \qw    & \gate{X} & \qw & \qw  
                                        }}                       
\vspace{0.2cm}                             
\caption{\label{fig:4.5} The output of the circuit is $H|\psi\rangle$; i.e., the input state has been {\em teleported} to the second qubit up to the application of a Hadamard gate. Recall that $|+\rangle \propto |0\rangle + |1\rangle$ is the $+1$ eigenstate of the Pauli operator $X$.
         }
\end{center}
\end{figure}

Quantum teleportation can be thought of as the answer to the question: Can we transfer the state of one qubit to another qubit?\footnote{We demand that we do not know the state to be transferred because, if we did, then the protocol is trivial: we could just prepare a second qubit in that exact state.} The answer is affirmative \cite{Bennett93} as illustrated by the circuit in figure \ref{fig:1.1.2}: It suffices to prepare two ancillary qubits in a {\em Bell state}, $|\Phi_0\rangle \propto |00\rangle + |11\rangle$, perform a {\sc cnot} gate from our first qubit to one of the two ancillary qubits and measure both our first qubit and this ancilla. The second ancillary qubit then acquires the state of the first qubit, $|\psi\rangle$, up to Pauli operators that depend on the outcomes of the two measurements. The entangled state $|\Phi_0\rangle$ is necessary if the qubits carrying the unknown state $|\psi\rangle$ in the beginning and in the end cannot interact directly. Else, simpler quantum teleportation circuits exist \cite{Zhou00} as, e.g., the one in figure \ref{fig:4.5}.

This circuit can alternatively be viewed as implementing a {\em teleportation} of the Hadamard gate. That is, instead of looking at this circuit as transferring the state $|\psi\rangle$ from the first to the second qubit, we can instead consider the Hadamard gate as being teleported. In general, the term {\em gate teleportation} denotes any procedure that achieves the simulation of some quantum gate using ancillary states and gates which are, depending on the context, easier to implement \cite{Shor96,Gottesman97,Gottesman99,Zhou00}. In chapter \ref{ch:lower-bounds}, we will discuss how the idea of gate teleportation is  used to obtain a universal set of gate gadgets satisfying properties  \ref{prop:2}; in this context, the gates that are easy to implement belong to the Clifford group. In measurement-based models \cite{Nielsen01,Raussen01,Leung03}, the set of easy to implement operations includes some fixed entangling gate (e.g., the {\sc cnot} or {\sc cphase} gate), the initialization of qubits in fixed Pauli eigenstates (e.g, the $+1$ eigenstates of $Z$ or $X$, $|0\rangle$ and $|+\rangle$ respectively) and single-qubit or two-qubit measurements along arbitrary orthonormal bases. 

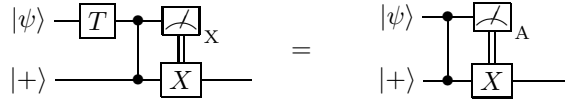
\begin{figure}[t]
\begin{center}
\hspace{-0.5cm}
\setlength{\unitlength}{1cm} 
\parbox{1cm}{
\Qcircuit @C=2.2ex @R=2.2ex @!R {  
            & \push{|\psi\rangle \hspace{0.1cm}} & \gate{T} & \ctrl{1} & \meterx \cwx[1] \\
            & \push{|+\rangle \hspace{0.1cm}}    & \qw      & \control \qw    & \gate{X} & \qw & \qw  
                                        }} \hspace{0.3cm} = \hspace{0.3cm}
\parbox{1cm}{
\Qcircuit @C=2.3ex @R=2.6ex @!R { 
            & \push{|\psi\rangle \hspace{0.1cm}} & \ctrl{1} & \meterobs \cwx[1] \\
            & \push{|+\rangle \hspace{0.1cm}}    & \control \qw    & \gate{X} & \qw & \qw  
                                        }}                          
\vspace{0.2cm}                             
\caption{\label{fig:4.6} On the left, the $T$ gate is first applied and, then, the circuit in figure \ref{fig:4.5} is executed; hence, the output is $HT|\psi\rangle$. On the right, an equivalent circuit where the $T$ gate has been removed and the measurement is now performed along the eigenbasis of the operator $A\equiv T^{\dagger} X T$ \cite{Leung01,Leung03}; again, the output is $HT|\psi\rangle$, i.e., the circuit {\em teleports} the gate $HT$.}
\end{center}
\end{figure}

More specifically, consider the {\em graph-state model} of quantum computation \cite{Raussen03} where the set,  $\mathcal{O}$, of available operations is $\mathcal{O}=\{${\sc cphase}$, \mathcal{P}_{|+\rangle}, \mathcal{M}_{A}\}$; here, $\mathcal{P}_{|+\rangle}$ denotes the preparation of a qubit in the state $|+\rangle$, and $\mathcal{M}_{A}$ denotes the measurement of a qubit along the eigenbasis of {\em any} single-qubit observable $A$. To show that $\mathcal{O}$ is universal for quantum computation, it suffices to show that any gate from the universal set $\mathcal{G} = \{${\sc cnot}$, H, T \}$ can be simulated (see proposition \ref{propo:1} in chapter \ref{ch:intro}). Indeed, the circuit in figure \ref{fig:4.5} shows how to simulate the Hadamard gate and, then, the {\sc cnot} can be simulated as {\sc cnot}$\,=(I\otimes H)\,${\sc cphase}$\,(I\otimes H)$. It remains to simulate the $T=\exp\left({-i{\pi \over 8}Z}\right)$ gate. The circuit in figure \ref{fig:4.6} shows how to implement the simulation of $HT$ using a measurement of the observable $T^{\dagger} X T$; then, $T$ can be simulated as $H\cdot HT=T$.

Since the only unitary interactions available in measurement-based models are entangling gates such as the {\sc cphase} gate, the Pauli operators that are conditioned on the measurement outcomes within each gate simulation cannot be directly applied; we then say that each gate is simulated up to {\em known} Pauli {\em corrections}. This is not a problem for the quantum computation as a whole, since we can easily propagate Pauli operators through subsequent gates---e.g., the {\sc cphase} gate belongs in the Clifford group and, so, conjugates Pauli operators to Pauli operators---and we can eventually absorb the Pauli corrections inside subsequent measurements similar to the trick in figure \ref{fig:4.6}. In this way, our knowledge of the Pauli corrections allows us to adapt some of the bases of subsequent measurements in order for the desired overall simulation to be executed.   

%----------------------------------------------------------------------%
\subsection{The Model-Conversion Lemma}
\label{sec:ModelReduction}

At first, it seems that the analysis of noisy measurement-based quantum computation is complicated by the fact that the measurement bases within each gate simulation depend on the outcomes of previous measurements. For example, if one measurement outcome is erroneous, then this will lead to adapting incorrectly some future measurement bases; hence, one fault can cause a series of other {\em correlated} faults. In order to obtain a proof of the quantum threshold theorem for measurement-based models, we will therefore need to show that such correlated faults are not an obstacle to executing fault-tolerant quantum computation.

Before proceeding further, it is helpful to develop a formalism of describing individual gate simulations in measurement-based models and how simulations compose together. For every simulated gate, let us define a {\em level-0 simulator}, or {\em 0-Sim}, as the union of operations required for simulating this gate. For example, the 0-Sim corresponding to the Hadamard gate consists of the preparation of an ancillary qubit in the state $|+\rangle$, a {\sc cphase} gate and a measurement along the eigenbasis of $X$ (Fig.~\ref{fig:4.5})---note that the conditional gate $X$ is not contained in the 0-Sim since Pauli corrections are never directly applied.

%We also need to formalize what we mean by the term simulation of a gate in measurement-based models. 
As discussed previously, every gate simulation in measurement-based models is achieved up to known Pauli corrections that are recorded in a classical memory and are used to adapt future measurement bases. To make the role of Pauli corrections more transparent, we will define an {\em ideal Pauli  corrector}, or {\em i-corrector}. An i-corrector is an ideal (i.e., faultless) device that conditioned on the classical record of the Pauli correction of a given qubit, applies this Pauli correction operator and discards the classical record. Like i-decoders and i-LRUs, i-correctors will be tools for our analysis and do not correspond to physical operations applied during a measurement-based computation. If we use a double line to denote the {\em classical} bits carrying the information about the Pauli correction of some qubit, then the i-corrector acting on this qubit is the operation

\vspace{1.2cm}
\setlength{\unitlength}{1pt}
\begin{picture}(292,24)
\put(0,12){\line(1,0){10}}
\put(0,35){\line(1,0){10}}
\put(0,33){\line(1,0){10}}
\put(10,0){\framebox(60,44){\shortstack{i-corrector}}}
\put(70,12){\line(1,0){10}}
\put(90,15){\makebox(20,12){=}}
\put(115,37){
\Qcircuit @C=2.3ex @R=2.6ex @!R { 
             & \ctrlc{1}  &  \\
             & \gate{P}  & \qw   
                                        }}    
\put(165,16){\makebox(20,12){,}}
\end{picture}
\vspace{0.3cm} 

\noindent where $P$ denotes the Pauli correction operator that the classical record indicates. 

We can now state what it means for a 0-Sim to correctly simulate some 0-Ga. \vspace{0.2cm} 

\begin{defi}[Correctness for 0-Sims]
\label{def:12}
A 0-Sim is {\em correct} if the 0-Sim followed by i-correctors is equivalent to i-correctors followed by the ideal 0-Ga that the 0-Sim simulates. For a single-qubit gate 0-Ga, correctness means schematically:
\end{defi}
\vspace{1cm}
\setlength{\unitlength}{1pt}
\begin{picture}(292,24)
\put(0,12){\line(1,0){10}}
\put(0,35){\line(1,0){10}}
\put(0,33){\line(1,0){10}}
\put(10,0){\framebox(48,44){\shortstack{correct\\0-Sim}}}
\put(58,12){\line(1,0){10}}
\put(58,35){\line(1,0){10}}
\put(58,33){\line(1,0){10}}
\put(68,0){\framebox(60,44){\shortstack{i-corrector}}}
\put(128,12){\line(1,0){10}}
\put(142,15){\makebox(20,12){=}}
\put(166,12){\line(1,0){10}}
\put(166,35){\line(1,0){10}}
\put(166,33){\line(1,0){10}}
\put(176,0){\framebox(60,44){\shortstack{i-corrector}}}
\put(236,12){\line(1,0){10}}
\put(246,0){\framebox(48,24){\shortstack{ideal\\0-Ga}}}
\put(294,12){\line(1,0){10}}
\put(308,15){\makebox(20,12){.}}
\end{picture}
\vspace{0.5cm}

We will also say that 

\begin{defi}[Goodness for 0-Sims]
\label{def:13}
A 0-Sim is {\em good} if it contains {\em no} faults, otherwise it is bad.
\end{defi}

Since a good 0-Sim contains no faults, it operates ideally; hence, the following lemma holds.

\begin{lem}[0-Sim-Correct]
\label{lem:6}
A good 0-Sim is correct.
\end{lem}

It is sufficient to construct 0-Sims for every gate 0-Ga in a finite universal gate set; e.g., such constructions for the graph-state model were discussed in the previous section. Definitions \ref{def:12} and \ref{def:13} also hold for 0-Sims simulating qubit preparation or measurement 0-Gas. The correctness of good measurement 0-Sims allows us to replace them by i-correctors followed by the corresponding ideal measurement 0-Gas, thereby creating i-correctors. Similarly, the correctness of good preparation 0-Sims allows us to annihilate i-correctors together with preparation 0-Sims, thereby obtaining the ideal preparation 0-Gas alone. We can now state our basic lemma.

\begin{lem}[Model Conversion]
\label{lem:7} 
Consider the measurement-based simulation of operations in a finite quantum universal set, $\mathcal{G}$, and let $C_{\rm sim}$ be the maximum number of locations in any 0-Sim. Then, if the measurement-based simulation of a quantum circuit expressed in terms of operations in $\mathcal{G}$ is subject to local noise of strength $\varepsilon$, there is a quantum circuit with operations in $\mathcal{G}$ which produces the same output probability distribution and is subject to local noise with strength
\[
\hat{\varepsilon} \leq  C_{\rm sim} \, \varepsilon \; 
\]

\noindent (or $\hat{p} \leq  C_{\rm sim} \, p$ if noise is stochastic).
\end{lem}

The proof of this lemma is essentially the same as the proof of the leakage-reduction lemma and will not be discussed in detail. Similar to coherent i-LRUs and their inverses, we can consider the coherent version of an i-corrector which retains the classical record of Pauli corrections. We can insert pairs of coherent i-correctors and their inverses preceding every bad 0-Sim in order to convert them to the simulation of some faulty operations (which may be correlated since they depend on the classical record of the Pauli corrections). Overall, by creating i-correctors out of 0-Sims simulating measurements, moving them to the left of 0-Sims simulating quantum gates, and annihilating them inside 0-Sims simulating qubit preparations, we can convert any noisy measurement-based simulation of an ideal quantum circuit to a direct noisy implementation of the quantum circuit. Then, local noise afflicting the measurement-based simulation will be equivalent to local noise acting on the equivalent quantum circuit, and the transformation of the noise strength can be estimated as in the proof of lemma \ref{lem:4}.

%----------------------------------------------------------------------%
\subsection{The Threshold Theorem for Quantum Computation by Measurements}
\label{sec:MBThreshold}

The model-conversion lemma tells us that noisy measurement-based quantum computation can be analysed in terms of an equivalent noisy quantum-circuit computation. In particular, if we consider the measurement-based simulation of a fault-tolerant quantum circuit, we expect that this simulation will also be fault tolerant. To prove the existence of an accuracy threshold for such a noisy measurement-based simulation, we can first use the model-conversion lemma to convert it to an equivalent noisy quantum circuit computation that implements the fault-tolerant quantum computation. Then, we can apply theorem \ref{theo:5} to complete the proof. We can thus prove the following theorem.

\begin{theo}[Accuracy Threshold for Measurement-Based Quantum Computation]
\label{theo:7}
Suppose gadgets can be constructed such that they satisfy properties \ref{prop:1} and \ref{prop:2} for a distance-$d$ quantum error-correcting code, and let $t=\lfloor{d-1\over 2}\rfloor$. Let $c$ be the largest size and $d$ be the largest depth in any 1-Rec, and let $C$ be the size of the largest 1-exRec. Consider the measurement-based simulation of all operations in our finite quantum universal set, $\mathcal{G}$, and let $C_{\rm sim}$ be the maximum number of locations in any 0-Sim. Suppose that elementary operations are afflicted by local noise of strength
\[
\varepsilon < \varepsilon_{\rm thr} \equiv \left( A' \right)^{-1/t} / C_{\rm sim} \;,
\]

\noindent (or $p< p_{\rm thr} \equiv (A')^{-1/t} / C_{\rm sim}$ if noise is stochastic) with $A'$ as in equation (\ref{3.1.15}) where $A$ is the largest number of malignant sets of $t{+}1$ faults in any 1-exRec and $B\geq (t+1) e^{(C-t-2)\varepsilon_{\rm thr}} {C\choose t+2}$ (or $B={C\choose t+2}$ if noise is stochastic). Then, for any fixed $\delta_0>0$, any ideal quantum circuit of size $L$ and depth $D$ can be simulated with error $\delta_0$ by a noisy measurement-based quantum computation of size $O\left(L(\log L)^{{\log_{t+1} c}}\right)$ and depth $O\left(D(\log L)^{{\log_{t+1} d}}\right)$.
\end{theo}

The value of $C_{\rm sim}$ depends on both our universal set and the particular construction of 0-Sims. For example, in the graph-state model, the simulation of the Hadamard gate requires $3$ elementary operations (one qubit preparation, one {\sc cphase} gate and one measurement as in figure \ref{fig:4.5}), the simulation of the {\sc cnot} gate requires $2\times 3+1=7$ elementary operations (two Hadamard simulations and one {\sc cphase} gate) and the simulation of the $T$ gate requires $2\times 3=6$ elementary operations (the three operations in figure \ref{fig:4.5} and one Hadamard simulation). Hence, if $\mathcal{G} = \{${\sc cnot}$, H, T \}$, then $C_{\rm sim}=7$ and the accuracy threshold for measurement-based quantum computation is at most an order of magnitude lower than for quantum-circuit computation. This conclusion is however overly pessimistic since the construction of 0-Sims can be optimized by carefully choosing our gate set; e.g., \cite{Nielsen05a} gives Monte-Carlo estimates of the accuracy threshold for measurement-based quantum computation which are very close to similar threshold estimates for quantum-circuit computation.

\section{History and Acknowledgements}

A discussion about leakage noise was informally given first in \cite{Knill96b} and \cite{Preskill97}. However, these discussions were not rigourous since physical leakage noise cannot be accurately captured by the stochastic noise models considered in all proofs of the quantum threshold theorem prior to the results in \cite{Terhal04} and \cite{Aliferis05b}. In fact, Terhal started thinking about proving the quantum threshold theorem for leakage noise shortly after the publication of \cite{Terhal04}, but she found it hard to adapt the proof in \cite{Aharonov99comb} to a setting where leakage faults occur. By using the new proof in \cite{Aliferis05b} which was much more transparent, Terhal and I analyzed the problem of fault-tolerant quantum computation subject to local leakage noise in \cite{Aliferis05c}. Later, and with the help of the idea of level reduction, I simplified the proof via the leakage-reduction lemma that was given here.   

The problem of fault-tolerant measurement-based quantum computation was first analyzed by Raussendorf in his thesis \cite{Raussen03b} which discussed cluster-state computation afflicted by a certain type of stochastic noise. A proof of a threshold theorem for measurement-based quantum computation for a more general noise model was later given by Nielsen and Dawson \cite{Nielsen04b}; this proof was based on the proof in \cite{Terhal04} for quantum-circuit computation that applies to coherent noise. In fact, Nielsen and Dawson emphasized in \cite{Nielsen04b} that noisy measurement-based simulations of quantum circuits induce effective noise correlations between the simulated quantum gates so that even if physical noise is independent and stochastic, a quantum threshold theorem that applies to coherent faults is necessary. Later, Leung and I gave a simpler proof \cite{Aliferis05a} by showing that noisy measurement-based quantum computation can be viewed as an equivalent noisy quantum-circuit computation where the quantum state is encoded in a random basis that corresponds to the Pauli correction operators---it is correlations due to this coding that Nielsen and Dawson were observing in \cite{Nielsen04b}, but the argument in \cite{Aliferis05a} shows that such correlations are no different than the correlations that are induced due to coding in an error-correcting code as explained in \S \ref{sec:BadRecs}. The proof given here is similar to the proof in \cite{Aliferis05a}, but it is phrased syntactically via the notion of correctness that is analogous to the semantic notion of {\em composable simulations} \cite{Childs04} used in \cite{Aliferis05a}.

%----------------------------------------------------------%

%-----------------------------------------%
%\chapter{The Threshold Theorem for Postselected Quantum Computation}
%\label{ch:postselection}
%\input{postselected-FTQC/postselection.tex}

%-----------------------------------------%
\chapter{Lower Bounds on the Quantum Accuracy Threshold}
\label{ch:lower-bounds}
%----------------------------------------------------------%
\section{Introduction}

In this chapter, I will present explicit gadgets satisfying properties \ref{prop:1} and \ref{prop:2} and I will use them to establish lower bounds on the quantum accuracy threshold. First, in \S \ref{sec:GadgetConstructions}, I will describe general gadget constructions that apply to any stabilizer code. Then, in \S \ref{sec:BS}, I will present  specific gadgets for fault-tolerant quantum computation based on the Bacon-Shor code.

In the gadget constructions that I will present, some additional assumptions to those in proposition \ref{propo:2} will be made. Although these assumptions are not essential for proving the quantum threshold theorem, they simplify the presentation and improve our accuracy threshold lower bounds. These assumptions are listed in the following proposition.

\begin{propo}
\label{propo:3}
For our gadget constructions and for deriving our lower bounds on the quantum accuracy threshold, we will make the following additional assumptions to those in proposition \ref{propo:2}:
\begin{enumerate}
\item We can perform quantum measurements during the computation and apply quantum gates conditioned on the measurement outcomes. Furthermore, the speed of quantum measurement is of the same order as the speed of quantum gates.
\item The classical processing of measurement outcomes can be done instantaneously and flawlessly.
\item Multi-qubit quantum gates can be applied on any subset of qubits irrespective of their physical separation.
\item There are no leakage faults.
\end{enumerate}
\end{propo}

As shown in \cite{Aliferis06b}, the assumptions about the speed of measurement and the speed of classical processing can be lifted without a significant effect on our accuracy threshold lower bounds. If classical processing is not flawless then, as noted in \S \ref{sec:Recursive}, the quantum threshold theorem can still be proved but the overall accuracy cannot be made arbitrarily small; instead, it will be limited by the accuracy of the gates implementing the classical processing. The quantum threshold theorem can also be proved if geometric constraints are imposed so that multi-qubit gates can only be applied on qubits that are geometrically local \cite{Gottesman00}; for a two-dimensional square lattice architecture with only nearest-neighbor gates, \cite{Svore06} shows that the accuracy threshold is decreased by a small factor (certainly less than an order of magnitude) compared to the case where no geometric constraints are imposed. Finally, as discussed in \S \ref{sec:Leakage}, leakage faults can be dealt with by inserting leakage-reduction units in the quantum circuits preceding every gate. The accuracy threshold in this case decreases by at most an order to magnitude compared to the case when there is no leakage; however, using contracted 0-Recs, it is expected that the accuracy thresholds with and without leakage faults can be brought to be closer. 

%----------------------------------------------------------%
%----------------------------------------------------------%
\section{Gadget Constructions}
\label{sec:GadgetConstructions}

%--------------------------------------------------------------%
\subsection{Fault-Tolerant Quantum Error Correction}
\label{sec:FTEC-Constr}

Let us first discuss how to construct 1-EC gadgets satisfying properties \ref{prop:1}. The key property guiding the construction is property 1(b) for $s=0$ and $r\leq t$; i.e., we must assure that when a 1-EC with at most $t$ faults has an input code block that is in the code space, it produces the same output after ideal decoding as if there were no faults inside the 1-EC.

Our 1-EC gadgets will consist of a syndrome measurement procedure which will be possibly followed by the application of some recovery operator. In fact, if the subsequent gates belong in the Clifford group, the recovery operators need not be explicitly applied but they can be kept in a classical memory similar to Pauli corrections in measurement-based quantum computation. Hence, our goal is to construct a quantum circuit that performs syndrome measurement and satisfies properties \ref{prop:1}. We will restrict our discussion to stabilizer codes as they are well-suited for fault-tolerant quantum computation; then the code generators are tensor products of Pauli operators.

%--------------------------------------------------------------%
\subsubsection{Syndrome Measurement with Cat States}
\label{sec:SyndrCatStates}

\begin{figure}[tb]
\begin{center}
\vspace{0.2cm}
\setlength{\unitlength}{1cm}  
\parbox{1cm}{
\Qcircuit @C=2.2ex @R=2.2ex @!R { \put(-1.2,-2.5){\small \shortstack{data\\block}}
            & \qw                                & \multigate{3}{U}    & \qw     & \qw  \\
            & \qw                                & \ghost{U}           & \qw     & \qw  \\
            & \vdots                             &                     & \vdots         \\
            & \qw                                & \ghost{U}           & \qw     & \qw  \\
            & \push{|+\rangle \hspace{0.1cm}}    & \ctrl{-1}   & \meterx &  
                                        }} 
\vspace{0.2cm}                             
\caption{\label{fig:5.1} A measurement of $U$ using one ancillary qubit. If the measurement outcome is ${+}1$ (respectively, ${-}1$), the data qubits are projected in the ${+}1$ (respectively, ${-}1$) eigenspace of $U$. The controlled-$U$ gate applies $U$ on the data block if the state of the ancillary qubit is $|1\rangle$, and applies the identity if the state of the ancillary qubit is $|0\rangle$. %Recall that $|+\rangle \propto |0\rangle + |1\rangle$.
         }
\end{center}
\end{figure}
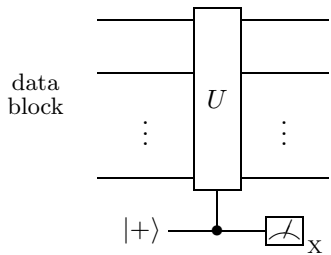

We first note that the code generators commute, so it is sufficient to measure each one separately in order to obtain the syndrome. A method for performing this measurement, which is due to Shor \cite{Shor96}, is based on the circuit in figure \ref{fig:5.1} that measures a unitary operator $U$ with eigenvalues $\pm 1$---since the eigenvalues of the code generators are $\pm 1$, they can be measured using this circuit. For instance, consider a code generator, $G_{a}$, of weight $w$; if we view $a$ as a length-$2n$ binary vector,
\begin{equation}
\label{5.1.1}
     G_a = \bigotimes_{j=1}^n \left( i^{a[j]\cdot a[j+n]} \; X^{a[j]} Z^{a[j{+}n]} \right) \, .
\end{equation}

\noindent Then, the circuit for measuring $G_{a}$ will consist of an ancillary qubit, a sequence of $w$ two-qubit gates (we apply $w$ controlled-$\left( i^{a[j]\cdot a[j+n]}X^{a[j]}Z^{a[j+n]} \right)$ gates with control the ancillary qubit and target qubit $j$ in the data block), and a final measurement of the ancillary qubit.

This circuit does {\em not} yet satisfy properties \ref{prop:1} because the ancillary qubit interacts with multiple qubits in the data block. This is a problem because a single fault acting on the ancillary qubit can give rise to an error that propagates to cause errors on {\em several} qubits in the data block; e.g., if the error acting on the ancillary qubit is $X$, then it propagates to become an $X$ error on any qubits in the data block that interact with this ancillary qubit via {\sc cnot} gates subsequent to the fault.\footnote{It is important to note how $X$ and $Z$ errors are propagated by {\sc cnot} gates. A {\sc cnot} propagates an $X$ error in its control qubit to $X$ errors in both control and target qubits; it also propagates a $Z$ error in its target qubit to $Z$ errors in both control and target qubits.} To prevent such propagation of errors to multiple qubits in the code block, we can replace the one ancillary qubit in figure \ref{fig:5.1} by multiple ancillary qubits. In particular, we can prepare $w$ ancillary qubits in the {\em cat state}, $|+\rangle_{\rm rep} \propto |0\rangle ^{\otimes w} + |1\rangle ^{\otimes w}$, i.e., the logical $|+\rangle$ state of the $w$-qubit quantum repetition code \cite{Shor96,DiVincenzo96}. Now, the interaction of ancillary and data qubits is bitwise or {\em transversal}: the $j$th ancillary qubit interacts via a controlled-$\left( i^{a[j]\cdot a[j+n]}X^{a[j]}Z^{a[j+n]} \right)$ gate with the $j$th qubit in the data block. The benefit of transversal interactions is that faults acting on, say, $r$ qubits in the cat state can cause errors on at most $r$ qubits in the data block, thus solving the problem of propagation of errors. The measurement in figure \ref{fig:5.1} needs now to be replaced by a measurement on all $w$ ancillary qubits. 
This measurement can also be performed {\em transversally} by separately measuring each of the $w$ ancillary qubits along the eigenbasis of $X$; then we can calculate the eigenvalue of $G_a$ by computing the parity of the transversal measurement outcomes. However, in general this eigenvalue cannot immediately be trusted: Since even a single fault might flip one of the tranversal measurements causing an error in the computation of the parity of the outcomes, each code generator must be measured repeatedly sufficiently many times depending on the distance of the code; the  eigenvalue of $G_a$ will then be obtained by taking a majority vote.

\begin{figure}[tb]
\begin{center}
\setlength{\unitlength}{1cm} \vspace{0.3cm}
\parbox{1cm}{
\Qcircuit @C=0.5ex @R=0.6ex @!R { \put(-1.3,-1.85){\rm \shortstack{data \\ block}}
            & \qw                                & \qw                 & \qw     & \qw & \qw  & \qw & \qw & \qw & \targ & \qw     & \qw   & \qw     & \qw & \qw     & \qw  \\
            & \qw                                & \qw                 & \qw     & \qw & \qw  & \qw & \qw & \qw & \qw   & \targ & \qw   & \qw     & \qw & \qw     & \qw  \\
            & \qw                                & \qw                 & \qw     & \qw & \qw  & \qw & \qw & \qw & \qw & \qw      & \targ & \qw     & \qw & \qw     & \qw  \\
            & \qw                                & \qw                 & \qw     & \qw & \qw  & \qw & \qw & \qw & \qw & \qw      & \qw   & \targ   & \qw & \qw     & \qw  \\
            & \\  
            &                                    & \push{|0\rangle \hspace{0.1cm}}    & \targ                          & \ctrl{4}            & \qw   & \qw     & \qw       & \qw & \ctrl{-5} & \qw & \qw & \qw & \meterx \\
            & \push{|+\rangle \hspace{0.1cm}}    & \ctrl{1}                           & \ctrl{-1}                     & \qw                 & \qw   & \qw     & \qw       & \qw & \qw & \ctrl{-5} & \qw & \qw & \meterx \\
            & \push{|0\rangle \hspace{0.1cm}}    & \targ                              & \ctrl{1}                      & \qw                 & \qw   & \qw     & \qw       & \qw & \qw & \qw & \ctrl{-5} & \qw & \meterx \\
            &                                    & \push{|0\rangle \hspace{0.1cm}}    & \targ                         & \qw                 & \ctrl{1} & \qw     & \qw    & \qw & \qw & \qw & \qw & \ctrl{-5} & \meterx \\  
            &                                    &                                    & \push{|0\rangle \hspace{0.1cm}}    & \targ   & \targ & \meterz & \cw & \rstick{{+}1}
                                        }} 
\end{center}  
\caption{ \label{fig:5.2}
           Example of syndrome measurement with cat states. In this case, the code generator $X^{\otimes 4}$ is measured. A four-qubit cat state is prepared and verified, and then it controls the application of the Pauli operator to the data block. Finally, all ancillary qubits are measured along the eigenbasis of $X$, and the parity of the measurement outcomes equals the measured eigenvalue of $X^{\otimes 4}$. 
           }
\end{figure}
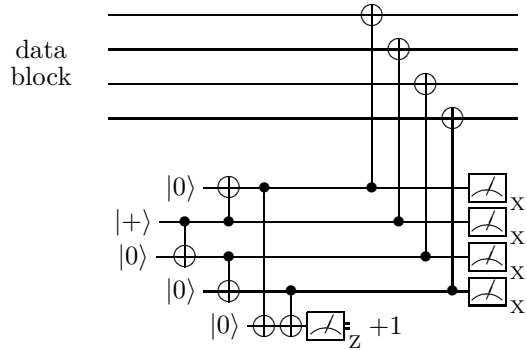

To finish the construction of the 1-EC gadget, it remains to specify how to prepare the cat state. Constructing an encoding circuit for $|+\rangle_{\rm rep}$ is straightforward: e.g., we may start with the state $|+\rangle \otimes |0\rangle^{\otimes (w-1)}$ and perform {\sc cnot} gates with control the first qubit and target every other qubit. To deal with the problem of the propagation of errors inside the encoding circuit, we may next {\em verify} the cat state \cite{Shor96}. Verification can be performed by preparing extra ancillary qubits and using them to measure operators in the stabilizer of the cat state in order to check for undesired fault patterns inside the encoding circuit---of concern are multiple $X$ errors since any pair of $Z$ errors acts trivially on $|+\rangle_{\rm rep}$. For instance, figure \ref{fig:5.2} shows the complete circuit for measuring the weight-$4$ operator $X^{\otimes 4}$. The verification step is a measurement of the operator $Z\otimes I \otimes I\otimes Z$ which belongs in the cat-state stabilizer. If the outcome of the verification measurement is ${-}1$, then a fault might have occurred in one of the two later {\sc cnot} gates during cat-state encoding which might have caused $X$ errors on two qubits of the cat state; in this case, the cat state is {\em discarded} and the encoding is repeated anew. If the outcome of the verification measurement is ${+}1$, at least two faults are required to introduce two $X$ errors in the cat state,  which can therefore be accepted. It is staightforward to adapt the principles underlying the construction of this circuit to measure code generators of higher weight for any stabilizer code.

Due to the transversal interaction between the verified cat state and the data block, property 1(b) is satisfied:  Indeed, if the input data block passes through an $s$-filter, then its state can be expanded as a sum of terms where, in each term, Pauli errors on at most $s$ qubits act on some state in the code space. Also, the $r$ faults inside the 1-EC gadget cannot cause errors acting on more than $r$ qubits in the data block output from 1-EC. Overall, the state of the data block that is output from 1-EC can be expanded as a sum of terms where, in each term, Pauli errors on at most $s+r$ qubits act on some state in the code space. Since $s+r\leq t$, all these Pauli errors can be corrected, which implies that their linear sum will also be corrected. %Hence, property 1(b) is satisfied. 
In \S \ref{sec:Prop1a}, we will discuss why property 1(a) is also satisfied. Finally, let us note that cat-state verification is {\em not} necessary for obtaining a 1-EC gadget satisfying properties 1; \cite{Aliferis06b} discusses a more efficient procedure in which verification is avoided and, instead, the cat state is decoded {\em after} interacting with the data. 

%--------------------------------------------------------------------------------------------------%
\subsubsection{Syndrome Measurement with Encoded Pauli Eigenstates}
\label{sec:SteFTEC}

A more efficient 1-EC gadget can be constructed for CSS codes \cite{Calderbank95,Steane97} for which, as discussed in \S \ref{sec:QCSSCodes}, each code generator can be chosen to be either $X$-type or $Z$-type. This 1-EC construction,  which is due to Steane \cite{Steane96b}, is based on the circuits

\begin{figure}[h]
\begin{center}
\setlength{\unitlength}{1.2cm}
\vspace{0.2cm} \hspace{0.05cm} 
\Qcircuit @C=0.7ex @R=2ex @!R {  
   & \push{|\psi\rangle_L \hspace{0.2cm}}  & \targ      & \qw  & \qw    \\
   & \push{|0\rangle_L \hspace{0.2cm}}     & \ctrl{-1}  & \meterx  
                                }
\hspace{0.5cm} \put(0.4,-0.35){;} \hspace{0.8cm} 
\Qcircuit @C=0.7ex @R=2ex @!R {  
   & \push{|\psi\rangle_L \hspace{0.2cm}}   & \ctrl{1}      & \qw   & \qw   \\
   & \push{|+\rangle_L  \hspace{0.2cm}}     & \targ         & \meterz  
                                }
\hspace{0.5cm} \put(0.4,-0.35){,}
\end{center}
\end{figure} \vspace{-0.3cm}

\noindent where $|\psi\rangle_L$ is the state of the data block and $|0\rangle_L$, $|+\rangle_L$ denote ancillary blocks encoded in the corresponding logical states in the {\em same} code as the data. For CSS codes with $k=1$ logical qubit, the logical {\sc cnot} gate can be implemented {\em transversally}, i.e., by performing bitwise a {\sc cnot} from each qubit in the control block to each qubit in the corresponding position in the target block \cite{Shor96,Gottesman98b}. The {\sc cnot} gates in the figure above denote such transversal {\sc cnot} gates acting between the data and ancillary blocks, and the measurements are understood as also being performed transversally, i.e., separately on each qubit in the ancillary block. 

Evidently, as we would expect for error-correction circuitry, these circuits act trivially on the logical state,  $|\psi\rangle_L$. However, their non-trivial content is that the outcomes of the transversal measurements reveal the {\em parity} of the code syndromes in the data and the ancillary blocks. Indeed, the transversal {\sc cnot} gates propagate any Pauli errors in the data block to the two ancillary blocks. Then, the $X$-type syndrome can be extracted by applying a classical parity check matrix to the measurement outcomes of those ancillary qubits measured along the eigenbasis of $X$. Specifically, for the $X$-type generator $X_a \equiv \otimes_{j=1}^n \; X^{a[j]}$, if the outcomes of the transversal measurements of $X$ are $x=(x_1,x_2,\dots , x_n)$, then the measured eigenvalue of $X_a$ is $a\cdot x$ (mod 2). The $Z$-type syndrome can be extracted similarly from the transversal measurements of $Z$. %Likewise, the $Z$-type syndrome can be extracted by applying a classical parity check to the measurement outcomes of those ancillary qubits measured along the eigenbasis of $Z$. For the $Z$-type stabilizer generator $Z_b \equiv\otimes_{j=1}^n \; Z^{b[j]}$, if the outcomes of the transversal measurements of $Z$ are $z=(z_1,z_2,\dots , z_n)$, then the measured eigenvalue of $Z_a$ is $b\cdot z$ (mod 2). 

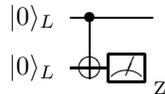
\begin{figure}[htb]
\begin{center}
\setlength{\unitlength}{1.2cm}
\vspace{0.2cm} \hspace{0.05cm} 
\Qcircuit @C=0.7ex @R=2ex @!R {  
   & \push{|0\rangle_L \hspace{0.2cm}}  & \ctrl{1}  & \qw  & \qw    \\
   & \push{|0\rangle_L \hspace{0.2cm}}  & \targ     & \meterz  
                                }
\end{center}  
 \caption{ \label{fig:anc-purif} A purification procedure against $X$ errors for logical $|0\rangle$ blocks encoded in some CSS code. Two blocks are encoded separately (the encoding circuits are not shown) and, then, they interact via a logical {\sc cnot} followed by measuring all qubits in the second block in the computation basis. Classical parity checks on the measurement outcomes allows to detect $X$ errors and also calculate the eigenvalue of the logical $Z$ operator on the first block. If no errors are detected and the eigenvalue of $Z_L$ is ${+}1$, then the first block is accepted; otherwise it is rejected. 
           }
\end{figure}

As for syndrome measurement with cat states, the logical $|0\rangle$ and $|+\rangle$ ancillary states need to be verified after encoding. Again, verification can be performed by using additional ancillary qubits and measuring suitable check operators. As an alternative to verification, a {\em purification} procedure can be used in which multiple copies of the logical states are encoded independently and, then, they interact pairwise in order to detect errors during encoding. For instance, one way to purify the $|0\rangle_L$ state against $X$ errors is to prepare another {\em verifier} $|0\rangle_L$ block, and perform a logical {\sc cnot} with the first block as control and the verifier block as target. Finally, each qubit in the verifier block is measured in the computation basis, and a parity check is applied to the measurement outcomes (Fig.~\ref{fig:anc-purif}). In this case, not just the $Z$-type stabilizer generators are extracted for the first block, but also, after performing classical error-correction on the measurement outcomes, the eigenvalue of the logical $Z$ operator (also a $Z$-type Pauli operator) can be calculated. The first ancillary block is then discarded if either the syndrome is nontrivial or the eigenvalue of $Z_L$ is $-1$. Purification against $Z$ errors can be conducted similarly by reversing the direction of the logical {\sc cnot} gate and measuring the verifier qubits in the eigenbasis of $X$. And, such purification schemes can be iterated if necessary by using the output copies from one purification round as the inputs to a following round. The $|+\rangle_L$ ancillary blocks can be purified similarly. 

Because of the transversal interaction between the data and ancillary blocks, this 1-EC gadget satisfies property 1(b); we will discuss why property 1(a) is also satisfied in \S \ref{sec:Prop1a}. Finally, we note that the purification procedure describe above, like the verification procedure for cat states, is nondeterministic: there are probabilistic fluctuations in the number of ancillary blocks that must be prepared before the first ancillary block is accepted. However, both methods can be adapted to become deterministic by increasing the measured check operators in verification procedures, or the number of verifier blocks in purification procedures. Alternatively, as for cat-state verification, the purification of $|0\rangle_L$ and $|+\rangle_L$ blocks can be avoided altogether and replaced by a suitable decoding step {\em after} the ancillary blocks interact with the data block \cite{Aliferis06b}. 

%--------------------------------------------------------------------------------------------------%
\subsubsection{Syndrome Measurement via Logical Teleportation}
\label{sec:KniFTEC}

\begin{figure}[tbh] \vspace{0.3cm}
\begin{center}
\hspace{-0.1cm} 
\Qcircuit @C=1.5ex @R=2ex @!R {  \put(4.7,0.6){$\otimes n$}
   & \push{|\psi\rangle_L \hspace{0.1cm}} & \qw & \qw      & \qw & \ctrl{1} & \meterx &  \\
   & & \push{|+\rangle_L \hspace{0.1cm}}      & \ctrl{1} & \qw & \targ    & \meterz &  \\
   & & \push{|0\rangle_L \hspace{0.1cm}}      & \targ    & \qw & \qw      & \qw     & \qw & \qw & \qw & \gate{P_L} & \qw & \qw \gategroup{1}{6}{2}{8}{2em}{--}
                                } \vspace{0.2cm}
\caption{\label{fig:A3} A schematic of Knill's 1-EC gadget for a code with $n$ physical qubits in the block. Two ancillary blocks are prepared in a logical Bell state and, then, transversal {\sc cnot} gates and measurements are performed. The measurement outcomes are constrained by parities which enable classical error correction. The {\em corrected} measurement outcomes allow us to determine the logical Pauli operator, $P_L$, necessary to complete the {\em logical teleportation} of the input data state, $|\psi\rangle_L$. }
\end{center}
\end{figure}
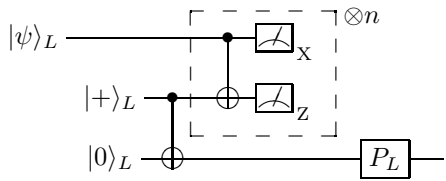

A third method for syndrome measurement that applies to any stabilizer code can be based on quantum teleportation. In this construction, which is due to Knill \cite{Knill05}, the 1-EC gadget implements the {\em logical} teleportation of the state of the input data block while at the same time extracting the code syndrome. Figure \ref{fig:A3} shows a schematic of Knill's 1-EC gadget: Two ancillary blocks are prepared in a logical Bell state, $|\Phi_0\rangle_L \propto |0\rangle_L |0\rangle_L + |1\rangle_L |1\rangle_L$, and then the data and one ancillary block interact via {\em transversal} {\sc cnot} gates followed by transversal measurements on all their qubits. Finally, a {\em logical} Pauli operator is applied on the second ancillary block in order to complete the {\em logical} teleportation of the state of the data block.

The outcomes of the transversal measurements reveal the parity of the code syndromes in the data and ancillary blocks. First, this syndrome information allows us to perform classical error correction on the measurement outcomes. Then, we can compute parities on the, now {\em corrected}, measurement outcomes in order to obtain the eigenvalues of the {\em logical} $X\otimes X$ and $Z\otimes Z$ operators acting on the first two blocks whose qubits were measured. Finally, by knowing the eigenvalues for these two logical operators, we can complete the {\em logical} teleportation of the state of the data block to the second ancillary block up to some appropriate logical Pauli correction operator. %As in measurement-based computation, this correction operator need not actually be applied but can be kept in a classical memory as long as the succeeding logical operations are in the Clifford group \cite{Knill05}. 
As in the previous two methods for syndrome measurement, the two ancillary blocks for this method need to be verified or purified before they are used. Overall, after a suitable verification or purification procedure is applied, the 1-EC gadget in figure \ref{fig:A3} satisfies property 1(b); again, property 1(a) will be discussed separately in \S \ref{sec:Prop1a}.

\setlength{\unitlength}{1cm}

A remarkable feature of this 1-EC gadget is that faults in the transversal {\sc cnot} gates and the subsequent transversal measurements---which combined can be viewed as performing transversal {\em Bell measurements}, i.e., measurements along the two-qubit basis $\{ (Z^{j_1} X^{j_2}\otimes I) |\Phi_0\rangle \, | \, j_1,j_2 = 0,1 \}$---can only result in an error in the logical Pauli correction operator. In other words, as long as the two ancillary blocks are prepared (and verified) without faults, the output code block is always in the code space independent of whether faults occurred in the transversal Bell measurements. The value of this observation comes when performing an analysis of malignant set of locations inside a 1-exRec where such 1-EC gadgets are used:  Consider the leading 1-EC gadgets inside the 1-exRec and let faults occur in the transversal Bell measurements inside them (while more faults may have occurred elsewhere in the 1-exRec). Then, since these faults cannot take the output code block outside the code space, these faulty locations cannot belong to malignant sets of locations for the 1-exRec. Therefore, for the purposes of an analysis of malignant sets of locations, the transversal Bell measurements in the {\em leading} 1-EC gadgets in all 1-exRecs can be taken to be effectively ideal!

%----------------------------------------------------%
\subsubsection{Satisfying Property 1(a)}
\label{sec:Prop1a}

Property 1(a) is the statement that independent of the state of the input block, a 1-EC with at most $r$ faults outputs a block which always passes through an $r$-filter for $0\leq r \leq t$. In particular, applying the property for $r=0$ requires that a 1-EC gadget without faults takes any input block to an output block which is in the code space. If each possible value of the syndrome corresponds to a correctable error (in which case the code is called {\em perfect}), then clearly the output of a faultless 1-EC will be in the code space after applying the necessary recovery operator. But, even if the code is not perfect, we may always assign to every syndrome that does not correspond to a correctable error some Pauli operator (chosen by convention) that returns the subspace labeled by the syndrome to the code space (see the last comment in \S \ref{sec:QStabCodes}). %Therefore, if $|\psi\rangle_L$ denotes a state in the code space, then for every Pauli error $E_a$, a faultless 1-EC maps $E_a|\psi\rangle_L$ to ${\cal O}_a|\psi\rangle_L$ for some {\em logical} operation ${\cal O}_a$ that may depend on $a$. 

At this point, it is convenient to include some reference system, $R$, that {\em purifies} the state of the input data block. This state can then be written as a sum of terms where, for each term, the state of the code block is a vector, $E_a|\psi_a\rangle_L$, and the state of the reference system is $|a\rangle_R$. Here $\{E_a\}$ are Pauli operators that label the orthogonal subspaces corresponding to all possible syndromes, $\{|\psi_a\rangle_L\}$ are some states in the code space (that may depend on $a$), and $\{|a\rangle_R\}$ need not be normalized nor mutually orthogonal. 
%Then, the state of the input code block can be expressed as $ \sum_{a} E_a |\psi\rangle_L \otimes |a\rangle_R $.
In this language, a faultless 1-EC realizes an isometry that acts on the data block and an ancillary system, $A$,  that is initialized in the state $|e_0\rangle_A$ as
\begin{equation}
\label{5.1.2}
  \sum_{a} E_a |\psi_a\rangle_L \otimes |a\rangle_R \otimes |e_0\rangle_A 
                 \stackrel{\rm{faultless}}{\stackrel{\rm{EC}}{\longrightarrow}} \sum_{a} {\cal O}_a |\psi_a\rangle_L \otimes |a\rangle_R \otimes |e_a\rangle_A \; ,
\end{equation}

\noindent where $\{\mathcal{O}_a\}$ are logical operators, i.e., they have support only on the code space, and $\{|e_a\rangle_A\}$ are also not assumed to be normalized nor mutually orthogonal. Thus, as desired, after tracing over the reference and ancillary systems, a faultless 1-EC gadget produces an output which is, in general, some density matrix in the code space. 

We can now show property 1(a) is satisfied even if $0<r\leq t$ faults occur inside the 1-EC gadget. By construction, in all 1-EC gadgets we have considered, the data and ancillary blocks interact via transversal gates. Moreover, the preparation and verification (or purification) of the ancillary blocks is done in a way that prevents $0< r' \leq t$ faults in this part of the 1-EC gadget from acting nontrivially on more than $r'$ qubits in an accepted ancillary block. Overall, each of the $r$ faults inside a 1-EC gadget can only act nontrivially on the {\em same} $r$ qubit positions in the data and ancillary blocks. 

Let $\mathcal{I}_r$ denote this set of at most $r$ qubit positions in the data or ancillary blocks that may have errors due to the $r$ faults inside the 1-EC gadget. Consider one term in the expansion of the state of the input data block, $\sum_{a} E_a |\psi\rangle_L \otimes |a\rangle_R $, where the input data block has Pauli error $E_a$. Also, expand the $r$ fault operators in terms of Pauli operators and pick one term in this expansion where the Pauli operator $E_b^{(\mathcal{I}_r)}$ acts on the data block and $E_c^{(\mathcal{I}_r)}$ acts on the ancillary blocks. (Here, our notation is that $E_*^{(\mathcal{I}_r)}$ is a tensor product of Pauli operators with support on the qubits in $\mathcal{I}_r$ alone.) For this particular term, the syndrome will correspond to the error operator $ E_c^{(\mathcal{I}_r)}E_a$, so that $E_{\rm rec}^\dagger E_c^{(\mathcal{I}_r)}E_a\equiv {\cal L}$ where $E_{\rm rec}^\dagger$ is the recovery operator and ${\cal L}$ is some logical operator. Then, after the recovery, $E_{\rm rec}^\dagger$, is applied to the data block that is in the state  $E_b^{(\mathcal{I}_r)} E_a |\psi\rangle_L \otimes |a\rangle_R$, the combined operator that acts on $|\psi\rangle_L \otimes |a\rangle_R$ is
\begin{equation}
\label{5.1.3}
 E_{\rm rec}^\dagger E_b^{(\mathcal{I}_r)} E_a = \mathcal{L} E_a^\dagger \left(E_c^{(\mathcal{I}_r)}\right)^\dagger E_b^{(\mathcal{I}_r)} E_a = \mathcal{L} E_d^{(\mathcal{I}_r)} \; ,
\end{equation}

\noindent where the crucial observation is that $E_d^{(\mathcal{I}_r)}$ is only supported on the qubits in $\mathcal{I}_r$. We conclude that, for each term in our expansion, the output of an $r$-good 1-EC deviates from the code space by the action of Pauli operators acting on at most $r$ qubits in the code block. Hence, since every term in our expansion passes through an $r$-filter, by linearity, the linear sum of all terms will also pass through an $r$-filter. 

This completes our general discussion of how 1-EC gadgets satisfying properties \ref{prop:1} can be constructed.

%----------------------------------------------------------------%

%----------------------------------------------------------------%
\subsection{Encoded Quantum Universality}

Let us now discuss how to construct a universal set of 1-Ga gadgets satisfying properties \ref{prop:2}. This construction is guided by the $C_k$ gate hierachy in definition \ref{def:1.2}. We have already discussed that the Clifford group, $C_2$, is {\em not} dense in $SU(2^n)$. In fact, the Gottesman-Knill theorem \cite{Gottesman98a} shows that quantum computation using qubits initialized in eigenstates of Pauli operators, Clifford-group gates, and measurements of Pauli operators can be simulated efficiently with a classical computer---let us call this set of operations {\em stabilizer operations} or $\mathcal{G}_{\rm stab}$. But, by adding any {\em one} non-Clifford gate to $\mathcal{G}_{\rm stab}$ is sufficient to give a new set of operations which {\em is} dense in $SU(2^n)$ \cite{nebe00} and, hence, is quantum universal; examples of such gate sets were given in proposition \ref{propo:1}.

%----------------------------------------------------------------%
\subsubsection{Encoded Clifford-Group Operations}
\label{sec:EncStabOper}

Although 1-Ga gadget constructions are known that apply to any stabilizer code \cite{Gottesman98b}, of practical interest are codes for which these constructions are efficient and give high accuracy thresholds. In fact, for some codes the construction of 1-Gas for the gates $\{H,S,\,${\sc cnot}$ \}$ that generate the Clifford group is particularly simple. Let us call a 1-Ga {\em transversal} if it can be realized by 0-Gas that act bitwise on the qubits in the code block (for a 1-Ga simulating a single-qubit gate), or bitwise between qubits at corresponding positions across different code blocks (for a 1-Ga simulating a multi-qubit gate). By construction, it is clear that transversal 1-Gas satisfy properties \ref{prop:2}.

CSS codes provide a family of codes with such useful transversality properties \cite{Shor96,Gottesman98b}: For any CSS code with $k=1$ logical qubit, the logical {\sc cnot} can be implemented by bitwise {\sc cnot} gates. Moreover, if the CSS code is constructed from a self-dual classical code (i.e., if the quantum check matrix in equation (\ref{1.3.7}) can be written so that $H=H'$), then the logical Hadamard gate can be implemented by bitwise Hadamard gates. If, in addition, this self-dual classical code is doubly even (i.e., if all its code words have Hamming weight multiple of 4), then the logical $S$ gate is transversal so that the logical Clifford group can be generated by 1-Gas which are transversal. An example of a code with these properties is Steane's $[[7,1,3]]$ code \cite{Steane96} discussed in \S \ref{sec:CodeExamples}.

\subsubsection{Quantum Software and Encoded Non-Clifford Gates}
\label{sec:software}

To complete our construction of 1-Gas satisfying properties \ref{prop:2} for a universal set of gates, it remains to construct a 1-Ga for a non-Clifford gate. This construction can be based on preparing a suitable ancillary state and then using a gate teleportation circuit \cite{Shor96,Gottesman97,Gottesman99,Zhou00}. 

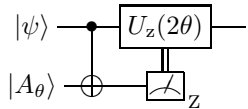
\begin{figure}[tbh]
\begin{center}
\vspace{0.2cm}
\parbox{1cm}{
\Qcircuit @C=1.5ex @R=1.5ex @!R {
            & \push{|\psi\rangle \hspace{0.1cm}}       & \ctrl{1} & \gate{U_{\rm z}(2\theta)} & \qw & \qw \\
            & \push{|A_{\theta}\rangle \hspace{0.1cm}} & \targ    & \meterz \cwx[-1]  
                                        }} 
\end{center}               
\caption{\label{fig:5.8}
        A circuit that {\em teleports} the gate $U_{\rm z}(\theta)$. An ancillary qubit is prepared in the state $|A_\theta\rangle \equiv U_{\rm z}(\theta)|+\rangle$, and then a {\sc cnot} gate is executed as shown. Finally, the ancillary qubit is measured in the computation basis and the correction gate $U_{\rm z}(2\theta)$ is applied if the measurement outcome is ${-}1$.
        }
\end{figure}

Consider the circuit shown in figure \ref{fig:5.8}: To realize the single-qubit rotation by angle $\theta$ around the $z$-axis, $U_{\rm z}(\theta) = \exp\left({-i {\theta \over 2}Z}\right)$, we first prepare an ancillary qubit in the state $|A_\theta\rangle \equiv U_{\rm z}(\theta) |+\rangle$ and, then, we perform a {\sc cnot} gate with the data qubit as control and the ancillary qubit as target. Finally, we measure the ancillary qubit in the computation basis. If the ${+}1$ eigenvalue is obtained, then we have successfully teleported the gate $U_{\rm z}(\theta)$; otherwise, when the outcome is ${-}1$, $U_{\rm z}(-\theta)$ has been teleported instead and the correction operator $U_{\rm z}(2\theta)$ needs to be applied.

\setlength{\unitlength}{1cm} 

%To obtain a 1-Ga for the gate $U_{\rm z}(\theta)$, we can consider the encoded version of the circuit in fig.~\ref{fig:5.8}. Using the methods discussed in the previous section, we can suppose we have already constructed 1-Ga's for the gates $H$, $S$, and {\sc cnot} that generate the Clifford group. Then, it is sufficient to construct a 1-Ga for the gate $T \equiv U_z(\pi/4)$ to obtain 1-Ga's for all gates in the first universal set in Proposition \ref{propo:1}. 
We observe that for $\theta=\pi/4$, the teleported gate is $T\equiv U_{\rm z}\left({\pi\over 4}\right) \not \in C_2$. Furthermore, in this case the correction gate is $S\equiv U_{\rm z} \left( {\pi\over 2} \right) \in C_2$, and, so, a 1-Ga for simulating it can easily be constructed. It remains to construct a procedure for preparing the {\em logical} $| A_{\pi/4}\rangle$ state such that the 1-Ga for the $T$ gate satisfies properties \ref{prop:2}. Since the logical $| A_{\pi/4}\rangle$ is just an ancillary state, its preparation can be done separate to the main computation; we can then view this quantum state as a form of ``software'' that we prepare and consume in order to implement fault-tolerant quantum computation. 

The logical $|A_{\pi/4}\rangle$ state is the ${+}1$ eigenstate of the {\em logical} Clifford-group operator $TXT^\dagger= T^2 X= SX$. Since we can construct a 1-Ga for the gate $SX$, we can also measure this operator using cat states: That is, we can measure the logical $SX$ by preparing a cat state, controlling each gate in the 1-Ga for $SX$ from a different cat-state qubit and, finally, measuring all qubits of the cat state along the eigenbasis of $X$ and computing the parity of the measurement outcomes. Since a single fault in this measurement can cause an error in the eigenvalue of the logical $SX$ operator that is measured, the measurement needs to be repeated sufficiently many times using different cat states with error-correction steps inserted between every two successive measurements.

With this procedure which applies to any stabilizer code, we can therefore construct a 1-Ga for the non-Clifford gate $T$ that satisfies properties \ref{prop:2}. An alternative construction for the non-Clifford Toffoli gate will be given in \S \ref{sec:BSUnivComp}. Together with the 1-Ga constructions for Clifford-group gates discussed in the previous section, we have constructed 1-Gas satisfying properties \ref{prop:2} for a quantum universal set of gates.

%------------------------------------------------------------%

%----------------------------------------------------------------%
\subsection{Fault-Tolerant Preparation and Measurement}
\label{sec:PrepMeasGas}

Finally, let us briefly discuss how 1-Ga gadgets for preparation of Pauli eigenstates and measurements in the eigenbases of Pauli operators can be constructed. As with Clifford-group gate 1-Ga constructions, efficient preparation and measurement 1-Gas can be constructed in the special case of CSS codes. Indeed, for CSS codes, the 1-Ga for the measurement of $Z$ consists of measurements of all qubits along the eigenbasis of $Z$, followed by classical postprocessing of the measurement outcomes. This postprocessing consists of classical error correction on the measurement outcomes (here, it is important that for CSS codes the stabilizer generators are only $X$- and $Z$-type), followed by computing the parity of the corrected outcomes to obtain the eigenvalue of the logical $Z$. A similar procedure can be used to construct a 1-Ga for measuring the operator $X$. 
Also, for CSS codes, there are efficient preparation 1-Gas for Pauli eigenstates. The construction of these 1-Gas can be based on a {\em purification} procedure in which multiple copies of the same logical state are encoded separately, and then they interact pairwise in order to detect errors similar to figure \ref{fig:anc-purif}. %The circuit in figure \ref{fig:5.9} shows such a 1-Ga for preparing the $|0\rangle$ state in a distance-3 code: In the first round of purification, pairs of logical $|0\rangle$ blocks interact via logical {\sc cnot} gates (which can be implemented transversaly) and the subsequent measurements detect $X$ errors during encoding. The second purification round detects $Z$ errors. The first ancillary block is only accepted if no errors are detected, otherwise the block is rejected and encoding followed by purification is attempted again. %We also note that, although the purification procedure in figure \ref{fig:5.9} is non-deterministic, it can become deterministic at the expense of using more ancillary blocks. %Also, for perfect distance-3 CSS codes, such as the Steane [[7,1,3]] code, there is no need to check for multiple $Z$ errors and only two ancillary blocks are needed.

%\begin{figure}[t]
%\begin{center}
%\vspace{0.2cm}
%\parbox{1cm}{
%\Qcircuit @C=1.8ex @R=1.8ex @!R {
%            & \push{|0\rangle_L \hspace{0.1cm}} & \ctrl{1} & \qw     & \targ     & \qw     & \qw \\
%            & \push{|0\rangle_L \hspace{0.1cm}} & \targ    & \meterz  \\
%            & \push{|0\rangle_L \hspace{0.1cm}} & \ctrl{1} & \qw     & \ctrl{-2} & \meterx \\
%            & \push{|0\rangle_L \hspace{0.1cm}} & \targ    & \meterz            
%                                        }} 
%\end{center}               
%\caption{\label{fig:5.9} A 1-Ga for the preparation of the $|0\rangle$ state in a distance-3 CSS code. Four ancillary blocks are encoded separately (the encoding circuits are not shown) and, then, they interact pairwise to detect $X$ and then $Z$ erros during encoding. For perfect CSS codes, such as the Steane [[7,1,3]] code, there is no need to check for multiple $Z$ errors and only two ancillary block are needed.
%        }
%\end{figure} 

For general stabilizer codes, preparation and measurement 1-Gas are more complex. First, 1-Gas for measuring Pauli operators can be constructed by repeated measurements of the corresponding logical Pauli operator using cat states. Then, the majority of the different measurement outcomes gives the eigenvalue of the Pauli operator that is measured by the 1-Ga. Similarly, 1-Gas for preparing Pauli eigenstates implement measurements of the code generators in order to transform any arbitrary initial state to a state in the code space, followed by a sufficient number of repeated measurements of the logical Pauli operator whose eigenstate we want to prepare. (For both preparation and measurement 1-Gas, error-correction steps need to be inserted between the repeated measurements in order to satisfy properties \ref{prop:2}.)

This completes our discussion of constructing preparation and measurement 1-Gas. At this point, we have shown how to obtain 1-Gas satisfying properties \ref{prop:2} for a universal set of operations (preparation of Pauli eigenstates, a quantum universal set of gates, and measurement of Pauli operators). 

%----------------------------------------------------------------%

%----------------------------------------------------------------%

%\input{lower-bounds/FTwithSteane.tex}

%----------------------------------------------------------%
\section{Quantum Fault Tolerance with the Bacon-Shor Code}
\label{sec:BS}

At last, it has come time to establish a numerical lower bound on the value of the quantum accuracy threshold. For simplicity, in the remaining discussion we will only consider {\em stochastic} local noise. Similar techniques can be used to give lower bounds of the same order for local noise that is not stochastic; however, the calculations are easier to perform and present in terms of  probabilities instead of quantum amplitudes. 

The fault-tolerant quantum circuits we will consider will be encoded using a subsystem code described by Bacon in \cite{Bacon05}---because of the close relation of this code with Shor's code \cite{Shor95b}, we will refer to it as the {\em Bacon-Shor code}. There is a different Bacon-Shor code for every integer $n>1$; for fixed $n$, the corresponding code, $\mathcal{C}_{BS}^{(n)}$, is a distance-$n$ stabilizer CSS code encoding one protected logical qubit into a code block of $n^2$ physical qubits.

\begin{figure}[ht]
\begin{center} 
%\leavevmode
%\epsfysize=5cm
%\epsfbox{pic01.eps}
%\includegraphics[width=5cm]{pic01}
%\epsfig{file=pic01.eps,width=5cm}
%
\vspace{2.8cm}
\begin{picture}(5,5)

\put(-1,5.4){$i$}
\put(0,3){\line(1,0){2.5}}
\put(3.5,3){\line(1,0){0.5}}
\put(3.5,7){\line(1,0){0.5}}
\put(-0.6,2.9){$n$}
\put(0,5){\line(1,0){2.5}}
\put(3.5,6){\line(1,0){0.5}}
\put(-0.6,4.9){3}
\put(0,6){\line(1,0){2.5}}
\put(3.5,5){\line(1,0){0.5}}
\put(-0.6,5.9){2}
\put(0,7){\line(1,0){2.5}}
\put(-0.6,6.9){1}

\put(1.4,7.8){$j$}
\put(0,4.5){\line(0,1){2.5}}
\put(0,3){\line(0,1){0.5}}
\put(-0.1,7.4){1}
\put(1,4.5){\line(0,1){2.5}}
\put(1,3){\line(0,1){0.5}}
\put(0.9,7.4){2}
\put(2,4.5){\line(0,1){2.5}}
\put(2,3){\line(0,1){0.5}}
\put(1.9,7.4){3}
\put(4,4.5){\line(0,1){2.5}}
\put(4,3){\line(0,1){0.5}}
\put(3.9,7.4){$n$}

\put(0,3){\color{Black} \circle*{0.4}}
\put(0,5){\color{lblue} \circle*{0.4}}
\put(0,6){\color{lblue} \circle*{0.4}}
\put(0,7){\color{Black} \circle*{0.4}}

\put(1,3){\color{Black} \circle*{0.4}}
\put(1,5){\color{lblue} \circle*{0.4}}
\put(1,6){\color{lblue} \circle*{0.4}}
\put(1,7){\color{Black} \circle*{0.4}}

\put(2,3){\color{Black} \circle*{0.4}}
\put(2,5){\color{lblue} \circle*{0.4}}
\put(2,6){\color{lblue} \circle*{0.4}}
\put(2,7){\color{Black} \circle*{0.4}}

\put(4,3){\color{Black} \circle*{0.4}}
\put(4,5){\color{lblue} \circle*{0.4}}
\put(4,6){\color{lblue} \circle*{0.4}}
\put(4,7){\color{Black} \circle*{0.4}}
\end{picture}
\vspace{-2.9cm}
\end{center}
\caption{\label{fig:5.3.1} Qubits in the \bs block are placed on the vertices of an $n{\times}n$ square lattice. An  element of the code stabilizer, the operator $X_{2,*}X_{3,*}$ applies $X$ on all qubits shown in blue.
         }
\end{figure}
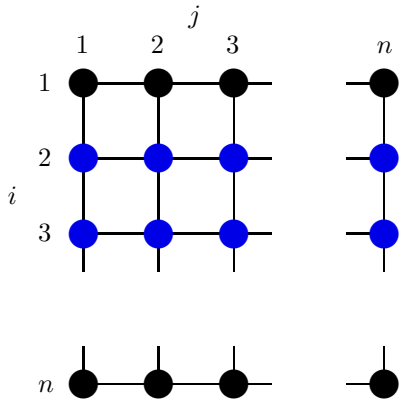

It is convenient to imagine placing the $n^2$ qubits in the \bs block on the {\em vertices} of an $n{\times}n$ square lattice  (Fig.~\ref{fig:5.3.1}). Then, the code's stabilizer group is 
\begin{equation}
\begin{array}{rcl}
\label{eq:stab}
\mathcal{S} = \langle X_{j,*}X_{j+1,*} \, ; \, Z_{*,j}Z_{*,j+1} \, | \, j\in \mathbb{Z}_{n-1} \rangle \, ,
\end{array}
\end{equation}  
\noindent where $O_{j,*}$, $O_{*,j}$ denote operators that act nontrivially as a tensor product of $O$ operators on all qubits in row or column $j$, respectively. Throughout this section, $\mathbb{Z}_{m}$ is understood to indicate the set $\{1,2,\dots,m \}$.

The code {\em syndrome}, $e$, induces a decomposition of the Hilbert space, $\mathcal{H}$, of the $n^2$ qubits in the code block into subspaces encoding $n^2-2(n{-}1)=(n{-}1)^2+1$ logical qubits. Therefore, within each subspace with fixed syndrome---and, in particular, within the {\em code space} corresponding to the trivial syndrome---we can define a subsystem decomposition
\begin{equation}
\label{eq:dec}
\mathcal{H}= \bigoplus_{e} \left( \mathcal{H}_L \otimes \mathcal{H}_{\rm T} \right) \; \vspace{-0.1cm},
\end{equation}
\noindent where we associate $\mathcal{H}_L$ with the one logical qubit protected by the full distance, $n$, of the code. The logical Pauli operators for this logical qubit can be defined as $X_L=X_{1,*}$ (i.e., a tensor product of $X$ operators applied on all qubits in the first row) and $Z_L=Z_{*,1}$ (i.e., a tensor product of $Z$ operators applied on all qubits in the first column).
The remaining $(n{-}1)^2$ logical qubits live in $\mathcal{H}_T$ and their logical Pauli operators can be chosen from the non-abelian group
\begin{equation}
\label{eq:t}
 T = \langle \; X_{j,i} X_{j+1,i} \; ; Z_{i,j} Z_{i,j+1} \; | \; i \in \mathbb{Z}_n \; ; j \in \mathbb{Z}_{n-1} \;  \rangle \; ,
\end{equation}
\noindent where $O_{i,j}$ denotes an operator $O$ acting on the qubit with coordinates $(i,j)$. Indeed, the operators in $T$ commute with every operator in the code stabilizer, they commute with the logical operators $X_L$ and $Z_L$ and, furthermore, they can be grouped into $(n{-}1)^2$ independent pairs of anticommuting operators with operators in different pairs commuting. 

Given some nontrivial syndrome value, error recovery for the logical qubit encoded in $\mathcal{H}_{L}$ proceeds in a similar manner as in the classical repetition code: The eigenvalues of the stabilizer generators $\{X_{j,*}X_{j+1,*}\}$ can be used to correct $Z$ errors on up to $\lfloor n/2 \rfloor$ rows. Moreover, only the parity of $Z$ errors in each row is relevant: an operator acting as $Z$ on a pair of qubits at the same row is an operator in $T$ and, therefore, has no effect on the protected information in $\mathcal{H}_L$. Error recovery for $X$ errors proceeds similarly along the columns. Further intuition on the Bacon-Shor code can be obtained by considering its derivation from other subspace codes; see chapter \ref{ch:bs-derivation} in the appendix.  

%--------------------------------------------------------------------------------------------------%

Since the logical Pauli operators for the logical qubits encoded in $\mathcal{H}_T$ act nontrivially on only two qubits in the code block, if we take into consideration all $(n{-}1)^2+1$ logical qubits then \bs has distance 2 and it is an {\em error-detecting} code. However, if we only consider the logical qubit encoded in $\mathcal{H}_{L}$, we effectively obtain a distance-$n$ code and errors with support on up to $\lfloor n/2 \rfloor$ qubits in the code block can be corrected---we will call this logical qubit the {\em protected} qubit. In fact, error recovery for the protected qubit may unavoidably result in applying at the same time nontrivial logical operations with support in $\mathcal{H}_{T}$. This is not a problem as long as we never encode useful information in $\mathcal{H}_{T}$. We can think of the $(n{-}1)^2$ logical qubits encoded in $\mathcal{H}_{T}$ as {\em gauge} qubits since they correspond to degrees of freedom for the logical information encoded in $\mathcal{H}_{L}$. In some cases it will be sufficient to completely disregard the state of the gauge qubits. More interestingly, we will next discuss how fault-tolerant error correction for the protected qubit can be simplified by taking advantage of the presence of gauge qubits.

%----------------------------------------------------------%
\subsection{Syndrome Measurement using the Gauge Qubits}
\label{sec:SyndrMeasGauge}

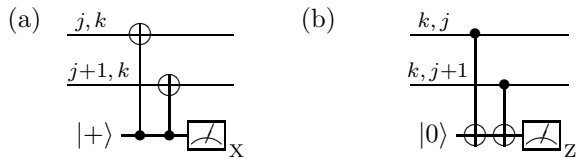
\begin{figure}[h]
\begin{center}
\begin{tabular}{ccc}
\put(-0.8,0.1){(a)}
\put(3.1,0.1){(b)}
\Qcircuit @C=0.7ex @R=2.0ex @!R {  \put(0.1,0.4){\footnotesize{$j,k$}}
   & \qw                             & \targ      & \qw       & \qw     & \qw \\
   & \qw  \put(-0.4,0.4){\footnotesize{$j{+}1,k$}}  & \qw        & \targ     & \qw     & \qw \\
   & \push{|+\rangle \hspace{0.1cm}} & \ctrl{-2}  & \ctrl{-1} & \meterx &
                                }    
%\parbox{2.5cm}{\epsfig{file=meas1.eps,width=2.35cm}} 
   &\hspace{1.2cm} & \hspace{0.2cm}
\Qcircuit @C=0.7ex @R=2.0ex @!R {  \put(0.1,0.4){\footnotesize{$k,j$}}
   & \qw                             & \ctrl{+2}  & \qw       & \qw     & \qw \\
   & \qw  \put(-0.4,0.4){\footnotesize{$k,j{+}1$}}  & \qw        & \ctrl{+1} & \qw     & \qw \\
   & \push{|0\rangle \hspace{0.1cm}} & \targ      & \targ     & \meterz &
                                }    
%\parbox{2.5cm}{\epsfig{file=meas2.eps,width=2.35cm}}
 \vspace{-0.2cm}
\end{tabular}
\end{center}
\caption{\label{fig:5.3.2} (a) A circuit for measuring the operator $X_{j,k}X_{j+1,k}$ using one ancillary qubit. (b) A similar circuit for measuring $Z_{k,j}Z_{k,j+1}$. }
\end{figure}

Let us first explain how we can extract the code syndrome indirectly by manipulating the state of the gauge qubits. Our first observation is that we can express the stabilizer generators as 
\begin{equation}
%\begin{array}{rcl}
\label{eq:gauge}
X_{j,*}X_{j+1,*}  =  \bigotimes _{k=1}^n \left( X_{j,k} X_{j+1,k} \right) \; ; \; {\rm and} \; \; Z_{*,j}Z_{*,j+1}  =  \bigotimes _{k=1}^n \left( Z_{k,j}\; Z_{k,j+1} \right) \; .
%\end{array}
\end{equation}  
\noindent What is remarkable about this decomposition is that the operators in parentheses are supported on $\mathcal{H}_T$; %as logical operators of the gauge qubits
hence, they commute with \emph{all} stabilizer generators and also commute with the logical operators for the protected qubit. Because of this, we can \emph{measure} each of them \emph{separately} and then equation (\ref{eq:gauge}) implies that the code syndrome can be computed by taking the appropriate parities of the measurement outcomes. Moreover, since these operators act nontrivially on only two qubits in the code block, measuring each one of them is especially easy; e.g., figure \ref{fig:5.3.2} shows simple circuits for measuring $X_{j,k}X_{j+1,k}$ and $Z_{k,j}Z_{k,j+1}$. 
\vspace{0.2cm}

Let us now check that these circuits can be used to construct a 1-EC gadget that satisfies properties \ref{prop:1}. Consider the circuit in figure \ref{fig:5.3.2}(a); similar observations apply to the other circuit. A fault in the preparation of the $|+\rangle$ state can only cause a $Z$ error that will propagate to flip the measurement outcome but which cannot harm the data. %In figure \ref{fig:5.3.2}(b), any errors due to faults in the subcircuit that prepares the ancillary Bell state, $|\Phi_0\rangle \propto |00\rangle + |11\rangle$, can always be attributed to one of the two qubits,\footnote{Since, $\forall \; 2{\times}2$ matrix $O$, $(O\otimes I)|\Phi_0\rangle = (I\otimes O^T)|\Phi_0\rangle$.} and, hence, an error can only propagate to at most one qubit in the data block. 
Consider now faults in the {\sc cnot} gates. %acting between the data block and the ancillary qubits. In figure \ref{fig:5.3.2}(b), the gates are transversal so errors cannot spread from one data qubit to the other. In figure \ref{fig:5.3.2}(a), 
We might worry that there is a problem if the first {\sc cnot} is faulty because an $X$ error can result in both data qubits. But we soon realize that $X_{j,k}X_{j+1,k}$ is not an error---it is exactly the operator we are trying to measure and so acts trivially after the measurement has taken place. Finally, we note that a single fault may cause one $X$ and one $Z$ error in {\em different} qubits in the data block which, therefore, will not pass through a 1-filter. However, as we will discuss in \S \ref{sec:BSLowerBounds}, the only extended rectangles that appear at the lowest levels of the fault-tolerant recursive simulation correspond to the logical {\sc cnot} and Hadamard operations. In this case, we may define the 1-filter in these levels as treating $X$ and $Z$ errors separately, and then our 1-EC gadgets satisfy properties \ref{prop:1} \footnote{Here it is important that {\sc cnot} gates do not mix $X$ and $Z$ errors upon conjugation and Hadamard gates transform $X$ errors to $Z$ errors and vice versa.}. Because of this observation, we will henceforth concentrate our discussion on the number of errors of the same ``type'' (i.e., either $X$ or $Z$) that faults may produce within the same code block. We should finally note that the measurement outcomes from the circuits in figure \ref{fig:5.3.2} cannot, in general, be trusted unless some of the measurements are repeated some number of times depending on the distance of the code. This redundancy is necessary since a single fault can cause errors in both the data block and the measurement outcomes. We will return to this point in \S \ref{sec:BSLowerBounds} by giving specific examples.

The benefit of this syndrome-measurement method is that it significantly reduces the qubit overhead for fault-tolerant error correction: Unlike in the standard 1-EC gadget constructions discussed in \S \ref{sec:FTEC-Constr}, this method does not require preparing and verifying entangled ancillary states---ancillary qubits initialized in the $|0\rangle$ or $|+\rangle$ state are sufficient as in figure \ref{fig:5.3.2}. For maximum qubit efficiency but at a cost of extra memory error, even a {\em single} ancillary qubit would suffice to sequentially measure all weight-two gauge-qubit operators necessary to extract the code syndrome. Furthermore, the specific gauge-qubit operators to be measured can be chosen to have support on neighboring physical qubits when qubits in the code block are arranged on a two-dimensional square lattice. For this reason, this method for syndrome measurement may prove to be especially advantageous for geometrically local quantum computing architectures such as, e.g., those envisioned for ion-trap or solid-state implementations.

\subsection{Standard Syndrome Measurement Procedures}
\label{sec:SyndrMeasStandard}

In settings without geometric locality constraints for the interaction of qubits or when qubit movement is much less noisy than quantum gates, syndrome measurement procedures that use encoded ancillary states and transversal interactions between the data and ancillary blocks will give the best accuracy thresholds. Since \bs is a CSS code, Steane's 1-EC gadget presented in \S \ref{sec:SteFTEC} can be used to extract the syndrome provided we can prepare and verify logical $|0\rangle$ ($|0\rangle_L$) and logical $|+\rangle$ ($|+\rangle_L$) states for the protected qubit. Alternatively, Knill's 1-EC gadget in \S \ref{sec:KniFTEC} requires preparing and verifying a logical Bell state ($|\Phi_0\rangle_L \propto |0\rangle_L|0\rangle_L + |1\rangle_L|1\rangle_L$). 
 
As is evident from the decomposition (\ref{eq:dec}), the distinctive feature of subsystem codes is that logical states in $\mathcal{H}_L$ are not uniquely encoded: after having specified a logical state in $\mathcal{H}_L$, the state in $\mathcal{H}_T$ can still be arbitrary. This freedom in choosing the state of the gauge qubits can be used to our advantage in the design of encoding circuits of logical states for the protected qubit. In particular, we will next discuss how, by exploiting this freedom, we can design remarkably simple encoding circuits for the logical ancillary states required for Steane's and Knill's 1-EC gadgets, thus also reducing the overhead associated with  the post-encoding verification of these states.

For concreteness, consider designing an encoding circuit for $|0\rangle_L$, i.e., a state in the code space which is the ${+}1$ eigenstate of $Z_L$. With the state in $\mathcal{H}_L$ specified, we can choose the  state in $\mathcal{H}_T$ to be the ${+}1$ eigenstate of the gauge-qubit operators $\{X_{i,j} X_{i+1,j}\}$. In order words, our encoding circuit prepares the $+1$ eigenstate of the operators in the following stabilizer group:
\begin{equation}
\label{eq:0-stab}
\mathcal{S}(|0\rangle_L) = \langle X_{i,j}X_{i+1,j} \; ; Z_{*,j} \; | \; i \in \mathbb{Z}_{n-1} ; j\in\mathbb{Z}_n \rangle \; .
\end{equation}
\noindent We recognize the state described by equation (\ref{eq:0-stab}) as a tensor product of $n$ {\em cat states} in the Hadamard-rotated basis, each one lying across a column in figure \ref{fig:5.3.1}. In other words, our $|0\rangle_L$ is the state
\begin{equation}
\label{eq:conj-cat}
 |0\rangle_L = \bigotimes\limits_{j=1}^n \left( {|++ \dots +\rangle_{\rm col:j} + |-- \dots -\rangle_{\rm col:j} \over \sqrt{2}} \right) \; ,
\end{equation}
\noindent where inside the parenthesis we have the state of the $n$ qubits in column $j$. 
  
We can obtain the state $|+\rangle_L$ by applying a logical Hadamard transformation to the state $|0\rangle_L$. We observe that applying Hadamard gates {\em bitwise} has the same effect as a logical Hadamard gate up to a 90-degree rotation of the square lattice in figure \ref{fig:5.3.1} (and, also, up to a nontrivial operation acting on the gauge qubits). We can therefore obtain $|+\rangle_L$ by first preparing $|0\rangle_L$, applying Hadamard gates bitwise and, finally, rotating the lattice by 90 degrees. The bitwise Hadamard gates will transform the Hadamard-rotated cat states to become usual cat states in the computation basis. And then, rotating the lattice by 90 degrees will align the $n$ cat states each to lie across a row. Hence, our $|+\rangle_L$ state is
\begin{equation}
\label{eq:cat}
 |+\rangle_L = \bigotimes\limits_{i=1}^n \left( {|00 \cdots 0\rangle_{\rm row:i} + |11 \cdots 1\rangle_{\rm row:i} \over \sqrt{2}} \right) \; .
\end{equation} 

%We observe that the $n$ Hadamard-rotated cat states in equation (\ref{eq:conj-cat}) need to be {\em verified} only against correlated $Z$ errors but not against correlated $X$ errors (any pair of $X$ errors acts trivially on a Hadamard-rotated cat state). This implies that instead of $Z$-error verification, {\em repetition} of the syndrome measurement using different $|0\rangle_L$ ancillary blocks is sufficient to satisfy properties \ref{prop:1}. This is because only $X$ errors can propagate from the $|0\rangle_L$ block to the data due to the direction of the {\sc cnot} gates, while here no correlated $X$ errors can appear in $|0\rangle_L$. Similar observations hold for $|+\rangle_L$ for which verification against correlated $X$ errors can be avoided if we repeat the syndrome measurement.

Finally, logical Bell states to be used in Knill's 1-EC gadget can be constructed by interacting two blocks encoded in the states $|+\rangle_L$ and $|0\rangle_L$ via a logical {\sc cnot} gate. Since the logical {\sc cnot} gate can be  implemented by transversal {\sc cnot} gates, this construction is also especially simple. Specific examples of Steane's and Knill's 1-EC gadgets will be given in \S \ref{sec:BSLowerBounds}.

%---------------------------------------------------------%
\subsection{Accuracy Threshold Lower Bounds}
\label{sec:BSLowerBounds}

We can now discuss how to construct 1-Gas for the Bacon-Shor code. As mentioned above, the logical {\sc cnot} and Hadamard gates have simple transversal implementations. Let us call {\em CSS operations} the operations in the set  
\begin{equation}
\mathcal{G}_{\rm CSS} = \{{\rm CNOT}, H, \mathcal{P}_{|0\rangle}, \mathcal{P}_{|+\rangle}, \mathcal{M}_X, \mathcal{M}_Z \} \; ,
\end{equation}

\noindent where $\mathcal{P}_{|\phi\rangle}$ denotes the preparation of the single-qubit state $|\phi\rangle$ and $\mathcal{M}_A$ denotes the measurement of the single-qubit operator $A$. Together with the operations in $\mathcal{G}_{\rm CSS}$, the phase gate $S$ is sufficient to generate the Clifford group and further adding, e.g., the Toffoli gate gives a quantum universal set of operations. To obtain a lower bound on the accuracy threshold for CSS operations, we will perform an analysis of malignant sets of locations in the {\sc cnot} 1-exRec. We note that we need not analyse the Hadamard 1-exRec since it contains fewer locations than the {\sc cnot} 1-exRec. The same observation is true for the preparation and measurement 1-exRecs. %Furthermore, by following the intuition developed in \S \ref{sec:Stretching}, preparation and measurement 1-exRecs can be contracted with {\sc cnot} 1-exRecs to form 1-conexRecs. These 1-conexRecs have fewer locations than the full {\sc cnot} 1-exRec and, as a consequence, the latter 1-exRec will be the one determining the accuracy threshold for CSS operations. 

The reason why the accuracy threshold for CSS operations can be analysed separately is that error correction for the Bacon-Shor code (and, more generally, for all CSS codes) can be implemented using CSS operations alone. Therefore, the extended rectangles corresponding to the logical phase and Toffoli gates need not appear at the lower levels of the recursive fault-tolerant simulation. It follows that if the physical noise strength is below the accuracy threshold for CSS operations, a recursive simulation of CSS operations can achieve any desired accuracy so that, effectively, CSS operations can be viewed as being {\em nearly ideal} at the highest levels of the recursion. Given highly reliable CSS operations, our next step will be to show that we can distill highly reliable quantum software (see \S \ref{sec:software}) that will enable the simulation of the logical phase and Toffoli gates that complete quantum universality. Overall, our strategy is to first determine the threshold for CSS operations which, by being significantly lower than the threshold for the distillation of quantum software, will set the accuracy threshold for all operations in our quantum universal operation set.

In the remaining of this section, I will first discuss the accuracy threshold for CSS operations for recursive simulations based on the Bacon-Shor code. Then, in \S \ref{sec:BSUnivComp}, I will give more details on the simulation of the logical phase and Toffoli gates which complete quantum universality.  

%------------------------------------------------------%
\subsubsection{The 9-Qubit Bacon-Shor Code}

%\vspace{0.2cm}
%------------------------------------------------%
\medskip \noindent {\bf The 1-EC Gadgets} \medskip

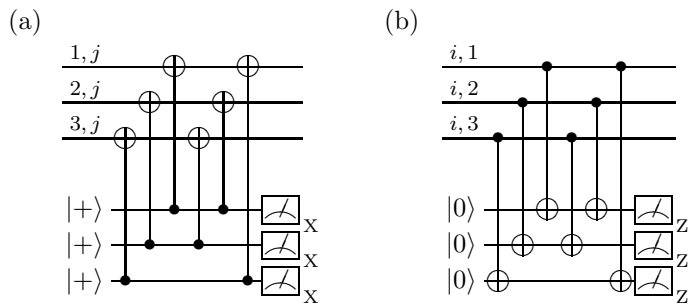
\begin{figure}[htb]
\begin{center}
\begin{tabular}{ccc}
\put(-0.2,0.5){(a)} \put(4.8,0.5){(b)}
\hspace{0.4cm}
\Qcircuit @C=0.3ex @R=0.7ex @!R {  \put(0.1,0.4){\footnotesize{$1,j$}}
   & \qw                             & \qw       & \qw       & \targ     & \qw       & \qw       & \targ  & \qw  & \qw \\
   \put(0.1,0.4){\footnotesize{$2,j$}}
   & \qw                             & \qw       & \targ     & \qw       & \qw       & \targ     & \qw    & \qw  & \qw \\
   \put(0.1,0.4){\footnotesize{$3,j$}}
   & \qw                             & \targ     & \qw       & \qw       & \targ     & \qw       & \qw    & \qw  & \qw \\
                                                                                              & \\
   & \push{|+\rangle \hspace{0.1cm}} & \qw       & \qw       & \ctrl{-4} & \qw       & \ctrl{-3} & \qw       & \meterx \\
   & \push{|+\rangle \hspace{0.1cm}} & \qw       & \ctrl{-4} & \qw       & \ctrl{-3} & \qw       & \qw       & \meterx \\
   & \push{|+\rangle \hspace{0.1cm}} & \ctrl{-4} & \qw       & \qw       & \qw       & \qw       & \ctrl{-6} & \meterx    
                                }  & \hspace{0.5cm} & \hspace{0.4cm} 
\Qcircuit @C=0.3ex @R=0.7ex @!R {  \put(0.1,0.4){\footnotesize{$i,1$}}
   & \qw                             & \qw       & \qw       & \ctrl{4}  & \qw       & \qw       & \ctrl{6}  & \qw  & \qw \\
   \put(0.1,0.4){\footnotesize{$i,2$}}
   & \qw                             & \qw       & \ctrl{4}  & \qw       & \qw       & \ctrl{3}  & \qw  & \qw  & \qw \\
   \put(0.1,0.4){\footnotesize{$i,3$}}
   & \qw                             & \ctrl{4}  & \qw       & \qw       & \ctrl{3}  & \qw       & \qw  & \qw  & \qw \\
                                                                                              & \\
   & \push{|0\rangle \hspace{0.1cm}} & \qw       & \qw       & \targ     & \qw       & \targ     & \qw  & \meterz \\
   & \push{|0\rangle \hspace{0.1cm}} & \qw       & \targ     & \qw       & \targ     & \qw       & \qw  & \meterz \\
   & \push{|0\rangle \hspace{0.1cm}} & \targ     & \qw       & \qw       & \qw       & \qw       & \targ & \meterz   
                                }
\end{tabular}
\end{center}
\caption{\label{fig:5.3.4} A fragment of the 1-EC gadget for \bsc{3} where the syndrome is measured using the gauge qubits. (a) Measuring operators in $T$ along column $j$. (b) Measuring operators in $T$ along row $i$. Identical circuits are run in parallel in all rows and columns. In both circuits, the third measurement is used to provide an extra redundancy bit that is needed to satisfy properties \ref{prop:1}.  }
\end{figure}

%\bsc{3} is the [[9,1,3]] code. %With reference to fig.~\ref{fig:5.3.1}, its stabilizer group generated by
%
%\begin{equation}
%\begin{array}{rcl}
%S(\mathcal{C}_{BS}^{(3)}) & = & \langle \; X_{{\rm row:}1,{\rm row:}2}\; ; X_{{\rm row:}2,{\rm row:}3}\; ; \; Z_{{\rm col:}1,{\rm col:}2}\; ; Z_{{\rm col:}2,{\rm col:}3}  \;  \rangle \; ,
%\end{array} 
%\end{equation}
%
%\noindent and the logical Pauli operators for the logical qubit protected in $\mathcal{H}_{\rm L}$ are $\bar{X}=X_{\rm row:1}$ and $\bar{Z}=Z_{\rm col:1}$. %The gauge-qubit operators acting nontrivially only in $\mathcal{H}_{\rm T}$ are generated by operators with an even number of $Z$ operators in each row and an even number of $X$ operators in each column (c.f., eq.$\,$(\ref{eq:t})).
%
Syndrome measurements using the gauge qubits for the 9-qubit Bacon-Shor code, $\mathcal{C}_{BS}^{(3)}$, can be done with the circuits in figure \ref{fig:5.3.4}. For this code, the measurement of one extra operator is necessary in order to satisfy properties \ref{prop:1}; in figure \ref{fig:5.3.4} we have chosen to measure the extra operator $X_{1,j}X_{3,j}$. This additional measurement provides the necessary redundancy in order to avoid the occurrence of more than one error in the data block due to a single fault\footnote{We are concerned with the case where, e.g., a $Z$ error occurs on both qubits on the support of the second {\sc cnot} operating on qubit $(2,j)$---this error will flip the eigenvalue of the measured operator $X_{1,j}X_{2,j}$, and it will also create a $Z_{2,j}$ error in the data. An error in the eigenvalue of $X_{1,j}X_{2,j}$ will lead us to incorrectly infer an error $Z_{1,j}$ in the data if we do not perform the extra measurement of $X_{1,j}X_{3,j}$. Combining $Z_{1,j}$ with $Z_{2,j}$ leads to an uncorrectable error in the data that is caused by a single faulty {\sc cnot}.}. %Similar observations hold for error correction against $X$ errors.

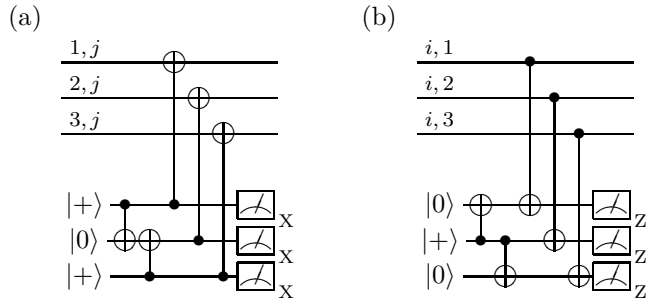
\begin{figure}[tb]
\begin{center}
\begin{tabular}{ccc}
\put(-0.4,0.5){(a)} \put(4.3,0.5){(b)}
\hspace{0.2cm} 
\Qcircuit @C=0.3ex @R=0.7ex @!R {  \put(0.1,0.4){\footnotesize{$1,j$}}
   & \qw                             & \qw       & \qw       & \targ     & \qw      & \qw        & \qw  & \qw \\
   \put(0.1,0.4){\footnotesize{$2,j$}}
   & \qw                             & \qw       & \qw       & \qw       & \targ    & \qw        & \qw  & \qw \\
   \put(0.1,0.4){\footnotesize{$3,j$}}
   & \qw                             & \qw       & \qw       & \qw       & \qw      & \targ      & \qw  & \qw \\
   & \\
   & \push{|+\rangle \hspace{0.1cm}} & \ctrl{1}  & \qw       & \ctrl{-4} & \qw       & \qw       & \meterx \\
   & \push{|0\rangle \hspace{0.1cm}} & \targ     & \targ     & \qw       & \ctrl{-4} & \qw       & \meterx \\
   & \push{|+\rangle \hspace{0.1cm}} & \qw       & \ctrl{-1} & \qw       & \qw       & \ctrl{-4} & \meterx        \\
                                } & \hspace{0.5cm} & \hspace{0.4cm} 
\Qcircuit @C=0.3ex @R=0.7ex @!R {  \put(0.1,0.4){\footnotesize{$i,1$}}
   & \qw                             & \qw       & \qw       & \ctrl{4}  & \qw      & \qw        & \qw  & \qw \\
   \put(0.1,0.4){\footnotesize{$i,2$}}
   & \qw                             & \qw       & \qw       & \qw       & \ctrl{4} & \qw        & \qw  & \qw \\
   \put(0.1,0.4){\footnotesize{$i,3$}}
   & \qw                             & \qw       & \qw       & \qw       & \qw      & \ctrl{4}   & \qw  & \qw \\
   & \\
   & \push{|0\rangle \hspace{0.1cm}} & \targ     & \qw       & \targ     & \qw       & \qw       & \meterz \\
   & \push{|+\rangle \hspace{0.1cm}} & \ctrl{-1} & \ctrl{1}  & \qw       & \targ     & \qw       & \meterz \\
   & \push{|0\rangle \hspace{0.1cm}} & \qw       & \targ     & \qw       & \qw       & \targ     & \meterz        \\
                                }
\end{tabular}
\caption{\label{fig:5.3.5} A fragment of Steane's 1-EC gadget for $\mathcal{C}_{BS}^{(3)}$. (a) Collecting the syndrome for $Z$ errors. The circuit in column $j$ is shown. (b) Collecting the syndrome for $X$ errors. Here, the circuit for row $i$ is shown. Identical circuits are run in parallel in all rows and columns. }
\end{center}
\end{figure}

Alternatively, Steane's 1-EC gadget requires preparing a logical $|0\rangle$ state for extracting the syndrome for $Z$ errors. Since our logical $|0\rangle$ state consists of three Hadamard-rotated cat states aligned across the three columns, the preparation circuit for column $j$ can be as in figure \ref{fig:5.3.5}(a). To obtain the syndrome for $X$ errors, we need to also prepare a logical $|+\rangle$ state which consists of cat states along the three rows; for row $i$, the preparation circuit is shown in figure \ref{fig:5.3.5}(b). We observe that the 1-EC gadget constructed with these circuits has the remarkable feature that it satisfies properties \ref{prop:1} {\em without} the need to verify the ancillary states. Indeed, consider a faulty {\sc cnot} inside the encoding circuit for these cat states. An $X$ error on both qubits on the support of this gate is either (i) an operator in the cat-state stabilizer or, (ii) is equivalent to a single $X$ error on the other qubit. Similarly for a $Z$ error acting on both qubits. Otherwise, an $X$ error acting on one qubit and a $Z$ error on the other cannot lead to more than one $X$ or one $Z$ error propagating from the cat state to the data block. Knill's 1-EC gadget for this code has the same feature as well; again, properties \ref{prop:1} are satisfied without the need to verify the ancillary logical Bell state which is prepared by encoding three-qubit cat states and then applying transversal {\sc cnot} gates. 

\vspace{0.2cm}
%------------------------------------------------%
\medskip \noindent {\bf Accuracy Threshold Lower Bounds} \medskip

A lower bound on the accuracy threshold can be obtained from theorem \ref{theo:4} by counting the total number of pairs of locations inside the {\sc cnot} 1-exRec since it is the largest among all 1-exRecs corresponding to CSS operations. Both Steane's and Knill's 1-EC gadgets contain $72$ qubit-preparation, measurement, gate or memory locations. Therefore, the {\sc cnot} 1-exRec contains $4\times 72 + 9 = 297$ locations, which implies $p_{\rm thr} \geq {297 \choose 2}^{-1} \geq 2.2\times 10^{-5}$. 

%Therefore, the {\sc cnot} 1-exRec contains $4\times 72 + 9 = 297$ locations, which implies the lower bound $\varepsilon_{\rm thr} \geq \left( e {297 \choose 2}\right) ^{-1} \geq 7.2\times 10^{-6}$ for the threshold strength of local noise (and $p_{\rm thr} \geq {297 \choose 2}^{-1} \geq 2.2\times 10^{-5}$ if local noise is stochastic). 

We can refine this lower bound by using the notion of malignant sets of locations described in \S \ref{sec:Malignant}. Let us denote by $p_i$ the noise strength for location type $i$, where locations are labeled as (1) memory, (2) $|0\rangle$ preparation, (3) $|+\rangle$ preparation, (4) measurement of $X$, (5) measurement of $Z$ and (6) {\sc cnot}. Assuming all sets of locations with at least three faults are malignant, analyzing all pairs of locations inside the {\sc cnot} 1-exRec gives for the effective noise strength at level $k$ of the recursive simulation:
\begin{equation}
\label{eq:error-rate}
p_{6}^{(k)} \leq \sum \limits_{j\leq i=1}^{6} \alpha_{ij}\, p_i^{(k{-}1)} p_j^{(k{-}1)} + B_{6}\, (p_{\rm max}^{(k{-}1)})^3 \; ,
\end{equation}
\noindent where $p_{\rm max}^{(k{-}1)} = \max_i \{ p_i^{(k{-}1)} \}$ (with our chosen set of CSS operations, $p_{\rm max}^{(k{-}1)} = p_6^{(k{-}1)}$). Third- or higher-order events contribute $B_{6} \equiv {297 \choose  3} \leq 4.33\times 10^{6}$ and the coefficients $\alpha_{ij}$ when Steane's 1-EC is used were found to be \cite{Aliferis06c}:
\begin{equation}
\label{eq:results}
\alpha = \left( \begin{array}{cccccc}
              52    & \\
              174	& 138 \\
              174	& 0	    & 138 \\
              234	& 0	    & 378	& 216 \\
              234	& 378	& 0	    & 0	     & 216 \\
              1154	& 1281	& 1281	& 1566	 & 1556	 & 3733 
              \end{array} \right) \, .
\end{equation}

\noindent Thus, by pessimistically taking all noise strengths equal, we obtain $A_{6}=12913$ malignant pairs, or
\begin{equation}
p_{6}^{(k)} \leq A'_{6} \left( p_{6}^{(k{-}1)} \right)^2 \; ,
\end{equation}
\noindent where $A'_{6} = {A_6 \over 2} \left( 1+ \sqrt{1 + {4B_6\over A_6^2}} \; \right) \leq 13241$. This indicates $7.55\times 10^{-5}$ as an improved lower bound on $p_{\rm thr}$. 

This lower bound can be improved further if we use the idea of contracting rectangles discussed in \S \ref{sec:Stretching}. In particular, we can contract state-preparation 1-exRecs and join them with the succeeding {\sc cnot} 1-exRecs and, similarly, we can contract measurement 1-exRecs to join them with the preceding {\sc cnot} 1-exRecs. That is, for all levels $k\geq 2$ of the recursive simulation, we need not consider separate preparation or measurement exRecs at level $k{-}1$. Instead, we can consider \parbox{1.8cm}{level-$(k{-}1)$} {\em contracted exRecs} or, simply, {\em conexRecs} that contain level-$(k{-}1)$ preparation (respectively, measurement) joined with the succeeding (respectively, preceding) level-$(k{-}1)$ {\sc cnot} operations and the intermediate \parbox{1.8cm}{level-$(k{-}1)$} EC gadgets \emph{omitted}. We observe that the effective noise strength acting on such conexRecs will be no worse than the effective noise strength acting on {\sc cnot} exRecs because (i) error correction following state preparation is superfluous as long as the preparation circuit satisfies properties \ref{prop:2}, and (ii) the  measurement of logical Pauli operators for CSS codes can be performed transversally on each qubit in the code block and, so, classical error correction on the measurement outcomes can replace quantum error correction preceding the measurements. 

Thus, for levels $k\geq 2$ of the recursive simulation, we can effectively set those coefficients in the matrix (\ref{eq:results}) that involve preparation or measurement locations to {\em zero}. Then, for levels $k\geq 2$, we compute $A_{6,{\rm str}}= 4939$, $B_{6,{\rm str}}={153 \choose 3} \leq 5.86\times 10^{5}$ and, so, $A'_{6,{\rm str}} \leq 5055$. Thus, we obtain the recursion equations
\begin{equation}
\begin{array}{rcl}
\label{eq:recursioneqs}
 p_{6}^{(1)} & \leq & A'_{6} \; \left( p_{6} \right)^2 \; , \\ 
 p_{6}^{(k)} & \leq & A'_{6,{\rm str}} \; \left( p_{6}^{(k{-}1)} \right)^2 \; , \; {\rm for} \; k>1 \; ,
\end{array}
\end{equation}

\noindent which imply the condition $p_{6}^{(1)} < 1.97 \times 10^{-4}$ and $p_{6} < 1.22\times 10^{-4}$. Hence, $1.22\times 10^{-4}$ is our new lower bound on $p_{\rm thr}$.

If we use Knill's 1-EC gadget instead, the analogous coefficients, $\alpha_{ij}$, of equation (\ref{eq:error-rate}) are \cite{Aliferis06c}:
\begin{equation}
\alpha = \left( \begin{array}{cccccc}
              60        & \\
              60	& 30 \\
              60	& 0	    & 30 \\
              180	& 0	    & 180	& 270 \\
              180	& 180	& 0	    & 0	     & 270 \\
              1104	& 552	& 552	& 1656	 & 1656	 & 4164 
              \end{array} \right) \, .
\end{equation}
\noindent Taking again all noise strengths equal, $A_6=11184$ and we find $A'_6\leq 11559$. For levels $k\geq 2$ of the recursive simulation, we can again use conexRecs; we then compute $A_{6,{\rm str}}=5328$. Furthermore, by using our observation in \S \ref{sec:KniFTEC}, we can take the transversal Bell measurements in the {\em leading} Knill's 1-EC gadgets inside the {\sc cnot} exRec to be ideal. This reduces the effective number of locations in this exRec from $297$ to $297-2\times 27=243$ at level 1, and from $153$ to $153-2\times 9=135$ at levels $k\geq 2$. The analogous recursion equations to equation (\ref{eq:recursioneqs}) give $1.26\times 10^{-4}$ as our lower bound on $p_{\rm thr}$ in this case.

%-----------------------------------------------%
\subsubsection{The 25-Qubit Bacon-Shor Code}

%\vspace{0.2cm}
%------------------------------------------------%
\medskip \noindent {\bf The 1-EC Gadgets} \medskip

\begin{figure}[htb]
\setlength{\unitlength}{1cm}
\begin{center} \vspace{0.2cm}
\begin{tabular}{ccc}
\put(-0.8,2.3){(a)} \put(4.4,2.3){(b)}
\parbox{1cm}{
\Qcircuit @C=0.1ex @R=0.6ex @!R {  \put(0.1,0.4){\footnotesize{$1,j$}}
   & \qw                             & \qw       & \qw       & \qw      & \qw       & \targ      & \qw      & \qw       & \qw       & \qw       & \qw     & \qw  & \qw \\
   \put(0.1,0.4){\footnotesize{$2,j$}}
   & \qw                             & \qw       & \qw       & \qw      & \targ     & \qw       & \qw      & \qw       & \qw       & \targ            & \qw     & \qw  & \qw \\
   \put(0.1,0.4){\footnotesize{$3,j$}}
   & \qw                             & \qw       & \qw       & \targ    & \qw       & \qw       & \qw      & \qw       & \targ  & \qw                 & \qw     & \qw  & \qw \\
   \put(0.1,0.4){\footnotesize{$4,j$}}
   & \qw                             & \qw       & \targ     & \qw      & \qw       & \qw       & \qw      & \targ  & \qw       & \qw                & \qw     & \qw  & \qw \\
   \put(0.1,0.4){\footnotesize{$5,j$}}
   & \qw                             & \qw       & \qw       & \qw      & \qw       & \qw       & \targ & \qw       & \qw       & \qw                & \qw     & \qw  & \qw \\
   & \push{|+\rangle \hspace{0.1cm}} & \qw       & \qw       & \qw      & \qw       & \ctrl{-5} & \qw      & \qw       & \qw       & \ctrl{-4}     & \qw     & \meterx \\
   & \push{|+\rangle \hspace{0.1cm}} & \qw       & \qw       & \qw      & \ctrl{-5} & \qw       & \qw      & \qw       & \ctrl{-4}     & \qw       & \qw     & \meterx \\
   & \push{|+\rangle \hspace{0.1cm}} & \qw       & \qw       & \ctrl{-5}& \qw       & \qw       & \qw      & \ctrl{-4}     & \qw       & \qw    & \qw        & \meterx \\
   & \push{|+\rangle \hspace{0.1cm}} & \qw       & \ctrl{-5} & \qw      & \qw       & \qw       &\ctrl{-4}    & \qw       & \qw       & \qw    & \qw        & \meterx %\\
%   & \push{|+\rangle \hspace{0.1cm}} & \ctrl{-5} & \qw       & \qw       & \qw      & \qw       & \qw       & \qw      & \qw       & \qw       & \ctrl{-9}   & \meterx      
                          }} 
& & \hspace{0.8cm} 
\parbox{1cm}{
\Qcircuit @C=0.1ex @R=0.6ex @!R {  \put(0.1,0.4){\footnotesize{$i,1$}}
   & \qw                             & \qw       & \qw       & \qw      & \qw       & \ctrl{5}  & \qw      & \qw       & \qw       & \qw       & \qw        & \qw  & \qw & \qw\\
   \put(0.1,0.4){\footnotesize{$i,2$}}
   & \qw                             & \qw       & \qw       & \qw      & \ctrl{5}  & \qw       & \qw      & \qw       & \qw       & \ctrl{4}  & \qw       & \qw  & \qw & \qw\\
   \put(0.1,0.4){\footnotesize{$i,3$}}
   & \qw                             & \qw       & \qw       & \ctrl{5} & \qw       & \qw       & \qw      & \qw       & \ctrl{4}  & \qw       & \qw       & \qw  & \qw & \qw\\
   \put(0.1,0.4){\footnotesize{$i,4$}}
   & \qw                             & \qw       & \ctrl{5}  & \qw      & \qw       & \qw       & \qw      & \ctrl{4}  & \qw       & \qw       & \qw       & \qw  & \qw & \qw\\
   \put(0.1,0.4){\footnotesize{$i,5$}}
   & \qw                             & \qw       & \qw       & \qw      & \qw       & \qw       & \ctrl{4} & \qw       & \qw       & \qw       & \qw       & \qw  & \qw & \qw\\
   & \push{|0\rangle \hspace{0.1cm}} & \qw       & \qw       & \qw      & \qw       & \targ     & \qw      & \qw       & \qw       & \targ     & \qw       & \qw & \meterz \\
   & \push{|0\rangle \hspace{0.1cm}} & \qw       & \qw       & \qw      & \targ     & \qw       & \qw      & \qw       & \targ     & \qw       & \qw       & \qw & \meterz \\
   & \push{|0\rangle \hspace{0.1cm}} & \qw       & \qw       & \targ    & \qw       & \qw       & \qw      & \targ     & \qw       & \qw       & \qw       & \qw & \meterz \\
   & \push{|0\rangle \hspace{0.1cm}} & \qw       & \targ     & \qw      & \qw       & \qw       & \targ    & \qw       & \qw       & \qw       & \qw       & \qw & \meterz \\
%   & \push{|0\rangle \hspace{0.1cm}} & \targ     & \qw       & \qw      & \qw       & \qw       & \qw      & \qw       & \qw       & \qw       & \targ   & \qw  & \meterz    
 }}  \vspace{0.2cm}                           
\end{tabular}
\caption{\label{fig:5.3.6} A fragment of the 1-EC gadget for \bsc{5} where the syndrome is measured using the gauge qubits. (a) Measuring  operators in $T$ along column $j$. (b) Measuring operators in $T$ along row $i$. Identical circuits are run in parallel in all rows and columns. Some additional measurements (not shown) are necessary in order  to satisfy properties \ref{prop:1}; e.g., the syndrome extraction could be repeated. }
\end{center}
\end{figure}
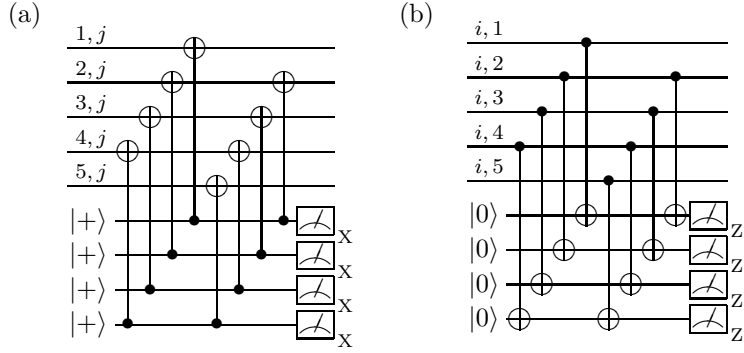

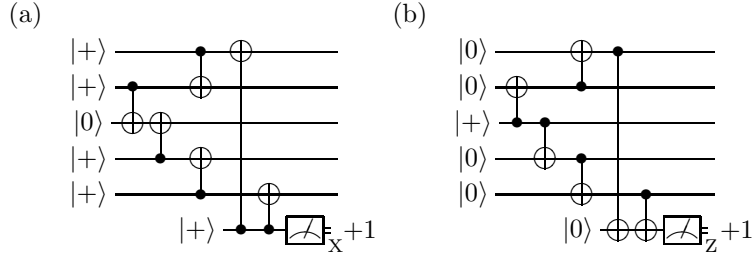
\begin{figure}[tb]
\begin{center}
\begin{tabular}{ccc}
\put(-0.5,0.4){(a)} \put(4.6,0.4){(b)}
\hspace{0.1cm}
\Qcircuit @C=0.6ex @R=0.7ex @!R {  
   & \push{|+\rangle \hspace{0.1cm}} & \qw       & \qw       & \ctrl{1}  & \targ     & \qw       & \qw & \qw       & \qw \\
   & \push{|+\rangle \hspace{0.1cm}} & \ctrl{1}  & \qw       & \targ     & \qw       & \qw       & \qw & \qw       & \qw \\
   & \push{|0\rangle \hspace{0.1cm}} & \targ     & \targ     & \qw       & \qw       & \qw       & \qw & \qw       & \qw \\
   & \push{|+\rangle \hspace{0.1cm}} & \qw       & \ctrl{-1} & \targ     & \qw       & \qw       & \qw & \qw       & \qw \\
   & \push{|+\rangle \hspace{0.1cm}} & \qw       & \qw       & \ctrl{-1} & \qw       & \targ     & \qw & \qw       & \qw \\
   &                                 &           &           & \push{|+\rangle \hspace{0.1cm}} & \ctrl{-5} & \ctrl{-1} & \meterx & \cw & \hspace{0.6cm} {+}1
                                } & & \hspace{0.5cm} 
\Qcircuit @C=0.6ex @R=0.7ex @!R {  
   & \push{|0\rangle \hspace{0.1cm}} & \qw       & \qw       & \targ     & \ctrl{5}  & \qw       & \qw & \qw       & \qw \\
   & \push{|0\rangle \hspace{0.1cm}} & \targ     & \qw       & \ctrl{-1} & \qw       & \qw       & \qw & \qw       & \qw \\
   & \push{|+\rangle \hspace{0.1cm}} & \ctrl{-1} & \ctrl{1}  & \qw       & \qw       & \qw       & \qw & \qw       & \qw \\
   & \push{|0\rangle \hspace{0.1cm}} & \qw       & \targ     & \ctrl{1}  & \qw       & \qw       & \qw & \qw       & \qw \\
   & \push{|0\rangle \hspace{0.1cm}} & \qw       & \qw       & \targ     & \qw       & \ctrl{1}  & \qw & \qw       & \qw \\
   &                                 &           &           & \push{|0\rangle \hspace{0.1cm}} & \targ & \targ & \meterz & \cw & \hspace{0.6cm} {+}1
                                } 
\end{tabular}
\end{center}
\caption{\label{fig:5.3.7} (a) A five-qubit cat state in the Hadamard-rotated basis with verification against $Z$ errors. (b) A five-qubit cat state with verification against $X$ errors. }
\end{figure}

%\bsc{5} is a [[25,1,5]] code. %With reference to fig.~\ref{fig:5.3.1}, its stabilizer group is generated by
%
%\begin{equation}
%\begin{array}{rcl}
%S(\mathcal{C}_{BS}^{(5)}) & = & \langle X_{{\rm row:}j,{\rm row:}j+1} \; ; \; Z_{{\rm col:}j,{\rm col:}j+1} \; | \; j\in\mathbb{Z}_4  \rangle \; ,
%\end{array}
%\end{equation}
%
%\noindent and the logical Pauli operators for the logical qubit protected in $\mathcal{H}_{\rm L}$ are again $\bar{X}=X_{\rm row:1}$ and $\bar{Z}=Z_{\rm col:1}$.
%
Syndrome measurements using the gauge qubits for the 25-qubit Bacon-Shor code, $\mathcal{C}_{BS}^{(5)}$, can be implemented with the circuits in figure \ref{fig:5.3.6}; for this code as well, some additional measurements to those shown must be performed in order to satisfy properties \ref{prop:1}. If instead we use Steane's 1-EC gadget, we need to prepare five-qubit cat states as is, e.g., shown in figure \ref{fig:5.3.7}. This time the cat states need verification to prevent a single fault from causing more than one $X$ or $Z$ error that will propagate to the subsequent transversal measurements. Alternatively, verification can be replaced by repeated syndrome extraction. %These cats states can be used to extract the syndrome similar to figure \ref{fig:5.3.5}.

\vspace{0.2cm}
%------------------------------------------------%
\medskip \noindent {\bf Accuracy Threshold Lower Bounds} \medskip

Now $t=2$ and, to lowest nontrivial order, we need to perform an analysis of malignant triples of locations inside the {\sc cnot} 1-exRec. Because of the size of this 1-exRec, the combinatorial analysis is significantly more time demanding than for \bsc{3}. The analysis in \cite{Aliferis06c} was performed by using a computer program running on a cluster of 20 Pentium III processors for approximately three months. The {\sc cnot} 1-exRec we analysed used Steane's 1-EC gadget, with cat-state verification as in figure \ref{fig:5.3.7}. Because this verification is nondeterministic, there is a statistical fluctuation in the number of cat-state preparation attempts that will need to be made before each cat state is accepted. For {\em stochastic} local noise, we can account for this nondeterminism in our estimate of the accuracy threshold by considering the probability of failure at level $k$ of the recursive simulation {\em conditioned on} having accepted all cat states inside all error-correction gadgets at {\em all} levels of the recursion. %Since it appears to be hard to perform a similar calculation when local noise is {\em not} stochastic, it will be preferrable in that case to use deterministic distillation procedures instead of non-deterministic verification (as discussed in \S \ref{sec:SteFTEC}).

%Since our analysis considered only non-deterministic verification procedures, our new lower bounds will only apply to local stochastic noise. 
The effective noise strength at the level-$k$ {\sc cnot} exRec if all of the verification tests contained inside this exRec are successful is
\begin{equation}
p_{6,{\rm joint}}^{(k)} \leq \sum_{\ell\leq j\leq i=1}^6 \alpha_{ij\ell}p_i^{(k-1)}p_j^{(k-1)} p_\ell^{(k-1)}+B_{6}(p_{\rm max}^{(k-1)})^4 \; ,
\end{equation}
\noindent where $B_6={1185\choose 4}$ gives the number of combinations of four locations inside the {\sc cnot} 1-exRec which contains 1185 locations in total. We now need to renormalize this noise strength to account for the fact that verification may fail. The probability of accepting all cat states inside Steane's 1-EC is
\begin{equation}
P_{{\rm accept}}^{(k)}\geq \left(1-p_{\rm max}^{(k-1)}\right)^{C_0},
\end{equation}
\noindent where $C_0=190$ is the total number of locations in the encoding and verification circuit for all cat states contained in this 1-EC gadget. Taking the noise strengths for all locations to be equal, the effective noise strength  at the level-$k$ {\sc cnot} exRec {\em conditioned on} the successful outcome in all verification tests contained inside it can be upper bounded by
\begin{equation}
\label{eq:cond}
p^{(k)} \leq \left(1-p^{(k-1)}\right)^{-4C_0}(A_6+B_6 p^{(k-1)})(p^{(k-1)})^3 \; ,
\end{equation}
\noindent where from our combinatorial analysis \cite{Aliferis06c}, $A_6=16625488$ is the number of malignant triples, and we have used the fact that the {\sc cnot} 1-exRec contains four 1-EC gadgets. Finally, as for \bsc{3}, for all levels $k\geq 2$ of the recursive simulation, we can consider contracted extended rectangles. Then, for levels $k\geq 2$ we only keep the coefficients $\alpha_{ij\ell}$ that do not involve preparation or measurement locations. Since $\alpha_{111}=287460$, $\alpha_{116}=1899918$, $\alpha_{166}=3911460$, and $\alpha_{666}=2554190$, we have $A_{6,{\rm str}}=8653028$. Also, $B_{6,{\rm str}}={705\choose 4}$ and $C_{0,{\rm str}}=120$. Thus, by solving for the fixed point of our recursion equations, we obtain $1.94\times 10^{-4}$ as our lower bound on $p_{\rm thr}$.

%-----------------------------------------------%
\subsubsection{Summary of Results}
\label{sec:summary-of-results}
 
Table \ref{table:1} summarizes our results \cite{Aliferis06c}. As already discussed, we carried out the combinatorial analysis for the concatenated \bsc{3} using both Steane's and Knill's 1-EC gadget and for the concatenated \bsc{5} using Steane's 1-EC gadget. Our best lower bound on the accuracy threshold for stochastic local noise is $1.94\times 10^{-4}$ and it was obtained with the concatenated \bsc{5}; this lower bound improves by nearly an order of magnitude the $2.73{\times}10^{-5}$ rigorous lower bound established in \cite{Aliferis05b} with the concatenated Steane [[7,1,3]] code. %The best lower bound on the threshold noise strength of general local noise is $\varepsilon_{\rm thr}\geq 1.14\times 10^{-4}$ and was obtained with the concatenated \bsc{3} using Knill's 1-EC gadget; this is the first rigorously established lower bound for local noise that is not stochastic.

\begin{table}[htb] \hspace{-0.4cm}
\begin{center}
\begin{tabular}{r|c|c|c|c|c}
  Code$\;\,\,$ & $\,$Parameters$\,$  & $\,$1-EC gadg.$\,$ & $\,$exRec locs.$\,$ & $\,p_{\rm thr}\, (\times 10^{-4})$  & $\, p_{\rm thr}^{\rm MC}\, (\times 10^{-4})$  \\
Steane  $\;$  &   [[7,1,3]]  &  Steane    &   575                     & $\, 0.27\,$ &  \\
\bsc{3} $\;$  &   [[9,1,3]]  &  Steane    &   297                     & $\,1.21\,$ & $1.21\pm 0.06$  \\
        $\;$  &              &  Knill     &   297                     & $\,1.26\,$ & $1.26\pm 0.05$  \\
\bsc{5} $\;$  &   [[25,1,5]] &  Steane    & 1,185                     & $\bf \,1.94\,$ & $1.92\pm 0.02$  \\
        $\;$  &              &  Knill     & 1,185                       &                              &                          $\bf\,2.07\pm 0.03$  \\
Golay   $\;$  &   [[23,1,7]] &  Steane    & 7,551                     &       & $\approx 1$    \\
\bsc{7} $\;$  &   [[49,1,7]] &  Steane    & 2,681                       &                              &                          $\,1.74\pm 0.01$  \\
        $\;$  &              &  Knill     & 2,681                       &                              &                         $\,1.91\pm 0.01$ 
\end{tabular} 
\caption{\label{table:1} Rigorous lower bounds on the accuracy threshold, $p_{\rm thr}$, for stochastic local noise with the concatenated Bacon-Shor code of varying block size and comparison with prior rigorous lower bounds using the concatenated Steane [[7,1,3]] code \cite{Aliferis05b} and Golay [[23,1,7]] code \cite{Reichardt06}. The fourth column gives the number of locations in the {\sc cnot} 1-exRec. The fifth column gives our lower bounds on $p_{\rm thr}$. The last column gives Monte-Carlo estimates for $p_{\rm thr}$ with $1\sigma$ uncertainties. Bold fonts indicate the best results in each column.
}
\end{center}
\end{table}

Analyzing codes of larger block size than \bsc{5} proved to be computationally difficult in this exact setting. In these cases, we may use a Monte-Carlo method and uniformly sample the set of fault paths with a fixed number of faulty locations inside an extended rectangle. By estimating what fraction, $\hat{f}$, of these sets is malignant, we obtain an estimate of the exact combinatorial coefficients with a standard error that can be determined by using $\sigma_{\hat{f}}=\sqrt{\hat{f}(1-\hat{f})/N}$ where $N$ is the sample size. We have also applied this Monte-Carlo method to the cases where we could extract the exact combinatorics in order to provide evidence that the Monte-Carlo estimates are accurate. Indeed, as can be seen in table \ref{table:1}, the exact lower bounds in those cases lie within $1\sigma$ of the estimated lower bounds. %Our Monte-Carlo results for the variety of codes we have analysed give evidence that the accuracy threshold achieves a maximum for the Bacon-Shor code with $n=5$ using Knill's 1-EC gadget. 

%----------------------------------------------------------%
\subsection{Universal Quantum Computation}
\label{sec:BSUnivComp}

It remains to discuss the fault-tolerant simulation of non-CSS operations that are required for quantum universality. We will see that as long as the physical noise strength is below the threshold for CSS operations, it will be possible to obtain arbitrarily reliable simulations of these additional operations as well. Therefore, the lower bounds presented in the previous section do indeed apply to universal quantum computation.  

Our first step will be to add to the set of CSS operations the phase gate, $S$, that allows us to generate the Clifford group. Next, to complete quantum universality, we will further add the non-Clifford Toffoli gate. Both the logical phase and Toffoli gates will be simulated by using CSS operations and by consuming special ``quantum software'' ancillary states; \S \ref{sec:logical-phase-gate} discusses the simulation of the logical phase gate and \S \ref{sec:logical-toffoli-gate} the simulation of the logical Toffoli gate. In both these sections, we will assume we are able to prepare these ancillary states with a fault probability that is upper bounded by some constant, $p_{\rm anc}$, and we will discuss how arbitrarily accurate copies can be distilled from them if $p_{\rm anc}$ is below some distillation threshold value. We will postpone until the final section, \S \ref{sec:bounds-software}, the discussion about how ancillary states with accuracy below this distillation threshold can be prepared for the 9- and 25-qubit Bacon-Shor codes.

\vspace{0.3cm}
%----------------------------------------------------%
\subsubsection{The Logical Phase Gate}
\label{sec:logical-phase-gate}

We first note that since the logical {\sc cnot} and Hadamard gates are transversal, the logical {\sc cphase} is also transversal; it can be implemented by doing a logical Hadamard on the target, followed by a logical {\sc cnot}, followed by another logical Hadamard on the target. %Thus, by imagining the qubits in each code block sitting on the vertices of the square lattice in fig.~\ref{fig:5.3.1}, the logical {\sc cphase} can be implemented transversally between two code blocks by i) rotating by 90 degrees the lattice of one, say the second, code block, ii) interacting the qubits in the two lattices bitwise with {\sc cphase} gates, and iii) performing another 90-degree rotation of the lattice of the second code block.

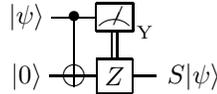
\begin{figure}[ht]
\begin{center} \vspace{0.2cm}
\begin{tabular}{c}
\Qcircuit @C=1ex @R=2.2ex @!R {  
   & \push{|\psi\rangle \hspace{0.1cm}} & \ctrl{1} & \metery \cwx[1]  \\
   & \push{|0\rangle \hspace{0.1cm}}    & \targ    & \gate{Z} & \qw & \qw & \push{S|\psi\rangle \hspace{0.1cm}} 
                                }  
\vspace{-0.1cm}
\end{tabular}
\end{center}
\caption{\label{fig:gate-telep} Circuit simulation of the phase gate, $S$, by using a measurement of $Y$. Contrary to the standard notation, here the correction operator $Z$ is applied when the measurement outcome is ${+}1$ and no correction is applied when the outcome is ${-}1$.}
\end{figure}

A destructive measurement of the logical $X$ (respectively, $Z$) operator can be performed by transversally measuring the operator $X$ (respectively, $Z$) on each qubit in the code block. (With the logical {\sc cnot} and {\sc cphase} transversal, we can also measure non-destructively the logical $X$ or the logical $Z$ operator by using as control an ancilla prepared in the logical $|+\rangle$ state.) If we were also able to easily measure the logical $Y$, $Y_L = i X_L Z_L$, we would be able to implement the logical phase gate by using, e.g., the circuit in figure \ref{fig:gate-telep}. %The logical {\sc cnot}, Hadamard and phase gates would then be sufficient for generating the logical Clifford group. 

Although $Y_L=\otimes_{i=1}^{n^2} Y^{(i)}$ when $n$ is odd, measuring transversally the operators $Y^{(i)}$ does not give a measurement 1-Ga which satisfies property \ref{prop:2}(b). This is because the Bacon-Shor code has no stabilizer operators that can be written as tensor products of $Y$ operators {\em alone}. Therefore, the problem is that we cannot perform error correction on the transversal measurement outcomes and, so, the eigenvalue we would deduce for the logical $Y$ could be erroneous even if a single one of the transversal measurements failed. 

We could instead measure the logical $Y$ operator nondestructively using cat states similar to our discussion in \S \ref{sec:SyndrCatStates}. Implementing this  measurement would require controlled-$Y$ gates, which at the next level of the recursive simulation would have to be implemented in an encoded form. But then, the problem is that the controlled-$Y$ gate is complex and all the transversal operations we have discussed so far ({\sc cnot}, {\sc cphase} and Hadamard) are real. Therefore, we do not have a direct transversal method for implementing the logical controlled-$Y$ gate.  

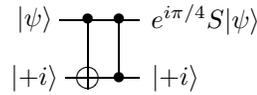
\begin{figure}[t]
\begin{center}
\begin{tabular}{c}
%
%\parbox{3cm}{\epsfig{file=S-simulation.eps,width=3.4cm}}
\Qcircuit @C=1ex @R=2.5ex @!R {
   & \push{|\psi\rangle \hspace{0.1cm}} & \ctrl{1} &\qw & \ctrl{1}    &\qw &\qw  & \push{e^{i\pi/4}S|\psi\rangle}  \\
   & \push{|{+}i\rangle \hspace{0.1cm}} & \targ    &\qw & \control \qw  &\qw & \qw  & \push{|{+}i\rangle \hspace{0.8cm}}
                                }  
\vspace{-0.2cm}
\end{tabular}
\end{center}
\caption{\label{fig:5.3.8} Circuit simulation of the phase gate $S$ using the ancillary state $|{+}i\rangle$.}
\end{figure}

Fortunately, there is a method for simulating the logical $S$ gate by using only logical {\sc cnot} and {\sc cphase} gates provided we can prepare a certain logical ancillary state: Consider the states $|{\pm}i\rangle \propto |0\rangle \pm i |1\rangle$ which are the $\pm 1$ eigenstates of $Y$. Given the state $|{+}i\rangle$, we can simulate $S$ (up to an irrelevant phase) with the circuit in figure \ref{fig:5.3.8}. Hence, the problem of constructing a 1-Ga for the logical phase gate reduces to the problem of preparing the {\em logical} ancillary state $|{+}i\rangle$.

\vspace{0.5cm}
%------------------------------------------------%
\medskip \noindent {\bf Simulating $\bf S$ and $\bf S^*$ in Superposition} \medskip

Based on the circuit in figure \ref{fig:5.3.8}, we will now first describe a simple procedure by which the simulation of the logical $S$ gate is still possible even if the logical ancillary state $|{+}i\rangle$ is replaced by another logical ancillary state which is easier to prepare \cite{Dennis02}. This procedure makes use the following two observations: (i) the $S$ gate is the only complex gate in our universal gate set, and (ii) if the state $|{-}i\rangle$ is used instead in the the circuit in figure \ref{fig:5.3.8}, then the operation $S^{\dagger} = S^*$ will be simulated. 

The idea is to use some standard state, say the state $|0\rangle$, instead of $|{+}i\rangle$ in the circuit in figure \ref{fig:5.3.8}. We can expand $|0\rangle \propto |{+}i\rangle + |{-}i\rangle$ and, for each of the two terms, we can consider the two paths of the subsequent computation which are executed in superposition. In one path $S$ is simulated and in the other $S^*$. Thus, if every time we want to simulate $S$ in our circuit we use the \emph{same} ancillary $|0\rangle$ state and because $S$ is the only complex gate in our gate set, the final state of the computation will be a linear superposition of one term where the desired computation unitary, $U$, has been implemented and a second term where $U^*$ has been implemented instead. In other words, if the initial computation state is $|\psi_{\rm initial}\rangle$, then the final computation state will be 
\begin{equation}
 |\psi_{\rm final}\rangle = {|{+i}\rangle \otimes U|\psi_{\rm initial}\rangle + |{-i}\rangle \otimes U^* |\psi_{\rm initial}\rangle \over \sqrt{2}} \; \; .
\end{equation}

In the end of the computation some operator $A$ will be measured which we can take to be real (and so, due to hermiticity, $A^T=A$). We now want to see that the expectation value for $A$ will be the same as if the desired $U$ had been simulated all along. Indeed, we compute
\begin{equation}
\langle A \rangle  =  \langle \psi_{\rm initial} | {U^\dagger A U + (U^*)^\dagger A U^* \over 2} |\psi_{\rm initial}\rangle = \langle \psi_{\rm initial} | U^\dagger A U |\psi_{\rm initial}\rangle \; ,
\end{equation}

\noindent since, $\forall |\psi\rangle$, $\langle \psi| (U^*)^\dagger A U^* |\psi\rangle =  \langle \psi| U^\dagger A^T U |\psi\rangle =  \langle \psi | U^\dagger A U |\psi\rangle $.

We can use this procedure at the logical level as well. The only penalty we pay for simulating the logical $S$ gate in this way is that we need to swap around the ancillary logical $|0\rangle$ block we use for the simulation if we are constrained to use only local interactions; but this will only give us a linear penalty in the size of the computation. Of course, we should emphasize that this trick works because the $S$ gate is not used in implementing error correction. As a consequence, we only need to implement logical $S$ gates at the {\em highest} level of the recursive fault-tolerant simulation and, moreover, it is not necessary that we execute different logical $S$ gates in parallel anywhere in our computation. Therefore, provided the physical noise strength is below the accuracy threshold for CSS operations, we may obtain any desired accuracy in the simulation of the logical $S$ gates at the highest level of our recursive simulation by choosing this highest level appropriately.

%------------------------------------------------%
\medskip \noindent {\bf Noisy $\mathbf{|{+}i\rangle}$ Distillation} \medskip

\begin{figure}[ht]
\begin{center} \vspace{0.2cm}
\parbox{1cm}{
\Qcircuit @C=1ex @R=2.5ex @!R {
   & \push{|{+}i\rangle \hspace{0.1cm}} & \qw & \ctrl{1} &\qw & \targ     &\qw  & \qw   & \push{\hspace{-0.4cm} |{+}i\rangle} \\
   & \push{|{+}i\rangle \hspace{0.1cm}} & \qw & \control \qw &\qw & \ctrl{-1} &\qw  & \meterx
                                }} 
\caption{\label{fig:5.3.9} Circuit for the nondeterministic distillation of the state $|{+}i\rangle$. Postselecting on the ${+}1$ measurement outcome, the output $|{+}i\rangle$ state has quadratically improved fidelity.}
\end{center}
\end{figure}
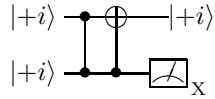

In cases when $S$ is not the only complex gate in our gate set, there is also a straightforward procedure for fault-tolerantly preparing the required ancillary ``quantum software:'' We can begin by preparing many noisy logical $|{+}i\rangle$ states and use them to progressively distill less and less noisy copies. A possible distillation circuit is shown in figure \ref{fig:5.3.9}.

The measurement outcome in this circuit is ideally ${+}1$, and an error on \emph{one} of the two $|{+}i\rangle$ states can be detected. We note that because $Y$ stabilizes the state $|{+}i\rangle$, we need only worry about $Z$ errors---we can write $X=iZY$ which is, up to the irrelevant phase $i$, equivalent to $Z$ when it acts on $|{+}i\rangle$. Postselecting on the ${+}1$ measurement outcome, the fidelity of the output $|{+}i\rangle$ state is increased quadratically relative to the fidelity of the input state since it takes errors in \emph{both} input $|{+}i\rangle$ states for an error in the output state to go undetected. 

To be concrete, assume noise is local and stochastic and let $p_{\rm anc}$ denote the probability that some fault has occurred during the preparation of one of the initial copies of the distillation protocol. Then by ``twirling,'' i.e., by applying at random with probability $1/2$ a $Y$ operator (and by doing nothing with probability $1/2$), we may describe the initial copies as having a $Z$ error with probability at most $p_{\rm anc}$. If the CSS operations of the protocol are executed ideally, then the probability of a $Z$ error on the output copy conditioned on getting the ${+}1$ measurement outcome and accepting the state is 
\begin{equation}
\label{eq:add-eq-1}
p_{\rm anc}^{(1)} \leq {p_{\rm anc}^2 \over p_{\rm anc}^2 + (1-p_{\rm anc})^2} \leq 2 p_{\rm anc}^2 \; ,
\end{equation}

\noindent since both input copies must have a $Z$ error in order for an erroneous output state to be accepted.\footnote{Here, we used the inequality, $\forall \eta_1,\eta_2\leq 1/2$, $\eta_1 \eta_2 + (1-\eta_1)(1-\eta_2)\geq 1/2$, which also implies that the bound (\ref{eq:add-eq-1}) is valid even if the error probabilities on the two input copies are not exactly equal.}

If we repeat the distillation protocol $l\geq 2$ times and since the input copies for each round are prepared independently at the previous round, we find $p_{\rm anc}^{(l)} \leq {1\over 2} \left( 2 p_{\rm anc}\right)^{2^l}$ 
%
%\begin{equation}
%\label{eq:add-eq-2}
%p_{\rm anc}^{(l)} \leq {1\over 2} \left( 2 p_{\rm anc}\right)^{2^l} \; ,
%\end{equation}
%
\noindent which indicates a distillation threshold, 
%\begin{equation}
$p_{\rm thr}^{\rm dist,|{+}i\rangle}={1/2}$.
\subsubsection{The Logical Toffoli Gate}
\label{sec:logical-toffoli-gate}

\begin{figure}[t]
\begin{center} \vspace{0.4cm}
\parbox{1cm}{
\Qcircuit @C=1.4ex @R=1.8ex @!R {
   & \push{|\psi_{1}\rangle \hspace{0.1cm}} & \qw      & \targ     & \qw       & \qw      & \meterz & \cw & \rstick{\hspace{-0.1cm} {j_1}} \\
   & \push{|\psi_{2}\rangle \hspace{0.1cm}} & \qw      & \qw       & \targ     & \qw      & \meterz & \cw & \rstick{\hspace{-0.1cm} {j_2}} \\
   & \push{|\psi_{3}\rangle \hspace{0.1cm}}  & \qw      & \qw       & \qw       & \ctrl{3} & \meterx & \cw & \rstick{\hspace{-0.1cm} {j_3}} \\
   & \push{|+\rangle \hspace{0.1cm}}         & \ctrl{2} & \ctrl{-3} & \qw       & \qw      & \qw & \qw   \\
   & \push{|+\rangle \hspace{0.1cm}}         & \ctrl{1} & \qw       & \ctrl{-3} & \qw      & \qw & \qw  \\
   & \push{|0\rangle \hspace{0.1cm}}         & \targ    & \qw       & \qw       & \targ    & \qw & \qw  \gategroup{4}{2}{6}{3}{1em}{--} \\
                                }}
\caption{\label{fig:5.3.11} Circuit simulation of the Toffoli gate using the ancilla $|{\rm Toffoli}\rangle$ shown in a box. Depending on the measurement outcomes, Clifford-group corrections (not shown) may need to be applied to the output qubits. To see that the circuit teleports the Toffoli gate, note that the Toffoli gate inside the box can be commuted to the right of the circuit thereby leaving behind three teleportation circuits for each of the three input qubits.}
\end{center}
\end{figure}
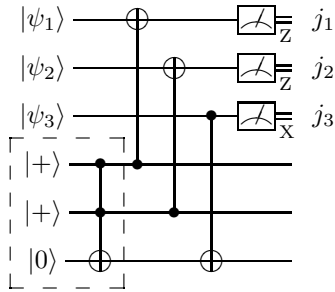

The simulation of the logical Toffoli gate uses the logical ancillary state 
\[
|{\rm Toffoli}\rangle\,{\equiv}\,\Lambda^2(X)\, \left( \, |+\rangle{\otimes}|+\rangle{\otimes}|0\rangle \, \right)
\] 
\noindent as an input to the circuit in figure \ref{fig:5.3.11} \cite{Shor96,Preskill97}. Depending on the $\pm1$  measurement outcomes, the post-measurement Clifford-group corrections are $S_1^{k_1} S_2^{k_2} S_3^{k_3}$ where $k_i=0$ if $j_i=+1$ and $k_i=1$ if $j_i=-1$; here,
\[ S_1=X^{(1)}\otimes {\rm CNOT}_{2\rightarrow 3}\;, S_2=X^{(2)}\otimes {\rm CNOT}_{1\rightarrow 3}\;, {\rm and} \; S_3 = {\rm CPHASE}_{1,2}\otimes Z^{(3)} \; ,
\]
\noindent and let us note that these operators are mutually commuting and $|{\rm Toffoli}\rangle$ is their simultaneous $+1$ eigenvector. Therefore, the problem of constructing a 1-Ga for the logical Toffoli gate reduces to preparing a logical ancillary state, the logical $|{\rm Toffoli}\rangle$.

%------------------------------------------------%
\medskip \medskip \noindent {\bf Recursive $|$Toffoli$\rangle$ Preparation} \medskip

%The state $|{\rm Toffoli}\rangle$ is the ${+}1$ eigenstate of the three operators 
%
%\[ S_1=X^{(1)}\otimes {\rm CNOT}_{2\rightarrow 3}\;, S_2=X^{(2)}\otimes {\rm CNOT}_{1\rightarrow 3}\;, {\rm and} \; S_3 = {\rm CPHASE}_{1,2}\otimes Z^{(3)} \; .
%\]

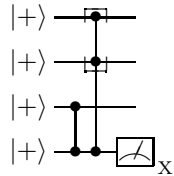
\begin{figure}[ht]
\begin{center} \vspace{0.2cm}
\parbox{1cm}{
\Qcircuit @C=0.9ex @R=1.5ex @!R {
   & \push{|+\rangle \hspace{0.1cm}} &\qw & \qw   &\qw    & \ctrl{1}  &\qw & \qw \\
   & \push{|+\rangle \hspace{0.1cm}} &\qw & \qw    &\qw   & \control \qw  &\qw & \qw \\
   & \push{|+\rangle \hspace{0.1cm}} &\qw & \control \qw &\qw & \qw   &\qw    & \qw \\
   & \push{|{\rm +}\rangle \hspace{0.1cm}} &\qw & \ctrl{-1} &\qw & \ctrl{-2} &\qw & \meterx  \gategroup{1}{6}{2}{6}{.8em}{--}
                                }}
\caption{\label{fig:5.3.12} A circuit that projects the input state $|+\rangle^{\otimes 3}$ onto the ${+}1$ eigenstate of the operator $S_3$ depending on the measurement outcome. For outcome $+1$ the output state is $|{\rm Toffoli}\rangle$. }
\end{center}
\end{figure}

Let us consider starting with the state $|+\rangle^{\otimes 3}$ which is clearly a ${+}1$ eigenstate of both $S_1$ and $S_2$; then, to obtain the state $|{\rm Toffoli}\rangle$, it remains to perform a measurement of the operator $S_3$, e.g., with the circuit in figure \ref{fig:5.3.12} \cite{Preskill97}.

%If this circuit were executed in an un-encoded form, the cat state would simply be one qubit in the state $|+\rangle$. When the logical circuit is executed instead, the cat state is $|{\rm cat}\rangle \propto |0\rangle^{\otimes n} + |1\rangle^{\otimes n}$ where $n$ the number of physical qubits in the code block. , the car state must either be verified against $X$ errors, or a decoding circuit can replace the transversal measurements in order to avoid verification \cite{Aliferis06b}. 

When this circuit is realized in an encoded form, the first three qubits become three code blocks each encoded in the logical $|+\rangle$ state and the fourth qubit is replaced by $n_c$ qubits in a cat state, where $n_c$ is the number of qubits in the code block. Since the measurement outcome can be erroneous due to even a single fault acting on the cat-state qubits, the measurement of the {\em logical} $S_3$ must be repeated several times to ensure properties \ref{prop:2} are satisfied. If the code has distance $d=2t+1$, $d$ repetitions are sufficient since then the majority vote is guaranteed to give the correct outcome if no more than $t$ of the individual measurement outcomes are erroneous, thereby reducing the probability of failure to order $t+1$. Furthermore, in order to correct faults in the noisy Toffoli gates, error-correction steps must be inserted on each one of the three code blocks between successive measurements.

Let us also comment on the second gate in figure \ref{fig:5.3.12}. Since the logical {\sc cphase} is transversal, applying the logical {\sc cphase} gate on the first two code blocks with a {\em control} by the cat-state qubits can be implemented as follows: We may first rotate the qubits in, say, the first code block by 90 degrees; next, we apply controlled-{\sc cphase} gates bitwise with control the cat state and target the first two code blocks; finally, we rotate back by 90 degrees the qubits in the first code block. By conjugating with bitwise Hadamard gates the cat-state qubits, the controlled-{\sc cphase}  gates become Toffoli gates so that the entire preparation circuit for the $|{\rm Toffoli}\rangle$ state uses only CSS operations and Toffoli gates. This is important because it allows us to apply this simulation procedure recursively. That is, we may start with ``level-0'' $|{\rm Toffoli}\rangle$ states that enable the simulation of noisy ``level-0'' Toffoli gates and, next, we may use these gates to prepare ``level-1'' $|{\rm Toffoli}\rangle$ states that will allow us to simulate less noisy ``level-1'' Toffoli gates, etc., until we eventually quickly reach our desired accuracy.

To be more concrete, let $p_{\rm anc}$ be the probability that a fault has occurred during the preparation of a ``level-0'' $|{\rm Toffoli}\rangle$ state, maximized over all our copies of such states. If we assume all CSS operations in the circuit in figure \ref{fig:5.3.12} are executed ideally, the noise strength on the output ``level-1'' $|{\rm Toffoli}\rangle$ state is 
\begin{equation}
\label{eq:add-eq-4}
p_{\rm anc}^{(1)} \leq d {n_c \choose t+1} p_{\rm anc}^{t+1} + {d\choose t+1} \left( n_c\; p_{\rm anc}\right)^{t+1} \; .
\end{equation}

\noindent Here, the first term is due to the fact that at least $t+1$ faults must have occurred in at least one of the $d$ sequential executions of the circuit in figure \ref{fig:5.3.12} in order for a logical error to be created---any number of faults less than $t+1$ will be corrected by the intermediate error-correction steps that are executed ideally since they only contain CSS operations. The second term is due to the possibility of having at least $t+1$ of the $d$ measurements of the logical $S_3$ give erroneous outcomes---for each individual measurement of $S_3$, an error can be produced due to a fault in any one of the $n_c$ Toffoli gates inside the measurement circuit. 

By repeating this simulation procedure recursively, we may achieve any desired accuracy for the simulation of the logical Toffoli gate as long as the noise strength acting on the initial ``level-0'' $|{\rm Toffoli}\rangle$ state is below a threshold value; equation (\ref{eq:add-eq-4}) implies that this threshold value is 
\begin{equation}
p_{\rm thr,determ}^{{\rm dist},|{\rm Toffoli\rangle}} = \left( d {n_c \choose t+1} + {d\choose t+1} n_c^{t+1} \right)^{-1/t} \; .
\end{equation}

\noindent For the $[[n_c=9,k=1,d=3]]$ Bacon-Shor code we find a threshold of at least $.28\%$; for the $[[n_c=25,k=1,d=5]]$ Bacon-Shor code, $.24\%$.

%------------------------------------------------%
\medskip \noindent {\bf Noisy $|$Toffoli$\rangle$ Distillation} \medskip

Alternatively, we can consider preparing several copies of the logical $|{\rm Toffoli}\rangle$ state and using them to distill less noisy output copies \cite{Gottesman04,Preskill-notes}. Let us first construct a circuit that takes as an input some number of copies of the $|{\rm Toffoli}\rangle$ state and outputs one $|{\rm Toffoli}\rangle$ state of improved fidelity.

\begin{figure}[tb]
\begin{center}
\begin{tabular}{ccccccc}
\hspace{0.05cm}
\Qcircuit @C=0.6ex @R=1ex @!R { 
   & \qw  & \qw      & \ctrl{2} & \qw      & \qw \\
   & \qw  & \ctrl{1} & \qw      & \ctrl{1} & \qw \\
   & \qw  & \ctrl{1} & \targ    & \ctrl{1} & \qw \\
   & \qw  & \targ    & \qw      & \targ    & \qw
                                }  &\;\; \put(0,-.8){=} \;\;\;\;& 
\Qcircuit @C=0.6ex @R=1ex @!R { 
   & \qw  & \ctrl{2} & \ctrl{1} & \qw \\
   & \qw  & \qw      & \ctrl{2} & \qw \\
   & \qw  & \targ    & \qw      & \qw \\
   & \qw  & \qw      & \targ    & \qw
                                } \vspace{0.7cm} & \hspace{1.5cm} &
\Qcircuit @C=0.6ex @R=1ex @!R { 
   & \qw  & \qw      & \targ     & \qw      & \qw \\
   & \qw  & \ctrl{1} & \qw       & \ctrl{1} & \qw \\
   & \qw  & \ctrl{1} & \qw       & \ctrl{1} & \qw \\
   & \qw  & \targ    & \ctrl{-3} & \targ    & \qw
                                } &\;\; \put(0,-0.8){=}  \;\;\;\; & 
\Qcircuit @C=0.6ex @R=1ex @!R { 
   & \qw  & \targ     & \targ     & \qw \\
   & \qw  & \qw       & \ctrl{-1} & \qw \\
   & \qw  & \qw       & \ctrl{-1} & \qw \\
   & \qw  & \ctrl{-3} & \qw       & \qw
                                }  \vspace{-0.8cm}
\end{tabular}
\end{center}
\caption{\label{fig-10} Commutation relations of {\sc cnot} and Toffoli.}
\end{figure}
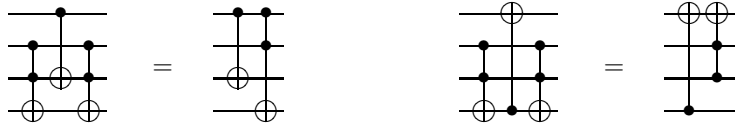

Consider the circuit in figure \ref{fig-9}(a). We start with one copy of the state $|{\rm Toffoli}\rangle$ and we measure the operator $S_1$ using one ancillary qubit initialized in the state $|+\rangle$. Since the state $|{\rm Toffoli}\rangle$ is an eigenstate of this operator, the effect of the measurement is trivial: the first three qubits remain in the state $|{\rm Toffoli}\rangle$ and the measurement on the ancilla always ideally gives outcome ${+}1$. The circuit in figure \ref{fig-9} includes two extra ancillary qubits. These qubits are useful because we now want to insert two cancelling Toffoli gates to obtain the equivalent circuit in figure \ref{fig-9}(b). Our motivation for inserting the two Toffoli gates is to create two copies of the $|{\rm Toffoli}\rangle$ state in the input. Thus, we next commute the leftmost of the two Toffoli gates to the left by using repeatedly the circuit identities in figure \ref{fig-10}. After the dust settles, the result is the circuit shown in figure \ref{fig-11}(a). 

\begin{figure}[tb]
\setlength{\unitlength}{1cm}
\begin{center}
  \begin{tabular}{ccc}
\put(-0.8,0.2){(a)} \put(6.3,0.2){(b)}  
\Qcircuit @C=0.6ex @R=1ex @!R {
   & \push{|{+}\rangle \hspace{0.1cm}} & \ctrl{2} & \qw       & \qw & \targ     & \qw       & \qw & \qw       & \qw       & \qw \\
   & \push{|{+}\rangle \hspace{0.1cm}} & \ctrl{1} & \qw       & \qw & \qw       & \ctrl{1}  & \qw & \ctrl{4}  & \qw       & \qw \\
   & \push{|0\rangle \hspace{0.1cm}}   & \targ    & \qw       & \qw & \qw       & \targ     & \qw & \qw       & \targ     & \qw \\
   & \\
   & \push{|{+}\rangle \hspace{0.1cm}} & \qw      & \qw       & \qw & \ctrl{-4} & \ctrl{-2} & \qw & \qw       & \qw       & \meterx \\
   & \push{|{+}\rangle \hspace{0.1cm}} & \qw      & \qw       & \qw & \qw       & \qw       & \qw & \targ     & \qw       & \qw \\
   & \push{|0\rangle \hspace{0.1cm}}   & \qw      & \qw       & \qw & \qw       & \qw       & \qw & \qw       & \ctrl{-4} & \qw \gategroup{1}{2}{3}{3}{1em}{--}
                                }    & \hspace{2.8cm} & 
\Qcircuit @C=0.6ex @R=1ex @!R {
   & \push{|{+}\rangle \hspace{0.1cm}} & \ctrl{2} & \qw       & \qw & \targ     & \qw       & \qw & \qw       & \qw       & \qw & \qw & \qw & \qw & \qw  \\
   & \push{|{+}\rangle \hspace{0.1cm}} & \ctrl{1} & \qw       & \qw & \qw       & \ctrl{1}  & \qw & \ctrl{4}  & \qw       & \qw & \qw & \qw & \qw  & \qw \\
   & \push{|0\rangle \hspace{0.1cm}}   & \targ    & \qw       & \qw & \qw       & \targ     & \qw & \qw       & \targ     & \qw & \qw & \qw & \qw  & \qw \\
   & \\
   & \push{|{+}\rangle \hspace{0.1cm}} & \qw      & \qw       & \qw & \ctrl{-4} & \ctrl{-2} & \qw & \qw       & \qw       & \qw & \ctrl{2} & \qw & \ctrl{2} & \qw & \meterx \\
   & \push{|{+}\rangle \hspace{0.1cm}} & \qw      & \qw       & \qw & \qw       & \qw       & \qw & \targ     & \qw       & \qw & \ctrl{1} & \qw & \ctrl{1} & \qw & \qw \\
   & \push{|0\rangle \hspace{0.1cm}}   & \qw      & \qw       & \qw & \qw       & \qw       & \qw & \qw       & \ctrl{-4} & \qw & \targ    & \qw & \targ    & \qw & \qw \gategroup{1}{2}{3}{3}{1em}{--}
                                }  \vspace{0.3cm}                             
  \end{tabular}
\caption{\label{fig-9} (a) A circuit that implements a measurement of $S_1$ on the input $|{\rm Toffoli}\rangle$ state shown in a box. The measurement outcome is ideally ${+}1$ since $|{\rm Toffoli}\rangle$ is an eigenstate of $S_1$. Note that the gates acting on the last two qubits operate trivially. (b) The same circuit after a pair of cancelling Toffoli gates has been inserted.  }
\end{center}
\end{figure}
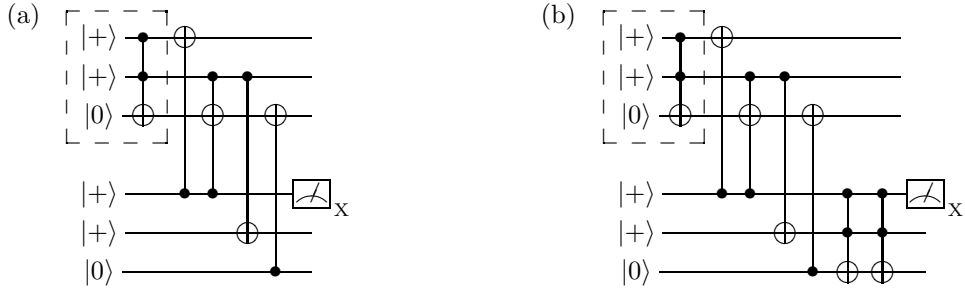 

%\begin{figure}[tb]
%\begin{center}
%\epsfig{file=lower-bounds/gadgets/T-ancilla-2.eps}
%\caption{\label{fig-9b} The circuit in figure \ref{fig-9} where a pair of cancelling Toffoli gates has been inserted. }
%\end{center}
%\end{figure}

%\setlength{\unitlength}{1cm}

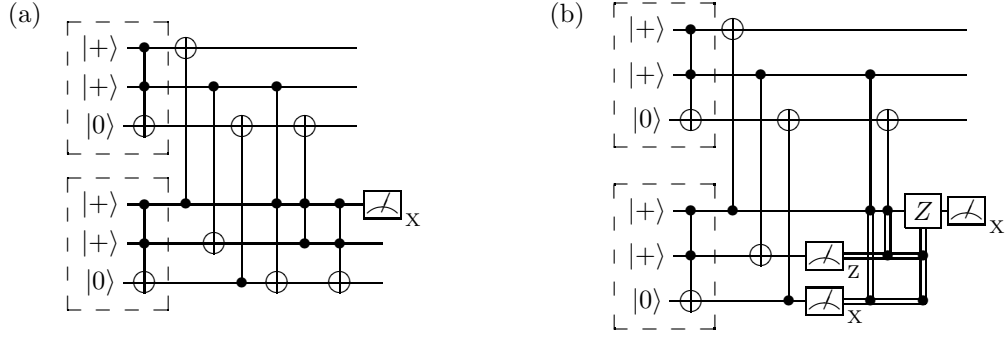
\begin{figure}[tb]
%\begin{center}
%\epsfig{file=lower-bounds/gadgets/T-ancilla-3.eps}
\setlength{\unitlength}{1cm}
\begin{center}
  \begin{tabular}{ccc}
\put(-0.8,2){(a)} \put(6.4,2){(b)} 
\parbox{3cm}{\Qcircuit @C=0.6ex @R=1ex @!R { 
   & \push{|{+}\rangle \hspace{0.1cm}} & \ctrl{2} & \qw       & \qw & \targ     & \qw       & \qw       & \qw & \qw      & \qw      & \qw & \qw & \qw  \\
   & \push{|{+}\rangle \hspace{0.1cm}} & \ctrl{1} & \qw       & \qw & \qw       & \ctrl{4}  & \qw       & \qw & \ctrl{4} & \qw      & \qw & \qw  & \qw \\
   & \push{|0\rangle \hspace{0.1cm}}   & \targ    & \qw       & \qw & \qw       & \qw       & \targ     & \qw & \qw      & \targ      & \qw & \qw  & \qw \\
   & \\
   & \push{|{+}\rangle \hspace{0.1cm}} & \ctrl{2} & \qw       & \qw & \ctrl{-4} & \qw       & \qw       & \qw & \ctrl{2} & \ctrl{-2}  & \qw & \ctrl{2} & \qw & \meterx \\
   & \push{|{+}\rangle \hspace{0.1cm}} & \ctrl{1} & \qw       & \qw & \qw       & \targ     & \qw       & \qw & \qw      & \ctrl{-1}  & \qw & \ctrl{1} & \qw & \qw \\
   & \push{|0\rangle \hspace{0.1cm}}   & \targ    & \qw       & \qw & \qw       & \qw       & \ctrl{-4} & \qw & \targ    & \qw      & \qw & \targ    & \qw & \qw \gategroup{1}{2}{3}{3}{1em}{--} \gategroup{5}{2}{7}{3}{1em}{--}
                                }}     & \hspace{2cm} & 
\parbox{3cm}{\Qcircuit @C=0.6ex @R=1ex @!R {
   & \push{|{+}\rangle \hspace{0.1cm}} & \ctrl{2} & \qw       & \qw & \targ     & \qw       & \qw       & \qw     & \qw & \qw & \qw     & \qw                   & \qw           & \qw  & \qw  \\
   & \push{|{+}\rangle \hspace{0.1cm}} & \ctrl{1} & \qw       & \qw & \qw       & \ctrl{4}  & \qw       & \qw     & \qw & \qw & \qw     & \ctrl{3}              & \qw           & \qw  & \qw \\
   & \push{|0\rangle \hspace{0.1cm}}   & \targ    & \qw       & \qw & \qw       & \qw       & \targ     & \qw     & \qw & \qw & \qw     & \qw                   & \targ         & \qw  & \qw \\
   & \\
   & \push{|{+}\rangle \hspace{0.1cm}} & \ctrl{2} & \qw       & \qw & \ctrl{-4} & \qw       & \qw       & \qw     & \qw & \qw & \qw     & \control   \qw        & \ctrl{-2}          & \gate{Z} & \meterx \\
   & \push{|{+}\rangle \hspace{0.1cm}} & \ctrl{1} & \qw       & \qw & \qw       & \targ     & \qw       & \meterz & \cw & \cw & \cw     & \cw                   & \control \cw \cwx[-1]  & \control \cw \cwx[-1] \\
   & \push{|0\rangle \hspace{0.1cm}}   & \targ    & \qw       & \qw & \qw       & \qw       & \ctrl{-4} & \meterx & \cw & \cw & \cw     & \control \cw \cwx[-2] & \cw                    & \control \cw \cwx[-1] \gategroup{1}{2}{3}{3}{1em}{--} \gategroup{5}{2}{7}{3}{1em}{--}
                                }}   \vspace{0.3cm}                             
  \end{tabular}
\caption{\label{fig-11} (a) An equivalent circuit to those in figure \ref{fig-9}. We identify the input to be two copies of the state $|{\rm Toffoli}\rangle$ shown in boxes. (b) After measuring the last two qubits and applying the appropriate corrections, the first three qubits are left in the state $|{\rm Toffoli}\rangle$. The fourth qubit is then measured and the outcome gives the product of the eigenvalues of $S_1$ acting in the two input $|{\rm Toffoli}\rangle$ states. }
\end{center}
\end{figure} 

This circuit includes, except for the two input $|{\rm Toffoli}\rangle$ states, additional Toffoli gates acting between these two states that we now want to eliminate---recall that our goal is to design a distillation circuit for the $|{\rm Toffoli}\rangle$ state that uses exclusively CSS operations. We observe that if we measure the last two qubits as shown in figure \ref{fig-11}(b), we can apply corrections depending on the measurement outcomes that will restore the state of the first three qubits to be the $|{\rm Toffoli}\rangle$ state. Furthermore, if we subsequently measure the fourth qubit, the eigenvalue will correspond to the \emph{product} of the eigenvalues of $S_1$ acting on the two input $|{\rm Toffoli}\rangle$ states\footnote{To see this, note that in figure \ref{fig-9} the measurement gives the eigenvalue of $S_1 \otimes X$ where $S_1$ acts on the first three qubits and $X$ on the fourth qubit. By  commuting $X$ to the right of the Toffoli gate inside the preparation circuit of the second $|{\rm Toffoli}\rangle$ copy, we see that we equivalently implement a measurement of the operator $S_1\otimes S_1$ on the two input $|{\rm Toffoli}\rangle$ states.}. Ideally this product is ${+}1$ because both input $|{\rm Toffoli}\rangle$ states are eigenstates of $S_1$. On the other hand, if the measured product is ${-}1$, we know that one of the input $|{\rm Toffoli}\rangle$ states had an error. Hence, if we postselect on the ${+}1$ measurement outcome, we expect the fidelity of the output $|{\rm Toffoli}\rangle$ state to improve quadratically.  

%\begin{figure}[tb]
%\begin{center}
%\epsfig{file=lower-bounds/gadgets/T-ancilla-4.eps}
%\caption{\label{fig-11b} After measuring the last two qubits in figure \ref{fig-11} and applying the appropriate corrections, the first three qubits are left in the state $|{\rm Toffoli}\rangle$. The fourth qubit is then measured with the outcome giving the product of the eigenvalues of $S_1$ acting in the two input $|{\rm Toffoli}\rangle$ states. }
%\end{center}
%\end{figure} 

In more detail, the $2^3$ eigenstates of the three commuting operators $S_1$, $S_2$ and $S_3$ form an orthonormal basis in the three-qubit Hilbert space. We can label the basis states by $|{\rm Toffoli}\rangle_{a,b,c}$ depending on their eigenvalues, $a,b,c={\pm}1$, with respect to the operators $S_1$, $S_2$ and $S_3$. Then, $|{\rm Toffoli}\rangle \equiv |{\rm Toffoli}\rangle_{+1,+1,+1}$ and we note that
\begin{equation}
|{\rm Toffoli}\rangle_{a,b,c} = Z^{\bar a} \otimes Z^{\bar b} \otimes X^{\bar c} |{\rm Toffoli}\rangle ,
\end{equation} 

\noindent where $\bar a=0$ if $a=+1$ and $\bar a=1$ if $a=-1$, and similarly for $\bar b$ and $\bar c$. Let us assume that $p_{\rm anc}$ is an upper bound on the probability of a fault during the preparation of a $|{\rm Toffoli}\rangle$ state that is input to the distillation protocol. Then, by flipping three coins and applying with probability $1/2$ each of $S_1$, $S_2$, and $S_3$ (and doing nothing otherwise in each of the three cases), we may assume that the input copies to the protocol are mixtures of the basis states $\{|{\rm Toffoli}\rangle_{a,b,c} \}$ where the probability of the state $|{\rm Toffoli}\rangle$ in the mixture is at least $1-p_{\rm anc}$; we may also say that the ``errors'' $Z \otimes I \otimes I$, $I \otimes Z \otimes I$, and $I \otimes I \otimes X$ appear with probabilities $p_a$, $p_b$, and $p_c$ such that $p_a, p_b, p_c \leq p_{\rm anc}$.

Let us know consider the circuit in figure \ref{fig-11}(b) and let the two input states to the distillation be the states $|{\rm Toffoli}\rangle_{a,b,c}$ and $|{\rm Toffoli}\rangle_{a',b',c'}$. In such a case the first three qubits at the output are in the state $|{\rm Toffoli}\rangle_{a,b\cdot b',c\cdot c'}$ and, moreover, the measurement outcome on the fourth qubit corresponds to $a\cdot a'$, i.e., it equals the product of the eigenvalues of $S_1$ of the two input copies. Therefore, postselecting on the $+1$ measurement outcome, we achieve a quadratic improvement of the error probability $p_a$ in the surviving state $|{\rm Toffoli}\rangle_{a,b\cdot b',c\cdot c'}$,
\begin{equation}
\label{eq:rec-eq-1}
p_a \rightarrow p'_a \leq {p_a^2 \over p_a^2 + (1-p_a)^2} \leq 2p_a^2 \;.
\end{equation} 

\noindent On the other hand, conditioned on accepting the output state, the output probabilities $p_b$ and $p_c$ get worse since if any one of the input copies was a ${-}1$ eigenstate of $S_2$ or $S_3$, the surviving copy will also be such a ${-}1$ eigenstate. In particular, 
\begin{equation}
\label{eq:rec-eq-2}
p_b \rightarrow p'_b \leq {2p_b(1-p_b) \over p_a^2 + (1-p_a)^2} \leq 4p_b \;,
\end{equation} 
\begin{equation}
\label{eq:rec-eq-3}
p_c \rightarrow p'_c \leq {2p_c(1-p_c) \over p_a^2 + (1-p_a)^2} \leq 4p_c \;.
\end{equation} 

Because the error probabilities with respect to the parameters $b$ and $c$ only get worse by a linear factor, if we repeat the distillation protocol using a different circuit that purifies with respect to $b$ and also with a third circuit that purifies with respect to $c$, we will eventually achieve a quadratic improvement in all three probabilities $p_a$, $p_b$, and $p_c$ \cite{Gottesman04,Preskill-notes}\footnote{Such circuits can be easily constructed using a similar trick as the one we used here to derive the circuit in figure \ref{fig-11}(b) from that in figure \ref{fig-9}.}. Overall, after these three distillation subprotocols, the error probabilities evolve as $(p_a,p_b,p_c) \rightarrow (32p_a^2,128p_b^2,512p_c^2)$, thus indicating a threshold value of at least $1/512\geq .19\%$. 

In fact, we can improve this value if we note that provided the error parameters never exceed $12\%$, we may use the bound $p_i^2 + (1-p_i)^2 \geq .75$ to replace equations (\ref{eq:rec-eq-1}) to (\ref{eq:rec-eq-3}) by $(p_a,p_b,p_c) \rightarrow (\beta p_a^2,\gamma p_b,\gamma p_c)$ where $\beta={4\over 3}$ and $\gamma={8\over 3}$. After the three distillation subprotocols, the error probabilities evolve as
\begin{equation}
(p_a,p_b,p_c) \rightarrow \left( \beta \gamma^2 p_a^2\; , \beta \gamma^3 p_b^2\; , \beta \gamma^4 p_c^2 \right) \;,
\end{equation}

\noindent indicating a distillation threshold,\footnote{We may check for consistency that $p_c$ after the second subprotocol remains below our $12\%$ upper bound; indeed, $\gamma^2 \times 1.45\% < 12\%$.} $p_{\rm thr}^{\rm dist,|Toffoli\rangle} \geq \left( \beta \gamma^{4} \right)^{-1} \geq 1.45\% $.
%
%\begin{equation}
%p_{\rm thr}^{\rm dist,|Toffoli\rangle} \geq \left( \beta \gamma^{4} \right)^{-1} \geq 1.45\% \; .
%\end{equation}

%-----------------------------------------------%
\subsubsection{Bounds on the Accuracy of Quantum Software}
\label{sec:bounds-software}

Let us finally discuss how we can prepare copies of the $|{+}i\rangle$ and $|{\rm Toffoli}\rangle$ states that are used as inputs to the distillation protocols described in the previous two sections. There are two issues to be addressed: The first one is related to the fact that our distillation procedures assumed that CSS operations were executed without faults. Although this assumption is never exactly satisfied, our idea was that we can {\em simulate} CSS operations to any desired accuracy provided the physical noise strength is below the accuracy threshold for CSS operations. Therefore our assumption about noiseless CSS operations is justified if we simulate the distillation protocols at a sufficiently high level, say $k$, of a recursive fault-tolerant simulation; this then implies that the input ancillary states to the distillation protocols must also be encoded in a level-$k$ code block. The second issue relates to the accuracy of these level-$k$ ancillary states: Not only must we show how such states can be prepared, but we must also be able to execute this preparation in such a way that the probability of a fault during preparation is below the distillation threshold.

Perhaps the simplest method for preparing a level-$k$ code block in a desired single-qubit state, $|\psi\rangle$, is to use quantum teleportation: We prepare a single qubit in the state $|\psi\rangle$ and some ancillary qubits in the state $|0\rangle |0\rangle_L + |1\rangle |1\rangle_L$ where, e.g., $|0\rangle |0\rangle_L$ denotes the $|0\rangle$ state on the first ancillary qubit and the level-$k$ logical $|0\rangle$ state on the remaining ancillary qubits. By performing a measurement along the Bell basis between the qubit in the state $|\psi\rangle$ and the first ancillary qubit, we teleport the level-$k$ $|\psi\rangle$ state to the remaining ancillary qubits. In chapter \ref{ch:injection} in the appendix we give more details on this procedure and we also derive the following upper bound on the accuracy of the level-$k$ code block, 
\begin{equation}
\label{eq:anc-accuracy-bound}
p_{\rm anc} \leq 3 p^{(k)} + D \sum\limits_{j=0}^{k-1} p^{(j)} + 4 p \; ,
\end{equation} 

\noindent where $p^{(k)}$ is the noise strength for CSS operations at level $k$ of the recursion ($p^{(0)}=p$ is the physical noise strength) and $D$ is the number of locations in a decoding circuit that performs a canonical mapping of the state of a level-1 code block to the state of one qubit\footnote{Here we assume that the same code is used at all levels of the recursive simulation; otherwise, $D$ would need to be maximized over all codes used at different levels.}. For the three-qubit $|{\rm Toffoli}\rangle$ state, we can similarly teleport the state of each qubit to a different code block. Then, we can upper bound the probability of a fault that is supported on a specific one of the three level-$k$ code blocks as in equation (\ref{eq:anc-accuracy-bound}) with $4 p$ replaced by $7 p$.

To show that our accuracy thresholds for CSS operations in \S \ref{sec:summary-of-results} do indeed apply to universal quantum computation, it remains to evaluate the upper bound in equation (\ref{eq:anc-accuracy-bound}) for our ``quantum software'' ancillary states and show that it is below the distillation thresholds. 

\vspace{0.3cm}
%---------------------------------------------------------------%
\medskip \noindent {\bf The 9-Qubit Bacon-Shor Code} \medskip

We first note that the first term in equation (\ref{eq:anc-accuracy-bound}) can be ignored since it can become arbitrarily small by choosing $k$ sufficiently large provided $p$ is below the accuracy threshold for CSS operations. To upper bound the second term, we recall the recursion equations (\ref{eq:recursioneqs}) that relate the accuracy of the level-$k$ simulation to the accuracy at level $k{-}1$. We have
\begin{equation}
\label{eq:expicit-bound-1}
D \sum\limits_{j=0}^{k-1} p^{(j)} \leq D \left( p + A'_6 p^2 + \sum\limits_{j=2}^{\infty } {1\over A'_{6,{\rm str}}} \left( A'_{6,{\rm str}} \left( A'_6 p^2 \right)  \right)^{2^{j-1}} \right) \; .
\end{equation} 

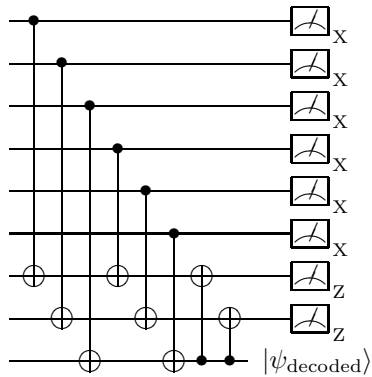
\begin{figure}[tb]
\begin{center}
\begin{tabular}{c}
\Qcircuit @C=0.6ex @R=1.3ex @!R {
   & \qw & \ctrl{6} & \qw      & \qw      & \qw      & \qw      & \qw      & \qw       & \qw       & \meterx \\
   & \qw & \qw      & \ctrl{6} & \qw      & \qw      & \qw      & \qw      & \qw       & \qw       & \meterx \\
   & \qw & \qw      & \qw      & \ctrl{6} & \qw      & \qw      & \qw      & \qw       & \qw       & \meterx \\
   & \qw & \qw      & \qw      & \qw      & \ctrl{3} & \qw      & \qw      & \qw       & \qw       & \meterx \\
   & \qw & \qw      & \qw      & \qw      & \qw      & \ctrl{3} & \qw      & \qw       & \qw       & \meterx \\
   & \qw & \qw      & \qw      & \qw      & \qw      & \qw      & \ctrl{3} & \qw       & \qw       & \meterx \\
   & \qw & \targ    & \qw      & \qw      & \targ    & \qw      & \qw      & \targ     & \qw       & \meterz \\
   & \qw & \qw      & \targ    & \qw      & \qw      & \targ    & \qw      & \qw       & \targ     & \meterz \\
   & \qw & \qw      & \qw      & \targ    & \qw      & \qw      & \targ    & \ctrl{-2} & \ctrl{-1} & \push{\hspace{0.2cm} |\psi_{\rm decoded}\rangle} \qw
                                }
\end{tabular}
\caption{\label{fig:9-bacon-shor-decoder} A decoding circuit for the 9-qubit Bacon-Shor code. The qubits in the code block are arranged starting with the qubit on the top left corner in the lattice in figure \ref{fig:5.3.1} and moving from left to right across successive rows. }
\end{center}
\end{figure}

\noindent To determine the value of $D$ we need an explicit construction of a decoding circuit such as the one shown in figure \ref{fig:9-bacon-shor-decoder}; for this decoder, $D=16$. %Since our decoder is assumed to fail even if a single one of the elementary operations it contains fails, we only need to count the {\sc cnot} gate locations; thus, we set $D=8$.  

We can now obtain a numerical upper bound on $p_{\rm anc}$ by explicitly calculating the first, say, 100 terms of the sum in equation (\ref{eq:expicit-bound-1}) and upper bounding the contribution of the remaining terms using 
\begin{equation}
\label{eq:expicit-bound-2}
\sum\limits_{j=102}^{\infty } p^{(j)} \leq {1\over A'_{6,{\rm str}}} \sum\limits_{j=0}^{\infty } \left( A'_{6,{\rm str}} p^{(102)} \right)^{2^j} \leq  {1\over  A'_{6,{\rm str}}} {1 \over 1- \left(A'_{6,{\rm str}} p^{(102)}\right)^2 }\; ,
\end{equation}

\noindent where we used the bound $\sum\limits_{j=0}^{\infty} x^{2^j} \leq \sum\limits_{j=0}^{\infty} \left( x^2 \right)^j \leq (1-x^2)^{-1}$. Using Knill's 1-EC gadget and by substituting the values for $A'_{6}$ and $A'_{6,{\rm str}}$ and setting $p= p_{\rm thr}=1.26\times 10^{-4}$, we find for single-qubit states, $p_{\rm anc} < 2.15\%$, which is safely below the $50\%$ $|{+}i\rangle$ distillation threshold. Similarly, for three-qubit ancillary states, we find that the probability of a fault that is supported on a specific one of the three code blocks is below $2.18\%$. Since $2.18\%$ is above our $1.45\%$ $|{\rm Toffoli}\rangle$ distillation threshold, we seem to need to revise our bound on $p_{\rm thr}$ in table \ref{table:1}. In fact, decreasing our bound by $10^{-5}$ to $p_{\rm thr} = 1.16\times 10^{-4}$ makes $p_{\rm anc}$ drop to $1.25\%$ for single-qubit states and to $1.29\%$ for three-qubit states which is now below the $1.45\%$ $|{\rm Toffoli}\rangle$ distillation threshold. 

However, it is not necessary that we revise our bound on $p_{\rm thr}$. The reason why we chose to describe the distillation of $|{\rm Toffoli}\rangle$ states in \S \ref{sec:logical-toffoli-gate} is because the distillation procedure was conceptually simple and the distillation threshold condition was easy to derive; however, much higher distillation thresholds are possible if we use more complex protocols for distilling certain ``magic'' single-qubit states as explained in \cite{Bravyi05}. In particular, the distillation threshold for the $|H\rangle$ magic state is at least $14\%$ \cite{Bravyi05} which is significantly above $2.15\%$ showing that if $p$ is below $1.26\times 10^{-4}$ universal quantum computation is still possible.

%(These bounds apply to the circuits using Steane's 1-EC gadgets; the bounds for Knill's 1-EC gadget are slightly better.) 

%---------------------------------------------------------------%
\medskip \noindent {\bf The 25-Qubit Bacon-Shor Code} \medskip
 
We may repeat the same calculation by using the recursion equations (\ref{eq:cond}) and also\footnote{A decoding circuit can easily be constructed similar to that in figure \ref{fig:9-bacon-shor-decoder}: We connect with {\sc cnot} gates all qubits in the first four rows to the qubits in the fifth row ($4\times5=20$ gates) and, then, we connect with {\sc cnot} gates the last qubit in the fifth row to every other qubit in the same row (another $4$ gates). Finally, we add 24 single-qubit measurements.} $D=48$. Instead of equation (\ref{eq:expicit-bound-2}) we now have
\begin{equation}
\label{eq:expicit-bound-3}
\sum\limits_{j=102}^{\infty } p^{(j)} \leq {1\over \Gamma} \sum\limits_{j=0}^{\infty } \left( \Gamma p^{(102)} \right)^{3^j} \leq  {1\over \Gamma}{1 \over 1- \left( \Gamma p^{(102)}\right)^3 }\; ,
\end{equation}

\noindent where 
\begin{equation}
\label{eq:expicit-bound-4}
\Gamma \equiv {A_{6,{\rm str}}+B_{6,{\rm str}} p^{(102)} \over \left( 1-p^{(102)} \right)^{4C_{0,{\rm str}}}} \; .
\end{equation}

\noindent For Steane's 1-EC gadget and by substuituting the values for $A_{6}$, $A_{6,{\rm str}}$, $B_{6}$, $B_{6,{\rm str}}$, $C_{0}$ and $C_{0,{\rm str}}$ and setting $p=p_{\rm thr}=1.94\times 10^{-4}$, we find for single-qubit states, $p_{\rm anc}<8.49\% $, which is below the $50\%$ $|{+}i\rangle$ distillation threshold. For three-qubit states, the probability of a fault that is supported on a specific one of the three blocks is below $8.55\% $. Again, $8.55\% $ is above our $1.45\%$ $|{\rm Toffoli}\rangle$ distillation threshold but safely below the $14\%$ $|H\rangle$ distillation threshold. Hence, universal quantum computation is possible if $p$ is below $1.94\times 10^{-4}$.

Finally, we should note that the sum $D \sum_{j=0}^{k-1} p^{(j)}$ in equation (\ref{eq:anc-accuracy-bound}) diverges as $k\rightarrow \infty$ when $p=p_{\rm thr}$ since $p_{\rm thr}$ is defined as the fixed point of our recursion equations (i.e., by definition, if we start with $p=p_{\rm thr}$ then $\forall j$, $p^{(j)}=p$). However, the lower bounds we give in table \ref{table:1} are slightly below this fixed point so that in fact the sum is convergent when we set $p=p_{\rm thr}$. 
 
%---------------------------------------------------------------%

%----------------------------------------------------------%
\section{History and Acknowledgements}

I first heard Bacon talk about the Bacon-Shor code at the QIP 2005 workshop, but it did not occur to me that this code could be useful for fault-tolerant quantum computation until after I read \cite{Bacon05} and I understood the relation to Shor's code (which, perhaps surprisingly, is not mentioned in \cite{Bacon05}). The analysis of malignant sets of locations for \bsc{3} and \bsc{5} was done by Cross and the results appear in \cite{Aliferis06c}.

The idea that the accuracy threshold for stabilizer operations (and CSS operations, in particular) sets the overall threshold for universal quantum computation has appeared in the work of Knill \cite{Knill04,Knill05} and has also been discussed by Bravyi and Kitaev \cite{Bravyi05} and by Reichardt \cite{Reichardt05,Reichardt06b}. 

%----------------------------------------------------------%

%-----------------------------------------%
\chapter{Epilogue}
\label{ch:conclusion}
%---------------------------------------------------------------%
\setlength{\unitlength}{1cm}
\begin{picture}(14,1)
\put(6,1){\small \parbox{8.3cm}{
%What is the greatest thing ye can experience? It is the hour of great contempt. The hour in which even your happiness becometh loathsome unto you, and so also your reason and virtue. The hour when ye say: ``What good is my happiness!  It is poverty and filth and wretched contentment. But my happiness should justify existence itself!'' The hour when ye say: ``What good is my reason!  Doth it long for knowledge as the lion for his food?  It is poverty and filth and wretched contentment!'' \\
{\em The most valuable intuitions are the last to be attained; the most valuable of all are those which determine methods.} \\
\hskip 6cm ---Nietzsche, {\em Antichrist}.}}
\end{picture}

\vskip 0.3cm

Level reduction is such a simple idea that it now appears to us to be completely natural. But as all valuable intuitions, the insights leading to a syntactic analysis of fault-tolerant simulations and the concept of level reduction have become possible based on a large body of work in the theory of fault-tolerant quantum computation during the last ten years. 

The value of level reduction is that it provides a simple methodology for analyzing recursive fault-tolerant simulations. In this thesis, this methodology was used to derive rigorous lower bounds on the accuracy threshold and to extend the proof of the quantum threshold theorem to new directions such as computation in the presence of coherent and leakage noise and measurement-based models of quantum computation. Level reduction is also used in a proof of the threshold theorem for postselected quantum computation \cite{Aliferis06a}.

Although the family of noise models I have called {\em local} includes models which are significantly more realistic than any of those considered in previous proofs, there is still a long road toward understanding and limiting the effect of physical noise on future quantum computers. In particular, it would be important to obtain new formulations of the quantum threshold theorem in terms of quantities such as spectral densities and correlation functions of the environment which are more easily accessible experimentally.

Let us hope that level reduction will become a new helpful tool among the many that will be needed to realize a formidable quantum computer.  

%---------------------------------------------------------------%

\begin{appendix}
%-----------------------------------------%
%\chapter{Designing Decoders/Encoders}
%\label{ch:design}
%\input{appendix/design.tex}

\chapter{Deriving the Bacon-Shor code}
\label{ch:bs-derivation}
%--------------------------------------------------------------------------------------------------%
\section{Starting from Shor's Code}

One natural way to derive the distance-$n$ Bacon-Shor code, \bsc{n}, is by starting with Shor's distance-$n$ code,  $\mathcal{C}_{Shor}^{(n)}$, as was done for the special case $n=3$ in \cite{Poulin05}. %Let us call $\mathcal{C}_Z^{(n)}$ the classical $n$-bit repetition code mapping the logical computation-basis state $|k\rangle$ to $|k\rangle^{\otimes n}\equiv |\bar{k}\rangle$ for $k=0,1$. The Hadamard-rotated $n$-bit repetition code, $\mathcal{C}_X^{(n)}$, maps the logical states $|\pm\rangle$ to $|\pm\rangle^{\otimes n}$, where $|\pm\rangle$ denote the ${\pm}1$ eigenstates of $X$. Then, \sh is the concatenated $\mathcal{C}_Z^{(n)} \circ \mathcal{C}_X^{(n)}$ mapping the logical state $|k\rangle$ to $|\bar{+}\rangle^{\otimes n} + (-1)^k |\bar{-}\rangle^{\otimes n}$, where $|\bar{\pm}\rangle \propto |0\rangle^{\otimes n} \pm |1\rangle^{\otimes n}$.    

If we place the $n^2$ qubits in the \sh block on the {\em vertices} of the $n{\times}n$ square lattice in figure \ref{fig:5.3.1}, then the code's stabilizer group is 
\begin{equation}
\begin{array}{rcl}
\label{eq:A2}
\hspace{-0.2cm} \mathcal{S}(\mathcal{C}_{Shor}^{(n)} ) & = & \langle X_{j,*}X_{j+1,*} \; ; \; Z_{i,j} Z_{i,j+1} \; | \; i\in \mathbb{Z}_n ; j\in \mathbb{Z}_{n{-}1} \rangle \; .
\end{array}
\end{equation}
\noindent Also, recall that the logical Pauli $X$ and $Z$ operators are $X_{1,*}$ and $Z_{*,1}$,  respectively.

By its construction, Shor's code treats $X$ and $Z$ errors asymmetrically: In each of the $n$ rows, up to $\lfloor n/2 \rfloor$ $X$ errors can be corrected because of the underlying classical repetition code. The code can also correct up to $\lfloor n/2 \rfloor$ $Z$ errors on different rows---pairs of $Z$ operators in the same row act trivially; the code is degenerate. However, we observe that pairs of $X$ operators in the same {\em column} commute with the logical $Z$ operator. It would therefore be sufficient for successful error correction if we could restore zero parity for $X$ errors in each given column instead of correcting $X$ errors in each row separately.

Since only the parity of $X$ errors in each column is relevant, we can replace the $n$ stabilizer generators $Z_{i,j} Z_{i,j+1}$ for $i\in \mathbb{Z}_{n}$ and {\em fixed} $j$ by their tensor-product, $Z_{*,j}Z_{*,j+1}$. Repeating for all $j\in \mathbb{Z}_{n-1}$, we reduce the stabilizer group of \sh in equation (\ref{eq:A2}) to the stabilizer group of $\mathcal{C}_{BS}^{(n)}$ in equation (\ref{eq:stab}).

%--------------------------------------------------------------------------------------------------%
\section{Starting from Bravyi and Kitaev's Surface Code}

\begin{figure}[hbt]
\begin{center}
%\leavevmode
%\epsfysize=5.2cm
%\epsfbox{pic12.eps}
%\includegraphics[width=4.6cm]{pic12}
%\epsfig{file=pic12.eps,width=4.6cm} 
%
\vspace{3cm} \hspace{1cm}
\begin{picture}(5,5)
%
%\put(-1,5.4){$i$}
%\put(0,2.5){\line(1,0){4}}
\put(3,6.5){\line(1,0){1}}
\put(-0.6,2.9){$n$}
%\put(0,4.5){\line(1,0){2.5}}
\put(3,5.5){\line(1,0){1}}
\put(-0.6,4.9){3}
\put(0,5.5){\line(1,0){2.5}}
%\put(3.5,4.5){\line(1,0){0.5}}
\put(-0.6,5.9){2}
\put(0,6.5){\line(1,0){2.5}}
\put(-0.6,6.9){1}

\put(0,3.5){\line(1,0){2.5}}
\put(3,3.5){\line(1,0){1}}
\put(4,2.5){\line(0,1){5}}

%\put(1.4,8){$j$}
\put(0,4.5){\line(0,1){3}}
\put(0,2.5){\line(0,1){1.5}}
\put(-0.1,7.7){1}
\put(1,4.5){\line(0,1){3}}
\put(1,2.5){\line(0,1){1.5}}
\put(0.9,7.7){2}
\put(2,4.5){\line(0,1){3}}
\put(2,2.5){\line(0,1){1.5}}
\put(1.9,7.7){3}
%\put(4,2.5){\line(0,1){5}}
\put(3.9,7.7){$n$}

\put(0,3){\color{Black} \circle*{0.4}}
\put(0,5){\color{lblue} \circle*{0.4}}
\put(0,6){\color{lblue} \circle*{0.4}}
\put(0,7){\color{Black} \circle*{0.4}}

\put(0.5,3.5){\thicklines \color{Black} \circle{0.4}}
\put(0.5,5.5){\thicklines \color{lblue} \circle{0.4}}
\put(0.5,6.5){\thicklines \color{Black} \circle{0.4}}

\put(1,3){\color{Black} \circle*{0.4}}
\put(1,5){\color{Black} \circle*{0.4}}
\put(1,6){\color{lred} \circle*{0.4}}
\put(1,7){\color{Black} \circle*{0.4}}

\put(1.5,3.5){\thicklines\color{Black} \circle{0.4}}
\put(1.5,5.5){\thicklines \color{lred} \circle{0.4}}
\put(1.5,6.5){\thicklines \color{lred} \circle{0.4}}

\put(2,3){\color{Black} \circle*{0.4}}
\put(2,5){\color{Black} \circle*{0.4}}
\put(2,6){\color{lred} \circle*{0.4}}
\put(2,7){\color{Black} \circle*{0.4}}

\put(3.5,3.5){\thicklines \color{Black} \circle{0.4}}
\put(3.5,5.5){\thicklines \color{Black} \circle{0.4}}
\put(3.5,6.5){\thicklines \color{Black} \circle{0.4}}

\put(4,3){\color{Black} \circle*{0.4}}
\put(4,5){\color{Black} \circle*{0.4}}
\put(4,6){\color{Black} \circle*{0.4}}
\put(4,7){\color{Black} \circle*{0.4}}
\end{picture}
\vspace{-2.6cm}
\end{center}
\caption{\label{fig:Abk} Qubits in the \bk block sit on the {\em edges} of a square lattice with different boundary conditions at top-bottom ({\em rough} edges) and left-right ({\em smooth} edges). Two elements of the code stabilizer are shown: $X$ is applied on qubits shown in blue ({\em site} operator); $Z$ is applied on qubits shown in red ({\em plaquette} operator). If all qubits shown as empty circles are measured in the computation basis, the remaining qubits will be encoded in $\mathcal{C}_{BS}^{(n)}$. }
\end{figure}
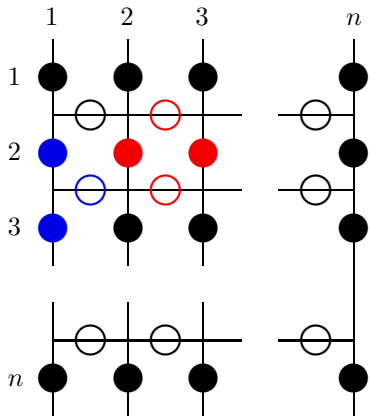

We will also give an alternative derivation of \bs starting from Bravyi and Kitaev's distance-$n$ surface code,  $\mathcal{C}_{BK}^{(n)}$ \cite{Bravyi98}. If we place the $n^2+(n{-}1)^2$ qubits in the \bk block on the {\em edges} of a square lattice as shown in figure \ref{fig:Abk}, the code's stabilizer group is generated by $X$ operators acting on all qubits neighboring a vertex ({\em site} operators) and $Z$ operators acting on all qubits neighboring a vertex of the {\em dual} lattice ({\em plaquette} operators). The logical Pauli $X$ and $Z$ operators are $X_L=X_{1,*}$ and $Z_L=Z_{*,1}$, respectively. It can easily be verified that if all qubits with half-integer coordinates are measured in the computation basis, the remaining qubits are left in a $+1$ eigenstate of the stabilizer generators in equation (\ref{eq:stab}). Moreover, the logical $X$ and $Z$ operators have no support on the measured qubits and match the logical Pauli operators for $\mathcal{C}_{BS}^{(n)}$. This implies that the same logical state that was encoded in the \bk block is, after the measurements, encoded in the protected qubit of $\mathcal{C}_{BS}^{(n)}$.

%--------------------------------------------------------------------------------------------------%

\chapter{Teleporting into Code Blocks}
\label{ch:injection}
%-------------------------------------------------------%

\begin{figure}[htb]
\begin{center}
\vspace{-0.4cm}
%\leavevmode
%\epsfysize=2.5cm
%\epsfbox{injection.eps}
%\includegraphics[width=6cm]{injection}
%\epsfig{file=injection.eps,width=6cm}
%
\begin{tabular}{c}
\Qcircuit @C=1.5ex @R=3ex @!R { 
   & \push{|\psi \rangle \hspace{0.1cm}}    & \qw      & \qw   & \qw                & \qw & \ctrl{1} & \meterx & \control \cw  \cwx[1] &  \\
   & \push{|+\rangle_L \hspace{0.1cm}}      & \ctrl{1} & \qw   & \gate{\vspace{0.1cm} \;\mathcal{D}\; \vspace{0.1cm}} & \qw & \targ    & \meterz &\control \cw \cwx[1] & \\
   & \push{|0\rangle_L \hspace{0.1cm}}      & \targ    & \qw   & \qw                & \qw & \qw      & \qw     & \gate{P_L} & \qw & \qw & \push{|\psi \rangle_L \hspace{0.1cm}}
                                }    
\end{tabular}
\vspace{-0.1cm}
\end{center}
\caption{\label{fig:C-1} Teleporting the single-qubit state $|\psi\rangle$ into the code block. First, two ancillary  blocks are prepared in a logical Bell state and, then, one of the two blocks is decoded using the decoder,  $\mathcal{D}$. Finally, a measurement along the {\em Bell basis} $\{(Z^{j_1}X^{j_2}\otimes I)|\Phi_0\rangle \, | \, j_1,j_2=0,1 \}$ is performed on an input qubit prepared in the state $|\psi\rangle$ and the qubit output from the decoder. After the {\em logical} Pauli  correction $P$ is applied, the state of the output code block is $|\psi\rangle_L$ as desired. }
\end{figure}
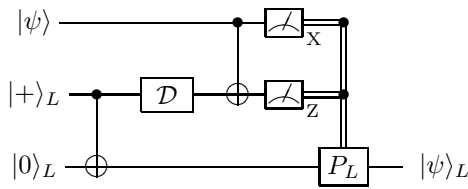

A general method for preparing a code block in the {\em logical} state $|\psi\rangle$ starting from a single-qubit state $|\psi\rangle$  is to {\em teleport} into the code block \cite{Knill05}; figure \ref{fig:C-1} shows the circuit. First, a {\em logical} Bell state is prepared by using logical $|+\rangle$ and $|0\rangle$ ancilla blocks and interacting them using a logical {\sc cnot} gate. Then, one of the two ancilla blocks is decoded by using the  decoding circuit, $\mathcal{D}$. Finally, the qubit output from the decoder interacts with a single input qubit prepared in the state $|\psi\rangle$ via a {\sc cnot} and both qubits are measured as shown in the figure. The {\em logical} Pauli correction operator, $P_L$, of the teleportation protocol restores the state of the output code block to be the {\em logical} state $|\psi\rangle$ as desired.  

We can use this method to prepare the logical $|{+}i\rangle$ or $|{\rm Toffoli}\rangle$ states that are input to the distillation protocols described in \S \ref{sec:BSUnivComp}. Consider the case where a single-qubit state $|\psi\rangle$ such as $|{+}i\rangle$ is teleported into the code block at level, say, $k$ of a recursive simulation. If noise is local and stochastic, the probability of a fault during the teleportation of the state into the code block can be upper bounded by
\begin{equation}
\label{eq:appB-1}
p_{\rm anc} \leq 3 p^{(k)} + p_{\rm dec}^{(k)} + 4 p \; .
\end{equation} 

\noindent Here $3 p^{(k)}$ is an upper bound on the probability of a fault in the level-$k$ logical Bell state (since it is prepared by using three level-$k$ operations), $p_{\rm dec}^{(k)}$ is the fault probability for decoding one half of the logical Bell state, and $4 p$ is an upper bound on the probability of a fault in the preparation of the input state $|\psi\rangle$ and in the three operations in the measurement along the Bell basis. In general, if a $m$-qubit state is teleported at level $k$ of a recursive simulation, we can upper bound the probability of a fault with support on a specific one of the $m$ level-$k$ code blocks by replacing the last term, $ 4 p$, in equation (\ref{eq:appB-1}) by $(3+s)p$ where $s$ is the number of locations in the preparation circuit of the state to be teleported (e.g., for the $|{\rm Toffoli}\rangle$ state, $m=3$ and $s=4$).

 %the analogous bound would be $p_{\rm anc} \leq m \left( 3 p^{(k)} + p_{\rm dec}^{(k)} + 3 p \right) + s\; p$, where $s$ is the number of locations in the preparation circuit of the state to be teleported (e.g., for the $|{\rm Toffoli}\rangle$ state, $m=3$ and $s=4$).

We can upper bound $p_{\rm dec}^{(k)}$ by decoding the level-$k$ block recursively starting from the highest coding level. That is, the decoding circuit, $\mathcal{D}$, first simulates the code decoder at level $k$ using logical operations at level $k{-}1$; this results in an output level-$(k{-}1)$ decoded block. Next, $\mathcal{D}$ simulates the code decoder at level $k{-}1$ using logical operations at level $k{-}2$ resulting in a output level-$(k{-}2)$ decoded block, etc., until the state is decoded to a single qubit. By this procedure and letting $D$ be the number of locations in the decoder, 
\begin{equation}
\label{eq:appB-2}
p_{\rm dec}^{(k)} \leq D \sum\limits_{j=0}^{k-1} p^{(j)} \; ,
\end{equation} 

\noindent where $p^{(0)} \equiv p$. As long as the strength of stochastic local noise, $p$, is below the accuracy threshold, $p^{(j)}$ decreases doubly exponentially with $j$ and, thus, our upper bound on $p_{\rm dec}^{(k)}$ approaches an asymptotic value of order $p$.

Actually, there is a detail regarding the presence of $D$ in equation (\ref{eq:appB-2}) that we need to discuss. The problem is that, at first, it seems that we need to define a level-1 extended rectangle for the decoding operation that includes a 1-EC followed by a noisy decoder. If this were indeed the case, the probability that this 1-exRec is bad is certainly higher than $Dp$ since we also need to consider the possible occurrence of faults in the leading 1-EC. However, it is easy to see that we do not really need to consider separate extended rectangles for the decoding operation. Instead, by following the intuition about  contracting extrended rectangles discussed in \S \ref{sec:Stretching}, we may have the decoder replace the trailing 1-EC of the preceding {\sc cnot} 1-exRec, thus creating a 1-conexRec. For example, to be more concrete, this means that in the second logical step in preparing, e.g., a level-1 logical Bell state as in figure \ref{fig:C-1}, we may consider a {\em single} 1-conexRec formed by (i) two leading 1-EC steps, (ii) a {\sc cnot} 1-Ga, (iii) the decoder acting on the first output block, and (iv) a 1-EC acting on the second output block. We then observe that the probability of badness of this 1-conexRec is upper bounded by $p^{(1)}+Dp$ since if the decoder is executed without faults it is equivalent to a faultless 1-EC followed by a faultless decoder. Indeed, we can obtain this upper bound by considering all fault paths for which the 1-conexRec is bad and the decoder is executed without faults (this probability is at most $p^{(1)}$) {\em plus} the probability of having even a single fault inside the decoder (this probability is at most $Dp$). Doing the same contraction of level-$k$ decoders with the preceding {\sc cnot} level-$k$ extended rectangles for all $k\geq 2$ implies that the upper bound we obtain by combining equations (\ref{eq:appB-1}) and (\ref{eq:appB-2}) is valid.

%-----------------------------------------------------------------------------------------------------------%

%-----------------------------------------%
\end{appendix}

\bibliographystyle{unsrt}
\bibliography{pa_bibliographia}

\end{document}